\documentclass[fleqn,twoside]{article}
\usepackage{espcrc1}

\usepackage{graphicx}
\usepackage{bm}
\usepackage{amsmath}
\usepackage{amssymb}

\renewcommand{\k}{\mbox{\boldmath$k$}}
\newcommand{\q}{\mbox{\boldmath$q$}}
\newcommand{\Q}{\mbox{\boldmath$Q$}}
\newcommand{\kk}{\mbox{\boldmath$k'$}}
\newcommand{\e}{\varepsilon}
\newcommand{\ee}{\varepsilon^{'}}
\newcommand{\s}{{\mit{\it \Sigma}}}

\newcommand{\LL}{\mbox{\boldmath$L$}}
\renewcommand{\SS}{\mbox{\boldmath$S$}}

\newcommand{\smallk}{\mbox{{\scriptsize \boldmath$k$}}}
\newcommand{\smallkk}{\mbox{{\scriptsize \boldmath$k'$}}}
\newcommand{\smallq}{\mbox{{\scriptsize \boldmath$q$}}}

\begin{document}

\title{\bf Theory of Superconductivity in Strongly Correlated Electron Systems}

\author{
Yoichi Yanase\address[Tokyo]{Department of Physics,
University of Tokyo, Hongo, Tokyo 113-0033, Japan},
Takanobu Jujo\address[Nara]{Nara Institute of Science and Technology,
Ikoma, Nara 630-0101, Japan},
Takuji Nomura\address[SPring8]{Synchrotron Radiation Research Center,
Japan Atomic Energy Research Institute, Mikazuki,
Hyogo 679-5148, Japan},
Hiroaki Ikeda\address[Kyoto]{Department of Physics, Kyoto University,
Kyoto 606-8502, Japan},
Takashi Hotta\address[Tokai]{Advanced Science Research Center,
Japan Atomic Energy Research Institute, Tokai,
Ibaraki 319-1195, Japan},
and
Kosaku Yamada\addressmark[Kyoto]
}

\maketitle

\begin{abstract}
\noindent {\bf Abstract} \\
In this article we review essential natures of superconductivity in
strongly correlated electron systems (SCES) from a universal point of view.
First we summarize experimental results on SCES by focusing on typical
materials such as cuprates, BEDT-TTF organic superconductors, and
ruthenate Sr$_2$RuO$_4$.
Experimental results on other important SCES, heavy-fermion systems,
will be reviewed separately.
The formalism to discuss superconducting properties of SCES is shown
based on the Dyson-Gor'kov equations.
Here two typical methods to evaluate the vertex function are introduced:
One is the perturbation calculation up to the third-order terms
with respect to electron correlation.
Another is the fluctuation-exchange (FLEX) method
based on the Baym-Kadanoff conserving approximation.
The results obtained by the FLEX method are in good agreement with
those obtained by the perturbation calculation.
In fact, a reasonable value of $T_{\rm c}$ for spin-singlet $d$-wave
superconductivity is successfully reproduced by using both methods
for SCES such as cuprates and BEDT-TTF organic superconductors.
As for Sr$_2$RuO$_4$ exhibiting spin-triplet superconductivity,
it is quite difficult to describe the superconducting transition
by using the FLEX calculation.
However, the superconductivity can be naturally explained
by the perturbation calculation, since the third-order terms
in the anomalous self-energy play the essential role
to realize the triplet superconductivity.
Another important purpose of this article is to review
anomalous electronic properties of SCES near the Mott transition,
since the nature of the normal state in SCES has been
one of main issues to be discussed.
Especially, we focus on pseudogap phenomena observed in
under-doped cuprates and organic superconductors.
A variety of scenarios to explain the pseudogap phenomena based on
the superconducting and/or spin fluctuations are critically reviewed
and examined in comparison with experimental results.
According to the recent theory, superconducting fluctuations,
inherent in the quasi-two-dimensional and strong-coupling superconductors,
are the origin of the pseudogap formation.
In these compounds, superconducting fluctuations induce a kind of resonance
between the Fermi-liquid quasi-particle and the Cooper-pairing states.
This resonance gives rise to a large damping effect of quasi-particles and
reduces the spectral weight near the Fermi energy.
We discuss the magnetic and transport properties as well as
the single-particle spectra in the pseudogap state
by the microscopic theory of the superconducting fluctuations.
As for heavy-fermion superconductors, experimental results are reviewed
and several theoretical analyses on the mechanism are provided
based on the same viewpoint as explained above.
\end{abstract}

\bigskip

\noindent {\it keywords}:
Unconventional superconductivity, Strongly correlated electron systems,
Fermi-liquid theory, Dyson-Gor'kov equations, Pseudogap phenomena,
Spin fluctuations, Superconducting fluctuations.

\bigskip

\noindent {\it PACS}:
74.20.Mn, 74.25.Fy, 74.72-h, 74.70.Kn, 74.70.Pq, 74.70.Tx.


\tableofcontents

\par\vfill
\eject

%
%
\section{Introduction}

In recent decades, elucidation of unconventional superconductivity
in strongly correlated electron systems (SCES) has been one of
the central issues both in experimental and theoretical research fields
of condensed matter physics.
As is well known, SCES form a vast category, including varieties of
materials such as transition metal oxides, molecular conductors,
and $f$-electron compounds.
Here we briefly introduce these superconducting materials in this order.

Among transition metal oxides, cuprate superconductors have certainly
attracted the most attentions of researchers in the condensed matter physics
since the discovery in 1986 \cite{rf:bednortz,rf:tanakahigh-tc},
due to the high superconducting transition temperature $T_{\rm c}$
as well as several kinds of anomalous behaviors in the electronic properties.
In fact, the high-$T_{\rm c}$ cuprate is one of main targets of this
review article.
It is emphasized here that varieties of superconducting materials
have been also discovered in other transition metal oxides.
Especially, Sr$_2$RuO$_4$ \cite{Rf:Maen.1} with isostructure of La$_2$CuO$_4$
(the parent compound of high $T_{\rm c}$ superconductors) exhibits
the superconductivity with triplet pairing, which has been confirmed
by nuclear magnetic resonance (NMR) measurements \cite{Rf:Ishi.1}.
The ruthenate has attracted attentions in spite of its low $T_{\rm c}$
as large as 1.5K, since it is one of rare materials that show triplet
superconductivity in the solid state.
The ruthenate will be also discussed in detail in this review article.

In addition to $d$-electron systems, materials composed of atoms with
lighter mass are also the member of SCES.
In general, those are called molecular conductors, including organics
and fullerides.
In 1980, (TMTSF)$_2$PF$_6$ has been discovered as the first organic
superconductor \cite{rf:jerome}, which has triggered the vigorous
experimental researches on these materials.
A characteristic issue of the molecular conductors is that it is
possible to construct artificial molecule by controlling the synthesis
in the atomic level.
Then, much effort has been made to elevate $T_{\rm c}$ by using
several techniques to synthesize new molecular superconductors.

Let us turn our attentions to $f$-electron materials
including rare-earth or actinide ions.
The pioneering discovery of superconductivity in the heavy fermion
material CeCu$_2$Si$_2$ \cite{rf:Steglich1}
has triggered the rapid increase of investigations on
exotic properties of $f$-electron superconductivity,
leading to a chain of further discoveries of superconductivity
in uranium and cerium compounds.
As we will review in Sec.~5, numbers of new superconducting materials
in heavy fermion systems have been discovered in recent years.
Here it is noted that new superconducting states such as coexisting
one with magnetic orders are increasing.

Even in the above brief survey, there are several kinds of
superconductors categorized in SCES.
It is quite natural for experimentalists to make the research field rich,
as a result of high activities to synthesize new superconducting materials.
However, the purpose in the theoretical research is not to pursue
the variety in materials, although at some stage it is necessary
to investigate a particular material as a typical example.
There should exist a universal picture to explain
the common essence in all SCES.
A task imposed on theoreticians is to unveil this universal concept, 
and to clarify the interesting aspects in materials.
We believe that we have arrived at a unified view on
the superconductivity in SCES.
Thus, the main purpose of this review article is to convey
this viewpoint by showing explanations for typical materials.

Before proceeding to the clarification of our unified viewpoint,
let us consider first the conditions on the theory of superconductivity
in SCES. Without any restrictions, the present review article may be
just the exhibition of previous theories for SCES.
The most exotic possibility for the superconducting mechanism was
superconductivity due to single-electron condensation, but the experimental
results on the unit of magnetic flux in high-$T_{\rm c}$ materials have
confirmed that the existence of Cooper-pair in high-$T_{\rm c}$ cuprates
\cite{Gough87}. Also in heavy-fermion superconductors, the existence of
Cooper-pair has been experimentally confirmed.
Those are quite important, since the basic point of the BCS theory for
superconductivity are invariant even for SCES.
Thus, in this review, issues on exotic superconductivity based
on the non Cooper-pair formation are simply ignored.

Furthermore, here we clarify our strategy to understand the
electric properties in the normal state before discussing
the pairing mechanism.
In the normal state of strongly correlated materials, ``anomalous''
metallic behaviors have been frequently observed.
Namely, behaviors of physical quantities are sometimes deviated from
those understood from the Fermi-liquid theory.
Especially in high-$T_{\rm c}$ cuprates, such ``anomalous'' bahaviors
have provided many challenging issues and stimulated much interests.
Indeed, the understanding of the ``anomalous'' metallic state is
one of the central issues in this review.
One way to explain such non Fermi-liquid behaviors is, of course,
to pursue the non Fermi-liquid ground state, appearing due to the
combined effects of strong correlation and low dimensionality.
In fact, as is well known in one dimension, there appears
the Tomonaga-Luttinger liquid state, essentially different from
the Fermi-liquid state, due to the restriction in the phase space of
one dimension.
No doubt is cast on the results in one dimension and we know that
there exist several approaches based on the non Fermi-liquid ground state
in order to understand the anomalous normal-state properties
of strongly correlated metals.
However, it should be noted that non Fermi-liquid behaviors do $not$
immediately indicate the breakdown of the Fermi-liquid theory in
strongly correlated materials with two or three dimensionality.
We believe that the Fermi-liquid state is a good starting
point to understand the non Fermi-liquid behaviors in some compounds.
If there is no discontinuity from the Fermi-liquid state,
the deviation from the Fermi-liquid behaviors should be derived
from such a starting point.
In this review, we will actually explain the non Fermi-liquid behaviors
from this point of view, but it is worth while to stress here
that the Fermi-liquid state is a robust concept
in two or three dimensions.
The essential issue of the Fermi-liquid state is the continuity
principle, which means that quasi-particles can be obtained
by the adiabatic continuations from the non-interacting systems.
Even if the overlap between bare electron state and dressed
quasi-particle state is vanishingly small like in $f$-electron materials,
the continuity principle can be still effective. 

Here readers may have a naive question in their minds:
Then, how to understand anomalous behaviors in the normal state?
First note that the conventional Fermi-liquid bahaviors may be
restricted to the very low-temperature region,
even if the ground state is a Fermi liquid.
Such a situation is considered to be realized in some heavy-fermion
compounds, owing to a strong renormalization of the Fermi energy.
Even if the quasi-particle renormalization is not so strong
like in high-$T_{\rm c}$ cuprates, this situation can be caused
by some kinds of fluctuations, which give corrections to
the Fermi-liquid behaviors.
The latter effect possibly appears more significantly,
when the long-range order exists in the ground state.
Since the Fermi-liquid state is eventually destroyed
by the long-range order, it is quite natural that
the Fermi-liquid behaviors are altered by its precursor,
namely, by the fluctuation.
We will attribute the non Fermi-liquid behaviors in the high-$T_{\rm c}$
cuprates and some organic superconductors to the latter origin.
In particular, the superconducting (SC) fluctuation as well as
the anti-ferromagnetic (AF) spin fluctuation should be
taken into account.
In those compounds, the effects of such fluctuations are
significantly enhanced owing to quasi-two-dimensionality.
The effective inclusion of fluctuations in the Fermi-liquid state
will lead to understandings of anomalous behaviors in the normal state.

A trial to take into account the effect of AF spin fluctuations
in the Fermi-liquid normal state has been developed for a long time,
since the AF insulating phase can be ubiquitously found in
strongly correlated materials.
The AF fluctuations are especially important near the phase boundary
between metallic and AF insulating phases.
For instance, the self-consistent renormalization (SCR) theory developed
by Moriya and co-workers is one of the powerful methods to include effects
of spin fluctuations, but the details will be simply skipped,
since readers can consult with the textbook \cite{rf:moriyatext}.
It is commented here that the AF spin fluctuation theory should be
interpreted as an extension of the Fermi-liquid theory \cite{rf:kohno},
which is called ``nearly anti-ferromagnetic Fermi-liquid theory'' 
at present.
The AF spin fluctuation is actually effective in high-$T_{\rm c}$ cuprates,
in which the NMR and neutron scattering measurements have clearly
observed the spin fluctuations.
For example, the anomalous temperature dependence of the
electric resistivity is explained by including the AF spin fluctuation.
Interestingly, the cross-over from the conventional $T$-square
to the anomalous $T$-linear resistivity has been observed
in Tl-based cuprate superconductors \cite{rf:kubo},
which occurs together with the enhancement of the AF spin fluctuation.
Then, a coherent understanding is obtained by extending the
Fermi-liquid theory to take into account the AF spin fluctuation.
Such a continuity clearly indicates that the theory based on
the Fermi-liquid picture is quite useful and effective
for strongly correlated systems.

Another effort has been devoted to include the effect of SC fluctuations.
This is closely related to the most challenging issue in the``anomalous''
metallic state, the pseudogap phenomenon, which is focussed in this review.
Among many theoretical proposals suggested before,
we will introduce an understanding based on the SC fluctuation.
Since high-$T_{\rm c}$ cuprates and organic superconductors have
the superconducting ground state, it is expected that the SC fluctuation
should be active in these materials at least near the phase transition.
It has been common knowledge that the SC fluctuation is usually negligible,
but the recent theoretical efforts, which has been stimulated by the recent
experimental results, have revealed the importance of the SC fluctuation
for the electronic properties.
We will clarify the condition for the appearance of the SC fluctuation and
show that the condition is actually satisfied in high-$T_{\rm c}$ cuprates
and some organic superconductors.
Then, the precursor of the phase transition destroys the Fermi-liquid
state and induces the excitation gap.
Many aspects of the anomalous properties, including the magnetic and
transport properties, are successfully explained by starting from
the Fermi-liquid state and taking into account appropriately
the effect of AF spin fluctuation and/or SC fluctuation
\cite{rf:yanaseFLEXPG}, as will be explained in Sec.~4 in details.
The scenario based on the Fermi-liquid theory is rather simple,
but the simple and unified scenario makes it easy to
understand anomalous behaviors in SCES.

Now our footing of this review article becomes clear, and the next point
is how to clarify the mechanism of the Cooper-pair formation.
First it is emphasized that a couple of quasi-particles form
the Cooper-pair. Typically, in heavy-fermion superconductivity,
the Cooper-pair is composed of heavy quasi-particles themselves,
as confirmed by the large jump in the specific heat at $T_{\rm c}$.
The investigation of pairing mechanism is reduced to the determination
of the residual interaction among quasi-particles.
As easily understood, non $s$-wave pairing should appear
for superconductivity in highly correlated systems
due to the effect of strong short-range Coulomb interaction.
In fact, anisotropic Cooper-pair has been found in common
in strongly correlated superconductors,
experimentally suggested by the power-law behavior of physical
quantities in the low-temperature region.
The value of the power sensitively depends on the node structure
of the gap function on the Fermi surface.
Thus, by analysing carefully the temperature dependence of physical
quantities in experiments, it is possible to deduce the symmetry of
the Cooper-pair under the group-theoretical restriction.
The phenomenological theory, which is not focused in this review, 
plays an important role for the determination of the pairing symmetry. 
Since the superconducting transition is the second-order phase
transition even for SCES, the Ginzburg-Landau theory has been still
applicable to those systems, by paying due attentions to the symmetry
of Cooper-pair as well as the group-theoretical restrictions
on the crystal structure.
In fact, there have been significant advances in the phenomenological
understandings on unconventional superconductivity without special
knowledge on the mechanism of Cooper-pair formation \cite{rf:sureview}.
In particular, when the degeneracy remains in the internal degree of
freedom in the triplet superconductor, the phenomenological Ginzburg-Landau
theory has been quite useful to identify the pairing symmetry.

When ones tried to discuss the microscopic aspects of the mechanism of
unconventional superconductivity, again AF spin fluctuations were
considered to play crucial roles to induce singlet $d$-wave superconductivity,
consistent with the node structure in high-$T_{\rm c}$ cuprates.
Indeed, the theory based on the AF spin fluctuations has provided many
important understandings in high-$T_{\rm c}$ superconductors, as will be
explained later. However, in order to arrive at the unified understanding
of superconductivity in SCES, more general viewpoint should be considered,
since the pairing potential originating from the residual interaction
among quasi-particles is $not$ always dominated by the spin fluctuations.
The most general conclusion is that the origin of the Cooper-pair
formation is the momentum dependence of the residual interaction.
We believe that
this point of view is important for the comprehensive understanding
of superconductivity in SCES including high-$T_{\rm c}$ cuprates.

A clear example for our belief can be found in the triplet superconductivity
confirmed in some $d$- and $f$-electron systems such as Sr$_2$RuO$_4$ and
UPt$_3$ \cite{Rf:Ishi.1,rf:Tou1}.
In order to understand the triplet pair formation, one may naively consider
the effect of ferromagnetic spin fluctuations, which was proposed for
the origin of spin-triplet superfluidity in $^3$He.
However, in the triplet superconductors, paramagnons are $not$ always
dominant in the spin fluctuation spectrum.
The ruthenate Sr$_2$RuO$_4$ is a typical example.
We have found that the incommensurate AF spin fluctuation is enhanced,
as observed in the neutron scattering experiment \cite{Rf:Sidi.1}.
Then, in contrast to the naive expectation, paramagnons do $not$ seem
to play a central role in the occurrence of triplet superconductivity.
In such a case, we have to consider the momentum dependence of
the residual interaction from the general point of view.
As we will discuss in detail in this review, it is possible to explain
triplet as well as singlet superconductivity, in addition to several
anomalous behaviors in the normal state, based on the microscopic
Hamiltonian, leading to the unified picture for unconventional
superconductivity in SCES.

For the purpose, it is indispensable to choose appropriately the method
for the calculation.
One is the application of numerical techniques such as
exact diagonalization and quantum Monte Carlo simulations.
There is a clear advantage that in principle we can include the effect
of electronic correlation correctly, but in the exact diagonalization,
the size of the model is severely restricted due to the limitation
in computer memory. Also in the quantum Monte Carlo simulations,
it is quite difficult to increase the strength of correlation because
of the negative sign problem and the system size is still restricted.
On the other hand, recently developed technique such as density matrix
renormalization group method is very powerful to analyze quantum systems
even with frustrations.
However, the target material is essentially limited to
the one-dimensional system, even though it may be possible to treat
ladder-like compounds.

Complementary to the numerical method, another traditional technique is
the quantum field theory, or more specifically, the Green's function
method, based on the assumption that it is allowed to perform
the perturbation expansion in terms of the interaction.
It is an advantage that we can calculate, in principle, physical quantities
in the thermodynamic limit, while it is inevitable to resort to
approximations for actual calculations, since it is quite difficult to
carry out the exact calculation based on the Green's function method.
Another important advantage is that the physical picture is to be clarified.
In this article, we focus on the Green's function technique and
the results obtained by this method.
It matches our principle based on the Fermi-liquid theory.
We will introduce an understanding obtained by the combination of
the numerical techniques and the Green's function approaches
\cite{rf:bulut1995}, but the techniques of numerical methods 
will not be introduced in this review article. 
Readers consult with other review articles regarding those issues 
(See, for instance, Ref.~\cite{Dagotto94}). 

Next, we need to define a microscopic model Hamiltonian.
In this paper, the Hubbard Hamiltonian will be focussed, since it is
widely recognized as a canonical model for SCES, although for
$f$-electron systems, we need to pay due attentions for its application.
We will also consider the multi-orbital Hubbard model, in which
orbital degree of freedom of $d$ electrons is explicitly included,
especially for the analysis of ruthenate.
For the purpose to study electronic properties of SCES, ones sometimes
considered the so-called $t$-$J$ model, obtained by the strong-coupling
expansion in the Hubbard or $d$-$p$ model.
The prohibition of double occupancy at each site is imposed on the model,
which is an essential point to include the strong correlation effect
in the $t$-$J$ model.
Since the Hilbert space becomes smaller than that of the Hubbard model,
the $t$-$J$ model has been frequently used for the analysis based
on numerical techniques. However, in the Green's function method or
even in simpler mean-field calculations, it is difficult to include
correctly the prohibition of double occupancy.
Thus, results and analysis on the $t$-$J$ model are out of the scope
of this review article.
Readers interested in the $t$-$J$ model can also consult with other
previous review papers \cite{Dagotto94}.

In this review article, then we will show the unified picture to
understand unconventional superconductivity in SCES based on
the Fermi-liquid framework by using the Hubbard (or Hubbard-like)
Hamiltonian and the Green's function method.
The calculations will be done based on the Dyson-Gor'kov equations
\cite{rf:AGD}, composed of normal and anomalous Green's functions
to characterize the normal and superconducting states. 
Formally those are the coupled equations, related by the irreducible
four-point vertex functions representing the interaction processes
among quasi-particles.
Depending on the effective strength of the interaction, we change
our approaches to evaluate the four-point irreducible vertex.

For the weak correlation systems,
we adopt the third-order perturbation theory with respect to
the on-site Coulomb interaction $U$ \cite{rf:hottasecond,rf:hottathird}.
By solving the linearized Dyson-Gor'kov equations, we can determine
$T_{\rm c}$, which will well reproduce the critical temperatures
for cuprates from overdoped to optimally doped regions.
Moreover, this perturbation theory gives reasonable values of
$T_{\rm c}$ also for organic superconductors
$\kappa$-type (BEDT-TTF)$_2$X \cite{rf:jujoTOP}.
The symmetry of these two superconductors  is considered to be
the $d$-wave one from various experimental results.
Theoretical calculations also predict the $d$-wave symmetry.

In addition to the above results, it should be stressed that
the perturbation theory gives also a reasonable explanation of
the mechanism of triplet superconductivity in Sr$_2$RuO$_4$
\cite{Rf:Nomu.3}.
The third-order terms in the vertex function play main roles
in realizing the triplet superconductivity.
Note that these terms are $not$ attributed to the contribution
from the spin fluctuations.
This fact is in sharp contrast with the spin fluctuation mechanism adopted
in cuprates, while it is consistent with the absence of paramagnon peak
in neutron diffraction experiments on ruthenate.

For the intermediate coupling systems such as optimally doped cuprates,
we adopt the fluctuation-exchange (FLEX) approximation \cite{rf:FLEX}
to evaluate the irreducible four-point vertex parts.
This theory based on the spin fluctuations has been developed
by many groups and used to give reasonable value of $T_{\rm c}$
for the $d$-wave superconductivity.
We confirm the reliability of these calculations by comparing the FLEX
and the third-order perturbation calculations with each other.
From the above calculation we conclude that
the momentum dependence in the effective interaction
between quasi-particles originating initially from the on-site Coulomb
repulsion $U$ induces the superconductivity.
The FLEX approximation is again used for the analysis of the anomalous
properties in the normal state. In order to take into account the effect
of the superconducting fluctuations, the FLEX approximation is modified
to be combined with the T-matrix approximation.
By using this FLEX + T-matrix approximation, we develop
a microscopic theory on the SC fluctuation and analyze
the under-doped cuprates. 

The organization of this review article is as follows.
In Sec.~2, experimental review will be provided regarding high-$T_{\rm c}$
cuprates, organic superconductors, and ruthenate.
The interests from the theoretical point of view will be also explained.
As is easily understood, those materials should include
millions of references, and it is almost impossible to cite all papers.
Thus, we will refer several papers, which will be relevant to
the later theoretical discussions.
In Sec.~3, theoretical results are reviewed based on the Green's function
techniques.
First we introduce some basic equations which will be needed
for the later analysis.
Then, the results for cuprates, organic superconductors, and ruthenate
will be discussed in this order.
In Sec.~4, as a typical result in the anomalous normal-state properties of
high-$T_{\rm c}$ cuprates, pseudogap phenomena will be discussed somewhat
in detail.
In Sec.~5, after the brief review of experimental results on $f$-electron
superconductors, we will discuss the mechanism of superconductivity
in heavy-fermion materials.
In Sec.~6, we summarize this review article.
In Appendix A, the correct derivation of the generalized
Fermi-liquid theory on the London penetration depth is presented and
we provide our picture to understand the so-called Uemura plot by taking
account of the quasi-particle interaction.
In Appendix B, the convergence of the perturbation expansion
with respect to $U$ is examined by calculating the forth-order terms.
The convergence is satisfactorily good for $d$-wave pairing cases,
while for $p$-wave pairing cases $T_{\rm c}$ shows oscillatory behavior
depending on the calculated order.
By summing up the ladder diagrams up to infinite order,
we can eliminate this oscillation and recover the convergence 
also for $p$-wave pairing case.

\par\vfill
\eject

%
%
\section{Experimental View}

\subsection{Cuprates}

The discovery of high-$T_{\rm c}$ cuprates
\cite{rf:bednortz,rf:tanakahigh-tc} was a trigger of the intensive
studies on superconductivity in strongly correlated electron systems (SCES).
The theoretical efforts stimulated by high-$T_{\rm c}$ superconductivity
have clarified many basic properties of strongly correlated materials.
Especially the understanding of the quasi-two-dimensional system
has been developed.

High-$T_{\rm c}$ cuprates have the perovskite-type crystal structure and
the parent compounds are the Mott insulators with anti-ferromagnetic (AF)
order.
Carrier doping into this Mott insulating state induces high-$T_{\rm c}$
superconductivity, which essentially occurs in the two-dimensional (2D)
CuO$_2$ plane.
Note here that both hole and electron dopings induce superconductivity
\cite{rf:electrondopesuper}.
From the early stage of the research, high-$T_{\rm c}$ cuprates have
been recognized to be one of SCES
\cite{rf:andersonscience,rf:andersonbook}.
Now the cuprate is confirmed to be the most established
unconventional superconductor.
Namely, it is predominantly believed that Cooper-pairs with
$d_{x^2-y^2}$-wave symmetry originates from the electronic mechanism.
We can understand also several normal-state anomalous properties
such as $T$-linear electric resistivity \cite{rf:iye} and
strongly enhanced AF spin correlation \cite{rf:NMR},
considering the effect of strong electron correlation.
In the following, we survey these experimental results on unconventional
superconductivity and anomalous normal-state properties
in high-$T_{\rm c}$ cuprates.

\subsubsection{Symmetry of Cooper-pair}

First let us see the results on pairing symmetry, since 
the symmetry of the Cooper-pair is an important issue 
to consider the superconducting mechanism. 
The non-$s$ wave pairing is a direct evidence for the unconventional 
superconductivity. 
According to the crystal symmetry of the square lattice composed of
copper and oxygen ions, the pairing symmetry is classified as
$s$-wave, $d_{\rm x^2-y^2}$-wave, $d_{\rm xy}$-wave, and so on.
Among them, in the theoretical point of view, the appearance of
the conventional $s$-wave superconductivity is suppressed, because
the strong on-site repulsion easily destroys the $s$-wave pairing.
In fact, the theoretical studies on the electron correlation have
predicted the $d_{\rm x^2-y^2}$-wave symmetry:
For instance, in the literatures, we find several works based on
the RPA \cite{rf:bickers1987,rf:shimahara},
scaling theory \cite{rf:shultz,rf:dzyaloshinskii,rf:lederer},
variational method \cite{rf:yokoyama,rf:ogata},
and quantum Monte Carlo simulation \cite{rf:white}.
In 80's, however, the many experimental results supported
the $s$-wave symmetry.
Only a few experiments including the NMR measurement have supported
the $d$-wave superconductivity \cite{rf:NMR}.
This discrepancy has been clearly resolved by the intensive
investigations in 90's, which have demonstrated the various evidence
for the anisotropic gap structure with line node 
\cite{rf:harlingenreview,rf:scalapinoreview,rf:ginsberg,rf:hardysymmetry}.
Some deviations from the clean $d$-wave superconductivity has been explained
by taking account of the strong impurity scattering \cite{rf:hottaimp}.
Finally, the phase-sensitive measurements have clearly confirmed
the $d_{\rm x^2-y^2}$-wave symmetry \cite{rf:sigrist,rf:harlingen,rf:tsuei}.
Nowadays the momentum dependence of the excitation gap is directly observed
in the angle resolved photoemission spectroscopy (ARPES), where
the line node appears in the diagonal direction of $k_{\rm x}=k_{\rm y}$
\cite{rf:shen1993,rf:shenrep,rf:ding1995,rf:norman1996}.

On the other hand, the pairing symmetry in electron-doped systems has not
been settled for a long time, because node-less behaviors were reported
by several groups \cite{rf:electron-dope_S}.
However, recent experiments have finally concluded that the electron-doped
systems are also the $d_{\rm x^2-y^2}$-wave superconductors
\cite{rf:hayashi,rf:electron-dope_D,rf:tsuei-el,rf:takahashi-el,rf:shen-el}.
Also in this case, crucial roles have been played by
the phase-sensitive measurement \cite{rf:tsuei-el} and
ARPES \cite{rf:takahashi-el,rf:shen-el}.
These results seem to be natural, since from the theoretical point of view,
the same pairing mechanism is expected both for hole- and electron-doped
materials. Note, however, that interesting particle-hole asymmetry has been
found in some aspects.
For instance, in electron-doped systems, we have observed that
(i) the transition temperature is relatively low,
(ii) the superconducting region in the phase diagram is narrow,
and (iii) the AF order is robust in comparison with hole-doped compounds.
Later in this review, this asymmetry will be discussed in our theoretical
approach by considering the detailed electronic structure.

Now we can conclude that the issues on the paring symmetry
in high-$T_{\rm c}$ cuprates have been settled.
In the next stage, the following two issues are especially important.
One is, of course, the pairing mechanism for high-$T_{\rm c}$ cuprates.
This theoretical subject will be discussed in Sec.~3.
Particularly we focus on the intensive studies from the microscopic
point of view, which have opened a new way to understand 
the superconductivity in SCES. 
Another issue is to explain the anomalous properties in the normal state,
which is one of the main points of this review.
This subject will be discussed in detail in Sec.~4 with our main interests
on the pseudogap phenomena.
As for the latter problem, intensive experimental investigations have been
performed and lots of important results have been piled up.
Thus, here we survey the experimental results on the normal-state properties
in the following subsections \cite{rf:timusk}.


\begin{figure}[t]
\begin{center}
\vspace{2.5cm}
\caption{The phase diagram of high-$T_{\rm c}$ superconductors.
The horizontal and vertical axes indicate the doping concentration
and the temperature, respectively.
``AF'', ``SC'', and ``PG'' denote anti-ferromagnetic, superconducting,
and pseudogap state, respectively.
The onset curve for the spin fluctuation ($T$=$T_0$) and that for
the pseudogap formation ($T$=$T^*$) show the typical cross-over
temperatures.}
\label{fig:high-Tcphasediagram2}
\end{center}
\end{figure}

\subsubsection{Phase diagram}

Before proceeding to the review of normal-state properties of
high-$T_{\rm c}$ cuprates, it is instructive to explain first
the outline of the $T$-$\delta$ phase diagram.
Both the theoretical and experimental investigations have been focussed
on the hole-doped systems, but recently, the electron-doped systems
also have been studied intensively.
Then, the whole phase diagram as shown
in Fig.~\ref{fig:high-Tcphasediagram2} has been clarified.
Let us see some characteristic points of this phase diagram in the following.

In general, the properties of high-$T_{\rm c}$ cuprates are controlled
by the carrier doping concentration $\delta$, which is defined as
$\delta$=1$-n$ and $n$ is the carrier number per copper site.
The system is the AF Mott insulator at half-filling ($\delta$=0) and
the AF order is easily destroyed by the slight amount of
hole-doping ($\delta$$>$0).
Then, there appears the metallic phase with the superconducting ground state.
The transition temperature $T_{\rm c}$ takes its maximum value
($\sim$100K) at the optimally-doped region ($\delta$$\sim$0.15) and
decreases in the over-doped region.

Here it is stressed that anomalous behaviors in the normal-state properties
are significant from the optimally-doped to the under-doped region,
although the meaning of ``anomalous behavior'' will be discussed in detail
in the following subsections.
As indicated by thin and thick dashed curves
in Fig.~\ref{fig:high-Tcphasediagram2},
there exist two characteristic temperatures, determined from
the several experimental results.
One is $T_0$, shown by the thin dashed curve, a temperature at which
the AF spin correlation begins to develop with decreasing temperature.
For instance, the NMR $1/T_{1}T$ and Hall coefficient $R_{\rm H}$
increase from the temperature $T_{\rm 0}$.
Another characteristic temperature is $T^{*}$,
indicated by the thick dashed curve.
This is a temperature for the onset of the opening of pseudogap,
observed in ARPES, NMR $1/T_{1}T$, tunnelling measurements, and so on.
The enormous investigations have been dedicated to the clarification
of the pseudogap because this issue has been regarded as
a central problem in high-$T_{\rm c}$ superconductors.
Note that these experiments have indicated some similarities between
the pseudogap and superconducting gap. We consider that 
the SC fluctuation becomes apparent around $T$=$T^*$. 
As shown in Sec.~4, our interests on the pseudogap phenomena are concerned
with this lower cross-over curve.
Note also that somewhat uncertainty may be included in the definitions of
$T_0$ and $T^*$, since two curves just denote the cross-over temperatures,
not the phase transition.
Nevertheless, we believe that the above classifications are useful
to understand the anomalous properties in a coherent way.

As already mentioned, the electron-doped systems ($\delta$$<$0) are
also superconducting.
However, some different features from hole-doped ones are observed, 
as explained in the previous subsection. 
For instance, the AF order is robust for the carrier doping in these systems
and superconductivity occurs in the narrow doping region with relatively
low transition temperature.
Especially, in the normal metallic state, the pseudogap phenomena and
the under-doped region have {\it not} been observed 
in electron-doped systems. 
In Secs.~3 and 4, we provide our understanding of the phase diagram
including this particle-hole asymmetry.

\subsubsection{Magnetic properties}  

As is well known, NMR and neutron scattering measurements are
the powerful methods to investigate spin correlations.
These experiments have played crucial roles to reveal that
the magnetic properties in the high-$T_{\rm c}$ cuprates
are ``anomalous'' in many aspects.
Here let us summarize these anomalous behaviors in the normal state.

\begin{figure}[t]
\begin{center}
\vspace{2.5cm}
\caption{The experimental data of the NMR $1/T_{1}T$ \cite{rf:yasuoka}.}
\label{fig:yasuoka}
\end{center}
\end{figure}

First we emphasize that the pseudogap phenomenon has been first
discovered in the NMR measurements \cite{rf:yasuoka}.
As shown in Fig.~\ref{fig:yasuoka}, the spin-lattice relaxation rate
over temperature $1/T_{1}T$ exhibits the peak around $T=T^{*}$ and
it begins to decrease below $T^{*}$. 
This is a typical pseudogap behavior observed in $1/T_{1}T$,
indicating the suppression of magnetic excitations in the low-energy part.
The detailed investigations have subsequently observed this phenomenon
in most of the under-doped compounds
\cite{rf:yasuoka,rf:warren,rf:takigawa1991,rf:itoh1992,rf:takigawa1994,rf:itoh1994,rf:julien,rf:itoh1996,rf:itoh1998,rf:ishida1998}
and the doping dependence of $T^{*}$ is found to be consistent with
Fig.~\ref{fig:high-Tcphasediagram2}.

In the conventional metal well described by the Fermi-liquid theory,
$1/T_{1}T$ should be a constant for the temperature higher than $T_{\rm c}$.
On the other hand, in high-$T_{\rm c}$ cuprates,
$1/T_{1}T$ is $not$ constant for $T > T_{\rm c}$.
Rather, the Curie-Weiss law $1/T_{1}T$$\propto$$(T+\theta)^{-1}$
has been observed for $T$$>$$T^*$ \cite{rf:NMR,rf:NMRCurie},
where $\theta$ is a Weiss temperature.
This fact clearly indicates that strong spin fluctuations exist
in high-$T_{\rm c}$ cuprates.
Note that this spin fluctuation has anti-ferromagnetic nature
\cite{rf:NMRKorringa}, since the ratio $(1/T_{1}T)/K^{2}$ is
larger by one order than the usual value of the Korringa law,
where $K$ is the NMR Knight shift.
For $T_{\rm c}<T<T^*$, as explained above, $1/T_{1}T$ begins to decrease
even before the system becomes superconducting.
Since those behaviors are significantly different from the results 
in the conventional Fermi-liquid theory, 
they are regarded to be anomalous normal-state properties.
Note that the Curie-Weiss law itself can be explained by
the self-consistent renormalization (SCR) theory \cite{rf:moriyatext},
which is one of the spin fluctuation theory.
However, it is difficult to understand the anomalous temperature
dependence including pseudogap behavior only from
the spin fluctuation theory.

The NMR Knight shift $K$ is proportional to the uniform spin susceptibility
$K \propto \chi({\bm 0},0)$.
Although the Knight shift satisfies the Korringa-relation
$K^{2} \propto 1/T_{1}T$ in usual metals,
in high-$T_{\rm c}$ cuprates, it shows a different temperature
dependence from the Korringa-relation.
Namely, the Knight shift gradually decreases below $T_{0}$
\cite{rf:takigawa1991,rf:alloul,rf:ishida1991}.
This behavior has been observed also in the uniform susceptibility
\cite{rf:johnston,rf:oda}.
Concerning the pseudogap, the Knight shift also shows an anomaly
around $T^{*}$; the decrease becomes rapid below $T^{*}$,
as observed in Fig.~\ref{fig:ishida1998} \cite{rf:ishida1998}.

\begin{figure}[t]
\begin{center}
\vspace{2.5cm}
\caption{The experimental data of the Knight shift
\cite{rf:ishida1998}.}
\label{fig:ishida1998}
\end{center}
\end{figure}

The spin-echo decay rate $1/T_{\rm 2G}$ measures the momentum summation
of the square of the static spin susceptibility (see Eq.~(\ref{eq:NMR-T2G})).
The experiments in early years have shown that the NMR $1/T_{\rm 2G}$
keep increasing below $T^{*}$ \cite{rf:itoh1992,rf:takigawa1994,rf:julien}.
However, the recent experiments have revealed the different behavior
depending on the number of CuO$_2$ layers
\cite{rf:itoh1994,rf:itoh1996,rf:itoh1998,rf:goto,rf:tokunaga}.
This difference is probably attributed to the effect of the
interlayer coupling \cite{rf:itoh1998,rf:goto}.
It is important to note that the $1/T_{\rm 2G}$ decreases below $T^{*}$
in the single layer compounds
\cite{rf:itoh1994,rf:itoh1996,rf:itoh1998,rf:goto}.
Then, the decrease of $1/T_{\rm 2G}$ is weaker than that of $1/T_{1}T$.

Now we turn our attentions to the neutron scattering experiments,
in which the wave vector of the spin fluctuation can be directly observed
by measuring the form factor in proportion to the imaginary part of
dynamical spin susceptibility, ${\rm Im}\chi({\bm q},\omega)$.
In high-$T_{\rm c}$ cuprates, the peak of the spin-spin correlation function
is located around ${\bm q}$=$(\pi,\pi)$ \cite{rf:neutronnormal}.
The peak position depends both on the temperature and doping concentration,
but interestingly enough, it can be incommensurate as
${\bm q}$$\neq$$(\pi,\pi)$ in some materials
\cite{rf:yamadalsco,rf:endohneutron,rf:yamadaneutron,rf:stripeEX,rf:arai,rf:yamadakurahashi,rf:lee}.
The interesting temperature dependence of this incommensurability has been
observed and discussed in relation with a stripe order
\cite{rf:stripeEX,rf:arai}, which is found in La-based compounds around
$\delta$=1/8 \cite{rf:tranquada}.
We will comment on this problem in Sec.~4.3.3.
Note that in the spin fluctuation theory, the detailed structure in
spin correlation function is not important, since the main part of the
magnetic excitation always locates around ${\bm q}$=$(\pi,\pi)$.

\begin{figure}[t]
\begin{center}
\vspace{2.5cm}
\caption{The experimental data of the dynamical spin susceptibility
at the anti-ferromagnetic wave vector ${\rm Im}\chi({\bm Q},\omega)$
\cite{rf:neutronPG}.}
\label{fig:neutronPG}
\end{center}
\end{figure}

The pseudogap is clearly observed also in the spectral weight of
the spin fluctuation, which is directly measured by
the neutron scattering experiments.
The spectral weight in the low frequency part is suppressed and
at the same time, it is shifted to the high frequency part,
as shown in Fig.~\ref{fig:neutronPG} \cite{rf:neutronPG}.
This frequency dependence is anomalous in comparison with the results
of the conventional Fermi-liquid theory and
also those of the spin fluctuation theory.
As will be discussed in Sec.~4.3.3, this behavior means that
a new energy scale appears in the pseudogap state.
We will attribute it to the energy scale due to the superconductivity.

\subsubsection{Transport properties}

In the previous subsection, anomalous behaviors in magnetic properties
have been briefly reviewed, but transport properties are also anomalous
in comparison with the conventional Fermi-liquid theory.
In this subsection, let us see anomalous features of in-plane electric
resistivity $\rho_{\rm ab}$, Hall coefficient $R_{\rm H}$, and
c-axis resistivity $\rho_{\rm c}$.

\begin{figure}[t]
\begin{center}
\vspace{2.5cm}
\caption{The experimental data of the in-plane resistivity
\cite{rf:odatransport}.}
\label{fig:odatransport}
\end{center}
\end{figure}

It is well known that the Fermi-liquid theory predicts the $T^2$-law
for the electric resistivity in the low-temperature region, but
in optimally- and under-doped region of high-$T_{\rm c}$ cuprates,
it has been observed that $\rho_{\rm ab} \propto T$, 
\cite{rf:iye,rf:takagi} as shown in Fig.~\ref{fig:odatransport}. 
However, when the amount of doping concentration is further increased,
the temperature dependence in the in-plane resistivity gradually changes
and eventually the $T^{2}$-law recovers in the over-doped region
\cite{rf:kubo}.
This crossover behavior in the temperature dependence of the in-plane
resistivity is well described by the nearly anti-ferromagnetic
Fermi-liquid theory, as will be explained in Sec.~4.3.4
\cite{rf:hlubina,rf:stojkovic,rf:yanaseTR}.

In the pseudogap state, a characteristic behavior can be observed 
in the in-plane resistivity.
Namely, it changes its slope at $T^{*}$ and slightly deviates downward
\cite{rf:ito,rf:mizuhashi,rf:odatransport}.
This rather weak deviation has been one of the puzzling issues among
anomalous normal-state properties of high-$T_{\rm c}$ cuprates.
The transport properties are generally insensitive to the pseudogap
in comparison with the magnetic properties \cite{rf:iye}.
Therefore, the pseudogap was sometimes called ``spin gap'',
in the sense that the gap occurs only in the spin excitations.
These properties were complicated and difficult to understand,
but it is now consistently explained by including simultaneously
the spin and superconducting fluctuations,
as will be described in Sec.~4.3.4 \cite{rf:yanaseTRPG}.

\begin{figure}[t]
\begin{center}
\vspace{2.5cm}
\caption{The experimental data of the Hall coefficient \cite{rf:ito}.}
\label{fig:ito}
\end{center}
\end{figure}

In the conventional Fermi-liquid, the Hall coefficient $R_{\rm H}$
is independent of $T$, while in high-$T_{\rm c}$ cuprates,
it strongly depends on the temperature, as show in
Fig.~\ref{fig:ito} \cite{rf:iye,rf:ito,rf:satoM,rf:hwang}.
Moreover, in the under-doped region, the Hall coefficient takes
much larger value than that expected from the 
band-structure calculation results.
As seen in the in-plane resistivity, these anomalous behaviors gradually
change to those in the conventional Fermi-liquid in the over-doped region.
The enhanced Hall coefficient was sometimes interpreted as an evidence for
the low-carrier density $n$, because the relation $R_{\rm H} \propto 1/n$
is derived in the isotropic electron system.
However, this interpretation is just a superficial expectation in the
anisotropic system such as the Hubbard model on the square lattice
(see Sec.~4.3.4).
In fact, the enhanced Hall coefficient has been explained within the nearly
AF Fermi-liquid theory \cite{rf:kontani,rf:kanki}.

In the pseudogap state, the Hall coefficient remarkably deviates downward
and decreases with temperature
\cite{rf:ito,rf:mizuhashi,rf:satoM,rf:xiong,rf:ong}.
The anomaly around $T^{*}$ is much clearer than that in the resistivity.
A theory based on the $d$-wave superconducting (SC)
fluctuation alone expects the
enhancement of the Hall coefficient \cite{rf:ioffe}.
We will also show that this discrepancy can be resolved by the
simultaneous consideration of AF and SC fluctuations (Sec.~4.3.4).

Another interesting property of high-$T_{\rm c}$ cuprates is the strongly
anisotropic transport, originating from the layered structure.
Typically the ratio of the c-axis and in-plane resistivity,
$\rho_{\rm c}/\rho_{\rm ab}$, increases in the under-doped
and/or low-temperature region \cite{rf:iye,rf:takenaka}.
Moreover, the pseudogap remarkably enhances the c-axis resistivity
\cite{rf:takenaka}.
This response to the pseudogap, qualitatively different from that
of the in-plane resistivity,
can be explained by considering the $d$-wave SC fluctuation
combined with the characteristic band structure
\cite{rf:ioffe,rf:okanderson} (see Sec.~4.3.4).

The c-axis optical conductivity $\sigma_{\rm c}(\omega)$
shows no Drude peak in the under-doped region
\cite{rf:homes,rf:basov,rf:tajima}, which is consistent with
the incoherent nature of the c-axis resistivity.
As will be explained in Sec.~4.3.4, owing to the momentum dependence of
the c-axis hopping matrix, the coherent transport along the c-axis is
disturbed by the pseudogap.
The gap structure appears in $\sigma_{{\rm c}}(\omega)$ in the pseudogap state.
This pseudogap smoothly changes to the SC gap and this behavior indicates
the close relation between the pseudogap and the SC gap.

\subsubsection{Spectroscopy}

The ARPES directly measures the single-particle spectral weight at
the selected momentum.
Therefore, many pieces of clear information on the single-particle excitation
can be obtained, while some cares are required for the resolution.
Especially, the ARPES has provided us an important suggestion on the
pseudogap phenomena \cite{rf:ding,rf:shenPG,rf:normanPG,rf:ARPESreview},
such as the leading edge gap observed above $T_{\rm c}$,
as shown in Fig.~\ref{fig:normanPG}.
This experimental result indicates the suppression of
the single-particle spectral weight near the Fermi energy,
which clarifies following two important natures of the pseudogap:
(i) The momentum dependence of the pseudogap is similar to that of the
SC gap, i.e., the pseudogap has the $d_{\rm x^2-y^2}$-wave form.
(ii) The magnitude of the pseudogap does $not$ change through
the superconducting transition.
Below $T_{\rm c}$, the coherent quasi-particle peak appears at the gap edge,
and the gap structure becomes sharp with keeping its magnitude.
These results have clearly suggested the close relation between the pseudogap
phenomena and superconductivity, leading to the pairing scenarios in which
the pseudogap is a precursor of superconductivity (see Sec.~4.1).

\begin{figure}[t]
\begin{center}
\vspace{2.5cm}
\caption{The results of the ARPES \cite{rf:normanPG}.}
\label{fig:normanPG}
\end{center}
\end{figure}

Note that the qualitatively same suggestion has been obtained from the
tunneling spectroscopy which measures the electronic density of states (DOS).
The suppression of the DOS near the Fermi level is observed above $T_{\rm c}$
\cite{rf:renner,rf:miyakawa,rf:dipasupil}.
The magnitude of the pseudogap in the DOS is nearly the same (slightly larger)
as that of the SC gap.
Again we observe that the gap structure becomes sharp below $T_{\rm c}$.

The measurements of the SC gap at $T \ll T_{\rm c}$ have revealed that
the SC gap increases in the under-doped region in spite of the decrease
in $T_{\rm c}$ \cite{rf:harris,rf:odatransport}.
If we ignore quantum fluctuations, the SC gap at $T=0$ is in proportion to
the transition temperature in the mean field theory $T_{\rm c}^{\rm MF}$,
which is increased with the decrease of the amount of doping even
in the under-doped region.
It is expected that the thermal SC fluctuation suppresses $T_{\rm c}$
in the under-doped region, since it should become strong with the decrease
of $\delta$.

Concerning the cross-over around $T=T_{0}$, both the tunneling spectroscopy
\cite{rf:renner,rf:miyakawa,rf:dipasupil} and 
the ARPES \cite{rf:sato} have observed the slight and broad suppression
of the DOS from $T=T_{0}$. This suppression is called ``large pseudogap'' 
and its energy scale is $3 \sim 4$ times larger than the SC gap. 
The large pseudogap should not be attributed to 
the precursor of superconductivity, and may have somewhat magnetic origin.
The pseudogap below $T^{*}$ is sometimes called ``small pseudogap'' 
in contrast with the large pseudogap. 

\subsubsection{Magnetic field penetration depth}

Finally in this subsection, let us discuss the magnetic field penetration
depth which is identical to the London penetration depth 
$\lambda_{\rm L}$ in the type II limit. 
This is not the quantity in the normal state, but it indicates an interesting
property in the SC state.
The doping dependence of the magnetic field penetration depth at $T$=0 is
known as ``Uemura plot'' \cite{rf:uemura,rf:bonnmagpen}.
The London constant $\Lambda$=$1/4\pi\lambda_{\rm L}^{2}$ is roughly
proportional to the hole-doping and $T_{\rm c}$
($\Lambda \propto \delta \propto T_{\rm c}$) in the under-doped region.
This constant shows a peak around the optimally-doping and decreases
with doping in the over-doped region \cite{rf:tallonmagpen}.
The London constant is sometimes expressed by the ``superfluid density''
$n_{\rm s}$ as $\Lambda$=$n_{\rm s}/m^{*}$,
where $m^*$ is the effective mass of carrier.
We focus on the London constant instead of the superfluid density
in order to avoid any confusion, since it will be shown that
the superfluid density is $not$ related to the electron density.
The London constant corresponds to the stiffness of the superconducting
phase variable, which is generally related to the SC fluctuation
in the ordered state.
The Uemura plot exhibits that the phase fluctuation is softened 
in the under-doped region.
It has been proposed along this line that the phase disordered state
is a possible pseudogap state \cite{rf:emery}.

The temperature dependence of the London constant has also attracted much
interests. The London constant exhibits $T$-linear dependence as
$\Lambda(T)$=$\Lambda(0)-aT$ in the low-temperature region
\cite{rf:hardysymmetry,rf:bonnmagpen,rf:panagopoulosmagpen},
which is a characteristic behavior of the $d$-wave superconductor.
This $T$-linear law reflects the nodal quasi-particles around
$\k = (\pi/2,\pi/2)$.
Interestingly, the coefficient $a$ is almost independent of the doping
concentration in sharp contrast to the drastic change of
$\Lambda(0)$ \cite{rf:bonnmagpen}.
These anomalous behaviors have much stimulated theoretical insights
\cite{rf:leewen,rf:millismagpen}.
In order to provide a systematic understanding for the anomalous behaviors,
we will explain the microscopic derivation of the London constant based on
the Fermi-liquid theory, as shown in Appendix
\cite{rf:jujomagpen1,rf:jujomagpen2}.
Then, it is emphasized that the Fermi-liquid correction as well as
the reduced symmetry in the square lattice plays an essential role
\cite{rf:millismagpen}.

The c-axis London constant is much smaller than the in-plane one and
the anisotropy $\Lambda_{\rm c}/\Lambda$ is reduced with
under-doping \cite{rf:bonnmagpen,rf:panagopoulosmagpen}.
These behaviors are consistent with the transport properties
in the normal state.
The power of the temperature dependence of $\Lambda_{\rm c}$ is larger
than unity in the low-temperature region
\cite{rf:bonnmagpen,rf:panagopoulosmagpen}.
The origin of this nature is in common with the incoherent c-axis
conductance in the pseudogap state.
Note that in the clean limit $T^{5}$-law is expected, but sample dependence
is frequently observed, probably owing to the effect of disorder. 

The London constant just below $T_{\rm c}$ shows a rapid growth rather
than that obtained in the BCS theory \cite{rf:panagopoulosmagpen,rf:kamal}.
This is regarded as an appearance of the critical fluctuation.
The critical behavior of the 3D-XY universality class has been confirmed
\cite{rf:kamal}.
This clear appearance of the SC fluctuation supports the pairing scenario
on the pseudogap phenomena.
The growth of $\Lambda_{\rm c}$ below $T_{\rm c}$ is more rapid than
that of $\Lambda$ \cite{rf:panagopoulosmagpen}.
This is also because of the momentum dependence of the inter-layer
hopping matrix.

\subsection{Organic Superconductor}

As mentioned in the introductory section, the discovery of
organic superconductor (TMTSF)$_2$PF$_6$ \cite{rf:jerome}
has made a great impact on the community of superconductivity.
The intensive exploitation in this field has revealed a universality of
superconducting phenomena and attached much interests on the physical
aspects of the organic materials \cite{rf:1}.
Among them, one of the most interesting superconductors is 
$\kappa$-type (BEDT-TTF)$_2$X.
In the following, it is expressed as $\kappa$-(ET)$_2$X for an abbreviation.
These compounds have quasi-two-dimensional electronic structures,
which have been confirmed by the Shubnikov-de Haas experiments \cite{rf:9}
and by the strong anisotropy in the electronic transport \cite{rf:28}.
The conduction band is mainly constructed from ET molecules.
A simplified model for the electronic state can be constructed
by noting the molecular $\pi$-orbitals (see Sec.~3.3.1),
although the ET molecule has a complicated structure.

\begin{figure}[t]
\begin{center}
\vspace{2.5cm}
\caption{Schematic phase diagram of $\kappa$-(ET)$_2$X compounds.}
\label{fig:organic}
\end{center}
\end{figure}

$\kappa$-(ET)$_2$X compounds are one of the central objects in this area,
because of their typical features as SCES.
The typical phase diagram is shown in Fig.~\ref{fig:organic}.
We have two ordered phases by tuning the pressure $P$.
One is the AF insulating state in the lower pressure region and the other
is superconducting phase in the higher pressure region \cite{rf:2,rf:3}.
It should be noticed that the superconductivity occurs when the AF order
disappears. The transition temperature decreases with increasing $P$.
It is true that this phase diagram includes similar aspects to that
of high-$T_{\rm c}$ cuprates, but several differences are observed.
First, the phase transition from the AF to the SC state is the first order.
Second, the carrier number in the conduction band is always half-filling
per dimer, independent of the control parameter $P$.
The pressure widens the band width $W$ and thus, it controls
the parameter $U/W$. The decrease in $U/W$ by the pressure has been
confirmed by many experimental facts.
For instance, the resistivity becomes smaller together with
the superconducting $T_{\rm c}$ \cite{rf:2,rf:3,rf:sushko}.
Therefore, the Mott transition in these compounds is regarded to be
``band width controlled'', which is contrasted with
the ``filling controlled'' Mott transition, as observed in cuprate
superconductors \cite{rf:imadareview}.

A series of these compounds have different properties at ambient pressure
due to the kinds of anions X:
$\kappa$-(ET)$_2$Cu[N(CN)$_2$]Cl is the AF insulator \cite{rf:4}.
The deuterated $\kappa$-(ET)$_2$Cu[N(CN)$_2$]Br is located
on the boundary between the two phases \cite{rf:5}.
$\kappa$-(ET)$_2$Cu[N(CN)$_2$]Br and $\kappa$-(ET)$_2$Cu(NCS)$_2$
show the superconductivity at the ambient pressure.
This systematic change due to X is regarded as an effect of the chemical
pressure, as described in the phase diagram of Fig.~\ref{fig:organic}.

It is predominantly believed with a few objections that the $d$-wave
superconductivity occurs in these compounds like high-$T_{\rm c}$ cuprates.
NMR experiments again have played an important role in the identification
of the pairing symmetry \cite{rf:mayaffreNMR,rf:sotoNMR,rf:kanodaNMR}.
The measurements below $T_{\rm c}$ have shown (i) no coherence peak,
(ii) the decrease of the Knight shift, and (iii) $T^3$-law of the NMR
$1/T_{1}$ in the low-temperature region. From these results,
the singlet pairing with line node has been suggested.
The same conclusion is obtained by the measurement of the electronic
specific heat, which is proportional to $T^2$ at low temperature \cite{rf:7}.
These results indicate that the pairing symmetry is not the $s$-wave,
but probably the $d$-wave.

The highest $T_{\rm c}$ among $\kappa$-(ET)$_2$X compounds is about 13K.
From the theoretical point of view, this value is the same order as that of
cuprate superconductors, when $T_{\rm c}$ is scaled by the band-width.
In general, organic materials have smaller band-width by an order,
because they are constructed from the molecular orbitals.
The scaling between $T_{\rm c}$ and $W$ have suggested a similar pairing
mechanism to the high-$T_{\rm c}$ cuprates. 
We will review in detail the theoretical results in Sec.~3.3.

Finally, we note that the similarity between organic and high-$T_{\rm c}$
superconductors should be extended to the anomalous properties
in the normal state.
In particular, the pseudogap has been also observed in the NMR $1/T_{1}T$
\cite{rf:mayaffrePG,rf:kanodaPG} with $T^{*}$$\sim$50K.
However, it is an important difference that $T^{*}$ is much higher than
$T_{\rm c}$ and the electronic state is almost incoherent above $T^{*}$.
Thus, the different nature around $T^{*}$ is expected.
As will be discussed in Sec.~4.2.4, however, the similar properties in the
electronic state indicate the manifestation of the pseudogap with the same
origin. Recently, Kanoda's group has given a clear understanding on this
problem \cite{rf:matsumotokanoda} by measuring the magnetic field dependence
of the NMR $1/T_{1}T$;
the SC fluctuation appears from the new cross-over temperature $T_{\rm c}^{*}$
which is between $T_{\rm c}$ and $T^{*}$.
The electronic state below $T_{\rm c}^{*}$ is regarded as the pseudogap state
induced by the SC fluctuation as in the under-doped cuprates. 
The details will be discussed in Sec.~4.2.4.

\subsection{Sr$_2$RuO$_4$}

Superconductivity in Sr$_2$RuO$_4$ has been discovered by Maeno {\it et al.}
in 1994 \cite{Rf:Maen.1} and the intrinsic $T_{\rm c}$ is now considered
to be as large as 1.5K in the high purity sample.
This compound possesses the same crystal structure as La$_{2-x}$Sr$_x$CuO$_4$,
one of the high-$T_{\rm c}$ superconductors, and it similarly has the quasi
two-dimensional nature.
For example, the resistivity exhibits a large anisotropy;
the ratio $\rho_{\rm c}$/$\rho_{{\rm ab}}$ is in the order of
several hundreds \cite{Rf:Maen.1,Rf:Maen.2}.
The quantum oscillation measurement has also clearly shown
the quasi-two-dimensional Fermi surfaces \cite{Rf:Mack.1}. 
Then, RuO$_2$ layers are expected to be essential for the metallic behavior
and superconductivity, as CuO$_2$ layers in high-$T_{\rm c}$ cuprates.

In contrast to cuprates, Sr$_2$RuO$_4$ is considered to be an ideal
two-dimensional Fermi-liquid \cite{Rf:Maen.2}, since there is no anomalous
behavior in the normal state.
Since $T_{\rm c}$ is much lower than that of cuprates and $\kappa$-(ET)$_2$X,
the superconductivity is easily destroyed
by the small perturbation or disorder.
For ruthenate it is difficult to find, at least at present,
a well-defined controlling parameter for the appearance of 
the superconductivity such as doping in cuprates. 
Thus, our interest is here focused only on
the superconductivity in Sr$_2$RuO$_4$. 

According to the first-principle band-structure calculations
\cite{Rf:Oguc.1,Rf:Sing.1}, it has been clarified that the electronic states
near the Fermi level mainly consist of the Ru4d$\varepsilon$ orbitals,
although the Ru4d$\varepsilon$ and O2p orbitals hybridize with each other.
Since it is expected that electrons strongly correlate through Coulomb
interactions at Ru sites, Sr$_2$RuO$_4$ belongs to a class of SCES.
In fact, the Mott insulating state has been found in the related compound
${\rm Ca}_{2}{\rm Ru}{\rm O}_{4}$ \cite{rf:nakatsuji},
suggesting the importance of strong correlation effect.

Compared to the other compounds focused in this review, Sr$_2$RuO$_4$ has
somewhat different nature both in the electronic structure and
in the superconducting properties.
One interesting issue is that this compound is a multi-band system.
The quantum oscillation measurement has shown three quasi-two-dimensional
Fermi surfaces, which are defined as $\alpha$, $\beta$, and $\gamma$ sheets
\cite{Rf:Mack.1}.
Note that the observed Fermi surfaces are in good agreement with the
first-principle band-structure calculation results \cite{Rf:Oguc.1,Rf:Sing.1}
and the recent ARPES measurement \cite{Rf:Dama.1}.
Another remarkable difference from cuprates and $\kappa$-(ET)$_2$X
is the electron filling. Since the valence of ruthenium ion is Ru$^{4+}$,
four electrons occupy Ru-site on the average.
Again according to the quantum oscillation measurement, about $4/3$ electrons
are included in the $\gamma$ band and remained part is included in $\alpha$
and $\beta$ bands.
Thus, it is regarded that this system is far from the half-filling.

The most outstanding and interesting difference is the pairing symmetry,
which has recently been clarified to be spin-triplet due to excellent
experiments \cite{Rf:Maen.3}.
The most important experimental evidence suggesting the spin-triplet pairing
has been obtained by NMR \cite{Rf:Ishi.1,Rf:Ishi.2}.
Ishida {\it et al.} have measured the $^{17}$O-NMR and Ru Knight shift
by applying the magnetic field parallel to the $ab$-plane,
and observed $no$ suppression in the spin susceptibility
below $T_{\rm c}$ \cite{Rf:Ishi.1,Rf:Ishi.2}.
This result excludes the possibility of the spin-singlet pairing,
and at the same time, indicates the $d$-vector 
along the $\hat{z}$-axis. The $d$-vector is an usual expression for 
the internal degree of freedom, which is an interesting subject 
in the triplet superconductivity \cite{rf:sureview,rf:leggett}. 
Recent inelastic polarized neutron scattering experiment
has suggested the same results \cite{Rf:Duff.1}.

Now the spin-triplet pairing in ruthenate has been experimentally confirmed.
Slight portion of non-magnetic impurities drastically suppresses
the superconductivity \cite{Rf:Mack.2}, in sharp contrast to 
the impurity effects in conventional $s$-wave superconductors.
The NMR and NQR relaxation rates exhibit no coherence peak just below
$T_{\rm c}$ \cite{Rf:Ishi.3,Rf:Ishi.4}.
The $\mu$SR measurement has shown that an internal magnetic field is
spontaneously turned on below $T_{\rm c}$ \cite{Rf:Luke.1}, indicating that
the time-reversal symmetry is broken in the SC state of ruthenate.
From this result, the chiral state
$\hat{d}=(k_{\rm x} \pm {\rm i} k_{\rm y})\hat{z}$
has been suggested \cite{Rf:Sigr.1}.
The temperature dependence of the critical current in
Pb/Sr$_{2}$RuO$_{4}$/Pb junction is also consistent with
the $p$-wave pairing state \cite{rf:jin}.


Theoretically, Rice and Sigrist pointed out a possibility of
spin-triplet superconductivity, immediately after the discovery of
superconductivity in Sr$_2$RuO$_4$ \cite{Rf:Rice.1}.
Their insights have been based on the following facts.
First some Fermi-liquid parameters are similar to those of $^3$He, which
is confirmed to be a spin-triplet $p$-wave superfluid \cite{rf:leggett}.
Second the three-dimensional analogous compound SrRuO$_3$ exhibits
the ferromagnetism with 
a Curie temperature $T_{\rm C}=160$K. 
They have considered that the Hund's rule coupling among Ru4d$\varepsilon$
orbitals stabilizes the spin-triplet pairing rather than the 
spin-singlet one. 
Although any microscopic justification for above two insights has not yet
been obtained up to now, their excellent prediction itself has obtained
a great success.

At the present stage, the theoretical interests on Sr$_2$RuO$_4$ are focused
on the two fundamental aspects of superconductivity,
namely the pairing symmetry and the pairing mechanism.
Concerning the pairing symmetry, the origin of the power-law behaviors is
a challenging subject.
As is mentioned above, the chiral state without time-reversal symmetry is
expected for the internal degree of freedom.
Then, assuming the simple momentum dependence of the SC gap,
for instance, $\Delta(\k) \propto \sin k_{\rm x}$ \cite{rf:miyakeRu},
the excitation gap opens on the whole Fermi surface.
On the contrary, the gap-less power-law behaviors, suggesting the existence
of the line node, have been observed in common among the several experimental
results on the specific heat \cite{Rf:Nish.1},
NMR $1/T_{1}T$ \cite{Rf:Ishi.4},
magnetic field penetration depth \cite{rf:bonalde},
thermal conductivity \cite{Rf:Tana.1,Rf:Izaw.1},
and ultrasonic attenuation rate \cite{Rf:Lupi.1}.
We should note that only the point node is derived from the symmetry argument,
even if the three-dimensional degree of freedom is taken into account 
\cite{rf:kusunose}. 
Thus, if we assume the chiral state, the line node should appear only
accidentally. Namely, the theoretical proposal on this problem has to rely
on somewhat an accidental reason.
Among them, the three-dimensional $f$-wave symmetry \cite{rf:hasegawa,rf:won}
has been supported by the thermal conductivity measurement
\cite{Rf:Izaw.1,rf:tanatar2}.
The more improved proposal based on the multi-band effect has been proposed
along this line \cite{rf:zhitomirski}.
The essential assumption of this proposal is that the zeros of the order
parameter corresponding to the symmetry $\Delta(k) \propto k_{\rm x}$
is parallel to the plane.
When the zeros have a slope, the point node is expected.
The pairing state assumed here is generally difficult in view of
the pairing mechanism, because the Fermi surface of Sr$_2$RuO$_4$ is
clearly two-dimensional.
Therefore, we consider that this problem should be resolved within
the two-dimensional model.
We will provide a different proposal along this line
on the basis of the microscopic theory (see Sec.~3.4.2).
Then, the node-like structure appears in the $\beta$ band.
This is also an ``accident'', but derived from the microscopic model.
The power-law behaviors can be explained,
although a fitting of the parameters is required.

The advances in the theoretical studies on the pairing mechanism will
be reviewed in Sec.~3.4.1. Since this issue is one of the main subjects
of this review, we will discuss it in detail.
Then, we show the results of the microscopic investigation based on
the perturbation theory in Sec.~3.4.2.
Subsequently, the microscopic mechanism of stabilizing the chiral state
will be investigated in Sec.~3.4.3.
The microscopic study on the internal degree of freedom becomes possible
owing to the relatively simple electronic state.
Finally we mention that such study was very difficult previously because
triplet superconductors were basically observed only in
the heavy-fermion compound.
We close this section by noting that the discovery of Sr$_2$RuO$_4$ has
accelerated the theoretical understanding on the triplet superconductivity.

\par\vfill
\eject

%
%
\section{Microscopic Mechanism of Superconductivity}

In this section, we review the theoretical investigations on the mechanism
of unconventional superconductivity based on the microscopic Hamiltonian
such as the single- and multi-band Hubbard model.
Before discussing the particular superconductors, in order to make
this review article self-contained, in Sec.~3.1 we briefly explain
the theoretical tools which will be used in the following subsections.
Readers who may not be interested in the theoretical formulation
can simply skip this subsection.
In Sec.~3.2, high-$T_{\rm c}$ superconductivity will be discussed in detail.
The basic properties and typical results of microscopic theories will be
reviewed.
The application to the organic superconductor is discussed in Sec.~3.3.
Then, the applicability of the microscopic theory for the molecular materials
are clearly shown. In Sec.~3.4, the microscopic theory is extended to
the triplet superconductivity in Sr$_{2}$RuO$_{4}$, where the qualitatively
different results are derived from the characteristic 
electronic structure. 

\subsection{Dyson-Gor'kov Equation}

In the following subsections, the Dyson-Gor'kov equation and
\'Eliashberg theory \cite{rf:eliashbergsuper} are used
to discuss superconductivity.
This formulation is suitable for the diagrammatic techniques in the quantum
field theory, although some approximations are usually needed for
the actual calculations.
However, this approach is effective to clarify the physical picture and
in principle, it is free from the finite size effect, compared with the
numerical methods.
This subsection is devoted to the introduction of the linearized
\'Eliashberg theory,
which is used to determine the superconducting transition.
For simplicity, we show the explicit expressions only for the single-band
case, but the extension to the multi-band system is straightforward.

In the Dyson-Gor'kov equation, the superconducting state is described
by introducing the normal and anomalous Green functions,
symbolically expressed as $G$ and $F$, respectively \cite{rf:AGD}.
In the homogeneous system, they are defined as
\begin{eqnarray}
 && G({\bm k},i\omega_n)=
 -\int_0^{\beta} d\tau e^{i\omega_n \tau}
 \langle T_{\tau} c_{{\bm k}\sigma}(\tau) c_{{\bm k}\sigma}^{\dag}
 \rangle, \\
 && F({\bm k},i\omega_n)=
 \int_0^{\beta} d\tau e^{i\omega_n \tau}
 \langle T_{\tau} c_{{\bm k}\uparrow}(\tau) c_{{\bm -k}\downarrow}
 \rangle, \\
 && F^{\dag}({\bm k},i\omega_n) =
 \int_0^{\beta} d\tau e^{i\omega_n \tau}
 \langle T_{\tau} c^{\dag}_{{\bm -k}\downarrow}(\tau)c^{\dag}_{{\bm k}\uparrow}
 \rangle,
\end{eqnarray}
where $c_{{\bm k}\sigma}(\tau)$=$e^{H\tau} c_{{\bm k}\sigma} e^{-H\tau}$
with a Hamiltonian $H$, $c_{{\bm k}\sigma}$ is an annihilation operator
for electron with spin $\sigma$ and momentum ${\bm k}$, $\beta$=1/$T$, and
$\omega_n$=$\pi T(2n+1)$ with an integer $n$ is a fermion Matsubara frequency.
The symbol $\langle \cdots \rangle$ means the operation to take statistical
average and $T_{\tau}$ is an ordering operator with respect to $\tau$.
Note that the following formulation is common to the singlet and
triplet pairing cases unless we explicitly mention.

The Green functions are expressed by normal and anomalous
self-energies through the Dyson-Gor'kov equation, diagrammatically
expressed in Fig.~\ref{fig:Dyson}.
Readers can find a clear derivation of the Dyson-Gor'kov equation
in Ref.~\cite{rf:AGD} and more general expression for
the inhomogeneous state in Ref.~\cite{rf:kopnin}.
The normal and anomalous self-energies are exactly obtained from
the Gor'kov equation \cite{rf:Gor'koveq,rf:kopnin}
in the homogeneous case through the Fourier transformation and
the Dyson-Gor'kov equation is written in the matrix form as
\begin{eqnarray}
 && \left(
 \begin{array}{cc}
   G(k) & F(k) \\
   F^{\dag}(k) & -G(-k)
 \end{array}
 \right)=
 \left(
 \begin{array}{cc}
   G^{(0)}(k)^{-1} - \Sigma_{\rm n}(k) & \Delta(k) \\
   \Delta^{*}(k) & -G^{(0)}(-k)^{-1} + \Sigma_{\rm n}(-k)
 \end{array}
 \right)^{-1}.
\end{eqnarray}
Here, $\Sigma_{\rm n}(k)$ and $\Delta(k)$ are the normal and anomalous
self-energies, respectively, $G^{(0)}(k)$ is the non-interacting
Green function, given by
$G^{(0)}(k)$=$[i\omega_{n}-\varepsilon({\bm k})]^{-1}$,
$\varepsilon({\bm k})$ is the one-electron dispersion energy,
and $k$ is a shorthand notation as $k$=$({\bm k},i\omega_n)$.
Note that a chemical potential is included in
$\varepsilon({\bm k})$ in our notation.

\begin{figure}[t]
\begin{center}
\vspace{2.5cm}
\caption{Diagrammatic representation of the Dyson-Gor'kov equation.}
\label{fig:Dyson}
\end{center}
\end{figure}

The normal and anomalous self-energies can be expressed by
the perturbation series, which is obtained in a usual manner 
\cite{rf:AGD}. 
Since an approximation is usually required for an explicit estimation, 
we will introduce three approximations such as
third-order perturbation (TOP),
random phase approximation (RPA),
and fluctuation-exchange (FLEX) approximation,
as will be explained later in detail.
In the other derivation, $\Sigma_{\rm n}(k)$ and $\Delta(k)$ are derived
from the functional derivatives using the Luttinger's functional
$\Phi[G,F^{\dag}]$ \cite{rf:luttinger,rf:eliashbergheat} as
\begin{eqnarray}
 \label{eq:variation}
 && \Sigma_{\rm n}(k) = \frac{1}{2}
 \frac{\delta \Phi}{\delta G(k)}, \hspace{5mm}
 \Delta(k) = -\frac{\delta \Phi}{\delta F^{\dag}(k)}.
\end{eqnarray}
Then, the variational conditions about the free energy $\Omega$ are
satisfied as
\begin{eqnarray}
 \label{eq:variational-condition}
 && \frac{\delta \Omega}{\delta \Sigma_{\rm n}(k)}=
 \frac{\delta \Omega}{\delta \Delta(k)}=0.
\end{eqnarray}
If above conditions are satisfied, the calculation is called
``conserving approximation'' \cite{rf:baym}.
Note that the perturbation scheme does not necessarily satisfy the above
conditions.

\begin{figure}[t]
\begin{center}
\vspace{2.5cm}
\caption{(a) The one-loop approximation for $\Delta(k)$ in the
electron-phonon system. The wavy line represents the phonon propagator.
Diagrammatic representation of (b) the anomalous vertex and
(c) the anomalous self-energy.}
\label{fig:anomalousvertex}
\end{center}
\end{figure}

For the conventional $s$-wave superconductivity, the anomalous self-energy
is obtained by the electron-phonon coupling.
Since the typical phonon frequency is smaller than the Fermi energy in
conventional metals, the Migdal's theorem holds and the vertex corrections
can be ignored.
Thus, the one-loop approximation shown in Fig.~\ref{fig:anomalousvertex}(a)
is valid. In the case of unconventional superconductivity arising from
electron correlations, the irreducible vertex $V_{\rm a}(k,k')$
in the particle-particle channel (Fig.~\ref{fig:anomalousvertex}(b))
is derived from the many-body effects.
In analogy with the electron-phonon mechanism, this vertex is regarded as
the effective interaction for the pairing.
Thus, the anomalous self-energy, represented formally by the diagram
in Fig.~\ref{fig:anomalousvertex}(c), is expressed by
\begin{eqnarray}
  \label{eq:anomalous-self-energy}
  && \Delta(k) = -\sum_{k'} V_{\rm a}(k,k') F(k').
\end{eqnarray}
Here the summation is defined as $\sum_{k} = (T/N)\sum_{{\bm k},n}$,
where $N$ is the number of sites.
It is considered that the unconventional superconductivity arises from
the momentum dependence in the effective interaction $V_{\rm a}(k,k')$.
The theoretical search for the pairing mechanism is then reduced to the
identification of the effective interaction.

The expression for the anomalous self-energy
Eq.~(\ref{eq:anomalous-self-energy}) is a self-consistent equation,
which is one of the mean-field equations.
We can reproduce the result of the weak-coupling BCS theory \cite{rf:BCS},
if we ignore the normal self-energy as well as
the frequency dependence of $V_{\rm a}(k,k')$.
The self-consistent equation is then transformed to the gap equation as
\begin{eqnarray}
 \label{eq:BCS-gap-equation}
 && \Delta({\bm k})=-\sum_{\bm k'} V_{\rm a}({\bm k},{\bm k'})
 \frac{\tanh (E({\bm k'})/2T)}{2 E({\bm k'})}  \Delta({\bm k'}), \\
 \label{eq:BCS-excitation-energy}
 && E({\bm k})=\sqrt{\varepsilon({\bm k})^2+\Delta({\bm k})^2}.
\end{eqnarray}
Here we further assume the BCS approximation for the interaction as
\begin{eqnarray}
 && V_{\rm a}({\bm k},{\bm k'}) = 
 \left\{
 \begin{array}{ll}
   -V & |\varepsilon({\bm k})|,|\varepsilon({\bm k'})|<\omega_{\rm c}, \\
    0 & {\rm otherwise},
 \end{array}
\right.
\end{eqnarray}
where $\omega_{\rm c}$ is a cut-off energy.
Then, we obtain the well-known formula for $T_{\rm c}$ as
\begin{eqnarray}
  && T_{\rm c} = 1.13 \omega_{\rm c} \exp(-1/\rho V),
\end{eqnarray}
where $\rho$ is the DOS at the Fermi level.

Also in the \'Eliashberg theory, the superconducting order is determined by
the non-trivial solution of the self-consistent equation.
In order to determine the critical temperature and corresponding pairing
symmetry, the Dyson-Gor'kov equation is linearized with respect to
$\Delta(k)$ as
\begin{eqnarray}
  && G(k)^{-1}=G^{(0)}(k)^{-1}-\Sigma_{\rm n}(k),\\
  && F(k) = |G(k)|^{2} \Delta(k).
\end{eqnarray}
where $G(k)$ is the dressed Green function, explicitly written as
$G(k)$=$[i\omega_n-\varepsilon({\bm k})-\delta\mu-\Sigma_{\rm n}(k)]^{-1}$.
Here the chemical potential shift $\delta\mu$ is determined
by the conservation for the particle number as
\begin{eqnarray}
  && \sum_{k}(G(k)-G^{(0)}(k))e^{{\rm i} \omega_{n} \eta}=0,
\end{eqnarray}
where $\eta$ is a positive infinitesimal.
The self-consistent equation for $\Delta(k)$ is expressed as
\begin{eqnarray}
 \label{eq:eliashberg-super2}
  && \Delta(k)=-\sum_{k'} V_{\rm a}(k,k') |G(k')|^{2} \Delta(k').
\end{eqnarray}
This linearized equation is called the \'Eliashberg equation,
which is valid just at $T$=$T_{\rm c}$.
The transition temperature is practically estimated by solving
the eigenvalue equation
\begin{eqnarray}
  \label{eq:eliashberg-super}
  && \lambda_{{\rm e}} \Delta(k) =
  -\sum_{k'} V_{{\rm a}}(k,k') |G(k')|^{2} \Delta(k').
\end{eqnarray}
The maximum eigenvalue becomes unity, $\lambda_{\rm e}$=1, at $T=T_{\rm c}$
and the anomalous self-energy $\Delta(k)$ plays a role of the eigenfunction.
The pairing symmetry is determined by the momentum dependence of $\Delta(k)$.
Since the pairing state with maximum $T_{\rm c}$ is usually realized
in the ground state, we can get knowledge on the behaviors
below $T_{\rm c}$ from the \'Eliashberg equation.
The momentum dependence of the quasi-particle energy gap is approximately
described by the absolute value $|\Delta(k)|$ at $\omega_{n} = \pi T_{\rm c}$.
Note that strictly the excitation gap $\Delta_{\rm ex}(\k)$ is renormalized as
$\Delta_{\rm ex}(\k)$=$z(\k)|\Delta(\k,\Delta_{\rm ex}(\k))|$, where
$z({\bm k})$=$(1-\partial{\rm Re}\Sigma_{\rm n}({\bm k},\omega)/
\partial \omega|_{\omega=0})^{-1}$ is the renormalization factor.

The \'Eliashberg theory includes the normal self-energy, which generally
induces de-pairing effects.
In particular, the quasi-particle damping gives rise to the pair-breaking
and suppresses $T_{\rm c}$. Note that the de-pairing effects are not expected
to alter the pairing symmetry.
The frequency dependence of the effective interaction represents
the retardation effect, which also reduces $T_{\rm c}$.
The \'Eliashberg theory is usually called ``strong-coupling theory''
in contrast to the weak-coupling theory in
Eqs.~(\ref{eq:BCS-gap-equation}) and (\ref{eq:BCS-excitation-energy}).
We should note that the ``strong-coupling superconductivity'' discussed
in Sec.~4 has a different meaning.

\begin{figure}[t]
\begin{center}
\vspace{2.5cm}
\caption{Diagrammatic representation of the normal self-energy within the TOP.}
\label{fig:3rdnormal}
\end{center}
\end{figure}

\begin{figure}[t]
\begin{center}
\vspace{2.5cm}
\caption{Diagrammatic representation of the anomalous self-energy
within the TOP.}
\label{fig:3rddiagrams}
\end{center}
\end{figure}

In the specific calculations, the normal and anomalous self-energies are
evaluated by using an approximation.
For the convenience in the following sections, we summarize the expressions
for $\Sigma_{\rm n}(k)$ and $V_{\rm a}(k,k')$ in several approximations.

First we show the perturbation series within the third order
in terms of $U$ \cite{rf:hottasecond,rf:hottathird},
which will be used in Secs.~3.2.3, 3.3.2, and 3.4.2.
The normal self-energy is separated into three terms as
\begin{eqnarray}
  \label{eq:thirdnormal}
  && \Sigma_{\rm n}(k)=
     \Sigma^{(2)}_{\rm n}(k)+
     \Sigma^{(3{\rm RPA})}_{\rm n}(k)+
     \Sigma^{(3{\rm VC})}_{\rm n}(k).
\end{eqnarray}
These terms are represented in Figs.~\ref{fig:3rdnormal}(a-c), written as
\begin{eqnarray}
\label{eq:thirdnormal1}
  && \Sigma^{(2)}_{\rm n}(k)=
  U^{2} \sum_{q} \chi_{0}(q) G^{(0)}(k-q),\\
  \label{eq:thirdnormal2}
  && \Sigma^{(3{\rm RPA})}_{\rm n}(k)=
  U^{3} \sum_{q} \chi_{0}(q)^{2} G^{(0)}(k-q),\\
  \label{eq:thirdnormal3}
  && \Sigma^{(3{\rm VC})}_{\rm n}(k)=
  U^{3} \sum_{q} \phi_{0}(q)^{2} G^{(0)}(q-k),
\end{eqnarray}
where $\chi_{0}(q)$ and $\phi_{0}(q)$ are, respectively, given by
\begin{eqnarray}
  \label{eq:irreducible-susceptibility}
  && \chi_{0}(q) = -\sum_{k} G^{(0)}(k+q) G^{(0)}(k),
\end{eqnarray}
and
\begin{eqnarray}
  \label{eq:irreducible-pair-susceptibility}
  && \phi_{0}(q) = \sum_{k} G^{(0)}(q-k) G^{(0)}(k).
\end{eqnarray}
Here $q$ denotes a shorthand notation as $q$=$({\bm q},i\Omega_n)$,
where $\Omega_n$=$2\pi T n$ is a boson Matsubara frequency.

Within the third-order perturbation, the effective interaction
in the singlet channel is expressed as
\begin{eqnarray}
  \label{eq:third-anomalous-singlet}
  && V^{{\rm s}}_{{\rm a}}(k,k') = U + V_{{\rm a}}^{(2)}(k,k') 
  +V_{{\rm a}}^{(3{\rm RPA})}(k,k') + V_{{\rm a}}^{(3{\rm VC})}(k,k'),
\end{eqnarray}
where the first, second, third, and fourth terms are represented
in the diagrams in Fig.~\ref{fig:3rddiagrams}(a), (b), (c)+(d),
and (e)+(f)+(g)+(h), respectively.
Explicitly, those terms are written as
\begin{eqnarray}
  \label{eq:third-anomalous-singlet1}
  V_{{\rm a}}^{(2)}(k,k') &=& U^{2} \chi_{0}(k-k'),\\
  \label{eq:third-anomalous-singlet2}
  V_{{\rm a}}^{(3{\rm RPA})}(k,k') &=& 2 U^{3} \chi_{0}(k-k')^{2},\\
  \label{eq:third-anomalous-singlet-vc}
  V_{{\rm a}}^{(3{\rm VC})}(k,k') &=&
  2 U^{3} {\rm Re} \sum_{q} G^{(0)}(k+q) G^{(0)}(k'+q)
  [\chi_{0}(q)+\phi_{0}(q)].
\end{eqnarray}
Note that the first three terms in $V^{{\rm s}}_{{\rm a}}(k,k')$ are
included in the RPA, while the last term is called ``vertex correction''.

The effective interactions for the triplet channel are given
by the same diagrams as Fig.~\ref{fig:3rddiagrams}.
Note here that these diagrams are obtained in case of
$\hat{d} \parallel \hat{z}$, where $\hat{d}$ is the $d$-vector
\cite{rf:leggett}, but the SU(2) symmetry ensures the same results
for the other $d$-vector.
The explicit expressions for the triplet channel are given by
\begin{eqnarray}
  \label{eq:third-anomalous-triplet}
  && V^{{\rm t}}_{{\rm a}}(k,k')=V_{{\rm a}}^{(2)}(k,k') 
  +V_{{\rm a}}^{(3{\rm VC})}(k,k'),
\end{eqnarray}
where
\begin{eqnarray}
  \label{eq:third-anomalous-triplet-rpa}
  && V_{{\rm a}}^{(2)}(k,k')  =  -U^{2} \chi_{0}(k-k'),\\
  \label{eq:third-anomalous-triplet-vc}
  && V_{{\rm a}}^{(3{\rm VC})}(k,k')  = 
  2 U^{3} {\rm Re} \sum_{q} G^{(0)}(k+q) G^{(0)}(k'+q)
  [\chi_{0}(q)-\phi_{0}(q)].
\end{eqnarray} 
Note that for the triplet channel, the third-order RPA terms cancel each other
and only the vertex correction terms remain.

\begin{figure}[t]
\begin{center}
\vspace{2.5cm}
\caption{Diagrammatic representation of the normal self-energy
within the RPA or FLEX approximation.
The bare (dressed) Green function is used in the RPA (FLEX).}
\label{fig:rpanormal}
\end{center}
\end{figure}

Next let us show the expressions for the RPA, which corresponds to
the partial summation, as shown in Figs.~\ref{fig:rpanormal} and
\ref{fig:rpavertex} \cite{rf:miyake,rf:scalapino}.
The normal self-energy is given as
\begin{eqnarray}
  \label{eq:rpa-normal}
  && \Sigma_{\rm n} (k) = \sum_{q} V_{\rm n} (q) G^{(0)}(k-q),
\end{eqnarray} 
where $V_{\rm n} (q)$ is given in the RPA as
\begin{eqnarray}
  \label{eq:rpa-normal-vertex}
  && V_{\rm n} (q)=U^{2} [\frac{3}{2} \chi_{{\rm s}} (q)
  +\frac{1}{2} \chi_{{\rm c}} (q)-\chi_{0} (q)].
\end{eqnarray}
Here $\chi_{{\rm s}}(q)$ and $\chi_{{\rm c}}(q)$ are the spin and charge
susceptibilities in the RPA, respectively, given by
\begin{eqnarray}
  \label{eq:rpa-susceptibility}
  && \chi_{{\rm s}}(q) = \frac{\chi_{0}(q)}{1 - U \chi_{0}(q)},
  \hspace{5mm}
  \chi_{{\rm c}}(q) = \frac{\chi_{0}(q)}{1 + U \chi_{0}(q)}.
\end{eqnarray}
The effective interaction for the singlet and triplet channel is given by
\begin{eqnarray}
  \label{eq:anomalous-rpa-singlet}
  && V^{{\rm s}}_{\rm a}(k,k')=U + \frac{3}{2} U^{2} \chi_{{\rm s}}(k-k') 
  -\frac{1}{2} U^{2} \chi_{{\rm c}}(k-k'),
\end{eqnarray}
and
\begin{eqnarray}
  \label{eq:anomalous-rpa-triplet}
  && V^{{\rm t}}_{\rm a}(k,k') = - \frac{1}{2} U^{2} \chi_{{\rm s}}(k-k') 
  -\frac{1}{2} U^{2} \chi_{{\rm c}}(k-k'),
\end{eqnarray}
respectively.
Because $\chi_{0}(q) > 0$ at $\Omega_{n}=0$, the spin part generally
gives larger contribution than the charge part.
In particular, the contribution from the charge susceptibility is suppressed
in the vicinity of the magnetic instability ($\chi_{0}(q) \sim 1$).
The closeness to the magnetic instability is an implicit assumption
for the RPA. Then, the RPA in the Hubbard model is a description
for the spin fluctuation theory (Sec~3.2.1).

\begin{figure}[t]
\begin{center}
\vspace{2.5cm}
\caption{The effective interaction within the RPA or FLEX approximation.}
\label{fig:rpavertex}
\end{center}
\end{figure}

The FLEX approximation \cite{rf:FLEX}, which is one of conserving
approximations, has been used very widely.
In this paper, Sec.~3.2.4 will be devoted to the review of the FLEX
approximation, but here we provide a short comment on the formulation.
The FLEX approximation can be also considered as one of the modifications of
the RPA, in the sense that the dressed Green function $G$ is used in
Eqs.~(\ref{eq:irreducible-susceptibility}),
(\ref{eq:irreducible-pair-susceptibility}),
(\ref{eq:rpa-normal})-(\ref{eq:anomalous-rpa-triplet}),
instead of the bare Green function $G^{(0)}$.
Then, the Green function, normal self-energy, spin and charge susceptibility
are determined self-consistently.

The numerical calculation is used to take a summation in these diagrammatic
techniques. The figures in this review show the results with 128$\times$128
points in the first Brillouin zone and 2048 Matsubara frequency.
In same cases, the calculations have been performed for smaller number
of meshes, when we have confirmed that there is no problem in accuracy
of the calculations.
Note that $T_{\rm c}$ should be zero in the strict two-dimensional system
due to the Mermin-Wagner theorem, although the estimation of $T_{\rm c}$
has been frequently performed in two dimensions.
This point will be briefly discussed later in this article.

\subsection{High-$T_{\rm c}$ cuprates}

\subsubsection{Overview}

A number of the pairing mechanisms have been proposed for
high-$T_{\rm c}$ superconductivity.
After intensive investigations over more than a decade, the magnetic
mechanism based on the spin fluctuation theory has been accepted most
predominantly.
This mechanism has been first discussed for the superfluidity in $^3$He,
where the ferromagnetic paramagnon mediates the $p$-wave
pairing interaction \cite{rf:leggett,rf:anderson1973}.
Next it has been pointed out that the $d$-wave superconductivity is
most favorable when the spin fluctuation is anti-ferromagnetic
\cite{rf:miyake,rf:scalapino}.

After the proposals on this mechanism for cuprates
\cite{rf:bickers1987,rf:millis} and the detailed investigations on the Hubbard
model \cite{rf:FLEX,rf:shimahara,rf:bickers1991}, the significant developments
have been given by the phenomenological theory
\cite{rf:moriya1990,rf:moriya1992,rf:monthoux1991,rf:monthoux1992,rf:monthouxST1992,rf:monthouxST1993,rf:moriyaST1994,rf:moriyaST1996}.
Subsequently, the microscopic theory using the FLEX approximation
\cite{rf:FLEX,rf:bickers1991,rf:monthouxFLEX,rf:paoFLEX,rf:dahmFLEX,rf:langerFLEX,rf:luo,rf:koikegamiFLEX,rf:takimotoFLEX,rf:takimotoFLEX2}
has clarified the detailed and important properties.
The calculated results have succeeded in the quantitative agreement on the
transition temperature. Moreover, the various properties both in the normal
and superconducting state have been explained
\cite{rf:moriyaAD,rf:chubukovreview},
except for the pseudogap phenomena.

Note that the $d_{\rm x^2-y^2}$-wave superconductivity is obtained also
in the $t$-$J$ Hamiltonian near the half-filling as a result of
the variational Monte Carlo simulation \cite{rf:yokoyama,rf:ogata},
exact diagonalization \cite{rf:dagottoexact}, and
Green function Monte Carlo simulation \cite{rf:sorella}.
Then, the pairing mechanism should be classified into the magnetic one,
since the $t$-$J$ Hamiltonian directly includes the anti-ferromagnetic
interaction.

At present, the magnetic mechanism should be regarded as one of
the limiting cases of the electronic mechanism, in which the momentum
dependence of the residual interaction among the quasi-particles
inevitably gives the superconducting ground state.
This idea has been first given by Kohn and Luttinger in 1965 \cite{rf:kohn}.
They discussed the possibility of the non-$s$-wave pairing state in the
three-dimensional fermion gas model within the second-order perturbation.
This general concept is particularly important for the comprehensive
understanding from the under-doped to the over-doped region.
This is because the spin fluctuation theory loses its justification
in the over-doped region, where the spin fluctuation is not clearly 
observed but $T_{\rm c}$ remains substantially. 

The results of the perturbation theory
\cite{rf:hottasecond,rf:hottathird} (see Sec.~3.2.3) 
have been quite instructive, in the sense that 
the roles of the RPA and non-RPA terms are clarified.
The spin fluctuation theory corresponds to the partial summation of the
perturbation series.
In other word, the RPA terms (including some renormalization) are included
in the spin fluctuation theory.
The perturbation theory has revealed that the RPA terms favor the
$d_{\rm x^2-y^2}$-wave superconductivity in the Hubbard model near
the half-filling, while it is suppressed by the non-RPA terms.
Therefore, it is expected that the RPA terms play a major role
in the pairing interaction, even far from the magnetic instability.
Again, the perturbation theory has revealed that the corrections
from the non-RPA terms are not important for the case of cuprates.
This is a microscopic justification why the concept of
the magnetic mechanism survives in the over-doped region.

Roughly speaking, the perturbation theory is appropriate to the
over-doped region, while the spin-fluctuation theory is appropriate
to the optimally-doped region.
Also in the under-doped region, the main pairing mechanism should be
in common with that in the optimally-doped region,
although the SC fluctuation remarkably reduces the transition temperature.
In our understanding, the SC fluctuation is the origin of the pseudogap
phenomena, which has been a challenging issue in the study of
high-$T_{\rm c}$ cuprates.
Therefore, we will describe in detail the theory including SC fluctuations
in Sec.~4 in order to understand the under-doped region.
In this section, we focus on the pairing mechanism in the optimally- and
over-doped region.

In Sec.~3.2.2, we provide a simple explanation on the
$d_{\rm x^2-y^2}$-wave superconductivity induced by the AF spin fluctuation.
The important scattering process for the pairing is explained.
In Sec.~3.2.3, we review the results on the perturbation theory which is
performed within the third order. The obtained results are qualitatively
similar to those of the spin fluctuation theory.
The corrections to the spin fluctuation theory are discussed.
In Sec.~3.2.4, we show the results of the FLEX approximation,
which is a microscopic description for the spin fluctuation theory.
The reasonable $T_{\rm c} \sim 100 {\rm K}$ is obtained
near the magnetic instability.
We discuss the effects of higher-order corrections in Sec.~3.2.5.
Some justifications for the spin fluctuation theory will be shown.

\subsubsection{Spin fluctuation-induced superconductivity}

First let us explain the phenomenological theory on
the spin fluctuation-induced superconductivity
\cite{rf:moriya1990,rf:moriya1992,rf:monthoux1991,rf:monthoux1992}.
The effective interaction $V_{{\rm a}}(k,k')$ is phenomenologically given
in this theory. Here we use the weak-coupling theory for simplicity.
This treatment is quantitatively insufficient, but it is enough to
grasp the basic idea of spin fluctuation-induced superconductivity.

A simple form of the effective Hamiltonian is described as
\begin{eqnarray}
  \label{eq:spin-fermion-model}
  && H_{\rm eff}=\sum_{{\bm k}\sigma} \varepsilon({\bm k}) 
  c_{{\bm k}\sigma}^{\dag}c_{{\bm k}\sigma}
  -g^{2} \sum_{\bm k,k',q}
  \chi_{\rm s}({\bm q}) {\bm \sigma}_{\alpha\beta}\cdot
  {\bm \sigma}_{\gamma\delta}
  c_{\bm{k+q}\alpha}^{\dag}c_{\bm{k'-q}\gamma}^{\dag} 
  c_{\bm{k'}\delta}c_{\bm{k}\beta},
\end{eqnarray}
where $\bm{\sigma}$=$(\sigma_x,\sigma_y,\sigma_z)$ are the Pauli matrices
and $g$ is the effective coupling constant for the interaction exchanging
the spin fluctuation. We denote the static spin susceptibility as
$\chi_{\rm s}(\q)=\chi_{\rm s}(\q,0)$.
The spin susceptibility $\chi_{\rm s}(\q,\Omega)$ near the magnetic
instability is phenomenologically expressed as \cite{rf:MMP,rf:barzykin}
\begin{eqnarray}
  \label{eq:phenomenological-susceptibility}
  && \chi_{{\rm s}}(\q,\Omega)=
  \frac{\alpha \xi^{2}}{1+\xi^{2} (\q-\Q)^{2}-{\rm i} \Gamma q^{2-z} \Omega},
\end{eqnarray}
where $\xi$ is the correlation length of the spin fluctuation,
$\bm{Q}$ is the wave vector defined as $\bm{Q}=(\pi,\pi)$ ($\bm{Q}=(0,0)$)
and $z$ is the dynamical exponent $z$=2 ($z$=3) for the anti-ferromagnetic
(ferromagnetic) case.
A large value of $\xi$ is a hypothesis of the spin fluctuation theory,
which is expected around the magnetic instability.
The strong enhancement of the static susceptibility around
$\bm{q} \sim \bm{Q}$ essentially induces the unconventional
superconductivity. 
The dissipation term $\Gamma$ describes the time scale of the spin fluctuation.
The diffusive dynamics of the spin fluctuation is a characteristic property
in the normal state. This is not the case in the SC state or in the AF state.
The $\Omega$-dependence induces the retardation effect, but it is not included
in the weak-coupling theory.

\begin{figure}[t]
\begin{center}
\vspace{2.5cm}
\caption{Effective interaction which corresponds to the single paramagnon
exchange. The wavy line represents the propagator of the paramagnon.}
\label{fig:phenomenovertex}
\end{center}
\end{figure}

In the weak-coupling theory, the gap equation is written as
\begin{eqnarray}
  && \Delta(\k)=-\sum_{\k'} V^{\rm s,t}_{\rm a}(\k-\k')
  \frac{\tanh (E(\k')/2T)}{2 E(\k')} \Delta(\k'),
\end{eqnarray}
where the pairing interaction $V^{\rm s,t}$ is given as
\begin{eqnarray}
  \label{eq:spin-fermion-singlet}
  && V^{\rm s}_{\rm a}(\k-\k')=\frac{3}{2} g^{2} \chi_{\rm s}(\k-\k'),
\end{eqnarray}
for the singlet pairing and
\begin{eqnarray}
 \label{eq:spin-fermion-triplet}
 && V^{\rm t}_{\rm a}(\k-\k')=-\frac{1}{2} g^{2} \chi_{\rm s}(\k-\k'),
\end{eqnarray} 
for the triplet pairing.
This is a result in the level of the one-loop approximation
for the effective model Eq.~(\ref{eq:spin-fermion-model}).
Here note that the effective interaction $V^{\rm s,t}_{{\rm a}}(\k-\k')$
corresponds to the single paramagnon exchange, which is diagrammatically
represented in Fig.~\ref{fig:phenomenovertex}.
The irreducible vertex estimated from the quantum Monte Carlo simulation
is in good agreement with the single paramagnon exchange at least
in the high-temperature region \cite{rf:bulut1993,rf:bulut1995}.
This form of the effective interaction can be derived from the RPA or FLEX
approximation for the Hubbard model, when $g$ is considered as
the on-site Coulomb interaction $U$.
In this sense, the RPA and FLEX approximations are regarded as
a microscopic description of the spin fluctuation theory.
An advantage of the phenomenological theory is its universality,
which does not depend on the microscopic details.
It should be considered that the renormalization from the high-energy
excitation and the higher-order corrections are effectively included
in the phenomenological parameters.

\begin{figure}[t]
\begin{center}
\vspace{2.5cm}
\caption{Typical Fermi surface of high-$T_{\rm c}$ cuprates.
The order parameter of the $d_{\rm x^2-y^2}$-wave superconductivity has
positive (negative) sign in the shaded (light) region. The scattering process
exchanging the AF spin fluctuation is shown by the dashed line.}
\label{fig:FS}
\end{center}
\end{figure}

By integrating the momentum perpendicular to the Fermi surface, introducing
the cut-off energy $\omega_{\rm c}$ and transposing the velocity
$v(\k)=|\partial \e(\k)/\partial \k|$ as the average on the Fermi surface,
the transition temperature is obtained as
\begin{eqnarray}
  && T_{\rm c}=1.13 \omega_{{\rm c}} \exp(-1/\rho |V_{{\rm sc}}|),
\end{eqnarray}
where the effective coupling constant $V_{\rm sc}$ is given by
\begin{eqnarray}
  && V_{\rm sc}=\int_{{\rm F}} \int_{{\rm F}} {\rm d}k {\rm d}k'
  \Delta(\k) V^{\rm s,t}_{{\rm a}}(\k-\k')  \Delta(\k').
\end{eqnarray} 
Here the gap function is normalized as
$\int_{{\rm F}} {\rm d}k |\Delta(\k)|^{2}=1$.
The integration $\int_{{\rm F}}$ is performed along the Fermi surface.

In actual systems, it is considered that the pairing symmetry with
maximum $T_{\rm c}$ is stabilized.
For example, the $d_{\rm x^2-y^2}$-wave superconductivity is stabilized
when $V_{{\rm sc}}$ is attractive ($V_{{\rm sc}} < 0$) and smallest
for the gap function
$\Delta(\k) = \Delta_{{\rm d}}(\k) \propto \cos k_{\rm x}-\cos k_{\rm y}$.
This is the case of high-$T_{\rm c}$ cuprates where anti-ferromagnetic
spin fluctuations are active.
We can easily understand this result from the typical Fermi surface,
as shown in Fig.~\ref{fig:FS}.
The scattering process from $\k_{1} \sim (\pi,0)$ to  $\k_{2} \sim (0,-\pi)$
is strongly enhanced and attractive for the $d_{x^{2}-y^{2}}$-wave symmetry
($\Delta_{\rm d}(\k_1) V_{\rm a}(\k_1-\k_2) \Delta_{\rm d}(\k_2)<0$).
It is essential that the order parameter changes its sign from
$\k_{1}$ to $\k_{2}$ and the effective interaction
$V_{{\rm a}}(\k_{1}-\k_{2})$ is enhanced around $\k_{1}-\k_{2} \sim \Q$.

In fact, it has been shown that the $d_{x^{2}-y^{2}}$-wave symmetry is
most favorable in the weak-coupling theory
\cite{rf:moriya1990,rf:moriya1992,rf:monthoux1991,rf:monthoux1992}.
After that, the quantitative estimation for the transition temperature was
performed on the basis of the strong-coupling theory
where the phenomenological parameters were determined from the results
of the NMR and resistivity.
Then, $T_{\rm c}$ was estimated to be a reasonable value as
$T_{{\rm c}} \sim 100 {\rm K}$
\cite{rf:monthouxST1992,rf:monthouxST1993,rf:moriyaST1994,rf:moriyaST1996}.
The quantitative agreement between theoretical and experimental $T_{\rm c}$'s
has strongly supported the validity of the spin fluctuation mechanism.

Before closing this subsection, let us comment on the ferromagnetic
spin fluctuation, where $\chi_{\rm s}(\q)$ is enhanced around $\q=(0,0)$.
Here we consider the three-dimensional fermion gas model as a typical
example. In this case, the triplet $p$-wave superconductivity is favored,
since the effective interaction in the triplet channel
Eq.~(\ref{eq:spin-fermion-triplet}) is strongly attractive
for the scattering process from $\k$ to $\k' \sim \k$.
Note that the detailed band structure is not important in this case.
It is widely believed that the ferromagnetic spin fluctuation stabilizes
the Anderson-Brinkman-Morel state in the superfluid $^3$He
\cite{rf:leggett,rf:anderson1973}.
The estimation by using the phenomenological theory \cite{Rf:Mont.1}
and the FLEX approximation \cite{rf:aritapd} shows that $T_{\rm c}$
in the ferromagnetic case is generally low compared with
the anti-ferromagnetic case.

The first proposal for the pairing mechanism in ${\rm Sr_2RuO_4}$ was
superconductivity mediated by ferromagnetic spin fluctuations
\cite{Rf:Mazi.1,Rf:Mazi.2} (see Sec.~3.4.1).
However, this naive expectation was denied by the experimental results
which did not observe the ferromagnetic spin fluctuation \cite{Rf:Sidi.1}.
This is not a mystery from the point of view on the electronic mechanism.
In such a case, we should restart the discussion from more general framework,
as will be shown in Sec.~3.4.

\subsubsection{Perturbation theory}

In the following subsections we explain the microscopic theories.
First let us review the perturbation theory, which is a method for
the systematic estimation of the effective interaction, certainly justified
in the weak-coupling region.
The expansion parameter here is $U/W$, where $U$ is the on-site Coulomb
repulsion and $W$ is the band width.
The principle of the adiabatic continuity in the Fermi-liquid theory
requires the regularity of the expansion.
For instance, a good convergence has been confirmed in the Kondo problem
based on the Anderson Hamiltonian, while the expansion with respect to
the Kondo exchange coupling $J$ is singular in the $s$-$d$ model
\cite{rf:Hewson1}.
Also in the periodic system, the perturbation method is expected to be
useful to understand the qualitative natures in the weak coupling region
$U$$<$$W$, as long as any long-range order does not occur.
Note that the low-dimensional system $d<2$ is beyond our scope
in this review ($d$ is the dimensionality).
We can see the applicability of the perturbation theory in the case of
$d=2$ in Refs.~\cite{rf:shanker} and \cite{rf:feldmann}.

As is mentioned before, the spin fluctuation theory can be 
described by the perturbation scheme.
The phenomenological form of the effective interaction in
Eqs.~(\ref{eq:spin-fermion-singlet}) and (\ref{eq:spin-fermion-triplet})
is derived from the partial summation of the RPA terms.
We should note that the partial summation is sometimes dangerous
as a microscopic estimation, since it is generally expected that
RPA terms are considerably canceled by the neglected terms,
e.g., vertex corrections.
In contrast to the spin fluctuation theory, the perturbation theory
has a well-defined basis.
Namely, all kinds of the terms are estimated on an equal footing.
The contribution from the non-RPA terms first appears in the third order.
Therefore, it is expected that the third-order perturbation (TOP) theory
clarifies the general tendency of the vertex corrections.

The application of the perturbation theory to the estimation of $T_{\rm c}$
in high-$T_{\rm c}$ superconductors has been first performed by Hotta.
He has performed both the second-order \cite{rf:hottasecond} and
third-order calculation \cite{rf:hottathird} for the $d$-$p$ model.
In the following, we show the calculated results for the Hubbard model,
because this is simpler than the $d$-$p$ model and it provides
the qualitatively same results. The Hubbard Hamiltonian is expressed as
\begin{eqnarray}
  \label{eq:Hubbard-model}
  && H=\sum_{{\bm k}\sigma} \varepsilon({\bm k}) 
  c_{{\bm k}\sigma}^{\dag}c_{{\bm k}\sigma}
  + U \sum_{\bm i} n_{{\bm i}\uparrow} n_{{\bm i}\downarrow},
\end{eqnarray}
where $n_{{\bm i}\sigma}$=$c_{{\bm i}\sigma}^{\dag}c_{{\bm i}\sigma}$
at site $\bm{i}$.
The two-dimensional dispersion relation $\varepsilon(\bm{k})$ is given by
the tight-binding model for the square lattice as
\begin{eqnarray}
  \label{eq:high-tc-dispersion}
  && \varepsilon(\bm{k})=-2t(\cos k_{x}+\cos k_{y})
  +4t'\cos k_{x} \cos k_{y}-\mu,
\end{eqnarray}
where $t$ and $t'$ represent the nearest-neighbor and next-nearest-neighbor
hopping amplitudes, respectively.
In Secs.~3.2 and 4, we take the energy unit as $2t$=1, indicating that
the band width $W$ is given by $W$=4.
If we choose $W$=4eV according to the band-structure calculation
\cite{rf:bandcalc}, the experimentally observed $T_{\rm c}$=100K
corresponds to $T_{\rm c}$=0.01 in our unit.
The next nearest-neighbor hopping, which corresponds to the $p$-$p$ hopping
in the $d$-$p$ model, is needed to reproduce the typical Fermi surface of
cuprates.
The reasonable value for $t'$ is considered to be $t'/t = 0.1 \sim 0.4$,
leading to the Fermi surface consistent with the experimental observation
\cite{rf:shenrep,rf:takahashi1989,rf:olson1989,rf:campuzano1990,rf:liu1992,rf:king,rf:dessau1993,rf:ino} and the band-structure calculation~\cite{rf:bandcalc}.

Following the \'Eliashberg theory (Sec.~3.1),
the superconducting transition
temperature is determined by solving the \'Eliashberg equation
Eq.~(\ref{eq:eliashberg-super}).
Within the TOP, the normal self-energy is estimated as
Eqs.~(\ref{eq:thirdnormal})-(\ref{eq:thirdnormal3}),
and the effective interaction for the singlet channel is estimated as
Eqs.~(\ref{eq:third-anomalous-singlet})-(\ref{eq:third-anomalous-singlet-vc}).
It should be again noted that the second-order term 
$V_{{\rm a}}^{(2)}(k,k')$ and a part of the third-order terms 
$V_{{\rm a}}^{(3{\rm RPA})}(k,k')$ contribute to the RPA terms.
The non-RPA terms $V_{{\rm a}}^{(3{\rm VC})}(k,k')$ exist in the third order,
leading to the lowest-order corrections to the spin fluctuation theory.

In the TOP, the most favorable pairing symmetry is the $d_{x^2-y^2}$-wave
around the half-filling.
Thus, we show only the results for the $d_{x^2-y^2}$-wave symmetry
in the following.
In Fig.~\ref{fig:various}, $T_{\rm c}$ is depicted as a function of $U$.
For comparison, we also show the results for the second-order perturbation
(SOP), third-order perturbation without vertex correction (TOPWOVC),
and RPA. The thin curves with labels ``P'' in Fig.~\ref{fig:various}
denote the results in which the normal self-energy is simply ignored.

\begin{figure}[t]
\begin{center}
\vspace{2.5cm}
\caption{$T_{\rm c}$ in the various approximations for $t'/t=0.15$ and
$\delta$=0.1.}
\label{fig:various}
\end{center}
\end{figure}

\begin{figure}[t]
\begin{center}
\vspace{2.5cm}
\caption{Irreducible susceptibility $\chi_{0}(\k)$ at the $10\%$, $15\%$,
and $19\%$ doping, respectively, for $t'/t=0.15$ and $T=0.01$.}
\label{fig:kai}
\end{center}
\end{figure}

We can see that the sufficiently high $T_{\rm c}$=0.01$\sim$100K is
obtained in the moderate coupling region.
Here the attractive interaction in the $d_{x^2-y^2}$-wave channel
is mainly from the RPA terms.
In Fig.~\ref{fig:kai}, we show the irreducible spin susceptibility,
which exhibits the peak around $(\pi,\pi)$.
Then, the momentum dependence of the RPA terms is moderate,
but qualitatively similar to the phenomenological theory
(see Eqs.~(\ref{eq:phenomenological-susceptibility}) and
(\ref{eq:spin-fermion-singlet})).
Therefore, the same scattering process as shown in Fig.~\ref{fig:FS} mainly
contributes to the $d_{x^{2}-y^{2}}$-wave superconductivity.
It should be stressed that so strong enhancement of the spin fluctuation
is not necessary for the appearance of superconductivity.
The perturbative renormalization group theory has consistently expected
the $d_{x^{2}-y^{2}}$-wave superconductivity arising from 
the same scattering process 
\cite{rf:shultz,rf:dzyaloshinskii,rf:lederer,rf:zanchi,rf:gonzalez,rf:furukawa,rf:halboth,rf:honerkamp}, where the scattering amplitude from
($\k_{1}$, $-\k_{1}$) to ($\k_{2}$, $-\k_{2}$) is enhanced
in the low energy effective action.

The comparison between the TOP(P) and the TOPWOVC(P) shows that
the vertex correction terms suppress the value of $T_{\rm c}$.
Thus, it is generally expected that the RPA terms are considerably
canceled by the vertex correction terms.
However, it is understood from the comparison between the TOP(P) and SOP(P)
that the RPA terms overcome the vertex corrections:
The third-order terms totally enhance the $d_{x^2-y^2}$-wave
superconductivity.
In other words, the RPA-terms are not completely suppressed by
the vertex corrections even far from the magnetic instability.
Thus, we conclude that the vertex correction is not so severe in this case.
Qualitatively the same mechanism as the spin fluctuation-induced
superconductivity is expected in the weak coupling region,
which corresponds to the over-doped cuprates.
Simultaneously, this continuity confirms the applicability of
the spin fluctuation theory in the optimally-doped region.

\begin{figure}[t]
\begin{center}
\vspace{2.5cm}
\caption{The imaginary part of the self-energy on the Fermi surface
${\rm Im}\Sigma^{\rm R}_{\rm n}(\k_{{\rm F}},0)$ for $t'/t$=0.15,
$\delta$=0.1, $U/t$=3, and $T$=0.01. The horizontal axis $\theta$ is shown
in Fig.~\ref{fig:FS}. The analytic continuation from the Matsubara frequency
to the real frequency is carried out by the Pad\'e approximation
\cite{rf:pade} through this review.}
\label{fig:damping}
\end{center}
\end{figure}

Next we discuss the effect of the normal self-energy.
We can see from Fig.~\ref{fig:various} that the value of $T_{\rm c}$
is reduced by the de-pairing effect.
Because the reduction factor is about a half in the TOP, the same order
of the magnitude of $T_{\rm c}$ still remains, even if the de-pairing
effect is considered.
The realistic value $T_{\rm c}$=0.01$\sim$100K is found around
$U/t$=3.2, where $U/W$=0.4.
On the other hand, $T_{\rm c}$ is remarkably reduced in the TOPWOVC,
since the RPA terms intensify each other and give rise to a large
self-energy.
Indeed, the RPA terms in the normal self-energy are considerably
compensated by the vertex correction.
This is the reason why the reduction of $T_{\rm c}$ is moderate
in the TOP.
The third-order terms in the normal self-energy completely cancels
each other in the particle-hole symmetric case, namely
$\Sigma^{(3{\rm RPA})}_{\rm n}(k)+\Sigma^{(3{\rm VC})}_{{\rm n}}(k)$=0
for $t'=0$ and $\delta=0$.
The remained contribution in the particle-hole asymmetric case rather
reduces the quasi-particle damping (see Fig.~\ref{fig:damping}).
As a result, the value of $T_{\rm c}$ in the TOP can be higher than
that in the TOPWOVC owing to the de-pairing effect.
Thus, it is generally expected that the normal self-energy is
overestimated in the spin fluctuation theory, and thus,
the magnitude of $T_{\rm c}$ is underestimated.

\begin{figure}[t]
\begin{center}
\vspace{2.5cm}
\caption{The doping dependence of $T_{\rm c}$ calculated by the TOP
for $U/t$=4. The normal self-energy is ignored for simplicity.}
\label{fig:Tc}
\end{center}
\end{figure}

Finally, we show the doping dependence of $T_{\rm c}$ in Fig.~\ref{fig:Tc}.
For simplicity, here we ignore the normal self-energy,
which is not important to see the qualitative behavior within the TOP.
If $t'/t=0$, $T_{\rm c}$ takes the maximum value just at half-filling,
because both the staggard susceptibility $\chi_{0}(\Q)$ and
the electronic DOS becomes largest at $\delta$=0.
By introducing the next nearest-neighbor hopping, the peak position is 
shifted to the hole-doped region. This is mainly due to the fact that
the DOS takes its maximum value in the hole-doping side, since
the Fermi surface crosses the van Hove singularity.
These features will again appear in the FLEX approximation.

\subsubsection{FLEX approximation}

It is considered that the spin fluctuation plays an important role
in the optimally- and under-doped region, where the strong enhancement
of the spin fluctuation is observed
\cite{rf:NMR,rf:NMRCurie,rf:NMRKorringa,rf:neutronnormal}.
Since the perturbation theory is not sufficient to describe
the strong spin fluctuation, some approximation beyond the TOP is required.
A simple microscopic theory on the spin fluctuation is the RPA, but
the tendency of the magnetic order is seriously overestimated in the RPA.
Thus, the strong spin fluctuation in the quasi-two-dimensional systems
is not correctly described, unfortunately.
Among the several modifications of the simple RPA,
fluctuation-exchange (FLEX) approximation has been used most widely.
This approximation includes the renormalization of the spin fluctuation
within the one-loop order, but the theory is surprisingly improved
at this level. The magnetic properties in the nearly AF Fermi-liquid
\cite{rf:moriya1990,rf:moriya1992,rf:monthoux1991,rf:monthoux1992,rf:moriyaAD,rf:chubukovreview} are appropriately reproduced,
since the mode coupling effect is partly included \cite{rf:moriyaAD}.
For instance, the Curie-Weiss law in the NMR $1/T_{1}T$ is obtained
in the wide temperature region, as shown in Fig.~\ref{fig:NMR}
(See also \cite{rf:dahmFLEX}).
Applying the FLEX approximation to the Hubbard model \cite{rf:FLEX,rf:bickers1991,rf:monthouxFLEX,rf:paoFLEX,rf:dahmFLEX,rf:langerFLEX} or
$d$-$p$ model \cite{rf:luo,rf:koikegamiFLEX,rf:takimotoFLEX,rf:takimotoFLEX2},
the doping dependences are properly reproduced at least from
the over- to optimally-doped region
Thus, the results of the FLEX approximation are reviewed in this subsection
and the following discussion is complementary with the perturbation
theory in Sec.~3.2.2. The higher-order corrections beyond the FLEX
approximation will be discussed in the next subsection.

It should be mentioned that the accuracy in the quantitative 
estimation from the microscopic level is still questionable, 
because the FLEX approximation is also a partial summation 
of the perturbation series.
However, we consider that the FLEX theory has a robust meaning as
a semi-phenomenological theory, since the effect of spin fluctuations
is qualitatively well grasped in this approximation.
Moreover, it is noted that unphysical results inherent
in the phenomenological theory are considerably excluded.

The FLEX approximation is one of the conserving approximation,
formulated in the scheme of Baym and Kadanoff \cite{rf:luttinger,rf:baym}
(see Eqs.~(\ref{eq:variation}) and (\ref{eq:variational-condition})).
The well-defined basis makes it possible to perform the systematic calculation
for the single- and two-particle properties in a coherent way.
In fact, various quantities in the normal state have been calculated
by using the FLEX approximation \cite{rf:kontani,rf:dahmFLEX,rf:deiszFLEX}.
In general, anomalous properties arising from the AF spin
fluctuation have been well explained within the FLEX approximation.
The applicable region of the FLEX approximation is roughly between
$T_{0}$ and $T^{*}$ in Fig.~\ref{fig:high-Tcphasediagram2},
since the SC fluctuation plays an essential role in the pseudogap state.
We review the results of the normal-state properties together with
the pseudogap phenomena in Sec.~4.3.
This subsection is devoted to the review on the superconducting
instability within the accuracy of mean field theory.

Let us move on to the formulation of the FLEX approximation.
In the FLEX approximation, the normal self-energy is expressed as
\begin{eqnarray}
  && \Sigma_{{\rm F}} (k)=\sum_{q}
  U^{2} [\frac{3}{2} \chi_{{\rm s}}(q) + \frac{1}{2} \chi_{{\rm c}}(q)
  -\chi_{0} (q)] G (k-q),
\end{eqnarray}
where $\chi_{{\rm s}} (q)$ and $\chi_{{\rm c}} (q)$ are given by 
Eq.~(\ref{eq:rpa-susceptibility}).
Note that $\chi_{0} (q)$ is the irreducible susceptibility,
which should be defined by using the dressed Green function as
\begin{eqnarray}
  && \chi_{0} (q) = - \sum_{k} G (k) G (k+q).
\end{eqnarray} 
Note that $G(k)$ is self-consistently determined with the self-energy
and susceptibilities. Throughout this self-consistent iterations,
the renormalization effect on the spin fluctuation is included
in the self-energy.
The spin fluctuation gives a large contribution to the self-energy
in the quasi-2D systems, and thus, the magnetic order is suppressed.
As a result, the FLEX approximation provides a wide critical region,
in which the spin fluctuation is strongly enhanced.
For an expression of the spin susceptibility
Eq.~(\ref{eq:rpa-susceptibility}) does not include the vertex corrections,
which should be required in the conservation scheme, but it is considered
that this correction is not severe \cite{rf:dahmFLEX}.
Note that even in the conservation scheme, this vertex correction is not
needed in the single-particle properties and superconducting $T_{\rm c}$.

The superconducting transition is determined by solving the
\'Eliashberg equation, Eq.~(\ref{eq:eliashberg-super}),
where the effective interaction in the singlet channel is given by
Eq.~(\ref{eq:anomalous-rpa-singlet}) with the self-consistently
determined susceptibility.
Near the magnetic instability, the second term in the right hand side of
Eq.~(\ref{eq:anomalous-rpa-singlet}) becomes dominant.
Then, the expressions are similar to those of the phenomenological theory
with $g$ replaced by $U$. It should be stressed again that the expressions
are determined from the microscopic model.

\begin{figure}[t]
\begin{center}
\vspace{2.5cm}
\caption{The momentum dependence of the static spin susceptibility in
(a) electron-doped ($10\%$), (b) under-doped ($10\%$), and
(c) optimally-doped ($16\%$) region.
Here, $t'/t=0.25$, $T=0.01$ and $U/t=3.2$ for hole-doped case,
$U/t=3$ for electron-doped case.}
\label{fig:kaiflex}
\end{center}
\end{figure} 

\begin{figure}[t]
\begin{center}
\vspace{2.5cm}
\caption{The momentum dependence of the order parameter
$\Delta(\k)=\Delta(\k,{\rm i} \pi T)$ on the Fermi surface, 
for $U/t=3$, $t'/t=0.25$ and $T=T_{{\rm c}}$.
The conventional form $\Delta(\k) \propto \cos k_{x}-\cos k_{y}$ is
also shown for the comparison. The horizontal axis $\theta$ is shown
in Fig.~\ref{fig:FS}.}
\label{fig:PhiFS}
\end{center}
\end{figure}

The FLEX approximation was first performed by Bickers et al.
\cite{rf:FLEX,rf:bickers1991} for the 2D Hubbard model with only
the nearest-neighbor hopping.
They have shown that the AF order exists near the half-filling and
the $d_{x^2-y^2}$-wave superconductivity occurs in the vicinity
to the AF state.
After that, intensive studies on the Hubbard model
\cite{rf:monthouxFLEX,rf:paoFLEX,rf:dahmFLEX,rf:langerFLEX}
and on the $d$-$p$ model
\cite{rf:luo,rf:koikegamiFLEX,rf:takimotoFLEX,rf:takimotoFLEX2}
have commonly obtained the $d_{x^2-y^2}$-wave superconductivity.
We show the momentum dependence of the static spin susceptibility
in Fig.~\ref{fig:kaiflex}, where the results for the optimally-doped,
under-doped, and electron-doped cases are shown.
It is observed that the spin fluctuation is strongly enhanced around
$\q=(\pi,\pi)$. Thus, the $d_{x^2-y^2}$-wave superconductivity is
mainly induced by the scattering process shown in Fig.~\ref{fig:FS}.
The order parameter is deformed, but similar to the conventional form
$\Delta(\k) \propto \cos k_{{\rm x}}-\cos k_{{\rm y}}$,
as shown in Fig.~\ref{fig:PhiFS}.
Since the scattering process between the zone boundary region 
(the circle in Fig.~\ref{fig:FS})
plays a dominant role, the order parameter takes large absolute value 
around $\theta$=0 or $\pi/2$.
This deformation is in agreement with the experimental results of ARPES
\cite{rf:mesot}. 

We show the results of $T_{\rm c}$ for the Hubbard model
and those for the $d$-$p$ model in Figs.~\ref{fig:electron-dope} and
\ref{fig:takimoto}, respectively.
Figure \ref{fig:electron-dope}(a) shows the results for
the electron-doped case, together.
In general, the transition temperature increases with the development of
the spin fluctuation, namely with decreasing $\delta$ and/or increasing $U$.
In contrast to the TOP, $T_{\rm c}$ tends to saturate near the magnetic
instability, since the de-pairing effect becomes stronger in that region.
The maximum value of $T_{\rm c}$ is commonly obtained as
$T_{\rm c} = 0.01 \sim 0.02$ \cite{rf:monthouxFLEX,rf:paoFLEX,rf:dahmFLEX,rf:langerFLEX,rf:koikegamiFLEX,rf:takimotoFLEX,rf:takimotoFLEX2},
which is consistent with the experimental value
$T_{\rm c} \sim 100 {\rm K}$ in order of magnitude.

It is straightforward to extend the FLEX approximation to the
superconducting state.
Actually, the calculation has been performed for the Hubbard model
\cite{rf:monthouxFLEX,rf:paoFLEX,rf:dahmFLEX} and for the $d$-$p$ model
\cite{rf:takimotoFLEX2}.
The ratio $2\Delta/T_{{\rm c}} = 10 \sim 12$ is commonly obtained,
which is much larger than the BCS value $2\Delta/T_{{\rm c}} = 3.5$.
This is mainly due to the feedback effect on the spin fluctuation:
Since the de-pairing effect is remarkably reduced by the finite 
excitation gap, the order parameter at $T=0$ exceeds the BCS value.
This feature is qualitatively consistent with the experimental value
$2\Delta/T_{{\rm c}} = 7 \sim 8$ in the optimally-doped region.
This value is over-estimated in the FLEX approximation probably
because the de-pairing effect around $T=T_{\rm c}$ is over-estimated
(see Sec.~3.2.3).

\begin{figure}[t]
\begin{center}
\vspace{2.5cm}
\caption{$T_{\rm c}$ in the FLEX approximation for the Hubbard model
\cite{rf:yanaseFLEXPG}. (a) Doping dependence and (b) $U$-dependence for
$t'/t=0.25$ and $\delta=0.1$.}
\label{fig:electron-dope}
\end{center}
\end{figure}

\begin{figure}[t]
\begin{center}
\vspace{2.5cm}
\caption{$T_{\rm c}$ in the FLEX approximation for the $d$-$p$ model
\cite{rf:takimotoFLEX2}.}
\label{fig:takimoto}
\end{center}
\end{figure}

The other interesting property in the superconducting state is
the resonance peak.
When the system is near the magnetic order, a sharp resonance peak appears
in the magnetic excitation with a smaller energy than the maximum gap
$2\Delta$ \cite{rf:dahmFLEX,rf:takimotoFLEX2,rf:RSphenomenology}.
It is considered that the resonance peak corresponds to the $41 {\rm meV}$
peak observed in the neutron scattering experiments in
YBa$_2$Cu$_3$O$_{6+\delta}$
\cite{rf:RSpeak1,rf:RSpeak2,rf:RSpeak3,rf:RSpeak4,rf:RSpeak5}.
The influence of the resonance peak appears in the dip-hump structure
in the quasi-particle spectrum and ``kink'' in the quasi-particle
dispersion around $(\pi,0)$
\cite{rf:takimotoFLEX2,rf:RSphenomenology,rf:diphumpnorman},
which is also consistent with the ARPES measurements
\cite{rf:shenrep,rf:ARPESreview,rf:campuzano1999}.

Finally let us review the application of the FLEX approximation to
the electron-doped cuprates which has been recently performed
\cite{rf:yanaseFLEXPG,rf:kontani,rf:kondo-el,rf:manske}.
The FLEX approximation also indicates the $d_{x^2-y^2}$-wave
superconductivity in the electron-doped region, which has been confirmed
by the recent experiments
\cite{rf:hayashi,rf:electron-dope_D,rf:tsuei-el,rf:takahashi-el,rf:shen-el}.
Moreover, the drastic particle-hole asymmetry in the phase diagram
\cite{rf:electrondopesuper} is explained simply by taking account of
the next-nearest-neighbor hopping $t'$ and choosing the electron number.
In general, the electron correlation is relatively weak
in the electron-doped region, because the electronic DOS decreases
with electron-doping.
Therefore, more conventional behaviors are expected 
in the normal state \cite{rf:yanaseFLEXPG}. 
The detailed discussion will be given in Sec.~4.3.2.
On the other hand, the AF order is robust for the carrier doping
(Fig.~\ref{fig:electron-dope}(a)) because of the nesting property
of the Fermi surface, as shown in Fig.~\ref{fig:fermi}.
An interesting nature of the AF state appears in the electron-doped
region \cite{rf:onose}, which has been well explained by the numerical
calculation on the $t$-$t'$-$J$ model \cite{rf:tohyama2001}.

More interestingly, the superconducting $T_{\rm c}$ is very low
and the doping region possessing the superconductivity is narrow
\cite{rf:yanaseFLEXPG,rf:manske}.
This is understood by the following two reasons:
One is the localized nature of the spin fluctuation in the momentum space
(see Fig.~\ref{fig:kaiflex}).
Then, the total weight of the spin fluctuation becomes small.
It should be noted that the effective interaction $|V_{{\rm sc}}|$ is
not determined by the magnetic correlation length. 
The other is the small DOS due to the fact that the Fermi level is apart 
from the flat dispersion around $(\pi,0)$. 
Thus, the effective coupling $\rho|V_{{\rm sc}}|$ decreases in the
electron-doped region, and thus, $T_{\rm c}$ becomes low.
These features are consistent with the experimentally observed
phase diagram \cite{rf:electrondopesuper}, and lead to the no appearance
of the pseudogap phenomena (see Sec.~4.3.2).

Another interesting feature is the strong modulation in order parameter.
Namely, the order parameter has its maximum magnitude around
the magnetic Brillouin zone.
This is an inevitable result of the spin fluctuation-induced superconductivity,
because the commensurate spin fluctuation (Fig.~\ref{fig:kaiflex})
induces the strongest interaction between the magnetic Brillouin zone.
Recent Raman scattering measurement has supported this modulation of
the order parameter \cite{rf:blumberg}.
Such a detailed consistency supports the importance of the spin fluctuation
in the electron-doped cuprates.

Summarizing, some detailed and interesting properties including the
particle-hole asymmetry are well reproduced by the microscopic theory
starting from the Fermi-liquid state.
We consider that this is an important suggestion for the wide applicability
and possibility of the microscopic theory for other strongly correlated
electron materials.
In fact, Secs.~3.3 and 3.4 are devoted to the application to the
organic superconductor and Sr$_2$RuO$_4$, respectively.

\begin{figure}[t]
\begin{center}
\vspace{2.5cm}
\caption{The Fermi surface obtained in the FLEX approximation
for the hole- (thick line) and electron-doped (thin line) cases.
Dashed lines show the non-interacting Fermi surface.
The nesting nature is enhanced by the AF spin fluctuation
\cite{rf:yanaseTR}.}
\label{fig:fermi}
\end{center}
\end{figure}

\subsubsection{higher-order corrections}

Thus far, we have not provided any explicit justification of
the spin fluctuation theory and/or the FLEX approximation.
They correspond to a partial summation in the perturbation series.
The role of the neglected terms is not clear.
Needless to say, an absolute work on this problem is very difficult.
However, several types of the higher-order corrections have been estimated
and the positive results have been obtained to some extent.

On the basis of the perturbation theory, the corrections first appear
in the third-order terms, which have been discussed in Sec.~3.2.3.
There are two types of the correction in the effective interaction.
One is represented by the Feynmann diagram in Figs.~\ref{fig:3rddiagrams}(e)
and (f). The other includes Figs.~\ref{fig:3rddiagrams}(g) and (h).
We have confirmed that both types of the diagram reduce the pairing
interaction, that is, both terms in Eq.~(\ref{eq:third-anomalous-singlet-vc})
are repulsive in the $d$-wave channel.
We furthermore find that the reduction from the two terms is nearly the same
magnitude.

\begin{figure}[t]
\begin{center}
\vspace{2.5cm}
\caption{The renormalization of the interaction corresponding to
the Kanamori theory.}
\label{fig:kanamori}
\end{center}
\end{figure}

This effect from the diagrams in Figs.~\ref{fig:3rddiagrams}(e) and
(f) is natural, because they are the lowest-order vertex correction
included in the Kanamori-type T-matrix diagram (Fig.~\ref{fig:kanamori}),
which represents the screening effect \cite{rf:kanamori}.
For instance, the higher-order correction represented 
by the T-matrix was approximately estimated 
by Bulut et al. \cite{rf:bulut1993}.
They have concluded that the bare interaction $U/t=4$ is renormalized to
$\bar{U}/t \sim 2$, where $\bar{U}$ denotes the renormalized interaction.
The coupling constant $g$ in the phenomenological theory, and further
$U$ in the FLEX approximation should be regarded as the renormalized
coupling constant including the screening effect.

The combination of the diagrammatic techniques and the quantum Monte Carlo
simulation has shown that the effective pairing interaction and
the spin susceptibility are consistently obtained by the generalized RPA,
where the renormalized particle-hole vertex $\bar{U}(q)$ is used
\cite{rf:bulut1995}.
The renormalized vertex in the intermediate coupling region $U/t$=4 is
estimated as about 80\% of $U$, namely $\bar{U}(q) \sim 0.8 U$.
Then, the expression for the effective pairing interaction, given by
\begin{eqnarray}
 && V^{{\rm s}}_{\rm a} (k,k')=
    U+\frac{3}{2} \bar{U}^{2} \chi_{{\rm s}} (k-k')
\end{eqnarray}
provides an appropriate estimation, when the renormalized coupling
constant $\bar{U}=0.8U$ is used with the spin susceptibility,
obtained by the quantum Monte Carlo simulation.
These results have indicated that the other scattering process,
such as the multi-paramagnon exchange (see Fig.~\ref{fig:vc}(b)),
is negligible at least in the high-temperature region $T/t \sim 0.25$.

\begin{figure}[t]
\begin{center}
\vspace{2.5cm}
\caption{(a) The lowest-order vertex correction in the phenomenological
spin fluctuation theory. (b) The effective interaction corresponding
to the multi-paramagnon exchange. Here, the solid and wavy lines are
the propagator of the fermion and spin fluctuation, respectively.}
\label{fig:vc}
\end{center}
\end{figure}

We find the other intensive investigations on the higher-order corrections,
namely, the vertex correction arising from the spin fluctuation
\cite{rf:chubukovreview,rf:bulut1993,rf:yonemitsu,rf:chubukov1995,rf:altshular1995,rf:stamp1996,rf:monthoux1997,rf:chubukov1997,rf:nakamura2000,rf:okabe2000}.
The lowest-order vertex correction from the spin fluctuation is shown in
Fig.~\ref{fig:vc}(a).
The diagrams in Figs.~\ref{fig:3rddiagrams}(g) and (h) are included in these
corrections, while only the higher-order terms than the forth-order are
included in the phenomenological spin fluctuation theory.

The strong cancellation of the effective vertex between the fermion and
spin fluctuation was suggested by Schrieffer \cite{rf:schrieffer1995}.
His suggestion raised a serious question on the spin fluctuation-induced
superconductivity, but it was just based on the mean-field theory
in the SDW state.
On the contrary, the studies in the normal state with diffusive spin
fluctuation have commonly concluded that the vertex correction represented in
Fig.~\ref{fig:vc}(a) furthermore enhances the effective vertex at
$\q=(\pi,\pi)$
\cite{rf:chubukovreview,rf:yonemitsu,rf:chubukov1995,rf:altshular1995,rf:monthoux1997,rf:chubukov1997,rf:nakamura2000,rf:okabe2000,rf:commentonvc}.
The numerical calculation has furthermore revealed that the momentum dependence
is not so altered by the vertex correction \cite{rf:monthoux1997}.
That is, we can appropriately include the vertex correction by renormalizing
the coupling constant as $g \rightarrow \alpha g$. 
Here the enhancement factor $\alpha$ increases with the development of
the spin fluctuation, but is expected not to exceed $2$ under
the reasonable parameters \cite{rf:monthoux1997,rf:chubukov1997}.
The effect of the vertex correction on the superconducting $T_{\rm c}$
has also been estimated \cite{rf:yonemitsu,rf:nakamura2000,rf:okabe2000}.
The explicit calculation is difficult, but the enhancement of $T_{\rm c}$
has been suggested commonly.

It should be noted that the diagrams in Figs.~\ref{fig:3rddiagrams}(g)
and (h) provide an opposite contribution, i.e., reduce the effective vertex.
This is because only the longitudinal component of the spin fluctuation
appears in the the vertex correction within the TOP.
The transverse component appears in the higher-order correction and changes
the sign. Combined with the corrections included in the T-matrix
(Fig.~\ref{fig:kanamori}), it is expected that the vertex correction
reduces the bare interaction $U$ in the weak-coupling region.
This expectation is consistent with the result in the high-temperature
region \cite{rf:bulut1993,rf:bulut1995}.
On the other hand, the enhancement of the effective vertex is expected
when the spin fluctuation is strongly enhanced.
The explicit calculation is an open problem up to now.

Another possible correction is the contribution from 
the multi-paramagnon exchange, which is shown in Fig.~\ref{fig:vc}(b).
This process has been proposed as the ``spin-bag'' mechanism
\cite{rf:schrieffer1988}.
However, the numerical estimations have concluded that the pairing interaction
arising from this process is almost negligible
\cite{rf:monthoux1997,rf:okabe2000}.

In short, from the above results, we expect that the theory based on
the single paramagnon exchange is qualitatively justified, although
the higher-order corrections are required in the quantitative estimations.

\subsection{Organic Superconductor $\kappa$-(ET)$_2$X}

In this section, we discuss $\kappa$-(ET)$_2$X compounds among many
organic superconductors.
The outline of the experimental results and interesting issues
on these compounds have been reviewed in Sec.~2.2.
The metallic conduction in this material is owing to the organic molecule,
which is complicated at first glance.
It is, however, shown that the simplified tight-binding model gives
a reasonable understanding for the superconductivity.
The successful application of the Hubbard model for the molecular systems
will extend the possibility of the microscopic theory.

\subsubsection{Electronic property and tight-binding model}

First we review the electronic properties and introduce an effective
Hamiltonian. 
This series of the organic materials are systematically described 
well by the Hubbard model with a tight-binding fitting \cite{rf:tamura,rf:8}.
Note that the site in the Hubbard model is not one atom, but corresponds to
one dimer of the organic molecules.
The procedure for this simplification is given in the following way.

The quasi-2D conduction band consists of the $\pi$-orbitals in the
ET-molecules. There are four ET-molecules and two holes in a unit cell,
as shown in Fig.~\ref{fig:kurokifig1}(a).
Then, the system is quarter-filling ($n=1.5$ per molecule).
In the first simplification, a dimer is regarded as a structural unit.
This procedure is justified because the transfer integral $t_{{\rm b}1}$
is twice larger than the other transfer integrals \cite{rf:8,rf:10}.
Then, the holes are contained in the anti-bonding orbitals in dimers.
The bonding orbitals are far below the Fermi level.
Focusing on the anti-bonding orbital, each dimer is connected
to the nearest neighbor sites by three kinds of transfer integrals 
(Fig.~\ref{fig:kurokifig1}(b)).
They are derived as $t_{{\rm b}}=-t_{{\rm b}2}/2$,
$t_{{\rm c}}=(t_{{\rm q}}-t_{{\rm p}})/2$,
and $t_{{\rm c'}}=(t_{{\rm q'}}-t_{{\rm p'}})/2$.
In this dimer model, the system is half-filling ($n=1$ per dimer).
The AF state in the lower pressure region is a Mott insulator constructed
from the dimers \cite{rf:8}.
Because the difference between $t_{{\rm c}}$ and $t_{{\rm c'}}$ is
negligible (only about $2\%$), the effective single-band model is
derived by setting $t_{{\rm c}}=t_{{\rm c'}}$.
Note that the obtained model is mapped to the anisotropic triangular
lattice shown in Fig.~\ref{fig:kurokifig1}(c).

\begin{figure}[t]
\begin{center}
\vspace{2.5cm}
\caption{(a) The lattice structure of $\kappa$-(ET)$_2$X. b$1$, b$2$, etc.
stand for the transfer integrals $t_{{\rm b}1}$, $t_{{\rm b}2}$, and so on.
The unit cell consists of the hatched molecules \cite{rf:commentkuroki}.
(b) The structure of the dimerized model. The difference between $t_{\rm c}$
and $t_{\rm c'}$ is usually neglected \cite{rf:commentkuroki}.
(c) The tight binding model corresponding to the 
anisotropic triangular lattice. (d) The typical Fermi surface.}
\label{fig:kurokifig1}
\end{center}
\end{figure}

Now superconductivity has been discussed on the basis of the Hubbard
model on the anisotropic triangular lattice.
Because the system is near the Mott transition, the on-site repulsion is
expected to be most effective. The Hubbard Hamiltonian has been given in
Eq.~(\ref{eq:Hubbard-model}), but the dispersion relation is now given as
\begin{eqnarray}
  && \varepsilon(\k)=-2t(\cos k_{x}+\cos k_{y})-2t'\cos (k_{x}+k_{y})-\mu.
  \label{eq:dispersion-organic}
\end{eqnarray}
Note that $t'/t$ defines the anisotropy of the system and typically,
$t'/t = 0.6 \sim 0.8$ according to the
extended H${\rm\ddot{u}}$ckel band calculation \cite{rf:9,rf:8}.
The Fermi surface in this model (Fig.~\ref{fig:kurokifig1}(d)) is
in good agreement with the result of the Shubnikov-de Haas effect.
This model has a strong frustration compared to the case of
high-$T_{\rm c}$ cuprates.
It should be noticed that the Coulomb repulsion $U$ is very small, 
because a lattice cite of this model is not an atom but a dimer of molecules.
However, this system is regarded as a strongly correlated electron system
since the transfer integral $t$ is also small.
The parameter $U/W$ is expected to be in the intermediate coupling region,
because the system is half-filling near the Mott transition.
The transfer integral has been estimated from the quantum chemistry
calculation \cite{rf:komatsu} as $t = 70 \sim 80$meV,
which is just about $1/10$ of the typical value in $d$-electron systems.
Thus, $T_{{\rm c}} \sim 10$K again corresponds to
$T_{{\rm c}}/t \sim 0.01$, which is theoretically comparable to
that for cuprate superconductors (Sec.~3.2).

\subsubsection{Application of the microscopic theories}

Microscopic calculations based on the Hubbard model have been performed
by using the RPA \cite{rf:vojta}, FLEX approximation
\cite{rf:kinoFLEX,rf:kondoFLEX,rf:schmalian}, TOP \cite{rf:jujoTOP},
and quantum Monte Carlo simulation \cite{rf:21}.
Note that strictly speaking, some small corrections on the model
(\ref{eq:dispersion-organic}) are included in several papers,
but they do not affect the results, qualitatively.
The above studies have concluded that
(i) the pairing symmetry is the $d_{x^{2}-y^{2}}$-wave,
(ii) qualitatively the same pairing mechanism to high-$T_{\rm c}$ cuprates
is expected, and
(iii) the comparable $T_{\rm c}/t \sim 0.01$ is obtained in the intermediate
coupling region.

\begin{figure}[t]
\begin{center}
\vspace{2.5cm}
\caption{The phase diagram obtained by the FLEX approximation
\cite{rf:kondoFLEX}. The inset shows $T_{\rm c}$ for the various $t'/t$.
Note that $\tau$ and $\tau'$ in this figure denote $t$ and $t'$,
respectively, in our notations.}
\label{fig:kondoFLEX}
\end{center}
\end{figure}

\begin{figure}[t]
\begin{center}
\vspace{2.5cm}
\caption{The spin susceptibility obtained by the FLEX approximation
at $T/t=0.02$. The value of $U/t$ is changed as $U/t=2.5$, $U/t=3.7$,
$U/t=5.5$, and $U/t=6.5$ with increasing $t'/t$.}
\label{fig:flexkai}
\end{center}
\end{figure}

Most of the above proposals are classified into the spin-fluctuation theory.
Although the phase transition from the AF to SC state is in the first order,
the strong AF spin fluctuation is observed experimentally \cite{rf:2,rf:3}.
Thus, it is expected that the spin fluctuation theory can be applied
in the metallic state.
The typical results of the FLEX approximation are shown in
Fig.~\ref{fig:kondoFLEX} \cite{rf:kinoFLEX,rf:kondoFLEX,rf:schmalian}.
The phase diagram in the plane of $t'/t$ and $U/t$ is shown.
The SC state appears in the neighborhood of the AF state.
By increasing $t'/t$, the introduced frustration destroys the AF state,
and then, the superconductivity appears.
We see that the static spin susceptibility in the normal state is
strongly enhanced around $\q=(\pi,\pi)$ and $\q=(\pi,-\pi)$
(Fig.~\ref{fig:flexkai}).
We have already shown that the scattering process exchanging 
the AF spin fluctuation is attractive for the $d$-wave 
superconductivity. 
The dominant scattering process is similar to that shown in Fig.~\ref{fig:FS}.
Thus, qualitatively the same mechanism to the high-$T_{\rm c}$ cuprates
is expected for the superconductivity in organic materials.
Note here that the spin fluctuation-induced superconductivity in organic
materials has been proposed in 80's for (TMTSF)$_2$X compounds 
\cite{rf:emeryorganic,rf:shimaharaorganic}. 
The transition temperature for the superconductivity is gradually reduced
by the frustration.
This is because the spin susceptibility becomes structure-less, and thus,
the spin fluctuation mechanism is ineffective.
In particular, the superconductivity almost disappears in case of
the isotropic triangular lattice ($t'/t=1$).

\begin{figure}[t]
\begin{center}
\vspace{2.5cm}
\caption{$T_{\rm c}$ obtained by the TOP. The results of the FLEX
approximation for $t'/t=0.7$ is shown for comparison.}
\label{fig:third}
\end{center}
\end{figure}

\begin{figure}[t]
\begin{center}
\vspace{2.5cm}
\caption{The bare spin susceptibility for various values of $t'/t$.}
\label{fig:jujofig7}
\end{center}
\end{figure}

As in high-$T_{\rm c}$ cuprates, qualitatively the same results are
obtained by the TOP \cite{rf:jujoTOP}.
For example, the $d_{x^2-y^2}$-wave superconductivity is stabilized
for $t'/t \leq 0.8$.
As shown in Fig.~\ref{fig:third}, $T_{\rm c}/t \sim 0.01$ is obtained
in the intermediate coupling region $U/t \sim 6$,
corresponding to $U/W \sim 0.7$.
The roles of the respective terms are the same as those for
high-$T_{\rm c}$ superconductors.
The superconductivity is mainly induced by the RPA-terms.
The irreducible susceptibility is moderate, but has a similar tendency
to the FLEX approximation, as is shown in Fig.~\ref{fig:jujofig7}.
The non-RPA terms in the effective interaction reduce $T_{\rm c}$.

\begin{figure}[t]
\begin{center}
\vspace{2.5cm}
\caption{$t'/t$-dependence of $T_{\rm c}$ in the TOP (circles).
The squares are the results when the vertex correction in the anomalous
self-energy is neglected.}
\label{fig:frustration}
\end{center}
\end{figure}

Figure \ref{fig:frustration} shows that $T_{\rm c}$ is reduced by
the frustration and almost disappears around $t'/t=0.8$.
This feature is also found in the result of the FLEX approximation.
Thus, the anisotropy in the triangular lattice plays an essential role
for the appearance of superconductivity.
The decrease in $T_{\rm c}$ for $t'/t < 0.3$ is caused by the normal
self-energy, which is enhanced by the nesting nature of the Fermi surface.
We have also shown the comparison with the result when we ignore
the vertex correction term $V_{{\rm a}}^{(3{\rm VC})}(k,k')$.
It is a natural conclusion that the vertex correction is more important
for the strongly frustrated case.
This is simply because the RPA terms become ineffective and then,
the non-RPA terms relatively become effective.
In general, the spin fluctuation theory such as SCR \cite{rf:moriyatext}
is justified when the Fermi surface is strongly nested.
Under the strong frustration, 
it is not allowed to ignore the vertex corrections simply.
Thus, the spin fluctuation theory is less appropriate to the
$\kappa$-(ET)$_2$X compounds for $t'/t = 0.6 \sim 0.8$ than to
the high-$T_{\rm c}$ cuprates.

Note that the TOP and FLEX approximation commonly show that $T_{\rm c}$
increases with $U$. This tendency is in good agreement with the phase diagram
in Fig.~\ref{fig:organic}.
The pressure increases the band width (equivalently, the transfer integral).
Then, the parameter $U/W$, and accordingly $T_{\rm c}$ decreases with
the applied pressure.
For the variation of anions, the frustration is enhanced from
X=Cu[N(CN)$_2$]Br to X=Cu(NCS)$_2$, namely in the right hand side of
the phase diagram in Fig.~\ref{fig:organic} \cite{rf:chisa}.
Therefore, the frustration furthermore reduces $T_{\rm c}$ in the metallic
state, combined with the chemical pressure.

Recent FLEX calculations have pointed out that the strong dimerization
is essential for the above results.
When the dimerization is not strong enough, the superconductivity is
severely suppressed, even if the bonding band is below the Fermi level
\cite{rf:commentkuroki,rf:kondo4band}.
In some cases, the $d_{\rm xy}$-wave symmetry is more favorable,
while $T_{\rm c}$ is very low \cite{rf:commentkuroki}.
The parameters obtained from the quantum chemistry calculation locate
in the boundary region between the strong and weak dimerization.
We mention that the reasonable $T_{\rm c}$ is obtained only
for the strongly dimerized region.
Some experimental efforts have been devoted to identify the symmetry.
Note that the $d_{x^{2}-y^{2}}$-wave symmetry has been supported by the
angle-dependence of the high-frequency conductivity \cite{rf:schrama},
while the $d_{\rm xy}$-wave symmetry has been supported by the thermal
conductivity measurement \cite{rf:izawaorganic} as well as
the tunnelling spectroscopy \cite{rf:araiorganic}.

Before closing this section, we comment on the other superconducting
materials in the molecular conductors.
Since the phonon excitation is generally strong in the organic conductors,
a class of the materials should be $s$-wave superconductor due to
the electron-phonon mechanism. 
For instance, superconductivity in alkali-metal-doped fullerides
A$_3$C$_{60}$ is considered to originate from the phonon-mediated
attractive interaction.
However, some class of the organic superconductor can be categorized 
into strongly correlated electron systems and
the unconventional superconductivity is expected. 
For example, (TMTSF)PF$_6$ is considered to be the case,
where the possibility of not only the $d$-wave pairing, but also
the triplet pairing has been proposed \cite{rf:leeTMTSF}.
The present microscopic calculations
\cite{rf:shimaharaorganic,rf:kinoTMTSF,rf:kurokiTMTSF,rf:nomuraTMTSF}
seem to indicate the predominance of the $d$-wave superconductivity.
We believe that the development of the microscopic theory for
organic materials is one of interesting future issues.

\subsection{Sr$_2$RuO$_4$}

Now let us review the theoretical investigations on Sr$_2$RuO$_4$
with main interests on the pairing mechanism, since interestingly
this material is confirmed to be a triplet superconductor (see Sec.~2.3).
The triplet superconductivity has been already discovered in heavy-fermion
compounds such as ${\rm UPt}_{3}$ \cite{rf:stewart}.
Probably, recently discovered uranium compounds
${\rm UGe}_{2}$ \cite{rf:saxena} and URhGe \cite{rf:Aoki} are
also the triplet superconductors,
since superconductivity coexists with ferromagnetism.
However, the microscopic investigation on these materials is generally
difficult due to their complicated electronic structure,
as will be shown in Sec.~5.

On the other hand, the electronic structure of ${\rm Sr_2RuO_4}$ is
relatively simple, deduced from such points as two-dimensional Fermi surfaces,
a few degenerate orbitals, weak spin-orbit coupling, and relatively
weak electron correlation.
Thus, both from experimental and theoretical points of view, 
${\rm Sr_2RuO_4}$ is considered to be a typical compound for 
the microscopic investigation on the triplet superconductivity, 
which will provide new understanding on the unconventional superconductivity.

\subsubsection{Overview}

Referring to the successes for the pairing mechanism in cuprates and $^3$He,
the spin fluctuation-induced superconductivity was examined for
Sr$_2$RuO$_4$ at the first stage.
Several authors discussed the ferromagnetic spin fluctuation, which can induce
the triplet $p$-wave superconductivity \cite{rf:leggett} 
(see the last in Sec.~3.2.2).
Along this line, Mazin and Singh have insisted that ferromagnetic spin
fluctuation is sufficiently strong to induce the triplet superconductivity
in this compound \cite{Rf:Mazi.1,Rf:Mazi.2}.
Monthoux and Lonzarich discussed the $p$-wave superconductivity
by phenomenologically introducing the ferromagnetic spin fluctuation
in a square lattice model \cite{Rf:Mont.1}.
However, it has become apparent that the situation is not so simple.
Neutron scattering measurement by Sidis {\it et al.} has revealed that
the ferromagnetic fluctuation is $not$ so enhanced.
Instead of ferromagnetic fluctuation, a sizeable incommensurate AF fluctuation
has been observed at the wave vector $\Q_{\rm IAF}=(0.6\pi, 0.6\pi, 0)$
\cite{Rf:Sidi.1}.
This incommensurate spin fluctuation is due to the Fermi surface nesting
effect, as predicted by the band calculation with the use of
local-density approximation (LDA) \cite{Rf:Mazi.2} and
also derived from simpler models \cite{Rf:Nomu.1,Rf:KKNg.1,Rf:Erem.1}.
Under those circumstances, it has been proposed that
the triplet superconductivity is possibly derived by assuming
the strong anisotropy of the AF spin fluctuation
\cite{Rf:Kuwa.1,rf:kohmoto,Rf:Kuro.1}.
The experimental support for the strong anisotropy has been obtained from
the NMR measurement \cite{rf:ishidaanisotropy}, but according to
the recent neutron scattering experiment by Servant et al. \cite{Rf:Serv.1},
the spin susceptibility is very isotropic at $\q=\Q_{\rm IAF}$.
This discrepancy should be resolved experimentally.
Theoretically, such anisotropy arises from the spin-orbit interaction
in the Ru ions \cite{Rf:KKNg.1,Rf:Erem.1}.
It should be commented that the significant exchange enhancement 
is needed to explain the anisotropy observed in NMR measurement.
Note also that the effective interaction in the Cooper channel
does not coincide with the observable spin susceptibility,
when the spin-orbit interaction is explicitly 
taken into account \cite{rf:ogataRu}. 
Thus, more explicit calculation is required for this scenario
in the present stage.

Takimoto has proposed that the orbital fluctuation is important for
the triplet superconductivity in Sr$_2$RuO$_4$ \cite{Rf:Taki.1}.
This type of fluctuation is similarly due to the nesting effect of
the Fermi surface.
He estimated the pairing interaction by using the RPA on the three-band
model and obtained the following results:
(i) The instability to the $f$-wave superconductivity is generally
derived by the RPA.
(ii) The $\alpha$ and $\beta$ bands are mainly superconducting.
(iii) The strong orbital fluctuation is necessary for the triplet pairing
to overcome the singlet one.
(iv) As a result, the triplet superconductivity is stabilized,
when the inter-orbital repulsion is larger than the intra-orbital one.
Among them, the condition for (iv) is difficult to be satisfied and thus,
somewhat convincing evidence may be needed for the scenario
based on orbital fluctuations.

The analysis based on the perturbation theory has been given 
on this issue \cite{Rf:Nomu.3,Rf:Nomu.2}. 
Nomura and Yamada have expanded the effective interaction 
with respect to the Coulomb interaction, as has been performed 
for high-$T_{\rm c}$ cuprates (Sec.~3.2.3) and 
$\kappa$-(ET)$_2$X (Sec.~3.3.2). 
Precedently, several works have discussed  
the two-dimensional Fermi gas model within the third-order
perturbation \cite{rf:chubukov-p-wave,Rf:Feld.1}.
They have concluded that the $p$-wave pairing state is expected to
provide the highest $T_{\rm c}$.
The important results obtained in Refs.~\cite{Rf:Nomu.3} and
\cite{Rf:Nomu.2} are summarized in the following:
(i) The non-RPA terms neglected in the fluctuation theory are 
significantly attractive in the $p$-wave channel.
These terms are the same ones discussed in the two-dimensional Fermi gas
model \cite{Rf:Feld.1}, while the singularity in the isotropic model 
disappears in the Hubbard model. 
(ii) The irreducible spin susceptibility derived from the $\gamma$ 
band shows rather weak momentum dependence.
This situation is in sharp contrast to high-$T_{\rm c}$ cuprates and
$\kappa$-(ET)$_2$X, but it is similar to the two-dimensional Fermi gas
model. The results of the perturbation theory and their differences from
the $d$-wave superconductors are discussed in the next subsection.
The failure of the spin-fluctuation theory will be clarified there.

Recently, Honerkamp and Salmhofer found the $p$-wave superconducting
phase in the single-band Hubbard model on the basis of the one-loop
renormalization group theory \cite{Rf:Hone.1}.
They have adopted a band structure similar to that of the $\gamma$ band.
According to their results, it seems that the momentum dependence of the
effective interaction is not dominated by the ferromagnetic spin fluctuation.
Rather it is similar to the one obtained within the naive third-order
perturbation theory, in spite of the adjacent ferromagnetic phase.
Note that this renormalization group analysis has an advantage
that the magnetic, pairing, and the other instabilities can be treated
on an equal footing.

\subsubsection{Perturbation theory for the triplet superconductivity}

In this subsection the results on the perturbation theory are reviewed.
First let us show the results on the single-band Hubbard model
corresponding to the $\gamma$ band \cite{Rf:Nomu.2}.
Referring to the fact that the $\gamma$ band has the largest DOS, we assume
that the superconductivity is mainly determined by the $\gamma$ band.
This assumption will be justified later from the results
on the three-band model.
The $\gamma$ band is mainly constructed from the Ru${\rm 4d}_{\rm xy}$
and O2p orbitals, which hybridize each other.
Since the correlation effect is dominated by the Coulomb interactions
at the Ru sites, the Hubbard model Eq.~(\ref{eq:Hubbard-model}) can be
a reasonable starting point.
This simplification of the model is similar to the case of high-$T_{\rm c}$
cuprates, where the $d$-$p$ model is reduced to the Hubbard model.
For ruthenate, the dispersion relation is given as
\begin{equation}
  \label{eq:ruthenium-dispersion}
  \e(\k)=-2t_1 (\cos k_{{\rm x}} + \cos k_{{\rm y}})-
  4t_2 \cos k_{{\rm x}} \cos k_{{\rm y}}-\mu.
\end{equation}
Due to the wave function of the 4d$_{\rm xy}$-orbitals, the sign of
the next-nearest neighbor hopping is opposite to the case of cuprates.
Note that the unit such as $t_{1}=1$ is used in this subsection.

\begin{figure}[t]
\begin{center}
\vspace{2.5cm}
\end{center}
\caption{The results for the single band model \cite{Rf:Nomu.2}.
The calculated $T_{\rm c}$ for the triplet $p$-wave and singlet
$d_{x^2-y^2}$-wave states.}
\label{Fg:pdtc}
\end{figure}

In Fig.~\ref{Fg:pdtc}, we show $T_{\rm c}$ obtained by the TOP
for the $p$- and $d_{\rm x^2-y^2}$-wave superconductivity.
We can see that the triplet pairing is more stable than the singlet one
for the electron filling $n=1.33$, corresponding to the case of
Sr$_2$RuO$_4$ \cite{Rf:Mack.1}.
On the other hand, the singlet pairing gives higher $T_{\rm c}$
near the half-filling.
This is because the AF component of the spin fluctuation is enhanced
and induces the $d_{\rm x^2-y^2}$-wave superconductivity.
The latter situation corresponds to the case of cuprate superconductors.
Thus, the electron number away from the half-filling is essential
for the appearance of the triplet superconductivity.
The transition temperature in the triplet channel increases with
the electron number.
This trend is consistent with the recent experiment \cite{rf:okuda}.
This $n$-dependence of $T_{\rm c}$ is dominated by the effective interaction
\cite{rf:yanaseRu}, not by the electronic DOS.

Let us explain the detailed roles of the respective terms
in the perturbation expansion \cite{rf:yanaseRu}.
It should be noted that the RPA terms cancel each other and do $not$ appear
in the third-order term of $V_{\rm a}^{\rm t}(k,k')$
(Eqs.~(\ref{eq:third-anomalous-triplet})-
(\ref{eq:third-anomalous-triplet-vc})).
That is, all of the third-order terms are non-RPA terms.
In the present case, both the second- and third-order terms are attractive
in the triplet channel. Namely, the RPA- and non-RPA terms cooperatively
induce the triplet superconductivity.
In contrast to high-$T_{\rm c}$ cuprates, however, the contribution from
the RPA term is not large, since the momentum dependence of $\chi_{0}(\q)$
is insignificant. On the other hand, the momentum dependence of
the third-order terms are particularly attractive.
Figure \ref{Fg:comparison-p-wave} shows the development of
the eigenvalue of the \'Eliashberg equation in various calculations.
We can see that the contribution from the third-order terms is important
even in the moderately weak-coupling region;
it overcomes the contribution from the second-order term around $U \sim 3$,
where the experimental $T_{\rm c} = 1.5{\rm K}$ is obtained.
Thus, an important role of the non-RPA terms is expected for the
appearance of the triplet superconductivity.
We have confirmed that the averaged magnitude of the third-order term
is still 1/3 of the second-order term in this region,
which does not contradict with the perturbative treatment adopted here.
The effect of the higher-order corrections will be discussed in Appendix B.

\begin{figure}[t]
\begin{center}
\vspace{2.5cm}
\caption{The eigenvalue $\lambda_{{\rm e}}$ in the triplet channel
\cite{rf:yanaseRu}. The temperature is fixed to $T=0.005$
($\sim 15 {\rm K}$). Then, $\lambda_{{\rm e}} \sim 0.4$ corresponds to
the experimental value of $T_{\rm c}$$\sim$1.5K by the extrapolation.
Here, the normal self-energy is ignored for simplicity.}
\label{Fg:comparison-p-wave}
\end{center}
\end{figure}

On the other hand, the RPA gives the observed value of
$T_{{\rm c}} \sim 1.5 {\rm K}$ only in the vicinity
of the mean-field magnetic instability ($U=2$).
This is mainly because the RPA term is not so suitable to the triplet pairing
and the magnetic order is significantly overestimated in the RPA.
It is also important that the contribution from the magnetic fluctuation
to the triplet pairing is 1/3 of that to the singlet pairing
(see Eqs.~(\ref{eq:anomalous-rpa-singlet}) and
(\ref{eq:anomalous-rpa-triplet})).
Note that this situation is not improved in the FLEX approximation,
as is shown in the figure.
This is mainly because the de-pairing effect is very strong,
when the spin fluctuation has a local nature, namely,
nearly $\q$-independent.
We have pointed out in Sec.~3.2 that the de-pairing effect is usually
overestimated in the FLEX approximation, which is quite serious for
the $\gamma$ band just due to the local nature of the spin fluctuation.
Also from this reason, the spin fluctuation theory is not suitable for
the $\gamma$ band.
In fact, the $\alpha$ and $\beta$ bands provide larger value of
$\lambda_{\rm e}$ in the FLEX approximation at $U>2.5$,
where the spin susceptibility has a sharp peak around $\q \sim \Q_{\rm IAF}$.
The obtained $\lambda_{\rm e}$ in this case is shown in
Fig.~\ref{Fg:comparison-p-wave}, but too small to lead to the observed
value such as $T_{{\rm c}} \sim 1.5 {\rm K}$.
These comparisons with the spin fluctuation theory indicate
the importance of the non-RPA term.
The results of the perturbation theory have supported this indication.
Note that the TOP is the lowest-order theory for the non-RPA terms.

Here we comment on the roles of the perturbative terms
in the singlet channel.
They are qualitatively the same as in the previous subsections
for high-$T_{\rm c}$ cuprates and organic superconductors.
However, there is one significant difference;
the third-order term is totally repulsive
in the $d_{x^{2}-y^{2}}$-wave channel.
That is, the vertex corrections completely suppress the RPA terms in
the third order. It is expected that the spin-fluctuation theory 
loses its justification in this case.
The electron filling far from the half-filling is also essential
for this conclusion.

We have compared in detail the stability of the $p$-wave,
$d_{x^{2}-y^{2}}$-wave,
and $d_{xy}$-wave superconductivity \cite{rf:yanaseRu}.
We found that the $d_{x^{2}-y^{2}}$-wave symmetry is most stable within
the SOP, but the $p$-wave symmetry becomes stable owing to the 
third-order terms even in the rather weak-coupling region ($U \geq 1.5$).
Note that if we increase the electron number furthermore,
the $p$-wave symmetry is most stable even in the SOP.
In other words, the parameter region stabilizing the triplet
superconductivity is obtained in the lowest-order theory and it is much
enlarged in the TOP. 

\begin{figure}[t]
\begin{center}
\vspace{2.5cm}
\end{center}
\caption{The results for the three-band model \cite{Rf:Nomu.3}.
The calculated $T_{\rm c}$ for the spin-triplet superconductivity
at $n_{l}=1.4$. (a) $|J|$=$|J'|$=0.667$U'$. (b) $|J|$=$|J'|$=$U'$.}
\label{Fg:tc3b}
\end{figure}

In the next stage, the discussion is extended to the three band model
\cite{Rf:Nomu.3}, given by 
\begin{eqnarray}
 \label{eq:three-band-model}
 && H=H_0+H',
\end{eqnarray}
where $H_0$ is written as
\begin{eqnarray}
 \label{eq:three-band-model-kinetic}
 H_0 = \sum_{\smallk,l,s}{\e}_{l}(\k)
 c_{\smallk,l,s}^{\dagger}c_{\smallk,l,s}
 + \sum_{\smallk,s}g(\k)
 (c_{\smallk,{\rm yz},s}^{\dagger}c_{\smallk,{\rm xz},s}+ {\rm h.c.})
 + H_{\rm LS},
\end{eqnarray}
with the spin-orbit coupling term
\begin{eqnarray}
 && H_{\rm LS} = 2 \lambda \sum_{i} \LL_{i} \cdot \SS_{i}. 
 \label{eq:three-band-model-ls}
\end{eqnarray}
The Coulomb interaction term $H'$ is given by
\begin{eqnarray}
 && H' = U \sum_{i,l} n_{i,l,\uparrow} n_{i,l,\downarrow} 
 +U' \sum_{i,l>l'} n_{i,l} n_{i,l'} 
 +J \sum_{i,l>l'} (2 \SS_{i,l} \SS_{i,l'} + \frac{1}{2} n_{i,l} n_{i,l'})
 \nonumber \\ 
 &&+J' \sum_{i,l \neq l'}c_{i,l,\downarrow}^{\dag}c_{i,l,\uparrow}^{\dag}
 c_{i,l',\uparrow}c_{i,l',\downarrow},
\label{eq:three-band-model-interaction}
\end{eqnarray}
where $l$ denote the Wannier states $({\rm xy},{\rm yz},{\rm xz})$
corresponding to the
Ru$({\rm 4d}_{\rm xy},{\rm 4d}_{\rm yz},{\rm 4d}_{\rm xz})$ orbitals.
The band dispersions are chosen as
\begin{eqnarray}
 {\e}_{xy}(\k) &= &-2t_1({\cos}k_x+{\cos}k_y)
 - 4t_2{\cos}k_x{\cos}k_y-\mu_{xy},\\
 {\e}_{yz}(\k) &= &-2t_3{\cos}k_y-2t_4{\cos}k_x-\mu_{yz},\\
 {\e}_{xz}(\k) &= &-2t_3{\cos}k_x-2t_4{\cos}k_y-\mu_{xz},\\
 g(\k) &= &4t_5{\sin}k_x{\sin}k_y.
\end{eqnarray}
For the present calculation, we take
$(t_1,t_2,t_3,t_4,t_5)=(1.0,0.4,1.25,0.125,0.2)$.
Under these parameters, the kinetic energy term $H_0$ well reproduces
the three Fermi surfaces observed by the de Haas-van Alphen effect
\cite{Rf:Mack.1}.
The last term in Eq.~(\ref{eq:three-band-model-kinetic}) is
the spin-orbit interaction in the Ru ions,
which plays an essential role in stabilizing the chiral superconductivity. 
We will discuss this subject in Sec.~3.4.3. 
For the time being, we ignore this term for simplicity.
The interaction term $H'$ represents the on-site Coulomb interactions
including the intra-band repulsion $U$, inter-band repulsion $U'$,
Hund's coupling term $J$, and pair hopping term $J'$.
The parameters satisfy the relation $U > U' > |J| \sim |J'|$ and
$J < 0$ in the ordinary situation.

\begin{figure}[t]
\begin{center}
\vspace{2.5cm}
\end{center}
\caption{
Contour-plots of $\Delta_{a}(\k,{\rm i}\pi T)$ for
(a) $\alpha$, (b) $\beta$, and (c) $\gamma$ band \cite{Rf:Nomu.3}.
The thick lines represent the Fermi surfaces.
The parameters are chosen as $U=3.385$, $U'=|J|=|J'|=0.5U$,
and $T=0.003$.
In (d), the gap magnitude on each Fermi surface is shown
as a function of the inplane azimuthal angle $\phi$
with respect to the Ru-O bonding direction
(The numerical data is from the work in Ref.~\cite{Rf:Nomu.4}).
Here we have assumed that the orbital symmetry
of the pairs is represented by $k_x \pm \rm{i} k_y$
and the unit of energy is about 4500K.}
\label{Fg:DVqtot}
\end{figure}

We similarly expand the normal and anomalous self-energies up to
the third order with respect to $H'$, and determine the superconducting
instability by solving the \'Eliashberg equation extended
to the multi-band system.
Then, the mixing term with coupling constant $t_5$ is approximately treated
for simplicity. As for details, readers can refer Ref.~\cite{Rf:Nomu.3}.
As will be mentioned below, this procedure does not affect seriously
the calculated results for $T_{\rm c}$, which is shown in Fig.~\ref{Fg:tc3b}.
We obtain the qualitatively same results as those for the single band model.
The inter-band interactions enhance $T_{\rm c}$ further, which takes 
the value $T_{\rm c} \sim 0.003 \sim 10{\rm K}$ in the moderate coupling
region $U = 3 \sim 4$.
The realistic value $T_{\rm c}$$=1.5$K should be obtained in the weaker
coupling region.
Note here that we restrict the calculation above $T \geq 0.003 t_1$
in order to avoid the finite size effects.

Figure~\ref{Fg:DVqtot} shows the momentum dependence of the anomalous
self-energy $\Delta_a(\k,{\rm i}\pi T)$ ($a=\alpha, \beta, \gamma$).
This figure clearly shows that the pairing symmetry is $p$-wave.
An important point to be noted is that the $\gamma$ band has the largest
magnitude of the anomalous self-energy. It is suggested that
the condensation energy is mainly gained in the $\gamma$ band.
We can show that this situation is very robust for Sr$_2$RuO$_4$
\cite{rf:yanaseRu}.
That is, the magnitude of the anomalous self-energy remarkably depends
on the band. Agterberg {\it et al.} have proposed this situation,
called ``orbital dependent superconductivity'' (ODS) \cite{Rf:Agte.1}.
In the present case, the $\gamma$ band is mainly superconducting,
because the DOS is largest in $\gamma$.
If we adopt the small $t_3$, the $\alpha$ and $\beta$ bands have larger DOS.
Then, the situation is converse.
In the realistic case, superconductivity is dominated by the $\gamma$ band
and the contribution from the other bands are small. 
This is the physical background by which the theoretical approach 
based on the single-band model is justified. 
The approximate treatment for the mixing term $g(\k)$ is simultaneously 
justified, because this term only couples the $\alpha$ and $\beta$ bands. 

\begin{figure}[t]
\begin{center}
\vspace{2.5cm}
\end{center}
\caption{Contour-plots of the effective interaction on the $\gamma$ band
for (a) the triplet channel, $V_{\gamma,\gamma}^{\rm t}(k,k')$ and
(b) the singlet channel, $V_{\gamma,\gamma}^{\rm s}(k,k')$ \cite{Rf:Nomu.3}.
Here the Matsubara frequency is chosen as $k_0=k'_0={\rm i}\pi T$ and
${\bm k}'$ is fixed as pointed by the arrow.
The parameters are chosen as $U=3.385$, $U'=|J|=|J'|=0.5U$ and $T=0.007$.}
\label{Fg:Vkq3b}
\end{figure}

In order to clarify the leading scattering process in the Cooper channel,
we show the contour-plots of the effective interactions on the $\gamma$-band
(Fig.~\ref{Fg:Vkq3b}).
It is shown that the effective interaction in the triplet channel 
$V_{\gamma,\gamma}^{\rm t}(k,k')$ takes the large value
around $\k \approx -\k'$. 
This characteristic momentum dependence mainly originates from the
third-order terms and is attractive for the triplet pairing. 
This nature is in common with that of the single-band model.
Figure \ref{Fg:Vkq3b}(b) shows that the characteristic peak appears
in the effective interaction in the singlet channel
$V_{\gamma,\gamma}^{\rm s}(k,k')$ around $\k=\k'+\Q_{\rm IAF}$.
These peaks arise from the nesting nature of the quasi-one-dimensional
Fermi surfaces, namely $\alpha$ and $\beta$ bands.
This momentum dependence appears through the inter-orbital interactions
and favors the $d_{x^2-y^2}$-wave state.
Thus, the singlet pairing is enhanced by the multi-band effect.
Although the triplet pairing is relatively suppressed by the multi-band
effect, such an effect is not significant when the inter-orbital interactions
are small compared with the intra-band one.

Finally let us comment on the understanding of the power-law behaviors
below $T_{\rm c}$
\cite{Rf:Ishi.4,Rf:Nish.1,rf:bonalde,Rf:Tana.1,Rf:Izaw.1,Rf:Lupi.1}
on the basis of the above results.
Below $T_{\rm c}$, the Bogoliubov quasi-particle energy is obtained as
$E_{a}(\k)=
z_{a}(\k)\sqrt{\varepsilon_{a}(\k)^{2}+\Delta_{a}(\k)^{2}}$.
Here $z_{a}(\k)$ is the renormalization factor for the $a$-band 
and $\Delta_{a}(\k)=|\Delta^{\rm R}_{a}(\k,E_{a}(\k))|$,
where $\Delta^{\rm R}_{a}(k)$ is the analytic continuation of $\Delta_{a}(k)$.
If we consider the weak coupling case, i.e.,
$(T_{\rm c}, \Delta_{a}(\k)) \ll W$, an approximation
$\Delta_{a}(\k) \simeq |\Delta_{a}(\k,{\rm i}\pi T)|$
becomes very accurate around the Fermi surface.
Therefore, the momentum and orbital dependence of the excitation gap is
obtained from $|\Delta_{a}(\k,{\rm i}\pi T)|$, except for the factor
arising from $z_{a}(\k)$.
Since we found that the $a$ and $\k$-dependence of $z_{a}(\k)$ is not
outstanding in Sr$_2$RuO$_4$, the qualitative nature of the excitation
spectrum below $T_{\rm c}$ is captured by Fig.~\ref{Fg:DVqtot}.
Strictly speaking, the momentum and orbital dependence of $\Delta_{a}(k)$
is deformed below $T_{\rm c}$.
However, this deformation is usually small in the weak-coupling region 
$T_{\rm c} \ll W$ \cite{rf:leggett}.
Note that the anisotropy of the excitation gap tends to be smeared below
$T_{\rm c}$. 

As shown in Fig.~\ref{Fg:DVqtot}, the momentum dependence of the anomalous
self-energy is highly anisotropic and cannot be fitted by the simple form
$\Delta(\k)=\sin k_{\rm x}$ and so on.
Recently, we have successfully shown that the gap structure derived
in the present formulation is consistent with the power-law behavior
of the specific heat \cite{Rf:Nomu.4}.
A node-like structure on the $\beta$ Fermi surface remains in the chiral
state $\hat{d}(\k)=(k_x \pm {\rm i}k_y)\hat{z}$ and results
in the power-law behavior at low temperature.
Combination with the ``orbital dependent superconductivity'' gives
a whole temperature dependence, in agreement with experiments,
although a fitting of the parameters is required.
Thus, the power-law behaviors and the time reversal symmetry breaking
can coexist within the two-dimensional model.
One of the remaining problems is to clarify whether or not
the other experimental results, such as the NMR $1/T_{1}T$ \cite{Rf:Ishi.4}, 
magnetic field penetration depth \cite{rf:bonalde}, 
thermal conductivity \cite{Rf:Tana.1,Rf:Izaw.1}, and 
ultrasonic attenuation rate \cite{Rf:Lupi.1} can be consistently 
explained.

\subsubsection{Identification of the internal degree of freedom} 

In this subsection, we discuss the subject concerning the internal
degree of freedom in the triplet superconductivity.
Since the triplet superconductivity has the spin degree of freedom,
an internal degeneracy remains under the crystal field.
The spin part of the order parameter is assigned by the $d$-vector as
\begin{eqnarray}
 \left(
 \begin{array}{cc}
 \Delta_{\uparrow\uparrow}(k) & \Delta_{\uparrow\downarrow}(k) \\
 \Delta_{\downarrow\uparrow}(k) & \Delta_{\downarrow\downarrow}(k) \\
 \end{array}
 \right)
 =\left(
 \begin{array}{cc}
 -d_{{\rm x}}(k)+{\rm i}d_{{\rm y}}(k) & d_{{\rm z}}(k) \\
 d_{{\rm z}}(k) & d_{{\rm x}}(k)+{\rm i}d_{{\rm y}}(k) \\
 \end{array}
 \right)
 ={\rm i} \hat{d}(k) \hat{\sigma} \sigma_{y}.
\end{eqnarray}
According to the tetragonal crystal symmetry, six eigenstates are degenerate
in the calculation in Sec.~3.4.2, because we have ignored the spin-orbit
interaction described in the last term of
Eq.~(\ref{eq:three-band-model-kinetic}).
This degeneracy is lifted by the spin-orbit interaction and classified into
four one-dimensional representations and a two-dimensional representation
\cite{rf:sureview,Rf:Sigr.1}.

The internal structure in the superconductivity is an attractive 
character which does not usually exist in the singlet pairing.
Since the SC state is characterized by this structure, both
theoretical and experimental interests are widely stimulated.
Owing to the internal degree of freedom, the multiple phase diagram
can appear, as has been found in $^{3}$He and UPt$_3$, to which
many theoretical investigations have been devoted
\cite{rf:sureview,rf:leggett,rf:ohmi,rf:sauls}.
For $^{3}$He, the weak dipole interaction works as the leading
spin-orbit interaction which stabilizes the A-phase near $T_{\rm c}$ 
and explains the interesting properties in the NMR shift \cite{rf:leggett}.
For the superconducting materials, the internal degree of freedom
has been discussed in the phenomenological level
\cite{rf:sureview,rf:ohmi,rf:sauls},
which classifies the possible eigenstates and provides some experimental
methods to identify the internal structure.
It has been difficult to develop the microscopic theory on this issue,
since the previous triplet superconductors are basically heavy-fermion systems.
We have expected that the relatively simple electronic structure makes 
Sr$_2$RuO$_4$ to be the first example of such microscopic study. 
Here we show that such investigation provides some interesting
conclusions \cite{rf:yanaseRu}. 

The experimental investigation on the multiple phase diagram
does not seem to be completed for Sr$_2$RuO$_4$. 
It has been discovered that the second phase appears 
in the low-temperature and high magnetic-field region 
with $H$ being {\it precisely} parallel to the plane \cite{rf:deguchi}. 
The property of this second phase is not identified up to now. 
Recently, NMR measurement has been performed to resolve the nature 
under the perpendicular field \cite{rf:ishidaprivate}. 
Then, the $d$-vector perpendicular to the $z$-axis is indicated under 
the high field. 
The stabilized state under the zero magnetic field 
has been identified experimentally. 
It has been revealed from the phenomenological 
argument \cite{Rf:Sigr.1} that only the chiral state 
$\hat{d}(k) = (k_{{\rm x}} \pm {\rm i} k_{{\rm y}}) \hat{z}$ 
is consistent with the present experiments 
\cite{Rf:Ishi.1,Rf:Ishi.2,Rf:Duff.1,Rf:Luke.1}. 
Here, our microscopic analysis is focused on the internal structure 
under the zero magnetic field. The microscopic mechanism of the chiral 
superconductivity has been one of important issues, 
and the resolution of the internal structure will indicate the phase 
diagram under the magnetic field. 

First we briefly explain the formulation.
The matrix representation of the kinetic energy term $H_0$ is described as
\begin{eqnarray}
 \label{eq:three-band-model-LS}
 H_{0} \!=\! \sum_{\smallk,s} 
 \left(
  \begin{array}{ccc}
   c_{\smallk,{\rm yz},s}^{\dag} &
   c_{\smallk,{\rm xz},s}^{\dag} &
   c_{\smallk,{\rm xy},-s}^{\dag}\\
   \end{array}
 \right)
 \left(
  \begin{array}{ccc}
   \varepsilon_{\rm yz}(\k) & {\rm i}g(\k)-s \lambda & -s \lambda\\
   -{\rm i}g(\k) -s \lambda & \varepsilon_{\rm xz}(\k) &  \lambda\\
   -s \lambda &  \lambda & \varepsilon_{\rm xy}(\k)\\
  \end{array}
 \right)
 \left(
  \begin{array}{c}
   c_{\smallk,{\rm yz},s}\\
   c_{\smallk,{\rm xz},s}\\
   c_{\smallk,{\rm xy},-s}\\
  \end{array}
 \right).
\end{eqnarray}
We denote the $3 \times 3$ matrix in Eq.~(\ref{eq:three-band-model-LS})
as $\hat{H}_{0}(\k,s)$.
Note that a constant phase factor is multiplied to the $4d_{{\rm xz}}$
orbital in order to simplify the notation.
This definition makes the pair hopping term $J'$ negative between the
$4d_{{\rm xz}}$ and other orbitals.
We ignore the hybridization term $g(\k)$ in the following, since this term
is not important when the $\gamma$ band is mainly superconducting.
We can show that the conclusions are not affected even if the $\alpha$ and
$\beta$ bands are mainly superconducting \cite{rf:yanaseRu}.

New quasi-particles are obtained by diagonalizing the matrix
$\hat{H}_{0}(\k,s)$ through the unitary transformation.
This procedure corresponds to the transformation of the basis as
$(a_{\bm{k},1,s}^{\dag}, a_{\bm{k},2,s}^{\dag}, a_{\bm{k},3,s}^{\dag})
=(c_{\bm{k},{\rm yz},s}^{\dag}, c_{\bm{k},{\rm xz},s}^{\dag},
c_{\bm{k},{\rm xy},-s}^{\dag})\hat{U}(\k,s)$.
New quasi-particles created by the operators
$a_{\mbox{{\scriptsize \boldmath$k$}},l,s}^{\dag}$ are
characterized by the pseudo-orbital $l$ and pseudo-spin $s$.
The Fermi surface of the new quasi-particle is consistent with the quantum
oscillation measurement \cite{Rf:Mack.1}.
The rotational symmetry in the spin space is violated by the spin-orbit
interaction. The anisotropy of the spin susceptibility is derived from
this effect \cite{Rf:KKNg.1,Rf:Erem.1,rf:yanaseRu}.

We calculate the scattering vertex
$\Gamma(k,k',a,a',b',b,s_{1},s_{2},s_{3},s_{4})$ in the Cooper channel
by using the old basis tentatively, and then, apply the unitary transformation
in order to obtain the effective interactions in the new basis,
\begin{eqnarray}
 && \tilde{\Gamma}(k,k',\alpha,\alpha',\beta',\beta,s_{1},s_{2},s_{3},s_{4})=
 \nonumber \\
 && \sum_{a,a',b,b'} u^{*}_{a,\alpha}(\k,s_{1}) u^{*}_{a',\alpha'}(-\k,s_{2}) 
 \Gamma(k,k',a,a',b',b,s_{1},s_{2},s_{3},s_{4}) 
 u_{b',\beta'}(-\k',s_{3}) u_{b,\beta}(\k',s_{4}),
\end{eqnarray}
where we redefine the up (down) spin in $4d_{\rm xy}$-orbital as $s=-1$ ($s=1$).
It should be noticed that many terms are added to the scattering vertex
$\Gamma(k,k',a,a',b',b,s_{1},s_{2},s_{3},s_{4})$ through
the off-diagonal Green functions.
Moreover, we have to calculate the off-diagonal part, because it contributes
through the unitary transformation.
The diagonal part with respect to the pseudo-orbital
$\tilde{V}_{\alpha,\beta}(k,k',s_{1},s_{2},s_{3},s_{4})
=\tilde{\Gamma}(k,k',\alpha,\alpha,\beta,\beta,s_{1},s_{2},s_{3},s_{4})$
contributes to the superconducting instability. 
As a result, the \'Eliashberg equation is extended in the following way:
\begin{eqnarray}
 \lambda_{{\rm e}} \Delta_{\alpha,s_{1},s_{2}} (k) =
 -\sum_{\beta,k',s_{3},s_{4}}
 \tilde{V}_{\rm \alpha,\beta} (k,k',s_{2},s_{1},s_{3},s_{4})
 |\tilde{G}_{\beta}(k')|^{2} \Delta_{\beta,s_{3},s_{4}}(k').
\end{eqnarray}
However, the full calculation for this equation is very tedious, 
since the effective interaction 
$\tilde{V}_{\rm \alpha,\beta} (k,k',s_{1},s_{2},s_{3},s_{4})$ 
includes lots of terms. 
Then, we simplify the calculation by using two additional approximations:
One is the perturbation with respect to the spin-orbit coupling 
$\lambda$ and the other is based on the ODS argument. 
Since the spin-orbit interaction in Sr$_2$RuO$_4$ is much smaller than
the band width, the perturbation with respect to $\lambda$ is very 
accurate. We find that the first-order term vanishes and 
the second-order term is the lowest order. 
We restrict the estimation within this order. 
The practical procedure has been explained in Ref.~\cite{rf:yanaseRu}.
We can show that the ODS is very robust in Sr$_2$RuO$_4$,
while the main band depends on the parameters \cite{rf:yanaseRu}.
Then, the eigenvalue $\lambda_{{\rm e}}$ is almost determined
by the interaction between the main orbital.
Therefore, it is sufficient to take into account only the diagonal part
of the interaction
$\tilde{V}_{\rm \alpha}(k,k',s_{1},s_{2},s_{3},s_{4})=
\tilde{V}_{\rm \alpha,\alpha}(k,k',s_{1},s_{2},s_{3},s_{4})$,
and investigate each case where the main pseudo-orbital is
$\alpha=1,2$ or $3$. 

Finally, the eigenstates are classified by using the $d$-vector
representation. We find the following eigenstates:
(1) $\hat{d}(k) = k_{{\rm x}} \hat{x} \pm k_{{\rm y}} \hat{y}$,
(2) $\hat{d}(k) = k_{{\rm x}} \hat{y} \pm k_{{\rm y}} \hat{x}$,
and
(3) $\hat{d}(k) = (k_{{\rm x}} \pm {\rm i} k_{{\rm y}}) \hat{z}$
with remained two-fold degeneracy.
Although other linear combinations are possible, we have chosen
the symmetric states, which are expected to be stabilized
in order to gain the condensation energy.
The degeneracy in (1) and that in (2) is finally lifted by the weak mixing
term $g(\k)$, but the main results are not affected by this effect
\cite{rf:yanaseRu}.

We estimate the eigenvalue $\lambda_{\rm e}$ for each eigenstate (1)-(3)
and regard the state with maximum $T_{{\rm c}}$ to be stabilized.
This procedure is surely correct around $T \sim T_{\rm c}$.
In order to estimate $\Gamma$, we use the perturbation method
with respect to the interaction term $H'$ and take into account
all of the second-order terms as well as the third-order term
with coupling constant $U^{3}$.
This approximation is justified when $|J'|$, $|J|$, $U' < U < W$.
We include the third-order term in the estimation because it is important
for the pairing mechanism, as is explained in Sec.~3.4.2. 
The results on the $d$-vector are almost not affected 
by the third-order terms since they do not lift the internal degeneracy.

\begin{figure}
\begin{center}
\vspace{2.5cm}
\end{center}
\caption{The stabilized state in the perturbation theory.
Circles and triangles represent the state (3) and (1), respectively.
The solid line shows the realistic value $n_{\gamma}=1.33$.
The Fermi surface is electron(hole)-like in the left (right) side
of the dashed line.
The parameters are chosen as $U=5$, $U'=0.3U$ and $|J'|=|J|=0.2U$.}
\label{fg:dvectorz1.25review}
\end{figure}

Now let us show the results by the fully microscopic calculation. 
We find that the ``symmetry breaking interaction'', which violates 
the SU(2) symmetry in the $d$-vector space, requires 
the Hund's coupling term $J$. 
Considering that the superconductivity is mainly induced by the 
intra-orbital repulsion $U$, we conclude that 
the triplet superconductivity and the chiral superconductivity 
have a quite different origin. 
In the present case, the cross-term with coefficient $UJ$ is the 
leading contribution to the ``symmetry breaking interaction'' 
and stabilizes the $d$-vector parallel to the {\it z}-axis.
Figure \ref{fg:dvectorz1.25review} shows the phase diagram with respect
to the parameter $t_2/t_1$ and the electron number in the $\gamma$ band.
We can see that the chiral state (3) is robustly stabilized in the
experimentally relevant region, $t_2/t_1 \sim 0.4$ and $n_{\gamma} \sim 1.33$.
The other state (1) also appears, but only in the experimentally irrelevant
case where the $\gamma$-Fermi surface is hole-like.
The difference of $T_{\rm c}$ for each eigenstate is estimated as 
$2 \sim 4\%$, if we put the parameters as $2\lambda =0.1{\rm eV}$ 
and $W =2{\rm eV}$. 
This value seems to be surprisingly small, but this is a natural result,
since the ``symmetry breaking interaction'' is in the second order
with respect to $\lambda$.
Note that the chiral state (3) is not stabilized,
if the $\alpha$ and $\beta$ bands are mainly superconducting.
We find that the state (2) is more stable there. 

Here we comment on the phase diagram under the magnetic field.
Since the $d$-vector along the $z$-axis is stabilized under zero
magnetic field, the parallel magnetic field does not alter the
$d$-vector. Thus, the second phase transition observed
under the parallel magnetic field is not the rotation of the $d$-vector.
This phenomenon may be a transition within the
two-dimensional representation
$\hat{d}(k) = (k_{{\rm x}} \pm {\rm i} k_{{\rm x}})\hat{z}$
or that of the vortex configuration.
Our estimation indicates that the $d$-vector can be altered
by the small magnetic field along the $z$-axis,
since the splitting of the degeneracy is small.
While the $H_{\rm c2}$ is small in this direction as
$H_{\rm c2} \sim 0.07$Tesla, our estimation for splitting of
the degeneracy corresponds to the magnetic field $H \sim 0.1$Tesla.
Since our estimation includes ambiguity in the order unity,
the $d$-vector perpendicular to the $z$-direction is possible
in the high-field region.

In the next stage, we compare several pairing states from the viewpoint
of the internal degree of freedom.
Assuming the pairing symmetry and the main band, we investigate
which pairing state is stabilized in the respective case.
Then, the ``symmetry breaking interaction'' is microscopically estimated
within the lowest order.
We believe that this semi-phenomenological comparison is meaningful,
for which the applicability of the perturbation theory may be wider.
This is because the perturbation series of the 
``symmetry breaking interaction'' has smaller coefficient 
than the total effective interaction. 
The obtained results will restrict the possible pairing symmetry, which
is consistent with the time-reversal symmetry breaking \cite{Rf:Luke.1}.
We restrict the discussion to the two-dimensional model,
because the three-dimensional pairing state is hardly promising.

\begin{table}[t]
\begin{tabular}{|c||c|c|} \hline
&
$\gamma$-band & $\alpha$-, $\beta$-band \\\hline\hline
$p$-wave & $(k_{{\rm x}} \pm {\rm i} k_{{\rm y}}) \hat{z}$ &
$k_{{\rm y}} \hat{x} \pm k_{{\rm x}} \hat{y}$ \\\hline
$p$-wave (with node) & none & $k_{{\rm y}} \hat{x} \pm k_{{\rm x}} \hat{y}$
\\ \hline
$f_{x^{2}-y^{2}}$-wave
&
$(k_{{\rm x}}^{2}-k_{{\rm y}}^{2})(k_{{\rm y}}\hat{x} \pm k_{{\rm x}}\hat{y})$
&
$(k_{{\rm x}}^{2}-k_{{\rm y}}^{2})(k_{{\rm x}}\hat{x} \pm k_{{\rm y}}\hat{y})$
\\ \hline
$f_{xy}$-wave
&
$k_{{\rm x}} k_{{\rm y}} (k_{{\rm x}}\hat{x} \pm k_{{\rm y}}\hat{y})$
&
none
\\ \hline
\end{tabular}
\caption{Stabilized state for each symmetry and main band.}
\label{tab:symmetry-d-vector-relation}
\end{table}

The obtained results are summarized in
Table~\ref{tab:symmetry-d-vector-relation},
where the stabilized state is shown for each pairing symmetry.
We see that the chiral state is stabilized {\it only when}
the symmetry is the $p$-wave and the main band is $\gamma$.
The other paring states including the $f$-wave symmetry
\cite{Rf:Taki.1,rf:dahmRu,rf:grafRu} provide other $d$-vector,
which is incompatible with the time-reversal symmetry breaking.
After all, it is concluded that the most favorable pairing state is
the $p$-wave on the $\gamma$-band, as expected in the perturbation theory.
In other words, the perturbation theory provides the probable
pairing symmetry and furthermore the compatible $d$-vector
along the $\hat{z}$-axis. 

Here we point out that our systematic treatment for the spin-orbit
interaction is necessary to discuss the internal degree of freedom.
The spin-orbit interaction contributes to the symmetry breaking 
interaction through
(i) the virtual process in the effective interaction and
(ii) the unitary transformation of the quasi-particles.
When the chiral state is stabilized, the dominant contribution comes
from the cross-term of (i) and (ii).
Thus, we have to treat the two effects on the same footing.
The estimation of the effect (i) only \cite{rf:ogataRu} or the effect
(ii) only \cite{rf:ngls} may be inadequate in the microscopic theory.
We should further mention that the effective interaction derived here is
quite different from the phenomenological assumption in Ref.~\cite{rf:ngls}.
Indeed, the chiral state is not stabilized, if we take account of only
the effect (ii).

At the last of this section, we comment on the heavy-fermion superconductor
${\rm UPt}_{3}$, which is a spin-triplet superconductor exhibiting
three different superconducting phases under the magnetic field.
The phase diagram has been explained by assuming the weak spin-orbit
interaction \cite{rf:ohmi} because this assumption is necessary
to explain the NMR Knight shift \cite{rf:Tou1}.
However, this assumption has raised a serious question, since the spin-orbit
coupling is generally strong in the heavy-fermion system \cite{rf:sauls}.
We think that the result obtained here gives a clue to this question.
It has been shown that the Hund's coupling term is required for
the violation of the SU(2) symmetry in the $d$-vector space.
On the contrary, the SU(2) symmetry in the real spin space is violated
even for $J=0$; the spin susceptibility is anisotropic, for example.
Therefore, the violation of the SU(2) symmetry can be much smaller
in the $d$-vector space than in the real spin space.
In other words, it is possible that the anisotropy is almost absorbed
in the character of the quasi-particles, and only a weak anisotropy
is remained in the residual interaction. The examination of this possibility
for ${\rm UPt}_{3}$ will be an interesting future issue.

\par\vfill
\eject

%
%
\section{Pseudogap Phenomena}

\subsection{Overview}

In this section, we discuss the normal-state properties of
high-$T_{\rm c}$ cuprates.
As we have noted in the review of the experimental results in Sec.~2.1
\cite{rf:timusk}, many anomalous aspects of the normal state have been
central issues of high-$T_{\rm c}$ superconductors.
We should clarify the nature of the normal state, because the comprehensive
understanding of the whole phase diagram is highly desirable.
Among them, the pseudogap phenomena, which are widely observed in the 
optimally- to under-doped region (see Fig.~\ref{fig:high-Tcphasediagram2}),
have attracted much interests.
Since the superconductivity arises through the pseudogap state in
optimally- and under-doped systems, the resolution of the pseudogap state
is an essential subject for the high-$T_{\rm c}$ superconductivity. 
Furthermore, the pseudogap phenomena have many interesting aspects, 
because they are in sharp contrast to the conventional Fermi-liquid theory
\cite{rf:nozieresbook}.
The unusual nature has indicated an appearance of a new concept
in the condensed matter physics.

The introduction of our understanding on this issue is one of the purposes
of this review. Among many theoretical proposals, we have adopted
the ``pairing scenario'' in which the pseudogap is a precursor of
the superconducting (SC) gap.
This scenario has been indicated by several experimental results,
as have been explained in Sec.~2.1.
Since there are several kinds of  ``pairing scenario'',
we will explain later them somewhat in detail to avoid confusion.
Before introducing the pairing scenarios, we briefly review
other scenarios, which have been widely investigated.

An interesting proposal is the appearance of the RVB state
\cite{rf:andersonscience,rf:andersonbook}.
Historically, the RVB theory was proposed for the quantum spin system with
frustration \cite{rf:andersonRVB}.
Motivated by the high-$T_{\rm c}$ superconductivity, the superconductivity
arising from the RVB state has been investigated intensively
\cite{rf:baskaranRVB,rf:yoshidaRVB,rf:ruckensteinRVB,rf:isawaRVB,rf:suzumuraRVB,rf:kotliar,rf:yoshiokaRVB,rf:fukuyamaRVBreview,rf:tanamoto,rf:ioffeRVB,rf:nagaosa,rf:nagaosaTR,rf:onodaTR,rf:ichinoseRVB,rf:leeRVB}.
Naively speaking, the spin-liquid state is the underlying state
which results in the superconductivity if holes are lightly doped. 
Among the several descriptions, the slave-boson method for the $t$-$J$ model
has been widely used \cite{rf:suzumuraRVB,rf:fukuyamaRVBreview,rf:tanamoto}.
In this theory, two essential excitations, spinon and holon, appear
in the mean-field level and couple through a gauge field.
The pseudogap state is regarded as a singlet pairing state of the spinons
and the superconductivity is described as Bose-Einstein condensation
(BEC) of the holons.
Some characteristics of the pseudogap state as well as the
whole phase diagram are reproduced within the mean-field theory.
The electric transport is explained by taking account of
the U(1) gauge field \cite{rf:nagaosaTR,rf:onodaTR}.
The gauge-field theory has been extended to the SU(2) symmetry
\cite{rf:leeRVB}.
These approaches are essentially from the Mott insulating state.
The metallic state is described as a ``doped Mott insulator''.
This starting point is contrastive to our approach from
the Fermi-liquid state.

Another class of the proposal is the ``hidden order scenario'',
where some long-range order exists in the pseudogap state.
For example, the anti-ferromagnetism \cite{rf:schrieffer1988,rf:pinesPG} and
the ``stripe'' order \cite{rf:ichiokaPG} have been proposed as candidates.
Recently, ``the $d$-density wave'' has been proposed as a new type of
the long-range order \cite{rf:chakravarty}.
If a finite value of some ``hidden'' order parameter is confirmed
experimentally, the corresponding scenario will be the best candidate,
but it is not the case up to now.

In other class of the scenarios, a fluctuation of the order parameter is
considered as an origin of the pseudogap.
The fluctuation-induced pseudogap has been investigated from the old days
\cite{rf:abraham}.
Then, the (quasi-) one-dimensional Peierls transition was mainly
investigated \cite{rf:leericeanderson,rf:mjrice,rf:sadovskii}.
Since any long-range order does not exist at finite temperature,
an extraordinary wide critical region is expected in one dimension.
The pseudogap in the spectrum is expected as a precursor
of the long-range order.
This mechanism may be generally expected in the low-dimensional system.
The pairing scenario is also classified into the case.
Another candidate is the AF spin-fluctuation \cite{rf:dahmFLEX,rf:chubukovreview,rf:deiszFLEX,rf:kampfPG,rf:chubukovPG,rf:vilkAFPG}.
The pseudogap in the single-particle spectrum can be derived from
the spin-fluctuation theory.
The pseudogap first opens at the ``hot spot'', which is qualitatively
consistent with experiments.
A naive and crucial problem is on the magnetic excitation
which is observed in the NMR and neutron scattering (Sec.~2.1.2).
It will be a subject how the decrease of the magnetic excitation is derived
from the magnetic fluctuation itself.
An alternative understanding for the comprehensive phase diagram is
proposed by Emery {\it et al.} \cite{rf:emerystripe} on the basis of
the fluctuating ``stripe'' state. 
In the Emery's proposal, the proximity effect from the insulating region
induces the spin gap.

Now let us introduce the pairing scenarios, which are classified into
some kinds. Because this concept is adopted in this review,
a detailed explanation is given, including the related theories.
The superconducting phase transition is usually described by the BCS
theory which is an established mean-field theory.
As is well known, the superconducting correlation does not appear above
$T_{\rm c}$ within the mean-field theory.
Therefore, we have to consider the breakdown of the BCS theory in order
to discuss the pairing scenario.
There are following two origins for the breakdown.
One is the strong-coupling superconductivity and the other is
the low-dimensionality.
The justification of the BCS theory is based on the long coherence length,
which is $\xi_{0}=10^{2} \sim 10^{3}$ in the conventional superconductors
and makes the fluctuation negligible.
However, the strong-coupling nature of the superconductivity results in the
short coherence length, which causes the softening of the fluctuation.
Here, the coupling of the superconductivity is indicated by the parameter
$T_{\rm c}^{\rm MF}/E_{\rm F}$, which determines the coherence length
in the clean limit.
Note that the non-$s$-wave superconductivity is always clean.
Here, $T_{\rm c}^{\rm MF}$ is the transition temperature
in the mean-field theory and $E_{\rm F}$ is the Fermi energy
renormalized by the electron correlation.
The fluctuation is generally enhanced by the low-dimensionality. 
In the strictly two-dimensional case, $T_{\rm c}$ is always zero
according to the Mermin-Wagner theorem, expect for the KT transition.
Then, wide critical region is expected even in the weak-coupling case.
While the three-dimensional long-range order occurs owing to the inter-layer
coupling, the quasi-two-dimensionality induces the strong fluctuation.
These two conditions are surely satisfied in the high-$T_{\rm c}$ cuprates,
where $\xi_{0}=3 \sim 5$.

The first proposal of the pairing scenario was based on 
the Nozi$\grave{{\rm e}}$res and Schmitt-Rink (NSR) theory 
\cite{rf:LeggettNSR,rf:Nozieres}, where the central subject 
is the cross-over problem from the BCS superconductivity to 
the Bose-Einstein condensation (BEC). 
In the weak-coupling region, the Cooper-pairing and phase coherence occur
at the same time according to the BCS theory.
In contrast to that, in the strong-coupling limit, the fermions construct
pre-formed bosons above $T_{\rm c}$ and the BEC occurs at $T_{\rm c}$.
The latter situation is called ``real space pairing'' in contrast to
the ``momentum space pairing'' in the BCS theory.
The cross-over of two regions was first formulated at $T=0$ by
Leggett \cite{rf:LeggettNSR} and extended to the finite temperature by
Nozi$\grave{{\rm e}}$res and Schmitt-Rink \cite{rf:Nozieres}.
After that, the application to the two-dimensional system has been investigated
\cite{rf:randeriaNSR,rf:schmitt-rink,rf:sereneNSR,rf:tokumitu}.
These two regions are continuously described by adjusting the chemical
potential. We note that this argument does not prohibit the phase transition.
The first-order phase transition with phase separation has been reported
in the dynamical mean-field theory \cite{rf:keller,rf:capone}.
The NSR theory at finite temperature takes the lowest-order correction to
the thermodynamic potential ($\Omega_{\rm B}$), which is shown 
in Fig.~\ref{fig:nozieres}. 
The particle number is obtained as $n=n_{\rm F}+n_{\rm B}$, 
where $n_{\rm F}$ is the usual Fermion contribution and $n_{\rm B}$
is the contribution from the fluctuation, namely
$n_{\rm B}=-\partial \Omega_{\rm B}/\partial \mu$.
The chemical potential is set below the conduction band
in the strong-coupling region.
Then, the fermionic excitation is fully gapped even in the anisotropic
superconductivity \cite{rf:LeggettNSR}.
In this limit, the system is regarded as a bosonic system with
residual interactions.
It should be noted that the NSR theory is basically justified
in the low-density system, since the shift of the chemical potential
is the leading effect in the low-density limit.

\begin{figure}[t]
\begin{center}
\vspace{2.5cm}
\caption{The correction to the thermodynamic potential in the NSR theory.}
\label{fig:nozieres}
\end{center} 
\end{figure} 

When the BCS-BEC cross-over was proposed for the pseudogap phenomenon
\cite{rf:micnasreview,rf:melo,rf:randeriareview},
the pseudogap state was regarded as a cross-over region.
The strong-coupling limit is evidently not relevant for cuprates,
because the excitation is clearly gapless along the diagonal direction.
Note here that this picture is adopted in the phenomenological
theories which assume the coexistence of the fermions and bosons
\cite{rf:ranninger,rf:geshkenbein,rf:perali}. 
By taking advantage of this proposal, intensive studies have been 
devoted to the cross-over problem 
\cite{rf:micnasreview,rf:melo,rf:randeriareview,rf:haussmann,rf:strinati,rf:engelbrecht,rf:stintzing,rf:janko,rf:koikegamiNSR,rf:kobayashiNSR,rf:hertog,rf:andrenacci,rf:maly,rf:chen}.
For example, the description beyond the original NSR theory has been 
given by the self-consistent T-matrix (SCT) approximation 
\cite{rf:haussmann} and by the pairing approximation \cite{rf:maly,rf:chen}.
The electronic spectrum has been calculated for the essentially 
low-density model by the T-matrix approach 
\cite{rf:janko,rf:fresard,rf:micnas,rf:perali3D,rf:kagan,rf:kyung,rf:hottaPG}. 
In particular, Janko {\it et al.} have shown that the pseudogap 
in the single-particle spectrum appears near the cross-over region 
in the 3D jellium model. Then, the importance of the self-energy correction 
has been pointed out. 
The quantum Monte Carlo simulation has been performed as a non-perturbative
method for the same problem \cite{rf:scalettar,rf:moreoQMC,rf:randeriaQMC,rf:trivedi,rf:singer,rf:vilk,rf:allen}.
Then, the results in the strong-coupling region commonly show the pseudogap
in the single-particle and magnetic excitations both in the high- and
low-density region.

We, however, believe that the BCS-BEC cross-over is not an essential
viewpoint for the high-$T_{\rm c}$ cuprates \cite{rf:yanasePG}. 
This point of view is especially important in order to clarify the
origin of the pseudogap. 
In Sec.~4.2, we show that the excitation gap is induced by the SC 
fluctuation even far from the cross-over region, if we adopt an 
appropriate model \cite{rf:yanasePG}. 
Then, the anomalous behaviors of the self-energy correction,
which is derived by the T-matrix approach, play an essential role.
This scenario is rather conventional in the fluctuation theory
\cite{rf:abraham,rf:leericeanderson,rf:mjrice,rf:sadovskii},
but several conditions are required in the SC fluctuation theory.

The self-energy correction arising from the SC fluctuation has been
investigated in 1970's \cite{rf:abraham,rf:varlamovreview}, which shows
the decrease of the electronic DOS due to the SC fluctuation.
However, it seems that this idea has not attracted much interests, 
because this is a very weak effect in the conventional superconductor.
We would like to stress that this common knowledge should be altered
for cuprates, where the SC fluctuation is remarkably enhanced
by the short coherence length and the quasi-two-dimensionality.
Then, the more pronounced phenomenon is derived;
the pseudogap appears in the single-particle spectral function.
The high-$T_{\rm c}$ cuprates are probably the first example
for the clear appearance of this phenomenon.

It should be noticed that the T-matrix approach is commonly used for
the NSR theory and for the estimation of the self-energy.
Actually, if we regard the Fig.~\ref{fig:nozieres} as a Luttinger's
functional Eq.~(\ref{eq:variation}), the self-energy should be estimated
by the SCT approximation.
Two aspects of the T-matrix approach, the BCS-BEC cross-over and 
the pseudogap, appear depending on the selected microscopic model. 
Indeed, the essential differences arise from the electron density and
the dimensionality.
If we use the 3D jellium model, the pseudogap appears from the cross-over
region \cite{rf:janko,rf:perali3D}.
Then, the leading scattering process is called ``resonance scattering''
since the SC fluctuation has a resonance nature in this region. 
On the other hand, if we use the quasi-2D square lattice model near the 
half-filling (the nearly half-filled system should be regarded as 
high density), the self-energy correction gives rise to the pseudogap
under more moderate condition. 

This is mainly because the quasi-two-dimensionality leads to the strong
fluctuation and induces the pseudogap for the relatively small value of
$T_{\rm c}^{\rm MF}/E_{\rm F}$.
Moreover, the high electron density makes the BCS-BEC cross-over difficult.
If we consider the repulsive Hubbard model as in Sec.~4.3, the BCS-BEC
cross-over is practically impossible.
These differences are schematically shown in Fig.~\ref{fig:modeldep}.
We see from this observation that the 3D jellium model is too simplified
to discuss the high-$T_{\rm c}$ cuprates, while several aspects are common.
In two dimensions, the resonance nature of the SC fluctuation is
not necessary for the pseudogap, while the renormalization effect induces
more propagating character \cite{rf:yanasePG}.
In the following part, however, we will commonly use the term
``resonance scattering'' keeping in mind that a kind of the resonance
between quasi-particle state and Cooper-pairing state is expected.

\begin{figure}
\begin{center}
\vspace{2.5cm}
\caption{The schematic figure for the relation between the NSR theory and
the pseudogap phenomena. (a) The three-dimensional jellium model and 
(b) two-dimensional square lattice model around the half-filling. 
The horizontal axis corresponds to the control parameter 
$T_{\rm c}^{\rm MF}/E_{\rm F}$ for the fixed electron number.
It should be noticed that the Fermi energy $E_{\rm F}$ is renormalized
in the SCES.} 
\label{fig:modeldep}
\end{center} 
\end{figure} 

Along this line, the T-matrix approach has been applied to
the two-dimensional lattice model including the $d$-wave symmetry 
\cite{rf:yanasePG,rf:dagotto,rf:yanaseMG,rf:kobayashi,rf:onodaPG,rf:metzner}. 
Engelbrecht {\it et al.} have applied the SCT approximation 
to the quarter-filled model \cite{rf:dagotto} and the present authors 
have performed both non-self-consistent and self-consistent 
T-matrix approximations for the nearly half-filled case using the
TDGL expansion \cite{rf:yanasePG,rf:yanaseMG}.
Subsequently, the microscopic theory starting from the repulsive Hubbard 
model has been developed \cite{rf:yanaseFLEXPG} (Sec.~4.3). 
Then, the pseudogap phenomena are systematically explained including 
their doping dependence. 
It should be stressed that this microscopic theory is a natural 
extension of the theories discussed in Sec.~3.2. 
Thus, we obtain a comprehensive understanding on the pairing mechanism 
and the normal state properties in Sec.~4.3. 

The T-matrix approximation is the lowest-order theory with respect to
the SC fluctuation.
Then, the pseudogap appears under the reasonable condition,
namely intermediate coupling region with a sufficient Fermi degeneracy.
Indeed, the chemical potential is almost not affected by the fluctuation.
Basically, the pairing symmetry is not important;
the $s$-wave model shows the pseudogap in a similar way \cite{rf:metzner}
except for the difference in the momentum dependence.
The SCT approximation is one of the methods to include 
the higher-order correction, which is taken into account within 
the renormalization of the Green function. 
Then, the pseudogap in the single-particle spectrum is suppressed,
while that of the DOS clearly remains. 
We will give a critical comment on the relevance of the partial 
summation performed in the SCT approximation (Sec.~4.2.5). 
The two-particle self-consistent (TPSC) approximation has been 
proposed for an improvement of the T-matrix approximation 
\cite{rf:kyungd-wave,rf:kyungTPSC}, 
where the coupling constants are adjusted to satisfy a sum-rule
while the renormalization for the single-particle Green function is
neglected.

A similar, but another approach is the ``phase fluctuation theory''
\cite{rf:emery,rf:franz,rf:kwon,rf:phase,rf:eckl,rf:tesanovic}. 
In this approach, the amplitude of the SC order parameter is fixed and
only the phase variable is taken into account. 
This approach has been indicated \cite{rf:emery} by the small London
constant $\Lambda$ in the under-doped region \cite{rf:uemura,rf:bonnmagpen}, 
which is usually attributed to the small superfluid density $n_{\rm s}$. 
In the case of the Kosterlitz-Thouless (KT) transition \cite{rf:KT},
which is the sole phase transition in the two-dimensional system,
the transition temperature $T_{\rm KT}$ is proportional to the
London constant. If we consider the actual three-dimensional
critical temperature is close to $T_{\rm KT}$, the $\delta$-dependence
of $T_{\rm c}$ is qualitatively consistent according to the Uemura plot
$\Lambda \propto \delta$.
The pseudogap state is described as the phase disordered state
which is expected above $T_{\rm KT}$.
The gap structure in the electronic spectrum is generally smeared
by the phase fluctuation in the SC state \cite{rf:yanaseSC}.
It has been, however, shown that the gap structure remains in the
phase disordered state \cite{rf:franz,rf:kwon,rf:phase,rf:eckl}.

We consider that the validity of the phase only theory on the 
electronic structure is not clear up to now. 
It is surely justified deeply below $T_{\rm c}$ where the phase fluctuation
is dominant but affects perturbatively, 
although the coupling to the density fluctuation should be considered
there \cite{rf:paramekantiFL}.
The description of the normal state may require further discussions;
the amplitude fluctuation certainly affects there.

Here we stress that the phase fluctuation theory is not intrinsically
incompatible with the T-matrix approach. 
We consider that the former is an approach from the low-temperature side
and the latter is that from the high-temperature side, 
which will be complementary. \
Because the T-matrix approach is a perturbation theory with respect to
the SC fluctuation, the precise description in the deeply critical region
is difficult \cite{rf:T-matrixKT}.
The two dimension is clearly below the upper critical dimension.
The phase only theory should be regarded as a phenomenological description
for it, while the (renormalized) Gaussian fluctuation region cannot be
described. 

An almost exact treatment is allowed for the one-dimensional model
in which the fermions couple to the classical field
(Lee-Rice-Anderson model) \cite{rf:leericeanderson}.
The classical field describes the fluctuating order parameter
for which the static approximation is done.
The statistical ensemble of the classical field is an essential assumption
in this model; the Gaussian fluctuation model has been investigated,
intensively.
While the Peierls transition has been focused in early years,
the AF spin fluctuation and SC fluctuation have been studied,
motivated by the high-$T_{\rm c}$ cuprates.
The pseudogap is commonly induced by the strong fluctuation
which is characteristic in the one-dimensional system.
In 70's, Sadovskii has given an ``exact'' solution where the fermion
self-energy is represented by the recursive continued fraction
\cite{rf:sadovskii}.
This method has been applied to the AF spin fluctuation with focus
on the high-$T_{\rm c}$ cuprates \cite{rf:mckenziePG}.
The extension to the two-dimensional case is difficult, but
it has been performed by Schmalian et al. \cite{rf:schmalianPG}
by using the separable form of the spin susceptibility. 
However, Tchernyshyov has pointed out the technical error of
the Sadovskii's solution, which is actually an approximation
\cite{rf:tchernyshyov1D}.
A sophisticated numerical method has shown that the Sadovskii's solution
is qualitatively a good approximation for the complex order
parameter \cite{rf:millisPG,rf:kopietzPG};
the commensurate CDW is not the case, but the SC is the case. 
Then, the accuracy of the approximations usable in higher dimensions
has been investigated \cite{rf:millisPG}.
It is shown that the SCT approximation properly reproduces the 
low-energy DOS, while the high-energy behavior $\omega \sim \Delta$ 
is not. 

Since the Gaussian fluctuation is inappropriate near the critical point,
a more sophisticated treatment for the statistical ensemble is required
for the critical behaviors.
The cross-over from the Gaussian fluctuation region to the phase fluctuation
region has been described on the basis of the Lee-Rice-Anderson model
with a careful treatment of the statistical ensemble \cite{rf:monienPG}.
Then, the excitation gap appears more clearly in the phase fluctuation region,
while the $\omega^{2}$-behavior appears in the Gaussian-fluctuation model
with complex order parameter.

Another solvable model is the BCS pairing model with
sufficiently long-range attractive interaction.
This model is almost exactly solvable in any dimension with the use of
the Sadovskii's method \cite{rf:fujimotoPG}.
Then, the calculation is technically reduced to the zero-dimensional problem
and quantum fluctuation plays a dominant role for the low-energy spectrum.
Note that the correction pointed out by Tchernyshyov \cite{rf:tchernyshyov1D}
does not exist in this model. 
The result shows that the non-self-consistent T-matrix approximation
appropriately reproduces the asymptotic behavior of the self-energy,
while the SCT approximation does not. 
This model is a clear example in which the partial summation of
the higher-order terms is dangerous. 

Although these simplified models cannot describe the phase space restriction
which is characteristic in higher dimensions, the importance of the
vertex correction will be common to the two-dimensional cases,
at least quantitatively.
It is expected that the exact solution will lie between the
non-self-consistent and self-consistent T-matrix approximations
which are used in the next section.
The higher-order correction in the two-dimensional case will be discussed
in the last subsection in Sec.~4.2. 

\subsection{General Theory}

In this section, we clarify the basic mechanism of the pseudogap phenomena
with the use of the attractive model.
It is explained how the pseudogap appears from the SC fluctuation.
The microscopic theory starting from the repulsive Hubbard model,
as will be given in Sec.~4.3, basically justifies this mechanism.

In Sec.~4.2.1, the drastic effect of the SC fluctuation on
the single-particle properties is demonstrated.
The basic idea of the ``resonance scattering'' is introduced.
The importance of the quasi-two-dimensionality is discussed in Sec.~4.2.2.
In Sec.~4.2.3, the pseudogap phenomena under the magnetic field are
investigated. The coherent understanding on the doping dependence
can be examined by the effects of the magnetic field.
The obtained results are qualitatively consistent with
the high field NMR measurements including their doping dependence.
From these results, $\kappa$-(ET)$_{2}$X compounds are suggested
to be another candidate for an appearance of the pseudogap.
Section 4.2.4 is devoted to the discussion on the $\kappa$-(ET)$_{2}$X
compounds. Basically, the theoretical analysis is performed with
the use of the T-matrix or self-consistent T-matrix approximation.
The higher-order correction beyond the T-matrix approximation is
discussed in Section 4.2.5.
It is shown that the vertex correction enhances the pseudogap furthermore.

\subsubsection{Basic mechanism of the pseudogap} 

As is explained in Sec.~2.1, several experimental results have indicated
the close relation between the SC gap and pseudogap.
Among them, rich information obtained from ARPES has led us to the
pairing scenario.
ARPES clearly shows the pseudogap in the single-particle spectrum.
In the theoretical point of view, the single-particle quantities
are simple compared with the two-particle quantities
such as the magnetic and transport properties.
Therefore, the study on the single-particle properties will
capture the basic mechanism of the pseudogap most clearly.
We focus on the estimation of the single-particle Green function 
in this section. 
It is simply expected that the pseudogap in the two-particle spectrum 
is derived from the pseudogap in the single-particle spectrum. 

It is instructive to adopt the attractive model in order to
investigate the fundamental roles of the SC fluctuation.
\begin{eqnarray}
  \label{eq:attractivemodel}
  && H=\sum_{{\bm k},s}\varepsilon(\k)c_{\bm{k},s}^{\dag}c_{\bm{k},s}
  +\sum_{\bm{k,k',q}}V_{\bm{k,k'}}
  c_{\bm{q}/2-\bm{k'},\downarrow}^{\dag}c_{\bm{q}/2+\bm{k'},\uparrow}^{\dag}
  c_{\bm{q}/2+\bm{k},\uparrow}c_{\bm{q}/2-\bm{k},\downarrow},
\end{eqnarray}
where $V_{\bm{k,k'}}$ is the $d_{x^2-y^2}$-wave pairing interaction,
given in the separable form as
\begin{eqnarray}
  \label{eq:d-wave}
  V_{\bm{k,k'}} = g \phi_{\bm{k}}\phi_{\bm{k'}}.
\end{eqnarray}
Here $g$ is negative and $\phi_{\bm{k}}$ is the $d_{x^2-y^2}$-wave form factor,
given by $\phi_{\bm{k}}=\cos k_{x}-\cos k_{y}$.
We use the tight-binding dispersion $\varepsilon(\k)$ defined in
Eq.~(\ref{eq:high-tc-dispersion}) for $t'/t=0.25$ and $\delta=0.10$.

The above Hamiltonian is an effective model in which the paring interaction
affects the renormalized quasi-particles.
Since the energy scale is renormalized, the magnitude of the gap and
$T_{\rm c}$ are relatively larger than those in the original model.
It should be noted again that the renormalized Fermi energy is used
in the control parameter for the SC fluctuation,
$T_{\rm c}^{\rm MF}/E_{\rm F}$.
This attractive model is a very simplified one, but we expect that
the fundamental features of the SC fluctuation and its effects are included.

\begin{figure}[t]
\begin{center}
\vspace{2.5cm}
\caption{(a) The scattering vertex represented by the ladder diagrams
(T-matrix). (b) The diagrammatic representation of the self-energy
in the T-matrix approach.}
\label{fig:vertexa}
\end{center}
\end{figure}

The SC fluctuation is diagrammatically described by the T-matrix
(Fig.~\ref{fig:vertexa}(a)) which is a propagator of the SC fluctuation.
The scattering vertex arising from the T-matrix is factorized into
$\phi_{\smallk-\smallq/2} t(q) \phi_{\smallk'-\smallq/2}$, where
\begin{eqnarray}
  \label{eq:t-matrix-small}
  && t(\mbox{\boldmath$q$},{\rm i}\Omega_{n})=
  [g^{-1}+\chi_{{\rm p0}}(\mbox{\boldmath$q$},{\rm i}\Omega_{n})]^{-1},
\end{eqnarray}
and
\begin{eqnarray}
  \label{eq:irreducible-pair-susceptibility-d-wave}
  && \chi_{{\rm p0}}(\mbox{\boldmath$q$},{\rm i}\Omega_{n}) =
  T \sum_{k}
  {\mit{\it G}} (\mbox{\boldmath$q/2$}+\mbox{\boldmath$k$},{\rm i}\omega_{m})
  {\mit{\it G}} (\mbox{\boldmath$q/2$}-\mbox{\boldmath$k$},
  {\rm i}\Omega_{n} - {\rm i}\omega_{m})
  \phi_{\mbox{\scriptsize \boldmath$k$}}^{2}.
\end{eqnarray}
The SC phase transition is determined by the divergence of
the SC susceptibility $t(\mbox{\boldmath$0$},0)$, namely
$1+g\chi_{{\rm p0}}(\mbox{\boldmath$0$},0)=0$.
This is called ``Thouless criterion'', which is equivalent to
the BCS theory in the weak coupling limit.

From early years, the effects of the SC fluctuation on the two-particle
correlation functions have been investigated intensively
\cite{rf:varlamovreview,rf:AL,rf:Maki}, 
which are observable in the conventional superconductors.
On the other hand, effects on the single-particle properties have not
attracted interests except for the early investigation
\cite{rf:abraham}, probably because this effect is very weak
in the conventional (low-$T_{\rm c}$ and 3D) superconductors.
However, the SC fluctuation seriously affects the electronic state
in the high-$T_{\rm c}$ superconductors in the following way. 

The anomalous contribution from the SC fluctuation generally originates 
from the enhanced T-matrix around $\bm{q}=\Omega=0$. 
Therefore, we tentatively expand the reciprocal of the T-matrix as 
\begin{eqnarray}
  \label{eq:TDGL}
  t(\mbox{\boldmath$q$},\Omega)=
  \frac{g}{t_{0}+b\mbox{\boldmath$q$}^{2}-(a_{1}+{\rm i}a_{2})\Omega}.
\end{eqnarray}
This procedure corresponds to the time-dependent Ginzburg-Landau (TDGL)
expansion. The TDGL parameters are expressed for the Gaussian fluctuation as
\begin{eqnarray}
  \label{eq:TDGLparameter}
  && t_{0} = 1 + g \int {\rm d}\varepsilon
  \frac{\tanh(\frac{\varepsilon}{2 T})}{2 \varepsilon}
  \rho_{{\rm d}}(\varepsilon)
  \cong |g| \rho_{{\rm d}}(0) \frac{T-T_{{\rm c}}}{T_{{\rm c}}}, \\
  \label{eq:TDGLparameterb}
  && b=|g| \int {\rm d}\varepsilon
  \frac{\rho_{{\rm d}}(\varepsilon) v_{{\rm F}}^{2}}{16 \varepsilon}
  \frac{\partial^{2} f(\varepsilon)}{\partial \varepsilon^{2}}
  \cong |g| \rho_{{\rm d}}(0) \frac{7 \zeta(3)}{32 (\pi T)^{2}}
  v_{{\rm F}}^{2}, \\
  && a_{1} = |g| \int {\rm d}\varepsilon
  \frac{\tanh(\frac{\varepsilon}{2 T})}{(2 \varepsilon)^{2}}
  \rho_{{\rm d}}(\varepsilon) \cong \frac{1}{2}
  \rho_{{\rm d}}'(0)/\rho_{{\rm d}}(0) \cong
  \frac{1}{2} |g|
  \frac{\partial \chi_{{\rm p0}}(\mbox{\boldmath$0$},0)}{\partial \mu},\\
  && a_{2} = |g| \rho_{{\rm d}}(0) \frac{\pi}{8 T},
\end{eqnarray}
where $\zeta(3)$ is the Riemann's zeta function and $v_{{\rm F}}$
is the averaged quasi-particle velocity on the Fermi surface.
Here the SC fluctuation is mainly determined by the electronic state
around $(\pi,0)$, owing to the $d$-wave symmetry. 
Then, $v_{{\rm F}}$ can be regarded as the velocity at the hot spot 
(see Fig.~\ref{fig:HotCold.tgif}). 
The flat dispersion observed around $(\pi,0)$ means that the effective
Fermi energy $E_{\rm F}=v_{{\rm F}}k_{{\rm F}}$ is small.
We have defined the effective DOS as
\begin{eqnarray}
  \label{eq:rhod}
  && \rho_{{\rm d}}(\varepsilon) = \sum_{\bm{k}}
  A(\bm{k},\varepsilon)\phi_{\bm{k}}^{2},
\end{eqnarray}
and $\rho_{{\rm d}}'(0)$ is its derivative at $\varepsilon=0$.
We have denoted the spectral function as
$A(\bm{k},\omega)=-(1/\pi){\rm Im}G^{{\rm R}}(\bm{k},\omega)$.

The mass term $t_{0}=1+g \chi_{0}(\bm{0},0)$ represents the closeness
to the critical point.
The fluctuation effect gradually emerges as $t_{0}$ is reduced.
The parameter $ b $ is generally related to the coherence length $\xi_{0}$
as $b \propto \xi_{0}^{2} \propto v_{{\rm F}}^{2}/T^{2}$,
and therefore sufficiently large in the weak-coupling superconductor.
This is the reason why the fluctuation is negligible in conventional
superconductors.
We have defined the parameter $T_{{\rm c}}^{{\rm MF}}/E_{{\rm F}} \sim
T_{\rm c}^{\rm MF}/v_{\rm F}k_{\rm F}$ as
``superconducting coupling'' which is an index of the fluctuation effect.
In case of high-$T_{\rm c}$ cuprates, $T_{\rm c}^{\rm MF}$ is large and
$v_{\rm F}$ is renormalized, and therefore the coherence length is
remarkably small $\xi_{0} = 3 \sim 5$.
This is an essential background of the pseudogap phenomena.

The parameter $a_{2}$ represents the dissipation and expresses
the time scale of the fluctuation.
This parameter is also small in the strong coupling region owing to the
high $T_{\rm c}$ and the renormalization effect by the pseudogap
\cite{rf:yanasePG,rf:tchernyshyov}.
The real part $a_{1}$ is usually ignored because this term is
in the higher order than the imaginary part $a_{2}$ with respect to
the small parameter $T_{\rm c}^{\rm MF}/E_{\rm F}$
in the weak-coupling region.
In the strong-coupling limit, the absolute value $|a_{1}|$ can exceed
$a_{2}$ and then, the SC fluctuation has a ``resonance nature''
\cite{rf:janko}.
While this situation is somewhat extreme, there is no justification
to ignore $a_{1}$ in the intermediate coupling region.
The sign of the parameter $a_{1}$ is related to the Hall conductivity
close to $T_{\rm c}$ \cite{rf:ebisawa}.
The anomalous sign of the Hall conductivity in high-$T_{\rm c}$ cuprates
has stimulated much interests \cite{rf:nagaoka} and
remains as an open problem.

The anomalous properties in the single-particle spectrum
\cite{rf:ding,rf:shenPG,rf:normanPG} are described by the
self-energy correction. Here, we estimate it within the one-loop order
(Fig.~\ref{fig:vertexa}(b)) as
\begin{eqnarray}
\label{eq:T-matrixap}
  && {\mit{\it \Sigma}}(\mbox{\boldmath$k$},{\rm i}\omega_{n})=
  T \sum_{\mbox{\boldmath$q$},{\rm i}\Omega_{m}}
  t(\mbox{\boldmath$q$},{\rm i}\Omega_{m})
  {\mit{\it G}} (\mbox{\boldmath$q$}-\mbox{\boldmath$k$},
  {\rm i}\Omega_{m} - {\rm i}\omega_{n})
  \phi_{\mbox{\scriptsize \boldmath$k$}-\mbox{\scriptsize \boldmath$q$}/2}^{2}.
\end{eqnarray}
This procedure corresponds to the T-matrix approximation.
In general, the T-matrix around $\mbox{\boldmath$q$}=0$ gives rise to the
anomalous properties and that far from $\mbox{\boldmath$q$}=0$ gives rise to
the Fermi-liquid properties.
The former process is very small in the weak-coupling case since $b \gg 1$.
Then, the self-energy shows the Fermi-liquid behaviors except for just
the vicinity to the critical point.
On the other hand, the former overcomes the latter in the intermediate-
or strong-coupling region, and then pseudogap appears. 
This criterion for the pseudogap is moderate in the 
quasi-two-dimensional system like high-$T_{\rm c}$ cuprates. 

We evaluate the anomalous contribution to the imaginary part 
by using the TDGL parameters \cite{rf:yanasePG}, 
\begin{eqnarray}
 \label{eq:imaginary-self-energy1D}
 {\rm Im} {\mit{\it \Sigma}}^{{\rm R}} (\mbox{\boldmath$k$},\omega)
 & = & -|g| \phi_{\mbox{\scriptsize \boldmath$k$}}^{2}
 \frac{T \xi_{\rm GL}}{4 \pi b v_{\mbox{\scriptsize \boldmath$k$}}}
 \frac{\xi_{\rm GL}^{-1}}
 {\alpha^{2}/v_{\mbox{\scriptsize \boldmath$k$}}^{2}+\xi_{\rm GL}^{-2}}
 \hspace{13mm}  ({\rm for \hspace{2mm} 1D}),
 \\
 \label{eq:imaginary-self-energy2D}
 & = & -|g| \phi_{\mbox{\scriptsize \boldmath$k$}}^{2} 
      \frac{T}{4 b v_{\mbox{\scriptsize \boldmath$k$}}} 
      \frac{1}{
      \sqrt{
      \alpha^{2}/v_{\mbox{\scriptsize \boldmath$k$}}^{2} + \xi_{\rm GL}^{-2}}}
 \hspace{12mm}  ({\rm for \hspace{2mm} 2D}),
 \\
 \label{eq:imaginary-self-energy3D}
 & = & 
    -|g| \phi_{\mbox{\scriptsize \boldmath$k$}}^{2} 
    \frac{T d}{8 \pi b v_{\mbox{\scriptsize \boldmath$k$}}}  
    \log [\frac{q_{\rm c}^{2}}
    {\alpha^{2}/v_{\mbox{\scriptsize \boldmath$k$}}^{2}}+\xi_{\rm GL}^{-2}] 
 \hspace{6mm}  ({\rm for \hspace{2mm} 3D}),
\end{eqnarray}
where we have defined as $\alpha=\omega+\varepsilon(\k)$ and
the GL correlation length is defined as $\xi_{\rm GL}=\sqrt{b/t_{0}}$.
These expressions are correct at
$|\alpha| \leq \sqrt{2 M v_{\mbox{\scriptsize \boldmath$k$}}^{2} T} \sim T$
where $1/2M=b/{\rm max}\{a_{1},a_{2}\}$.
We have estimated for the one- and three-dimensional cases for a comparison.
Equations (\ref{eq:imaginary-self-energy2D}) and
(\ref{eq:imaginary-self-energy3D}) clearly show that the anomalous part 
of the self-energy increases when the parameter $b$ is small and/or 
the system is two-dimensional.
The real part can be obtained by the Kramers-Kronig relation as
\begin{eqnarray}
  \label{eq:real-self-energy1D}
  {\rm Re} {\mit{\it \Sigma}}^{{\rm R}} (\mbox{\boldmath$k$},\omega)
  & = &
  |g| \phi_{\mbox{\scriptsize \boldmath$k$}}^{2}
  \frac{T \xi_{\rm GL}}{4 \pi b v_{\mbox{\scriptsize \boldmath$k$}}}
  \frac{\alpha/v_{\mbox{\scriptsize \boldmath$k$}}}
       {\alpha^{2}/v_{\mbox{\scriptsize \boldmath$k$}}^{2}+\xi_{\rm GL}^{-2}}
\hspace{55mm}  ({\rm for \hspace{2mm} 1D}),
\\
  \label{eq:real-self-energy2D}
  {\rm Re} {\mit{\it \Sigma}}^{{\rm R}} (\mbox{\boldmath$k$},\omega)
  & = &
  |g| \phi_{\mbox{\scriptsize \boldmath$k$}}^{2}
  \frac{T}{4 \pi b v_{\mbox{\scriptsize \boldmath$k$}}}
      \frac{1}{
      \sqrt{
      \alpha^{2}/v_{\mbox{\scriptsize \boldmath$k$}}^{2}+\xi_{\rm GL}^{-2}}}
  \log [\frac{\alpha/v_{\mbox{\scriptsize \boldmath$k$}}
                +\sqrt{\alpha^{2}/v_{\mbox{\scriptsize \boldmath$k$}}^{2}
                       +\xi_{\rm GL}^{-2}}}
               {\alpha/v_{\mbox{\scriptsize \boldmath$k$}}
                -\sqrt{\alpha^{2}/v_{\mbox{\scriptsize \boldmath$k$}}^{2}
                       +\xi_{\rm GL}^{-2}}}]
\hspace{5mm}  ({\rm for \hspace{2mm} 2D}),
\\
  \label{eq:real-self-energyasinptotic}
  & = &
  \left\{
    \begin{array}{ll}
    |g| \phi_{\mbox{\scriptsize \boldmath$k$}}^{2}
       \frac{T}{2 \pi b} \frac{1}{\alpha}
       \log [\frac{2 \alpha}{\xi_{\rm GL}^{-1}
       v_{\mbox{\scriptsize \boldmath$k$}}}]
    & (|\alpha| \gg \xi_{\rm GL}^{-1}v_{\mbox{\scriptsize \boldmath$k$}})
\\
    |g| \phi_{\mbox{\scriptsize \boldmath$k$}}^{2}
      \frac{T \xi_{\rm GL}^{2}}
      {2 \pi b v_{\mbox{\scriptsize \boldmath$k$}}^{2}} \alpha
    & (|\alpha| \ll \xi_{\rm GL}^{-1}v_{\mbox{\scriptsize \boldmath$k$}})
   \end{array}
  \right.
\end{eqnarray}
These expressions correspond to the classical approximation,
which is justified near the critical point.
We note that the expression for the real part is not so accurate
in the low-frequency region; the dynamics of the SC fluctuation affects
on it. However, the qualitative behaviors are correctly grasped by
Eqs.~(\ref{eq:real-self-energy1D}) and (\ref{eq:real-self-energy2D}).

\begin{figure}[t]
\begin{center}
\vspace{2.5cm}
\caption{The characteristic behaviors of (a) the real part and 
(b) imaginary part of the self-energy induced by the 
``resonance scattering''. (c) The spectral function.}
\label{fig:selfreal}
\end{center}
\end{figure}

We show the typical behaviors in Fig.~\ref{fig:selfreal}.
It should be noticed that the real part of the self-energy shows the positive
slope around $\omega=0$, and there, the imaginary part shows the sharp peak
in its absolute value.
Both features are very anomalous compared with the conventional
Fermi-liquid theory.
This is caused by the ``resonance scattering'' from the SC fluctuation,
which is identified to be the origin of the pseudogap phenomena.
Owing to the anomalous properties of the self-energy, the spectral function
clearly shows the pseudogap (Fig.~\ref{fig:selfreal}(c)). 

\begin{figure}[t]
\begin{center}
\vspace{2.5cm}
\caption{(a) The real and (b) imaginary part of the self-energy 
obtained by the T-matrix approximation. 
(c) The spectral function at $\k=(\pi,0.15\pi)$. 
The inset in (a) and (b) shows the enlarged results for $g/t=-0.5$.
The inset of (c) shows the relation between $2\Delta$ and
$T_{{\rm c}}^{{\rm MF}}$. (d) DOS.
The temperature is chosen as $T=1.2 T_{\rm c}$ in all figures.}
\label{fig:selfr}
\end{center}
\end{figure}

Let us provide a more simple explanation on these anomalous features.
If we disregard the detailed structure around $\omega=0$, 
the self-energy is simply approximated as 
\begin{eqnarray}
  \label{eq:staticap}
  {\mit{\it \Sigma}}^{{\rm R}}(\mbox{\boldmath$k$},\omega) =
  \frac{\Delta^{2} \phi_{\mbox{\scriptsize \boldmath$k$}}^{2}}
  {\omega + \varepsilon(\k) + {\rm i}\delta}, \\
  \Delta^{2} = - \sum_{q} t(\q,{\rm i}\Omega_{n}).
\end{eqnarray}
This approximation is obtained by ignoring the $q$-dependence of the
Green function and that of the form factor in Eq.~(\ref{eq:T-matrixap}).
This form of the self-energy is assumed in the pairing approximation
\cite{rf:maly,rf:chen}.
Also, the ARPES experiments have been analyzed by assuming the similar
form of the self-energy \cite{rf:norman2}.

If the self-energy is expressed as Eq.~(\ref{eq:staticap}),
the Green function has the same momentum and frequency dependence as 
the normal Green function in the SC state as 
\begin{eqnarray}
  \label{eq:Greenfunctionsuper}
  G^{{\rm R}}(\mbox{\boldmath$k$},\omega)=\frac{\omega+\varepsilon(\k)}
  {(\omega+\varepsilon(\k))
  (\omega-\varepsilon(\k))-\Delta^{2}
  \phi_{\mbox{\scriptsize \boldmath$k$}}^{2}}.
\end{eqnarray}
Then, the energy gap appears in the single-particle spectrum, 
where the quasi-particle energy is obtained as 
\begin{eqnarray}
  \label{eq:gap}
  E_{\mbox{\scriptsize \boldmath$k$}}
  = \pm \sqrt{\varepsilon(\k)^{2}
  +\Delta^{2} \phi_{\mbox{\scriptsize \boldmath$k$}}^{2}}.
\end{eqnarray} 
The amplitude of the gap is determined by the total weight of the (thermal)
SC fluctuation, while the gap is related to the order parameter
in the mean-field theory.

\begin{figure}[t]
\begin{center}
\vspace{2.5cm}
\caption{The static part of the T-matrix $|t(q,0)|$ at $T=1.2 T_{{\rm c}}$.
(a) $g/t=-0.5$. (b) $g/t=-1.5$.}
\label{fig:rephi0.25}
\end{center}
\end{figure}

The above expression Eq.~(\ref{eq:staticap}) is somewhat extreme and
inaccurate around $\omega=0$ (see Eqs.~(\ref{eq:imaginary-self-energy2D})
and ~(\ref{eq:real-self-energy2D})).
We calculate Eqs.~(\ref{eq:t-matrix-small}),
(\ref{eq:irreducible-pair-susceptibility-d-wave}), and (\ref{eq:T-matrixap})
without using the TDGL expansion and show the self-energy, spectral function,
and DOS in Fig.~\ref{fig:selfr}.
Here $\phi_{\mbox{\scriptsize \boldmath$k$-\boldmath$q$/2}}$ is replaced
with $(\phi_{\mbox{\scriptsize \boldmath$q$-\boldmath$k$}}+
\phi_{\mbox{\scriptsize \boldmath$k$}})/2$ in order to restore
the periodicity, but the results are not affected by this procedure.
We see that the Fermi-liquid behaviors appear in the weak-coupling 
region ($|g|/t=0.5$) and the anomalous behavior emerges in the
intermediate-coupling region ($|g|/t=1 \sim 2$).
In the latter case, the pseudogap appears both in the spectral function
(Fig.~\ref{fig:selfr}(c)) and in the DOS (Fig.~\ref{fig:selfr}(d)).
The inset of Fig.~\ref{fig:selfr}(c) shows that the magnitude of the
pseudogap is approximately scaled by the mean-field transition temperature
$T_{{\rm c}}^{{\rm MF}}$.
The pseudogap in the two-particle excitations including the NMR $1/T_{1}T$
is derived from the pseudogap in the single-particle excitations
(see Secs.~4.2.3 and 4.3.3). 
It should be noted that the shift of the chemical potential is negligible
in this region. Thus, the scenario based on the NSR theory is irrelevant 
in the two-dimensional high-density system. 

As is mentioned above, the drastic change from the weak- to
the strong-coupling region mainly arises through the parameter $b$.
It is clearly understood from Fig.~\ref{fig:rephi0.25} that
the total weight of the SC fluctuation is very small
in the weak-coupling region, owing to the large value of $b$.
In this case, the anomalous contribution from SC fluctuation is smeared by
the Fermi-liquid contribution. This is why the pseudogap does not appear
in the weak-coupling superconductor.

The momentum dependence of the spectral function is shown in
Fig.~\ref{fig:momentum}.
The anomalous contribution to the self-energy is smeared around
the gap node.
Therefore, the ``Fermi arc'' appears around the diagonal direction
and disappears as approaching to the critical point.
These features are observed behaviors in the ARPES measurements
\cite{rf:normanPG}.

\begin{figure}[t]
\begin{center}
\vspace{2.5cm}
\caption{The momentum dependence of the spectral function
for $g/t=-1.5$ and $T=1.2 T_{{\rm c}}$.
The momentum and the Fermi surface are shown in the inset.}
\label{fig:momentum}
\end{center}
\end{figure}

Figure \ref{fig:spectrtemp} shows the temperature dependence.
It is shown that the pseudogap is smeared as the temperature increases.
The pseudogap disappears around $T^{*}$ which is estimated as 
$T^{*} \sim 2 T_{\rm c}^{\rm MF}$ at $g/t=-1.5$.
It should be stressed that the pseudogap is closed by the broadening,
but not by the disappearance of the gap amplitude.
This feature is in contrast to Eq.~(\ref{eq:staticap})
which is revealed to be a very rough estimate.
This nature of the pseudogap closing is also consistent with the
experiments \cite{rf:ding,rf:shenPG,rf:normanPG,rf:renner,rf:miyakawa}.

\begin{figure}[t]
\begin{center}
\vspace{2.5cm}
\caption{The temperature dependence of (a) the spectral function and
(b) DOS. Here, $g/t=-1.5$ and the momentum is A in the inset of
Fig.~\ref{fig:momentum}.}
\label{fig:spectrtemp}
\end{center}
\end{figure}

The transition temperature suppressed by the fluctuation has been calculated
by the SCT approximation \cite{rf:yanasePG}, where the T-matrix,
self-energy, and Green function are determined self-consistently.
Here, $T_{\rm c}$ is determined by the condition $t_{0}=0.02$ by taking
account of the weak three-dimensionality.
If we choose a strictly two-dimensional model, $T_{\rm c}$ is always zero.
We see from Fig.~\ref{fig:T-matrixphase} that $T_{\rm c}$ is remarkably
suppressed as the superconducting coupling increases,
while the suppression rapidly disappears in the weak-coupling region.
Thus, the wide critical region is expected in the strong-coupling
superconductor. Since the under-doped cuprates correspond to the case,
the pseudogap is widely observed.

\begin{figure}[t]
\begin{center}
\vspace{2.5cm}
\caption{The phase diagram of the attractive model.
The transition temperatures in the mean field theory and
that in the SCT approximation are shown.}
\label{fig:T-matrixphase}
\end{center}
\end{figure}

The pseudogap is basically obtained also in the SCT approximation.
For example, the DOS is shown in Fig.~\ref{fig:sctDOS}.
Exceptionally the single-particle spectral function shows no
(or very weak) gap structure. 
On this point, the quantum Monte Carlo simulation for the attractive
Hubbard model gives more similar results to those of
the non-self-consistent T-matrix approximation,
where the pseudogap is clearly observed in the spectral function.
It seems that the estimation of the spectral function is not improved
by the higher-order corrections within the SCT.
We will discuss this point in Sec.~4.2.5.

\begin{figure}[t]
\begin{center}
\vspace{2.5cm}
\caption{DOS in the SCT approximation.}
\label{fig:sctDOS}
\end{center}
\end{figure}

The extension of the SCT approximation to the superconducting state
\cite{rf:yanaseSC,rf:jujoSC} has shown that the SC fluctuation is rapidly
suppressed in the ordered state while the phase-mode remains gapless.
Then, the order parameter grows more rapidly than the BCS theory.
This property is a general consequence of the critical fluctuation.
The rapid growth of the order parameter is commonly obtained in the FLEX
approximation \cite{rf:monthouxFLEX,rf:paoFLEX,rf:takimotoFLEX2}
in which the feedback effect is its origin.
These two effects additively contribute in high-$T_{\rm c}$ cuprates.
It should be noticed that we cannot directly measure the order parameter
in high-$T_{\rm c}$ cuprates because the pseudogap already exists
in the spectrum.
Instead, the London constant $\Lambda$, which is proportional to the
inverse square of the magnetic field penetration depth $\lambda_{\rm L}$ 
($\Lambda = 1/4\pi\lambda_{\rm L}^{2}$), 
is a better probe of this property \cite{rf:yanaseSC}.
This quantity reflects the long-range coherence of the order parameter
and rapidly develops just below $T_{\rm c}$ with
the increase of the order parameter.
Experimental results have confirmed the rapid growth, which precisely
corresponds to the universality class of the 3D XY model
\cite{rf:panagopoulosmagpen,rf:kamal}.

\subsubsection{Effect of three-dimensionality}

So far, we have stressed that the strong SC fluctuation gives rise to
pseudogap phenomena. In that case, it is necessary that the total weight
of the SC fluctuation $\sum_{q} |t(\q,{\rm i}\Omega_{n})|$ is large compared
with the conventional superconductor.
The previous discussion in Sec.~4.2.1 has pointed out two conditions for
the pseudogap phenomena. One is the strong-coupling superconductivity,
namely the short coherence length. Another is the quasi-two-dimensionality.
The former point has been clarified in the previous subsection.
In this subsection, we discuss the latter point in more detail.

First, we show the expression for the imaginary part of the self-energy
in case of the layered superconductor,
\begin{eqnarray}
 \label{eq:imaginary-self-energyaniso}
 {\rm Im} {\mit{\it \Sigma}}^{{\rm R}} (\mbox{\boldmath$k$},\omega)=
 -|g| \phi_{\mbox{\scriptsize \boldmath$k$}}^{2}
 \frac{T d}{8 \pi v_{\mbox{\scriptsize \boldmath$k$}} b \gamma}
 \log [\frac{
     \gamma q_{\rm c} + \sqrt{\gamma^{2} q_{\rm c}^{2}+ \xi_{\rm GL}^{-2}
     +\alpha^{2}/v_{\mbox{\scriptsize \boldmath$k$}}^{2}}}
     {-\gamma q_{\rm c} + \sqrt{\gamma^{2} q_{\rm c}^{2}+ \xi_{\rm GL}^{-2}
      +\alpha^{2}/v_{\mbox{\scriptsize \boldmath$k$}}^{2}}}], 
\end{eqnarray}
where $\gamma$ is the anisotropy of the coherence length
$\gamma=\xi_{{\rm c}0}/\xi_{0}$ and $q_{\rm c}=\pi/d$ is the cut-off
momentum along the {\it c}-axis. $d$ is the inter-layer spacing.
We can easily confirm that Eq.~(\ref{eq:imaginary-self-energyaniso}) is
reduced to Eq.~(\ref{eq:imaginary-self-energy2D}) and
Eq.~(\ref{eq:imaginary-self-energy3D}) in the limit $\gamma \rightarrow 0$
and $\gamma \rightarrow 1$, respectively.
It is clearly shown that the quasi-particle damping
$-{\rm Im} {\mit{\it \Sigma}}^{{\rm R}} (\mbox{\boldmath$k$},0)$ shows
power-law anomaly in 2D case ($\propto 1/\sqrt{t_{0}}$), but
only weak logarithmic anomaly in 3D case ($\propto -\log t_{0}$).
This is evidently due to the restriction of the phase space by the
dimensionality.
Thus, the quasi-two-dimensionality generally enhances the anomaly.
It is necessary for the two-dimensional behaviors that the c-axis coherence
length is smaller than the inter-layer spacing $\xi_{{\rm c}0} \leq d$,
which is well satisfied in high-$T_{\rm c}$ cuprates.
Then, the critical behavior shows the dimensional cross-over at
\begin{eqnarray} 
  \label{eq:dimensional-cross-over}
  \xi_{\rm GL}^{c}=\gamma \xi_{\rm GL} \sim d/2,
\end{eqnarray} 
where $\xi_{\rm GL}^{c}$ is the GL correlation length along the c-axis.
Note that the critical behaviors are always three-dimensional
near the critical point.
Even in the three-dimensional region, the anisotropy enhances the self-energy
through the factor $\gamma^{-1}$.
We note that the dimensional cross-over occurs for the variation of the
frequency $\omega$; two-dimensional behavior is more robust at finite
frequency.

\begin{figure}[t]
\begin{center}
\vspace{2.5cm}
\caption{The self-energy obtained by the T-matrix approximation for
$t_{\rm z}/t=0.1$ ($T_{\rm c}^{\rm MF}/t=0.337$) and
$t_{\rm z}/t=1$ ($T_{\rm c}^{\rm MF}/t=0.195$) \cite{rf:jujo}.
(a) real part and (b) imaginary part at $g/t=-2$.
The decrease of mean field $T_{\rm c}^{\rm MF}$
by the three-dimensionality is not essential for the results.}
\label{fig:jujofig1a}
\end{center}
\end{figure}

We have actually taken into account the weak three-dimensionality and
numerically estimated the single-particle properties 
like Sec.~4.2.1 \cite{rf:jujo}. 
The anisotropy is represented by the parameter $t_{\rm z}/t$, 
where $t_{\rm z}$ is the hopping matrix along the $c$-axis. 
We show the results of the T-matrix approximation for the different value
of $t_{\rm z}/t$. The self-energy is shown in Fig.~\ref{fig:jujofig1a}.
We see that the anomalous behaviors which are shown in Fig.~\ref{fig:selfr}
appear in the quasi-two-dimensional case, $t_{\rm z}/t=0.1$.
Thus, the pseudogap is robust for the weak three-dimensionality.
This robustness also results from the strong coupling nature of
the superconductivity, as is explained later. 
On the contrary, typical Fermi-liquid behaviors appear in the 
isotropic three-dimensional case, $t_{\rm z}/t=1$. 
In this case, the contribution from the SC fluctuation shows
only a weak anomaly (Eq.~(\ref{eq:imaginary-self-energy3D})), 
which is almost smeared by the Fermi-liquid contribution.
To make sure, the pseudogap phenomena are possible even
in the three-dimensional system \cite{rf:preosti}.
However, somewhat hard condition, such as very large coupling constant
or just vicinity to the critical point, is needed.
Recent calculation for the three-dimensional jellium model has concluded
that the pseudogap appears in the vicinity of the BCS-BEC cross-over
region \cite{rf:perali3D}.
It is concluded that the appearance of the pseudogap is quantitatively
difficult for the three-dimensional systems like heavy-fermion compounds.

Furthermore we comment on the dimensional cross-over
in the quasi-two-dimensional systems.
The two-dimensional behaviors generally appear
in the high-temperature region and three-dimensional ones are expected
below the cross-over temperature.
The condition for the dimensional cross-over has been given in
Eq.~(\ref{eq:dimensional-cross-over}).
Then, the cross-over temperature ($T_{\rm cr}$) is estimated as
\begin{eqnarray}
  \label{eq:dimensional-cross-over-temp}
  (T_{\rm cr}-T_{\rm c})/T_{\rm c} \sim 4 (\xi_{{\rm c}0}/d)^{2}
  \sim 4 \xi_{0}^{2} (t_{\rm z}/t)^{2},
\end{eqnarray}
since $\gamma/d \sim t_{\rm z}/t$.
If we choose $t_{\rm z}/t = 0.05$ and $\xi_{0} =4$,
the above expression results in $(T_{\rm cr}-T_{\rm c})/T_{\rm c} \sim 0.04$.
Thus, two-dimensional behaviors are widely expected in high-$T_{\rm c}$ 
cuprates. This is also owing to the short coherence length.
The strong coupling nature of the superconductivity plays an essential
role also in this stage.

From the results in Secs.~4.2.1 and 4.2.2, the doping dependence is
understood in the following way: 
According to the tunnelling experiment \cite{rf:odatransport,rf:harris}, 
the gap amplitude at $T \ll T_{\rm c}$ increases as $\delta$ decreases. 
This result indicates that $T_{\rm c}^{\rm MF}$ increases 
with decreasing $\delta$.
Combined with the renormalization of the quasi-particle velocity,
which is clearly shown by ARPES \cite{rf:shenrep,rf:ARPESreview},
the superconducting coupling increases with under-doping.
Then, two-dimensionality simultaneously becomes enhanced \cite{rf:iye}.
As a result, the clear pseudogap is expected in the under-doped region,
while the pseudogap gradually disappears with doping.
Since the pseudogap onset temperature is scaled by 
$T_{\rm c}^{\rm MF}$ in the intermediate-coupling region, 
$T^{*}$ increases with decreasing $\delta$.
These features are quite consistent with the phase diagram
in high-$T_{\rm c}$ cuprates.
The microscopic knowledge is necessary for $T_{\rm c}$, which is
scaled by the renormalized Fermi energy.
We will show the results of the microscopic calculation in Sec.~4.3, 
where the appropriate doping dependence is reproduced. 

\subsubsection{Effect of the magnetic field}

It is generally expected that superconducting phenomena show characteristic
behaviors under the magnetic field.
The quantum phase of the order parameter is modulated by the magnetic field,
and then the characteristic value of the magnetic field is very small
in the conventional superconductors.
Such a remarkable response is contrastive to the magnetic properties
where the magnetic field dependence is usually small,
because the Zeeman term ($\sim 10 {\rm K}$) is much smaller than
the exchange coupling $J (\sim 1000 {\rm K}$).
Therefore, the behaviors under the magnetic field are one of the
important tests for the theoretical scenarios based on the SC fluctuation.
Actually, the magnetic field dependence of the pseudogap has been
investigated by recent experiments. 
Among them, we discuss here the NMR $1/T_{1}T$ which has been
measured for high-$T_{\rm c}$ superconductors 
\cite{rf:zheng,rf:gorny,rf:mitrovic,rf:zheng2} and
organic superconductors \cite{rf:matsumotokanoda}.
Important knowledge has been obtained for both systems.

In this subsection, we focus on the high-$T_{\rm c}$ cuprates.
The experimental results are summarized as follows.
The magnetic field dependence is little observed in the under-doped
region \cite{rf:zheng,rf:gorny}. 
In particular, the onset temperature $T^{*}$ does not depend on
the magnetic field up to 20 Tesla.
The effect of the magnetic field is observed only close to $T=T_{\rm c}$.
It is also observed that the transition temperature is suppressed
by the magnetic field.
On the contrary, the effects of the magnetic field are clearly visible
around the optimally-doped region in which only the weak pseudogap phenomenon
is observed in the narrow temperature region \cite{rf:mitrovic,rf:zheng2}.

The theoretical results for the magnetic field dependence are obtained
in the following way \cite{rf:yanaseMG}.
Although the following explanation basically considers the Gaussian
fluctuation, qualitatively the same tendency is expected for the critical
fluctuation.
We consider the magnetic field applied along the c-axis
($B \parallel \vec{c}$) in accordance with the experimental condition.
The main effect of the magnetic field is the Landau level quantization of the
SC fluctuation, which is quasi-classically expressed by the replacement of
the quadratic term in the T-matrix as
$\mbox{\boldmath$q$}^{2} \Rightarrow 4{\rm e} B (n+1/2)$
\cite{rf:dorin,rf:eschrig}.
The Landau quantization of the quasi-particle is ignored in the
usual condition $\omega_{{\rm c}} \tau \ll 1$,
where $\omega_{{\rm c}}$ is the cyclotron frequency.

The Landau quantization has the following two important effects.
One is the suppression of superconductivity.
The mass term is increased by the zero-point oscillation energy as
$t_{0} \Rightarrow t_{0} + 2 b {\rm e} B $,
which corresponds to the lowest Landau level.
The SC fluctuation is suppressed by this effect.
The other is the Landau degeneracy which generally enhances the fluctuation
because the effective dimension is reduced.
The transition temperature is further reduced by the enhanced fluctuation.
Considering at the fixed temperature in accordance with the experiments,
the dominant effect is the former.
Then, we see that the characteristic magnetic field $B_{{\rm ch}}$
for the SC fluctuation is expressed as
$B_{{\rm ch}} \sim t_{0}/b=\xi_{{\rm GL}}^{-2}$.
That is, the effects of the magnetic field are scaled by
the magnetic flux penetrating the correlated area $\xi_{{\rm GL}}^{2}$.

As we have stressed above, pseudogap appears in the short coherence
length superconductors.
Then, a large value of the characteristic magnetic field is expected.
Moreover, the parameter $t_{0}$ around the onset temperature $T^{*}$ is
not so small in the strong-coupling superconductors.
Then, the characteristic magnetic field $ B_{{\rm ch}} $ around
$T=T^{*}$ is in the same order as $H_{{\rm c2}}(0)$ which is 
over 100Tesla in the under-doped region. 
This is the simple but robust reason why the magnetic field dependence
of the onset temperature $T^{*}$ is not observed in the under-doped region.
Because the characteristic magnetic field is generally small in the
conventional superconductors, this magnetic field independence was sometimes
interpreted as a negative evidence for the pairing scenario.
However, this interpretation is inappropriate.

The magnetic field effects are apparent in the temperature
region close to $T_{{\rm c}}$, since the GL correlation length
$\xi_{{\rm GL}}$ diverges at the critical point.
This effect is actually observed even in the under-doped region
\cite{rf:zheng}. 
These behaviors are confirmed by the numerical calculation based
on the T-matrix approximation \cite{rf:yanaseMG}, 
where the magnetic field with experimentally relevant order 
$B$$\sim$10 Tesla is considered. 
Then, the Kubo formula is used to estimate the correlation function 
and the effect of the SC fluctuation is included in 
the self-energy correction. 
The vertex correction corresponding to the Maki-term is usually
not important for the $d$-wave superconductor \cite{rf:eschrig,rf:kuboki}.
Figure \ref{fig:figure0.8} shows the results for $g/t=-1.6$
where the pseudogap is clearly observed (Sec.~4.2.1). 
We see the magnetic field independence below $T^{*}$ and 
magnetic field dependence around $T_{\rm c}$. 

It should be noted that the magnetic field dependence of $T_{\rm c}$
is explained by taking account of the critical fluctuation.
The latter effect ``Landau degeneracy'' significantly enhances the
fluctuation and suppresses the $T_{\rm c}$.
Therefore, the magnetic field dependence of $T_{\rm c}$ is more
drastic than that expected from the mean-field theory.
This is a qualitative explanation \cite{rf:yanaseMG,rf:levinMG} of
the different behaviors between $T^{*}$ and $T_{{\rm c}}$
\cite{rf:zheng,rf:gorny}.

\begin{figure}[t]
\begin{center}
\vspace{2.5cm}
\caption{The calculated results for $1/T_{1}T$ under the magnetic field.
The relatively strong coupling case $g/t=-1.6$. The horizontal axis is
the temperature scaled by the $T_{\rm c}$ under the zero magnetic field.}
\label{fig:figure0.8}
\end{center}
\end{figure}

On the contrary, larger magnetic-field dependence is expected
in the relatively weak-coupling case. 
In this case, the onset temperature $T^{*}$ may be reduced by the 
magnetic field. The result of the T-matrix approximation for $g/t=-1.0$ 
shows a clear magnetic field dependence below $T^{*}$ 
(Fig.~\ref{fig:figure0.5}). 
This behavior corresponds to the optimally-doped region.
These doping dependences are in good agreement with the experimental results.
The detailed estimation on the magnetic field dependence in the 
optimally-doped region has been performed in Ref.~\cite{rf:eschrig}. 

\begin{figure}[t] 
\begin{center}
\vspace{2.5cm}
\caption{The calculated results for $1/T_{1}T$ under the magnetic
field. The weak coupling case $g/t=-1.0$.}
\label{fig:figure0.5}
\end{center}
\end{figure}

The pseudogap phenomena should be comprehensively understood from the
under- to optimally-doped region.
Based on this belief, we consider that the magnetic field dependence of
NMR $1/T_{1}T$ gives a strong support for the pairing scenario.
The usual behaviors of the SC fluctuation in the optimally-doped region
and the strong-coupling behaviors in the under-doped region are consistent
with the understanding of the phase diagram
(Fig.~\ref{fig:high-Tcphasediagram2} and Sec.~4.2.2).

Here note that we have ignored the Zeeman coupling term.
This procedure is justified because the Zeeman term gives only a higher-order
correction in the fluctuating regime \cite{rf:yanaseMG}.
The role of the Zeeman term can be clarified by changing the direction
of the magnetic field because the ``orbital effect'' is suppressed
when the magnetic field is applied along the plane. 

Recently, the magnetic field dependence of the c-axis conductivity is
measured for the Bi-based compounds.
Owing to the strong two-dimensionality, it is expected that 
the incoherent process dominates the c-axis conductivity in these compounds.
Then, the c-axis DC conductivity basically measures the electronic DOS
at $\omega=0$.
Therefore, this probe additively detects the large pseudogap below $T_{0}$ as
well as the small pseudogap below $T^{*}$
(Fig.~\ref{fig:high-Tcphasediagram2}).
The c-axis resistivity increases below $T_{0}$ especially in the
under-doped region.
This upturn of the c-axis resistivity coincides with the decrease of 
the uniform spin susceptibility \cite{rf:watanabe2000}. 
Then, the magnetic resistance is negative \cite{rf:yan} 
owing to the destruction of the pseudogap. 
Shibauchi {\it et al.} have reported the measurements for Bi-2212 compounds
up to 60 Tesla \cite{rf:shibauchi}. 
The result shows that the Zeeman term suppresses the large pseudogap
in the over-doped region where the typical magnetic field is small.
This result may be an evidence for the magnetic origin of the large pseudogap.
Lavrov et al. have reported the measurements for Bi-2201 compounds
from over-doped to heavily under-doped (insulating) region \cite{rf:lavrov}.
Then, two distinct responses to the magnetic field are observed
in the under-doped (superconducting) region.
This result implies two distinct origins of the suppression of the DOS 
which have been discussed in Sec.~2.1.5. 
In particular, the anisotropy of the magnet resistance
will be a clear evidence for the appearance of the SC fluctuation.
The sensitivity of the ``orbital effect'' to the direction of
the magnetic field is clearly observed in the vicinity of
$T_{\rm c}$, which is consistent with the pseudogap induced
by the SC fluctuation.
It should be noted again that the orbital effect is not clear around 
$T=T^{*}$ because the characteristic magnetic field is too large.
Therefore, we have a question about the interpretation in 
Ref.~\cite{rf:lavrov}, which concluded from the weak magnetic field
dependence that $T^{*}$ decreases with under-doping.
Taking account of the short correlation length $\xi_{\rm GL}$ at $T=T^{*}$,
the $T^{*}$ in the under-doped region will be higher than that determined
in Ref.~\cite{rf:lavrov}.
The analysis in the higher magnetic field is desirable to estimate
$T^{*}$ more precisely.
Another interesting observation in Ref.~\cite{rf:lavrov} is that the
magnetoresistance is positive in the non-superconducting region.
This result implies the qualitative difference of the electronic state 
between the insulating and superconducting regions,
which supports our theoretical approach to the high-$T_{\rm c}$
superconductivity from the metallic side.

\subsubsection{Another candidate: organic superconductor
$\kappa$-(ET)$_{2}$X}

In this subsection, we suggest another candidate for the pseudogap
state induced by the SC fluctuation. The organic superconductor
$\kappa$-(ET)$_{2}$X is a clear two-dimensional system with
$T_{\rm c} \sim 10$K.
As is noted in Sec.~3.3, this value of $T_{\rm c}$ is comparable 
to that of the high-$T_{\rm c}$ cuprates when normalized 
by the Fermi energy. 
Thus, two important conditions explained in Sec.~4.2.2 can be satisfied
in this compound. Therefore, we expect a similar appearance of the pseudogap.

Concerning the pressure dependence, the pseudogap is expected in the
lower pressure region because the $T_{\rm c}$ is large and $E_{\rm F}$
is small. By applying the pressure, the superconductivity gradually becomes
the weak-coupling one and therefore the pseudogap will disappear.
This variation is similar to the doping dependence of high-$T_{\rm c}$
cuprates. The coherence length has been estimated from the $H_{\rm c2}$ 
as $\xi=28\pm 5 (\overset{\circ}{\rm A})$ and
$\xi=53\pm 6 (\overset{\circ}{\rm A})$
for $\kappa$-(ET)$_2$Cu[N(CN)$_2$]Br and $\kappa$-(ET)$_2$Cu(NCS)$_2$,
respectively \cite{rf:29}.
This variation is consistent with the above expectation because
$\kappa$-(ET)$_2$Cu[N(CN)$_2$]Br corresponds to the
lower pressure region (Fig.~\ref{fig:organic}).
These values of the coherence length are comparable to the length scale of
the structural unit. This is just a condition for the pseudogap phenomena.

The NMR measurements have actually observed the pseudogap in
$\kappa$-(ET)$_{2}$X with $T^{*} \sim 50$K \cite{rf:mayaffrePG,rf:kanodaPG}.
However, this $T^{*}$ should not be identified to the onset of the
SC fluctuation, because the pressure dependence is not consistent.
The nature around $T^{*}$ probably corresponds to the metal-insulator
cross-over; the electronic state is almost incoherent above $T^{*}$.
Then, the manifestation of the SC fluctuation is expected
in the lower temperature region.
This expectation is clearly supported by the recent NMR measurement
by the Kanoda's group \cite{rf:matsumotokanoda}.
They have measured the magnetic field dependence of the NMR $1/T_{1}T$ and
shown that the SC fluctuation appears below the new cross-over temperature
$T_{\rm c}^{*}$ which is between $T_{\rm c}$ and $T^{*}$.

It is an advantage of the organic materials that the superconductivity is
destroyed by the magnetic field about $10$ Tesla which is experimentally
practical.
The experimental result shows that the NMR $1/T_{1}T$ increases {\it above}
$T_{\rm c}$ and below $T_{\rm c}^{*} \sim 20$K by applying the magnetic field
$H > H_{{\rm c2}}$ along the c-axis.
On the contrary, the NMR $1/T_{1}T$ is not affected by the same
magnetic field along the plane.
$H_{{\rm c2}}$ is much larger than the applied field in this direction.
The anisotropy is a clear evidence for the ``orbital effect''
which is a characteristics of the SC fluctuation.
Therefore, the further decrease of $1/T_{1}T$ below $T_{\rm c}^{*}$
is attributed to the SC fluctuation.
These results have revealed the hidden phenomena under
the zero magnetic field.
This effect has been unclear since it is masked by the sizable decrease
from 50 K.
The magnetic field dependence has played a key role to identify
the contribution from the SC fluctuation.
Thus, the pseudogap induced by the SC fluctuation is a
universal phenomenon for the quasi-two-dimensional short
coherence length superconductors. 

\subsubsection{higher-order corrections}

So far, we have discussed the effect of the SC fluctuation within the
1-loop order (T-matrix approximation).
At the last of Sec.~4.2, we briefly discuss the effect of
the higher-order corrections.
The SCT approximation, which is an extended version of
the T-matrix approximation, shows the pseudogap in many aspects but
only a weak one in the single-particle spectral function.
This result implies that the qualitative role of SC fluctuation
is robust but the higher-order corrections quantitatively suppress
the pseudogap.
However, because the SCT is a partial summation of the higher-order terms,
particular effects tend to be overestimated.
Therefore, the effect of the vertex correction is interesting
to capture the qualitative tendency more precisely.

\begin{figure}[t]
\begin{center}
\vspace{2.5cm}
\end{center}
\caption{The loop expansion of the self-energy within the 3-loop order.}
\label{fig:SC-higher-order}
\end{figure}

In this subsection, we carry out the loop expansion 
within the 3-loop order \cite{rf:yanaseSCFVC}.
The corresponding diagrams are shown in Fig.~\ref{fig:SC-higher-order}.
The 1-loop term (Fig.~\ref{fig:SC-higher-order}(a)) is included
in the T-matrix approximation.
The 2-loop term (Fig.~\ref{fig:SC-higher-order}(b)) and a part of
the 3-loop term (Fig.~\ref{fig:SC-higher-order}(c,d)) are included
in the SCT approximation.
Importantly, the lowest-order vertex correction term appears
in the 3-loop order (Fig.~\ref{fig:SC-higher-order}(e)).

Here, we ignore the higher-order corrections on the T-matrix.
The corrections to the T-matrix are well expressed by the renormalization
of the TDGL parameters (Eq.~(\ref{eq:TDGL})) and qualitatively not important.
The renormalization effect further reduces the parameter $b$ from
Eq.~(\ref{eq:TDGLparameterb}), while $T_{\rm c}$ is reduced as is shown
in Fig.~\ref{fig:T-matrixphase}.
It is more important that the pseudogap in the spectral function is smeared
in the SCT approximation through the terms like
Fig.~\ref{fig:SC-higher-order}(b-d).
This kind of the correction is not included in the
two-particle self-consistent (TPSC) approximation
\cite{rf:kyungd-wave,rf:kyungTPSC}, where the spectral function has
a similar feature to the non-self-consistent T-matrix approximation.

Here we use the static approximation where only the classical part
of the T-matrix ($t(\q,{\rm i}\Omega_{n}=0)$) is taken into account.
This approximation is appropriate near the critical point because
the singular contributions from the SC fluctuation
(Eqs.~(\ref{eq:imaginary-self-energy1D})-(\ref{eq:real-self-energy2D}))
are derived from the classical part.
At least, this approximation is sufficient to capture the qualitative
bahaviors. In the following figure, we show the results in the $s$-wave
case, namely $\phi_{\smallk}=1$ for simplicity. Almost the same results
are expected in the $d$-wave case.

\begin{figure}[t]
\begin{center}
\vspace{2.5cm}
\caption{(a) The imaginary part of the self-energy and
(b) the spectral function at $\k=(\pi,0.13\pi)$. Here, $g/t=-1.4$ and
$T=0.049=1.04T_{\rm c}$. In (a), each contribution from the diagram in
Fig.~\ref{fig:SC-higher-order}(a), (b), (d), and (e) is shown, respectively.}
\label{fig:higher-order-selfi}
\end{center}
\end{figure}

We show the imaginary part of the self-energy in
Fig.~\ref{fig:higher-order-selfi}(a).
It is shown that the $1$-loop term has similar features to the results
of the T-matrix approximation, although the structure around $\omega=0$
is broadened. The 2-loop term gives an oppsite contribution to the
self-energy. Therefore, the spectral function within the 2-loop order
becomes very sharp (see Fig.~\ref{fig:higher-order-selfi}(b)),
which indicates a well-defined quasi-particle. 
To the contrary, the $3$-loop terms enhance the 1-loop term,
and therefore, pseudogap clearly appears in this order.
This recurrence occurs in the SCT approximation and finally results
in Fig.~\ref{fig:sctDOS}.
It is important to note that the lowest-order vertex correction term
enhances the pseudogap furthermore. Therefore, it is expected that 
the smearing of the spectral function and DOS in the SCT approximation 
is considerably canceled by the vertex correction.
In other words, the effects of the SC fluctuation are
underestimated in the SCT approximation.
Roughly speaking, the correct result will lie between
the non-self-consistent and self-consistent T-matrix approximations.
Then, $T_{\rm c}$ is reduced more remarkably than that in
Fig.~\ref{fig:T-matrixphase}.
This conjecture is consistent with the Quantum Monte Carlo simulation
in the attractive Hubbard model \cite{rf:vilk,rf:allen}.

Interestingly, the three kinds of the 3-loop term has
qualitatively same feature.
In particular, the comparison between the term in
Fig.~\ref{fig:SC-higher-order}(d) and that in
Fig.~\ref{fig:SC-higher-order}(e) may be interesting because
they are equivalent in the Sadovskii's meathod for the one-dimensional
model \cite{rf:sadovskii} and for the BCS pairing model
\cite{rf:fujimotoPG}.
This equivalency is not satisfied in two dimensions, but
Fig.~\ref{fig:higher-order-selfi}(a) shows that
it is qualitatively justfied.
We can see that the vertex correction term has larger contribution.
This tendency is remarkable around the ``hot spot'' and insignificant
around the ``cold spot''.
The difference increases with increasing the superconducting coupling
$T_{\rm c}^{\rm MF}/E_{\rm F}$. 
It should be noticed that the asymptotic behavior of the specral 
function in the Sadovskii's meathod \cite{rf:fujimotoPG} is more similar 
to the non-self-consistent T-matrix approximation than to the SCT one.
From the above observations, we expect that the correct result in the
two-dimensional model will behave as in the T-matrix approximation.
Then, the pseudogap clearly appears in the spectral function 
(see Sec.~4.2.1). 

\subsection{Microscopic Theory: FLEX+T-matrix approximation}

In the previous section, we have discussed the general aspect of
the pseudogap phenomena induced by the SC fluctuation.
Then, a model with attractive interaction has been used as an
effective model.
Of course, the relevant Hamiltonian for cuprates is the repulsive model,
where the pairing interaction should be derived from
the many-body effects as performed in Sec.~3.
Then, the microscopic theory starting from the repulsive model
is highly desired for the pseudogap phenomena.
It should be confirmed that the effective coupling of the superconductivity
$T_{\rm c}^{\rm MF}/E_{\rm F}$ is strong enough to lead to
the pseudogap phenomena.
The doping dependence will be an interesting consequence of such microscopic
theory. We will see that the microscopic treatment is essential for
the coherent understanding of the magnetic and transport properties,
which have been stimulated much interests. 
The purpose of this section is a review of the microscopic theory
based on the FLEX+T-matrix approximation, by which the SC fluctuation is
microscopically taken into account. 
The Hubbard model (Eqs.~(\ref{eq:Hubbard-model})
and (\ref{eq:high-tc-dispersion})) is chosen as a microscopic model.
We similarly fix the parameter as $2t=1$ and $t'=0.25t$. 

The FLEX+T-matrix approximation was first proposed in
Ref.~\cite{rf:dahmpseudogap} where the NMR $1/T_{1}T$ is calculated
by using the phenomenological form of the T-matrix.
Fully microscopic treatment has been proposed in \cite{rf:koikegamiFLEXPG}
and developed in Ref.~\cite{rf:yanaseFLEXPG}.
We explain how the anomalous properties in the normal state are understood
from this approach.

\begin{figure}[t]
\begin{center}
\vspace{2.5cm}
\caption{(a) The real part and (b) imaginary part of the self-energy
in the FLEX approximation. (c) The spectral function.
The momentum A, B and C is shown in the inset of (b).
Here, $U/t=3.2$, $\delta=0.10$ and $T=0.01$.}
\label{fig:FLEXselfr}
\end{center}
\end{figure}

\subsubsection{Single-particle properties}

Let us briefly explain the basic formulation of the FLEX+T-matrix
approximation.
The characteristic behaviors of the single-particle properties are
discussed in parallel. We show that the resonance scattering gives rise to
the pseudogap in the under-doped region.
The doping dependence including the electron-doped cuprates is consistently
explained in Sec.~4.3.2.
The magnetic and transport properties are discussed in Secs.~4.3.3 and
4.3.4, respectively.

First we describe the quasi-particles and AF spin fluctuation
by using the FLEX approximation.
The characteristic properties of the quasi-particles in the nearly AF
Fermi-liquid \cite{rf:stojkovic,rf:yanaseTR} are summarized
in the following way.
Figure \ref{fig:FLEXselfr} shows the typical results of the self-energy. 
The local minimum of the absolute value of the imaginary part at $\omega=0$
assures the Fermi-liquid behaviors. 
There clearly appear notable features of the nearly AF Fermi-liquid.
The first is the $\omega$-linear dependence of the imaginary part,
which is in contrast to the $\omega$-square dependence in the
conventional Fermi-liquid theory. 
The origin of the $\omega$-linear dependence is the 
low-energy spin excitation, which is also the origin of 
the $T$-linear law of the resistivity \cite{rf:stojkovic,rf:yanaseTR} 
(see Sec.~4.3.4). 
The second is the strong momentum dependence of the self-energy.
For example, the damping rate of quasi-particles,
$1/\tau(\k)=
-{\rm Im}{\mit{\it \Sigma}}_{{\rm F}}^{{\rm R}}(\mbox{\boldmath$k$},0)$
is large around $(\pi,0)$ (``hot spot'') and is small around $(\pi/2,\pi/2)$
(``cold spot''). This momentum dependence plays a crucial role
in the transport properties, especially in the pseudogap state (Sec.~4.3.4).
The renormalization factor
$z_{\mbox{{\scriptsize \boldmath$k$}}}^{-1}=
1 - \partial {\rm Re}\Sigma_{\rm F}^{{\rm R}}
(\mbox{\boldmath$k$}, \omega)/\partial \omega|_{\omega =0}$
has the qualitatively same momentum dependence.
The quasi-particle velocity is considerably small at the hot spot
because of the van Hove singularity and the large renormalization factor.
This is just the flat dispersion observed in ARPES
\cite{rf:shenrep,rf:ARPESreview}.
We have commented in Fig.~\ref{fig:fermi} that the AF spin fluctuation 
transforms the Fermi surface \cite{rf:yanaseTR}.
The Fermi surface is pinned to the flat dispersion owing to the
transformation of the Fermi surface.
Because the effective Fermi energy of the superconductivity is
determined by the quasi-particle velocity around the hot spot as
$E_{\rm F} \sim v_{\rm F}k_{\rm F}$, the effective coupling of
the superconductivity $T_{\rm c}^{\rm MF}/E_{\rm F}$ is enhanced by
the flat dispersion.
Then, the coherence length $\xi$ and the TDGL parameter $b$ are much reduced.

The spectral function $A(\k,\omega)$ shows the single peak structure
(Fig.~\ref{fig:FLEXselfr}(c)), which justifies the picture of
the quasi-particles in the nearly AF Fermi-liquid.
The spectral function is remarkably broad at the hot spot and sharp
at the cold spot reflecting the momentum dependence of the quasi-particle
damping. This is also observed in ARPES around the optimally-doping
\cite{rf:ARPESreview}. However, the pseudogap appears neither
in the spectral function nor in the DOS.
Thus, the FLEX approximation is insufficient for the description of
the pseudogap state.

\begin{figure}[t]
\begin{center}
\vspace{2.5cm}
\caption{(a) The T-matrix. (b) The self-energy arising from the
SC fluctuation, $\Sigma_{{\rm S}}(k)$.}
\label{fig:vertex}
\end{center}
\end{figure}

\begin{figure}[t]
\begin{center}
\vspace{2.5cm}
\caption{(a)The real part and (b) the imaginary part of the self-energy
in the FLEX+T-matrix approximation. (c)The spectral function and (d) DOS.
The inset in (d) shows the same result in the large energy scale.
The momentum A, B and C is shown in the inset in Fig.~\ref{fig:FLEXselfr}(b).
Here, $U/t=3.2$, $\delta=0.10$ and $T=1.2 T_{{\rm c}}$.}
\label{fig:FLEX+Tselfr}
\end{center}
\end{figure}

The SC fluctuation is derived from the effective pairing interaction
described in Eq.~(\ref{eq:anomalous-rpa-singlet}).
We consider the T-matrix which is expressed by the ladder diagrams
in the particle-particle channel (Fig.~\ref{fig:vertex}(a)).
By restricting to the $d$-wave channel, the T-matrix is estimated
by extending the \'Eliashberg equation \cite{rf:yanaseFLEXPG},
\begin{eqnarray}
  && T(k_{1},k_{2}:q)=
  \frac{g\lambda(q)\phi(k_{1})\phi^{*}(k_{2})}{1-\lambda(q)},
  \label{eq:FLEX+T-matrix1} \\
  && \lambda(q)=
  -\sum_{k,p}\phi^{*}(k) V_{\rm a}(k-p) G(p) G(q-p) \phi(p).
  \label{eq:FLEX+T-matrix2}
\end{eqnarray}
Here, the expression of the FLEX approximation is used for the effective
interaction $V_{\rm a}(k)$ and the Green function $G(k)$.
$\phi(k)$ is the eigenfunction of the \'Eliashberg equation
Eq.~(\ref{eq:eliashberg-super}) with its maximum eigenvalue
$\lambda_{{\rm e}}$. This function corresponds to the wave function
of the fluctuating Cooper-pairs.
This procedure is justified when the correlation length is long enough,
namely in the vicinity of the critical point $T=T_{{\rm c}}$.
Because this procedure is based on the \'Eliashberg equation with fully including the
momentum and frequency dependence, the characteristics of 
the spin-fluctuation-induced superconductivity are included 
in the T-matrix. 
In Eqs.~(\ref{eq:FLEX+T-matrix1}) and (\ref{eq:FLEX+T-matrix2}),
the wave function is normalized as
\begin{eqnarray}
\label{eq:normalization}
  \sum_{k} |\phi(k)|^{2} = 1.
\end{eqnarray}
Then, the constant factor $g$ is obtained as
\begin{eqnarray}
\label{eq:defg}
  g=\sum_{k_{1},k_{2}}\phi^{*}(k_{1}) V_{\rm a}(k_{1}-k_{2}) \phi(k_{2}).
\end{eqnarray}
It should be noted that the parameter $\lambda(0)$ is equivalent 
to the maximum eigenvalue of the \'Eliashberg equation, 
namely $\lambda(0)=\lambda_{{\rm e}}$. 
Then, the divergence of the T-matrix is equivalent to the criterion 
in the \'Eliashberg equation $\lambda_{{\rm e}}=1$. 

The self-energy correction arising from the SC fluctuation is obtained
within the 1-loop order (FLEX+T-matrix approximation),
\begin{eqnarray} 
\label{eq:FLEX+T-matrixap}
 \Sigma_{{\rm S}}(k)=\sum_{q} T(k,k:q) G(q-k).
\end{eqnarray}
The total self-energy is obtained by adding it to the contribution
from the spin fluctuation $\Sigma(k)=\Sigma_{\rm F}(k)+\Sigma_{\rm S}(k)$.

\begin{figure}[t]
\begin{center}
\vspace{2.5cm}
\caption{(a) The doping dependence of the DOS at $T=1.2T_{{\rm c}}$ and
$U/t=3.2$. (b)$U$-dependence of the DOS at $T=1.2T_{{\rm c}}$ and
$\delta=0.10$.}
\label{fig:FLEX+TDOSdoping}
\end{center}
\end{figure}

The self-energy in the under-doped region is shown in
Figs.~\ref{fig:FLEX+Tselfr}(a) and (b).
We clearly see the anomalous behaviors which arises from the SC fluctuation.
The similar structure to that in the attractive model (Fig.~\ref{fig:selfr})
appears in the low-frequency region.
In particular, the extremely large damping around the Fermi level
gives rise to the clear pseudogap in the spectral function and in the DOS
(Figs.~\ref{fig:FLEX+Tselfr}(c) and (d)). 
This is a result of the ``resonance scattering''. 
Thus, the superconductivity described by the FLEX approximation is 
strong coupling enough to induce the pseudogap 
in the under-doped region. 
It should be noted that the SC fluctuation is still over-damped,
while an asymmetric structure appears owing to not so small
$a_{1}/a_{2}$ (see Eq.~\ref{eq:TDGL}).
Because of the momentum dependence of the wave function $\phi(k)$,
the pseudogap has a $d$-wave form which is similar to the SC gap.
The anomalous behaviors are smeared around the cold spot and
the ``Fermi arc'' appears in the pseudogap region.
This is an important character of the pseudogap observed by
ARPES measurements \cite{rf:ding,rf:shenPG,rf:normanPG}.

\begin{figure}[t]
\begin{center}
\vspace{2.5cm}
\caption{(a) The doping dependence of the TDGL parameter $b$ at
$T=T_{{\rm c}}$ and $U/t=3.2$.
(b) The $U$-dependence at $T=T_{{\rm c}}$ and $\delta=0.10$.
The inset shows the result at the fixed temperature $T=0.0082$.}
\label{fig:bdopingTc}
\end{center}
\end{figure}

It is an advantage of the microscopic calculation that the reasonable
energy scale of the pseudogap is obtained, while much larger value is
obtained in the attractive model.
Here, the gap magnitude $\Delta_{{\rm pg}}$ and $T_{\rm c}$ are smaller
by an order.
This is because the superconductivity and the pseudogap take place in the
renormalized quasi-particles near the Fermi surface.
As a result, the energy scale takes a realistic value
$2 \Delta_{{\rm pg}} \sim 100 {\rm meV}$.
We again stress that the effective Fermi energy is renormalized
by the electron correlation and therefore the effective coupling
$T_{{\rm c}}^{{\rm MF}}/E_{{\rm F}}$ is enhanced.

Before closing this subsection, we discuss the relation to the SC gap.
The ratio $2\Delta_{{\rm s}}/T_{{\rm c}}^{\rm MF} \sim 12$ has been
obtained by the FLEX approximation in the optimally-doped region
\cite{rf:monthouxFLEX,rf:paoFLEX,rf:takimotoFLEX2},
where $\Delta_{{\rm s}}$ is the maximum value of the SC gap.
A larger value of this ratio is expected in the under-doped region.
The experimental results have shown that the energy scale of the
pseudogap is slightly larger than the SC gap
\cite{rf:ding,rf:shenPG,rf:normanPG,rf:renner}.
Our result indicates the ratio $2 \Delta_{{\rm pg}}/T_{{\rm c}} \sim 20$
in the under-doped region.
Thus, the pseudogap obtained in the FLEX+T-matrix approximation
has a relevant magnitude compared with the SC gap.

\subsubsection{Doping dependence}

The FLEX+T-matrix approximation appropriately reproduces 
the doping dependence. In the previous subsection, a clear pseudogap 
is shown in the under-doped region. The anomalous properties gradually 
disappear as the doping concentration is increased.
Consequently, the pseudogap in the DOS is filled up with hole-doping
(Fig.~\ref{fig:FLEX+TDOSdoping}(a)).

\begin{figure}[t]
\begin{center}
\vspace{2.5cm}
\caption{Temperature dependence of the NMR $1/T_{1}T$ and
$T_{1}T/(T_{2{\rm G}})^{2}$ in the FLEX approximation for $U/t=3.2$.}
\label{fig:NMR}
\end{center}
\end{figure}

This closing of the pseudogap is basically understood from the
doping dependence of the TDGL parameter $b$ (Fig.~\ref{fig:bdopingTc}).
The large value of $b$ generally means the weak SC fluctuation, which is
realized in the over-doped region.
Owing to the relation $b \propto (v^{*}_{\rm F}/T)^{2}$,
the parameter $b$ rapidly develops in the weak-coupling region.
Here $v^{*}_{\rm F}$ is the quasi-particle velocity around $(\pi,0)$.
In our notation, the parameter $b$ includes the coupling constant $g$
as $b \propto |g| \rho_{{\rm d}}(0) (v^{*}_{{\rm F}}/T)^{2}$.
Because the decrease of $v^{*}_{{\rm F}}/T$ is considerably canceled
by the increase of $g$, the parameter $b$ saturates in the under-doped region.
This feature does not contradict with the development of the SC fluctuation.

The similar closing of the pseudogap is caused by decreasing $U$
(Figs.~\ref{fig:FLEX+TDOSdoping}(b) and ~\ref{fig:bdopingTc}(b)).
Thus, the pseudogap phenomena are one of the characteristics of
the strongly correlated electron systems,
where superconductivity tends to be strong coupling.
The inset of Fig.~\ref{fig:bdopingTc}(b) shows the TDGL parameter
at the fixed temperature.
The result clarifies the role of the electron correlation, namely the
renormalization of the quasi-particle velocity.
We see that the TDGL parameter $b$ and the coherence length
is reduced by the electron correlation.

At last, we discuss the electron-doped cuprates.
Owing to the next-nearest neighbor hopping term $t'$,
the electronic DOS is relatively small in the electron-doped region.
In particular, the Fermi level is not pinned to the flat dispersion
around $(\pi,0)$.
This feature means that the electron correlation is effectively weak
in the electron-doped cuprates.
Therefore, more conventional behaviors are expected in the normal state.
The calculated results based on the FLEX approximation confirm this
naive expectation.
The imaginary part of the self-energy shows the $\omega^{2}$-dependence,
which is a conventional Fermi-liquid behavior.
This feature indicates the $T^{2}$-law of the resistivity which is
confirmed in Sec.~4.3.4 and experimentally observed in the wide 
temperature region \cite{rf:electrondopesuper}. 

Concerning the pseudogap, the SC fluctuation is very weak in the
electron-doped region.
This is because the quasi-particle velocity $v^{*}_{{\rm F}}$ is 
large around $(\pi,\pi/4)$, and more importantly, because
$T_{\rm c}$ is low.
We have already explained the reason of low $T_{\rm c}$ in Sec.~3.1.4.
Thus, the superconductivity is the weak-coupling one,
which is indicated by the large TDGL parameter $b \sim 30$.
The comparable value to the over-doped region indicates that 
the pseudogap is very weak even if it is observed by some experiments.
The large value of the coherence length $\xi$ is consistent with the 
small $H_{\rm c2} = 5 \sim 10$Tesla \cite{rf:zheng_el}. 
These results are consistent with the recent experiments of the ARPES
\cite{rf:almitage}, neutron scattering \cite{rf:yamadakurahashi},
NMR $1/T_{1}T$ \cite{rf:zheng_el}, and tunnelling spectroscopy
\cite{rf:tunneling_el}.
These measurements do not show the ``small pseudogap''
in the electron-doped cuprates.

Other authors have suggested the pseudogap in the electron-doped cuprates
${\rm Nd}_{2-x}{\rm Ce}_{x}{\rm Cu}{\rm O}_{4-y}$ from the optical
measurements \cite{rf:basov-el}.
They have shown that the characteristic structure of the frequency dependent
scattering rate $1/\tau(\omega)$.
The suggested pseudogap has much larger magnitude than the SC gap,
and appears from the very high temperature $T > 300 {\rm K}$.
It is therefore considered that this phenomenon is not attributed to
the same origin as that of the small pseudogap in the hole-doped 
cuprates. 
Because a similar structure is observed even in the AF state, 
this phenomenon may be attributed to the ``large pseudogap'' 
with magnetic origin. 
The observation of the ARPES \cite{rf:almitage} indicates 
the broad suppression of the single particle spectrum 
around the magnetic Brillouin zone. 
Since this suppression has also large energy scale, the existence of the 
large pseudogap and importance of the AF spin fluctuation have been 
suggested. 

\subsubsection{Magnetic properties}

In the following subsections, we discuss the magnetic and transport
properties which are calculated by using the Kubo formula.
The estimation of the two-particle correlation functions are
required for this purpose.
First we focus on the magnetic properties, which have been
the central issue among the several pseudogap phenomena. 

Many authors have used the FLEX approximation in order to investigate
the magnetic properties in high-$T_{\rm c}$ cuprates \cite{rf:dahmFLEX}.
The vertex correction is needed in the context of the
conserving approximation \cite{rf:FLEX,rf:bickers1991},
but usually it is ignored.
The estimation in the early stage \cite{rf:dahmFLEX} has indicated
that the vertex correction reduces the spin susceptibility
but does not alter the qualitative behaviors.
The characteristic behaviors of the nearly AF Fermi-liquid are well
reproduced within the FLEX approximation.
The NMR $1/T_{1}T$ shows the Curie-Weiss law (Fig.~\ref{fig:NMR})
and its relation to the NMR $1/T_{2{\rm G}}$ shows the magnetic scaling
with the dynamical exponent $z=2$ (see the inset in Fig.~\ref{fig:NMR}).
It is microscopically shown that the AF spin correlation increases
with decreasing $\delta$.
These features have been assumed in the phenomenological theory
\cite{rf:moriya1990,rf:moriya1992,rf:monthoux1991,rf:monthoux1992}.
It should be, however, noted that the pseudogap is not derived 
within the FLEX approximation. 

The pseudogap in the NMR (see Sec.~2.1.3) is explained by taking 
account of the SC fluctuation, as is shown in the following way 
\cite{rf:yanaseFLEXPG}. 
In the FLEX+T-matrix approximation, the dynamical spin susceptibility
$\chi_{{\rm s}}^{{\rm R}}(\mbox{\boldmath$q$},\Omega)$ is obtained
by extending the FLEX approximation
\begin{eqnarray}
 \label{eq:FLEX+T-matrixkai}
 && \chi_{{\rm s}}(q) = \frac{\chi_{0}(q)}{1 - U \chi_{0}(q)}, \\
 && \chi_{0}(q) = \sum_{k} G(k) G(k+q),
\end{eqnarray}
where the effects of the SC fluctuation are included in the self-energy
$\Sigma_{{\rm S}}(k)$.
The NMR spin-lattice relaxation rate $1/T_{1}$ and spin-echo decay rate
$1/T_{2{\rm G}}$ are obtained by the following formula
\cite{rf:NMR,rf:pennington1991}
\begin{eqnarray}
 \label{eq:NMR}
 1/T_{1}T & = & \sum_{\mbox{\boldmath$q$}} F_{\perp}(\mbox{\boldmath$q$})
             [\frac{1}{\omega}
             {\rm Im} \chi_{{\rm s}}^{{\rm R}} (\mbox{\boldmath$q$}, \omega)
             \mid_{\omega \to 0}],
 \\
 \label{eq:NMR-T2G}
 (1/T_{2{\rm G}})^{2} & = & \sum_{\mbox{\boldmath$q$}}
                     [F_{\parallel}(\mbox{\boldmath$q$})
        {\rm Re} \chi_{{\rm s}}^{{\rm R}} (\mbox{\boldmath$q$}, 0)]^{2}
                   - [\sum_{\mbox{\boldmath$q$}}
                      F_{\parallel}(\mbox{\boldmath$q$})
        {\rm Re} \chi_{{\rm s}}^{{\rm R}} (\mbox{\boldmath$q$}, 0)]^{2}.
\end{eqnarray}
Here $F_{\perp}(\mbox{\boldmath$q$}) = \frac{1}{2}\{A_{1} + 2 B
(\cos q_{x}+\cos q_{y})\}^{2} + \frac{1}{2}
\{A_{2} + 2 B (\cos q_{x}+\cos q_{y})\}^{2}$
and $F_{\parallel}(\mbox{\boldmath$q$}) =
\{A_{2} + 2 B (\cos q_{x}+\cos q_{y})\}^{2}$ \cite{rf:milla}.
The hyperfine coupling constants $A_{1}, A_{2}$, and $B$ are evaluated
as $A_{1} = 0.84 B$ and $A_{2} = -4 B$ \cite{rf:barzykin}.

We show the calculated results for the NMR $1/T_{1}T$, $1/T_{2{\rm G}}$,
and static spin susceptibility in Fig.~\ref{fig:magnetic}.
We see that the pseudogap clearly appears in the NMR $1/T_{1}T$
(Fig.~\ref{fig:magnetic}(a)).
The decrease of the NMR $1/T_{1}T$ above $T_{{\rm c}}$ is 
the first observation of the pseudogap phenomenon \cite{rf:yasuoka}.
The $1/T_{1}T$ is reduced with approaching to the critical point
because the dissipation of the spin fluctuation is suppressed
by the pseudogap in the DOS.

\begin{figure}[t]
\begin{center}
\vspace{2.5cm}
\caption{(a) The temperature dependence of the NMR $1/T_{1}T$
and $1/T_{2{\rm G}}$ in the FLEX+T-matrix approximation.
(b) The static spin susceptibility
$\chi_{{\rm s}}^{{\rm R}} (\mbox{\boldmath$q$},0)$
at $\mbox{\boldmath$q$} = (0,0)$ (open squares) and
at $\mbox{\boldmath$q$} = (\pi,\pi)$ (closed circles).
Here, $\delta=0.10$ and $U/t=3.2$.
The inset of (b) shows the results of the FLEX approximation.}
\label{fig:magnetic}
\end{center}
\end{figure}

The NMR $1/T_{2{\rm G}}$ also shows the pseudogap with the onset
temperature $T^{*}$ close to that in the $1/T_{1}T$
(Fig.~\ref{fig:magnetic}(a)). 
This is also an effect of the SC fluctuation.
So far, different behaviors of $1/T_{2G}$ have been reported for several
high-$T_{\rm c}$ compounds \cite{rf:itoh1992,rf:takigawa1994,rf:itoh1994,rf:julien,rf:itoh1996,rf:itoh1998,rf:goto,rf:tokunaga}.
At present, they are attributed to the effects of the interlayer coupling
\cite{rf:itoh1998,rf:goto}. 
The experimental data on the single layer compounds show 
the decrease of $1/T_{2G}$ in the pseudogap state 
\cite{rf:itoh1996,rf:itoh1998}, which is consistent with our result 
in Fig.~\ref{fig:magnetic}(a). 
It is interesting that the pseudogap in the NMR $1/T_{2{\rm G}}$
is moderate compared with the NMR $1/T_{1}T$.
This is because the NMR $1/T_{2{\rm G}}$ measures the static part
which reflects the total weight of the spin fluctuation.
It should be noticed that the pseudogap suppresses only the low-frequency
component of the spin fluctuation, because the superconductivity has
a smaller energy scale than that of the spin fluctuation.
Thus, it is no wonder that the scaling relation of the spin
fluctuation is violated in the pseudogap state (see the inset in
Fig.~\ref{fig:magnetic}(a) and experimental result \cite{rf:itoh1998}).
These features are qualitatively consistent with the experimental results
\cite{rf:itoh1996,rf:itoh1998,rf:goto,rf:tokunaga}.
The obtained behavior of the $1/T_{2{\rm G}}$ is expected 
from the result in the superconducting state \cite{rf:bulut}.
Then, the $1/T_{2{\rm G}}$ remains even at low temperature,
although the $1/T_{1}T$ rapidly decreases.
This is a characteristic behavior of the $d$-wave superconductivity
and has played an important role for identifying the pairing symmetry
\cite{rf:itoh1992}. 
Because the pseudogap is a precursor of the $d$-wave superconductivity,
the above results in the pseudogap state are expected ones. 

Contrary to the other magnetic properties, experiments have shown that
the uniform spin susceptibility decreases from much higher
temperature than $T^{*}$ \cite{rf:alloul,rf:oda}.
The decrease becomes more rapid below $T^{*}$ \cite{rf:ishida1998}.
We show the calculated results of the uniform susceptibility
$\chi_{{\rm s}}(\mbox{\boldmath$0$},0)$ and staggard susceptibility
$\chi_{{\rm s}}(\mbox{\boldmath$Q$},0)$ in Fig.~\ref{fig:magnetic}(b).
While the staggard susceptibility shows the pseudogap with the
onset temperature $T^{*}$ determined from the $1/T_{1}T$,
the decrease of the uniform susceptibility begins
from much higher temperature and becomes rapid below $T^{*}$.
Thus, the qualitatively different behavior of the uniform susceptibility
is consistently obtained by simultaneously taking account of 
the spin and SC fluctuations. 

\begin{figure}[t]
\begin{center}
\vspace{2.5cm}
\caption{The dynamical spin susceptibility
$\chi_{{\rm s}}^{{\rm R}}(\mbox{\boldmath$Q$},\Omega)$
in the FLEX+T-matrix approximation at $\delta=0.10$ and $U/t=3.2$.
(a) The real part. (b) The imaginary part.}
\label{fig:T-matrixch0.4rekaiomega}
\end{center}
\end{figure}

The frequency dependence of the spin susceptibility well characterizes
the magnetic properties in the pseudogap state.
The results for the dynamical spin susceptibility
at $\mbox{\boldmath$q$} = \mbox{\boldmath$Q$}$ is shown
in Fig.~\ref{fig:T-matrixch0.4rekaiomega} \cite{rf:yanaseFLEXPG}.
We see the suppression of the real part in the low-frequency region,
that is, the magnetic order is suppressed by the SC fluctuation.
The imaginary part in the low-frequency region is more
remarkably suppressed in the pseudogap state.
At the same time, the spin excitation develops
in the higher-frequency region.
In other words, the pseudogap transfers the spectral weight
of the spin fluctuation from the low- to high-frequency region.
This is just the pseudogap phenomenon observed in the
inelastic neutron scattering measurements \cite{rf:neutronPG}.
This character in the frequency dependence indicates that
the pseudogap has the smaller energy scale than that
of the magnetic excitation.
We stress that this is a natural consequence of the pairing scenario.
These features are consistent with the above discussion on
the NMR $1/T_{2{\rm G}}$.
It is notable that the spin fluctuation in the relatively wide frequency
region contributes to the pairing interaction.
The low-frequency component rather gives a de-pairing effect through the
self-energy. Therefore, the $d$-wave pairing is not suppressed
by the pseudogap in the spin fluctuation.
We have actually confirmed that the feedback effect rather enhances
the $d$-wave superconductivity \cite{rf:yanaseFLEXPG}.

\begin{figure}[t]
\begin{center}
\vspace{2.5cm}
\caption{Doping dependence of the incommensurability
$\delta$ \cite{rf:kurokiFLEX}, which is defined by the peak of the magnetic
excitation at $(\pi,(1\pm\delta)\pi)$ and $((1\pm\delta)\pi,\pi)$.
The inset shows the Fermi surface in the electron-doped region.} 
\label{fig:kurokifig3} 
\end{center}
\end{figure}

The detailed momentum dependence of the magnetic excitation is a current
interest brought by the recent experiments.
There are detailed analysis by using the FLEX approximation
\cite{rf:dahmFLEX,rf:kurokiFLEX}.
Then, the magnetic excitation is commensurate in the under-doped region and
becomes incommensurate with hole-doping. It is robustly commensurate 
in the electron-doped case (Fig.~\ref{fig:kurokifig3}).
The long-range hopping term $t'$ plays an essential role in this structure.
The qualitatively same results have been obtained in the RVB theory
including the long-range hopping \cite{rf:tanamoto}.
These features are qualitatively consistent with the inelastic neutron
scattering for ${\rm La}_{2-x}{\rm Sr}_{x}{\rm Cu}{\rm O}_{4}$
\cite{rf:endohneutron,rf:yamadaneutron}
and ${\rm Nd}_{2-x}{\rm Ce}_{x}{\rm Cu}{\rm O}_{4-y}$ 
\cite{rf:yamadakurahashi}. 
The same result for the electron-doped cuprates has been reported
by the numerical studies on the Hubbard model
\cite{Dagotto94,rf:dagottoreview} and $t$-$J$ model \cite{rf:tohyama}.

\begin{figure}[t]
\begin{center}
\vspace{2.5cm}
\caption{The momentum dependence of the dynamical spin susceptibility
${\rm Im}\chi_{{\rm s}}^{{\rm R}}(\mbox{\boldmath$q$},\Omega)$
at $\delta=0.10$, $U/t=3.2$, and $T=0.01$.
(a) The FLEX approximation at $\Omega=0.01$.
(b) The FLEX+T-matrix approximation at $\Omega=0.01$.
(c) The FLEX+T-matrix approximation at $\Omega=0.1$.}
\label{fig:Imkai_under_FLEX}
\end{center}
\end{figure}

Recently, interesting temperature and frequency dependence has been
reported for ${\rm Y}{\rm Ba}_{2}{\rm Cu}_{3}{\rm O}_{6+\delta}$,
${\rm Bi}_{2}{\rm Sr}_{2}{\rm Ca}{\rm Cu}_{2}{\rm O}_{8+\delta}$,
and ${\rm La}_{2-x}{\rm Sr}_{x}{\rm Cu}{\rm O}_{4}$
\cite{rf:stripeEX,rf:arai,rf:lee}.
The commensurate magnetic excitation is observed at high temperature
and the incommensurability develops below the onset temperature
above $T_{{\rm c}}$.
Here it is shown that these features can be explained by taking account
of the pseudogap induced by the SC fluctuation \cite{rf:yanaseFLEXPG}.
First we point out that the SC fluctuation generally enhances
the incommensurability.
Although the commensurate peak is obtained by the FLEX approximation
(Fig.~\ref{fig:Imkai_under_FLEX}(a)),
it becomes incommensurate owing to the SC fluctuation
(Fig.~\ref{fig:Imkai_under_FLEX}(b)).
These features result from the $d$-wave momentum dependence of
the pseudogap in the spectral function.
The spectral gap around $(\pi,0)$ reduces the spin excitation
at $\q=\Q$ more remarkably than the incommensurate component.
The similar effect is expected in the superconducting state.
Thus, the incommensurate structure in the low-temperature region
can be explained as an effect of the pseudogap or SC gap.
If so, the incommensurate structure disappears 
in the high-frequency region as is shown
in Fig.~\ref{fig:Imkai_under_FLEX}(c), which has been observed
in the experiments \cite{rf:stripeEX,rf:arai,rf:lee}.
This is simply because the pseudogap appears only
in the low-frequency region.
The incommensurate structure disappears with increasing the temperature,
owing to the closing of the pseudogap.
Thus, the whole temperature and frequency dependences are consistent
with the incommensurability induced by the pseudogap.
It should be stressed that the incommensurate structure observed in
inelastic neutron scattering does not necessarily mean the stripe order.

\subsubsection{Transport properties}

In this subsection, we discuss the transport phenomena which have been
one of the central issues of the cuprate superconductors.
The understanding for the anomalous behaviors consistent with the
magnetic properties has been a fundamental problem for a long time.
A solution is provided by the FLEX+T-matrix approximation
in the following way. Then, the anomalous properties above $T^{*}$ 
are explained by taking the spin fluctuation into account 
\cite{rf:hlubina,rf:stojkovic,rf:yanaseTR,rf:kontani,rf:kanki}, 
and those below $T^{*}$ are explained by {\it simultaneously} 
taking account of the spin and SC fluctuations \cite{rf:yanaseTRPG}.
In the RVB theory, the electric conductivity around $T^{*}$ is explained
by taking account of the coupling between the spinons and holons through
the gauge field \cite{rf:ioffeRVB,rf:nagaosaTR,rf:onodaTR}.
The singlet pairing of the spinon weakly affects the charge transport
carried by the holon.
On the contrary, our understanding is based on the characteristic momentum
dependences of the quasi-particle properties.

Before describing the microscopic theories, we introduce the general formula
for the electric transport on the basis of the Fermi-liquid theory.
According to the Kubo formula, the electric conductivity is expressed as
the current-current correlation function
\begin{eqnarray}
 \label{eq:kuboformula}
 \sigma_{\mu\nu} & = & e^{2} \lim_{\omega=0}
 \frac{{\rm Im}K_{\mu\nu}^{{\rm R}}(\omega)}{\omega}, \\
 K_{\mu\nu}(\omega_{{\rm n}}) & = & \int_{0}^{\beta} {\rm d}\tau
 \langle T_{\tau} J_{\mu}(\tau) J_{\nu}(0) \rangle
 e^{{\rm i} \omega_{{\rm n}} \tau},
\end{eqnarray}
where $\omega_{{\rm n}}=2 n \pi T$ is the bosonic Matsubara frequency.

Since the conductivity is infinite in the non-interacting system,
some procedure of the renormalization is required  
in the perturbation expansion. 
Then, the expression for $K^{{\rm R}}(\omega)$ is generally complicated
in the process of the analytic continuation.
However, \'Eliashberg has given a compact formula
for the longitudinal conductivity
$\sigma_{xx}$ by taking the most divergent terms with respect to the
quasi-particle lifetime
$\tau(\mbox{\boldmath$k$})=
1/|{\rm Im}{\mit{\it \Sigma}}^{{\rm R}}(\mbox{\boldmath$k$},0)|$
\cite{rf:eliashbergtransport}.
This procedure is correct in the coherent limit
$l_{\rm max}=
v(\mbox{\boldmath$k$})\tau(\mbox{\boldmath$k$})|_{{\rm max}} \gg 1$,
which is justified in the low-temperature region.
The exceptional case is a system close to the phase transition.
For example, the Aslamazov-Larkin (AL) term \cite{rf:AL}
in the SC fluctuation theory is higher order with respect to $1/l_{\rm max}$,
but divergent at the critical point.
We will estimate the AL-term afterward and conclude that its contribution
is not important \cite{rf:yanaseTRPG}.

Using the Green function, the \'Eliashberg formula is expressed as
\begin{eqnarray}
  \label{eq:conductivity}
  \sigma_{\mu\mu} = e^{2} \sum_{\k} \int \frac{{\rm d}\e}{\pi} (-f'(\e))
  |G^{{\rm R}}(\k,\e)|^{2} \tilde{v}_{\mu}(\k,\e) J_{\mu}(\k,\e),
\end{eqnarray}
where
$\tilde{v}_{\mu}(\k,\e) = v_{\mu}(\k)+
\partial {\rm Re} \s(\k,\e)/\partial k_{\mu}$
is the velocity including the $k$-mass renormalization.
Note that the $k$-mass renormalization corresponds to a part of
the vertex correction.
The total current vertex $\vec{J}(\k,\e)$ is obtained
by solving the Bethe-Salpeter equation
\begin{eqnarray}
\label{eq:currentvertex}
  J_{\mu}(\k,\e) = \tilde{v}_{\mu}(\k,\e) + \sum_{\kk}
  \int \frac{{\rm d}\ee}{4 \pi {\rm i}} \Im_{22}(\k,\e:\kk,\ee)
  |G^{{\rm R}}(\kk,\ee)|^{2} J_{\mu}(\kk,\ee).
\end{eqnarray}
The vertex function $\Im_{22}(\k,\e:\kk,\ee)$ is obtained by the
analytic continuation of the irreducible four-point particle-hole vertex
\cite{rf:eliashbergtransport}.
The renormalization of the total current vertex $\vec{J}(\k,\e)$
is usually the main contribution of the vertex correction.
In the following, we use ``the vertex correction'' as the correction
arising from the vertex function $\Im_{22}(\k,\e:\kk,\ee)$.

The expression for the Hall conductivity $\sigma_{\mu\nu}$ corresponding
to Eq.~(\ref{eq:conductivity}) has been given by Kohno and Yamada
\cite{rf:kohnohall}: 
\begin{eqnarray}
  \label{eq:Hallconductivity}
  \sigma_{\mu\nu} &=&
  -H e^{3} \sum_{\k} \int \frac{{\rm d}\e}{\pi} (-f'(\e))
  |{\rm Im} G^{{\rm R}}(\k,\e)| |G^{{\rm R}}(\k,\e)|^{2}
  \nonumber \\
  && \times \tilde{v}_{\mu}(\k,\e)
  [J_{\mu}(\k,\e) \partial J_{\nu}(\k,\e)/ \partial k_{\nu}-
  J_{\nu}(\k,\e) \partial J_{\mu}(\k,\e)/ \partial k_{\nu}].
\end{eqnarray}
In case of the Hall conductivity, the most divergent term with respect to
$\tau(\mbox{\boldmath$k$})$ is the quadratic term.
In this section, the magnetic field is fixed to be parallel to the
{\it c}-axis $H \parallel {\it c}$, and the current $J$ is fixed to
be perpendicular to the {\it c}-axis $J \perp {\it c}$.

The vertex correction is generally required in order to satisfy
the Ward identity which corresponds to the momentum conservation law
\cite{rf:yamada,rf:toyoda,rf:okabe,rf:maebashi}.
When the current operator commutes with the Hamiltonian,
the infinite conductivity is generally expected.
However, the finite conductivity is obtained when we consider only the
self-energy correction.
The infinite conductivity is derived by taking into account
the vertex correction according to the scheme of Baym and Kadanoff.
In the lattice system with Umklapp processes, however, 
the current operator does not commute with the Hamiltonian.  
Then, the vertex corrections are usually taken into account
by only multiplying a constant factor, 
and have no important role \cite{rf:yamada}. 
This argument is based on the assumption that the temperature dependence of
the four-point vertex is negligible, which is justified in the conventional
Fermi-liquid. However, the vertex correction sometimes gives 
a significant effect, when a collective mode induces a temperature dependence
of the four-point vertex \cite{rf:kontani,rf:kanki}.
This case is realized in the under-doped region. 

The above expressions (Eqs.~(\ref{eq:conductivity}) and
(\ref{eq:Hallconductivity}))
are rewritten in the Fermi-liquid limit $z(\k) \gamma(\k) \ll T$ as
\begin{eqnarray}
  \label{eq:eliashberg-Kohno}
  \sigma_{xx} & = & e^{2} \sum_{\k} z(\k) (-f'(\e^{*}(\k)))
  \tilde{v}_{x}(\k,\e^{*}(\k)) J_{x}(\k,\e^{*}(\k))/\gamma(\k,\e^{*}(\k)),
  \\
  & \cong & e^{2} \int_{{\rm FS}} \frac{{\rm d}k}{(2 \pi)^{2}}
  \frac{\tilde{v}_{x}(\k)}{\tilde{v}(\k)} J_{x}(\k) \tau(\k),
  \\
  \sigma_{xy} & = & -\frac{H e^{3}}{2} \sum_{\k} z(\k) (-f'(\e^{*}(\k)))
  \tilde{v}_{x}(\k,\e^{*}(\k))
  \nonumber \\
  & & \times
  [J_{x}(\k,\e^{*}(\k)) \frac{\partial J_{y}(\k,\e^{*}(\k))}{\partial k_{y}}-
  J_{y}(\k,\e^{*}(\k)) \frac{\partial J_{x}(\k,\e^{*}(\k))} {\partial k_{y}}]
  /\gamma(\k,\e^{*}(\k))^{2},
  \\
  \label{eq:angle-rep}
  & \cong &  \frac{H e^{3}}{4} \int_{{\rm FS}} \frac{{\rm d}k}{(2 \pi)^{2}}
  |\vec{J}(\k)|^{2} (\frac{\partial \varphi(\k)}{\partial k_{\parallel}})
  \tau(\k)^{2}.
\end{eqnarray}
We have used the definition
$\tilde{v}_{\mu}(\k)=\tilde{v}_{\mu}(\k,0)$, $J_{\mu}(\k)=J_{\mu}(\k,0)$,
and so on.
Above expressions are similar to the consequence of the Boltzmann equation;
the velocity is replaced by the total current vertex $J_{\mu}$.
In Eq.~(\ref{eq:angle-rep}), we have defined the angle of the current vertex as
$\varphi(\k)={\rm Arctan}(J_{x}(\k)/J_{y}(\k))$
(see Fig.~\ref{fig:HotCold.tgif}).
It should be noticed that the Hall conductivity depends on the differential
of the angle $\varphi(\k)$ with respect to the momentum along the Fermi
surface $k_{\parallel}$.
In the isotropic systems, the current vertex is always perpendicular to
the Fermi surface, and therefore the relation $R_{{\rm H}} \propto 1/n$
is proved. This expression is, however, very delicate.
It is clearly understood from the above expressions
(Eqs.~(\ref{eq:conductivity})-(\ref{eq:angle-rep}))
that the Hall coefficient is not directly related to the carrier number $n$.
The electric transport is determined by the quasi-particles near the
Fermi surface, not by the excitations deeply below the Fermi level.

\begin{figure}[t]
\begin{center}
\vspace{2.5cm}
\caption{The schematic figure of the band velocity $\vec{v}$ and
total current vertex $\vec{J}$. Here the hot and cold spots are shown.
The dashed line indicates the magnetic Brillouin zone boundary,
while the dotted line denotes the wave vector $\Q=(\pi,\pi)$.
Note that the angle $\theta$ is also used in other figures.}
\label{fig:HotCold.tgif}
\end{center}
\end{figure}

In the following, we use
Eqs.~(\ref{eq:conductivity})-(\ref{eq:Hallconductivity}),
although Eqs.~(\ref{eq:eliashberg-Kohno})-(\ref{eq:angle-rep}) are expected
to give qualitatively same results.
The resulting resistivity $\rho$ and Hall coefficient $R_{{\rm H}}$
are obtained by the formula, $\rho = 1/\sigma_{xx}$ and
$R_{{\rm H}}=\sigma_{xy}/\sigma_{xx}^{2} H$, respectively.
Hereafter, the constant factor arising from 
the unit of charge $e$ is omitted.

It should be noted that the coherent transport is assumed in the above
expressions which are based on the Fermi-liquid theory.
This assumption seems to be incompatible with the pseudogap induced
by the ``resonance scattering'', where the extremely large damping
is the origin of the pseudogap.
However, this difficulty is removed by the characteristic momentum
dependence; the pseudogap occurs at the hot spot, while the in-plane
transport is determined by the cold spot as explained below.
Because the coherent nature of the quasi-particles is sufficiently maintained
at the cold spot (Figs.~\ref{fig:HotCold.tgif} and \ref{fig:LifeFS}),
the above formula are justified even in the pseudogap state. 

Now let us explain the microscopic theories. 
First the results of the spin fluctuation theory are reviewed 
\cite{rf:yanaseTRPG,rf:kontani}. 
We show the calculated results of the resistivity and Hall coefficient
in the FLEX approximation. 
The four-point vertex in the FLEX approximation includes three terms
(see Fig.~\ref{fig:AFMT}).
Because the spin-fluctuation Maki-Thompson (SPMT) term shown
in Fig.~\ref{fig:AFMT}(a) is dominant among them, 
only the SPMT-term is taken into account
\cite{rf:yanaseTRPG,rf:kontani,rf:kanki} 

\begin{figure}[t]
\begin{center}
\vspace{2.5cm}
\caption{The four-point vertex in the FLEX approximation.
The wavy dashed line represents the spin fluctuation.
(a) The ``SPMT-term'' which is dominant.
The diagrams (b) and (c) are usually negligible.}
\label{fig:AFMT}
\end{center}
\end{figure}

We provide a detailed explanation because a part of the following
understanding is obtained in recent years.
The $T$-linear law of the electric resistivity and the enhancement of
the Hall coefficient are discussed. These behaviors are the
characteristics of the nearly AF Fermi liquid and observed in 
the experimental results above $T^{*}$. 

We point out three important properties from which the unconventional
transport above $T^{*}$ is explained.
The first is the momentum dependence of the quasi-particle lifetime
$\tau(\k)$. The typical results are shown in Fig.~\ref{fig:LifeFS}.
We see that the lifetime is long at the cold spot $\theta \sim \pi/4$ 
and short at the hot spot $\theta \sim 0$ or $\sim \pi/2$
\cite{rf:hlubina,rf:stojkovic,rf:yanaseTR}.
When the exchange of the AF spin fluctuation is the dominant scattering 
process, the quasi-particle damping $\gamma(\k)=-{\rm Im}\s^{\rm R}(\k,0)$
is almost determined by the low-energy DOS around $\kk = \k+\Q$. 
Then, the lifetime is short around the magnetic Brillouin zone boundary 
and/or around the van Hove singularity. 
The electric transport is practically carried
by the quasi-particles at the cold spot which is located around
$\k=(\pi/2,\pi/2)$ (see Fig.~\ref{fig:HotCold.tgif}).
This momentum dependence of the lifetime affects the spectral function,
as is explained in Sec.~4.3.1, which is actually observed in the ARPES
measurements \cite{rf:shenrep,rf:ARPESreview,rf:shenschrieffer,rf:bogdanov}.
Also the magnetic-transport measurement has confirmed this momentum
dependence \cite{rf:hussey}. 

\begin{figure}[t]
\begin{center}
\vspace{2.5cm}
\caption{The momentum dependence of the lifetime $\tau(\k)$ on the Fermi
surface. The circles and triangles correspond to $T=0.005$ and $T=0.009$,
respectively and $U/t=3.2$. The solid curve is a result in the electron-doped
region ($\delta=-0.10$, $U/t=3$ and $T=0.018$).}
\label{fig:LifeFS}
\end{center}
\end{figure}

\begin{figure}[t]
\begin{center}
\vspace{2.5cm}
\caption{The results of the FLEX approximation at $U/t=3.2$.
(a)The resistivity and (b) Hall coefficient in under-doped
($\delta=0.09$, circles), optimally-doped ($\delta=0.15$, squares),
and electron-doped ($\delta=-0.1$, stars) region.}
\label{fig:FLEX-R}
\end{center}
\end{figure}

The second is the $T$-linear dependence of the damping rate
$\gamma_{{\rm c}}= \gamma(\k_{{\rm c}})$, 
where $\k_{{\rm c}}$ is the momentum at the cold spot.
The $T$-linear resistivity originates from this $T$-linear dependence
\cite{rf:hlubina,rf:stojkovic,rf:yanaseTR}. 
It have been pointed out that the $T^{2}$-resistivity is always
obtained in the low-temperature limit even in the quantum critical point
unless the Fermi surface is perfectly nested \cite{rf:yanaseTR}.
This is because the quasi-particles at the cold spot are not directly
scattered by the AF spin fluctuation at $\q=\Q$.
Note that the anomalous power-law can be induced by the slight impurity
scattering near the quantum critical point \cite{rf:rosch}.
However, the crossover temperature from $T$-square to $T$-linear resistivity
is sufficiently small in the under-doped region (Fig.~\ref{fig:FLEX-R}(a)).
The transformation of the Fermi surface plays a role to reduce
the crossover temperature \cite{rf:yanaseTR}.

The third is the temperature dependence of the 
vertex correction \cite{rf:kontani,rf:kanki}. 
We will see that the contribution from the SPMT-term to the total 
current vertex $\vec{J}(\k)$ significantly enhances the Hall coefficient.
The schematic figure (Fig.~\ref{fig:HotCold.tgif}) shows that
the total current vertex $\vec{J}(\k)$ is significantly altered
from the band velocity $\vec{v}(\k)$.
This is an effect of the AF spin fluctuation.
Then, the four-point vertex $\Im_{22}(\k,\e:\kk,\ee)$ is significantly
enhanced around $\k - \kk =\Q$.
In the vicinity of the magnetic instability, the differential
$\partial \varphi(\k)/\partial k_{\parallel}$ increases around the cold spot.
Therefore, the Hall coefficient increases with the development of
the spin fluctuation (Fig.~\ref{fig:FLEX-R}(b)). 
The vertex correction is not so important for the resistivity. 
The resistivity is enhanced by the SPMT-term, however, the qualitative
temperature dependence is not altered (Fig.~\ref{fig:FLEX-R}(a)).

The nearly AF Fermi-liquid theory explains the transport phenomena
also in the electron-doped region. 
Then, the cold spot is located around $(\pi,\pi/4)$ and $(\pi/4,\pi)$, 
as is shown in Fig.~\ref{fig:LifeFS}, where the Fermi surface is far
from the magnetic Brillouin zone (see Fig.~\ref{fig:fermi}). 
The large damping around the magnetic Brillouin zone is consistent with 
the observation in the ARPES measurements \cite{rf:almitage}, 
which confirms this picture. 
Then, the electric transport shows considerably different behaviors.
Combined with the sharp commensurate structure of the
magnetic excitation (Fig.~\ref{fig:kaiflex}),
the quasi-particle lifetime at the cold spot follows
the usual $T$-square law.
Note that the lifetime is much longer than the hole-doped case.
Therefore, the resistivity shows the $T$-square law with a small magnitude
\cite{rf:yanaseFLEXPG}.
More interestingly, the Hall coefficient changes its sign because the
differential of the angle
$\partial \varphi(\k)/\partial k_{\parallel}$
has negative sign around $(\pi,\pi/4)$ \cite{rf:kontani,rf:kanki}.
These behaviors are consistent with the experimental results
\cite{rf:electrondopesuper,rf:iye,rf:satoM}.

In the following part, we take account of the SC fluctuation and
discuss the electric transport in the pseudogap state \cite{rf:yanaseTRPG}.
In order to make the discussion clear,
we classify the effects of the SC fluctuation in the following way.
\begin{enumerate}
\item  The pseudogap in the single-particle properties.
\item  The feedback effects through the AF spin fluctuation.
\item  The vertex corrections from the SC fluctuation.
       The AL-term is classified into them.
\end{enumerate}
The following calculation identifies the effect (2) as a main contribution.
That is, the coupling between the spin and SC fluctuations plays an essential
role for the electric transport. 
Therefore, we perform the self-consistent FLEX+T-matrix (SCFT) 
approximation \cite{rf:yanaseFLEXPG} in which the dynamical 
spin susceptibility, T-matrix, and the Green function are 
determined self-consistently. 

We first discuss the resistivity.
The effect (1) obviously reduces the longitudinal and transverse
conductivities. 
However, the increase of the resistivity is not significant because
the pseudogap occurs at the hot spot which is not important for the 
transport phenomena. It should be noted that the self-energy 
at the cold spot is always dominated by the spin fluctuation.
Then, the effect (2) increases the conductivity and
gives the larger contribution than (1). 
The quasi-particle damping arising from the spin fluctuation
$\gamma_{{\rm F}}(\k)=-{\rm Im} \s_{{\rm F}}^{{\rm R}}(\k,0)$
is reduced by the pseudogap in the spin excitation. 
Thus, the SC fluctuation induces the downward deviation of 
the resistivity around $T=T^{*}$ (Fig.~\ref{fig:SC-ResistivityU=2.0}(a))
through the feedback effect.
This downward deviation becomes more remarkable as decreasing the hole-doping,
which is consistent with the experimental results
\cite{rf:ito,rf:mizuhashi,rf:odatransport,rf:takenaka}.

\begin{figure}[t]
\begin{center}
\vspace{2.5cm}
\caption{(a) The temperature dependence of the resistivity $\rho$
at $U/t=4.0$ without the effect (3).
The closed and open symbols show the results with and without the SPMT-term,
respectively. (b) The effects of the vertex correction. The results with
the SCMT-term and with the SCMT- and AL-terms are shown.}
\label{fig:SC-ResistivityU=2.0}
\end{center}
\end{figure}

It is another important property that the deviation of the resistivity
is only a slight one, while the NMR $1/T_{1}T$ is remarkably reduced by
the SC fluctuation.
This is because the cold spot is not so sensitive to spin fluctuation
at $\q=\Q$. As shown in Fig.~\ref{fig:Imkai_under_FLEX}, the suppression
of the magnetic excitation is moderate for the incommensurate component
$\q \neq \Q$. This is an origin of the incommensurate peak in the
pseudogap state. Since the quasi-particle damping at the cold spot 
is determined by some average of the incommensurate component, 
the enhancement of the lifetime is not so significant. 
Thus, the weak response of the resistivity to the pseudogap 
is explained by the detailed analysis of the momentum dependence. 
This explanation is in sharp contrast with that in the RVB theory 
\cite{rf:ioffeRVB,rf:nagaosaTR,rf:onodaTR}. 

\begin{figure}[t]
\begin{center}
\vspace{2.5cm}
\caption{(a) The four-point vertex in the lowest order with 
respect to the SC fluctuation (SCMT-term).
(b) The Feynmann diagram representing the AL-term. 
The wavy dashed line represents the SC fluctuation.}
\label{fig:SCMT2}
\end{center}
\end{figure}

We have clarified the role of the vertex correction arising
from the SC fluctuation \cite{rf:yanaseTRPG}. 
Then, we have estimated the lowest-order term within the \'Eliashberg theory
(``SCMT-term'') and the AL-term \cite{rf:AL} beyond the \'Eliashberg theory 
(Fig.~\ref{fig:SCMT2}). 
Both contributions increase the conductivity. 
It should be noted that the SCMT term is not equivalent to 
the Maki-Thompson term \cite{rf:Maki} which has been investigated 
in the fluctuation theory 
\cite{rf:varlamovreview,rf:dorin,rf:bieri}. 
The SCMT term includes the lowest order contribution with respect to 
$1/\tau(\k)$, and take into account the vertex correction iteratively. 
Note that the diffusion propagator, which is important in the $s$-wave 
superconductor, is negligible in the $d$-wave superconductor. 

Figure \ref{fig:SC-ResistivityU=2.0}(b) shows that the SCMT-term does not
affect the temperature dependence, qualitatively.
This is mainly because of the momentum dependence of
the $d$-wave order parameter.
Although the contribution from the SCMT-term increases as the
temperature decreases, it is not visible in the temperature
dependence of the resistivity because the other contribution to
the conductivity also increases. 

The role of the AL-term may be interesting because this contribution 
has been intensively investigated in the fluctuation theory 
\cite{rf:varlamovreview,rf:dorin,rf:bieri,rf:IAV}. 
The AL-term is interpreted as the conductivity carried by the
fluctuating Cooper-pairs~\cite{rf:tinkham}, and should be written as the
superconducting part $\sigma_{{\rm s}}$ in contrast to the normal part
$\sigma_{{\rm n}}$ included in the \'Eliashberg theory.
We have concluded that the AL-term is almost negligible
in the wide temperature range (see Fig.\ref{fig:SC-ResistivityU=2.0}(b)).
This is simply because the AL-term is higher order
with respect to the parameter 
$1/l_{\rm max}=1/v(\k_{{\rm c}})\tau(\k_{{\rm c}})$. 
The AL-term becomes dominant just above $T_{\rm c}$ because this term is
more divergent with respect to $1/t_{0}=1/(1-\lambda(0))$. 
However, this temperature range is narrow because of the long lifetime 
of the quasi-particle at the cold spot. 
The detailed analysis on the fluctuation conductivity in the vicinity of
the critical point has been performed within the weak-coupling theory 
\cite{rf:varlamovreview,rf:dorin,rf:bieri,rf:IAV}. 
Here we stress that the downward deviation of the resistivity is not
attributed to the AL-term, but to the feedback effect through the 
suppression of the spin fluctuation. 

Here we discuss the role of the AL-term furthermore,
since some characteristics of the high-$T_{\rm c}$ cuprates manifest.
There are two important points in the above discussion on the AL-term.
One is the strong-coupling superconductivity.
While the resonance scattering is strong in the superconductor with
short coherence length, the AL-term does not directly depend on the
coherence length in two dimensions \cite{rf:AL,rf:tinkham}.
This is because the short coherence length means the small velocity
of the fluctuating Cooper-pairs.
Then, the pseudogap occurs under not so small value of $t_{0}$
where the AL-term is still small. 
The other is the characteristic momentum dependences 
in high-$T_{\rm c}$ cuprates. 
For example, in case of the strong-coupling $s$-wave superconductors,
the AL-term will be much more important. This is because the pseudogap opens
on the whole Fermi surface and therefore the normal part $\sigma_{{\rm n}}$
is remarkably suppressed.

Let us comment on the c-axis transport, which shows qualitatively
different behaviors from the in-plane transport.
That is, the {\it c}-axis transport is strongly incoherent in the pseudogap
state, while the in-plane transport is sufficiently coherent
\cite{rf:takenaka}.
We can understand these qualitatively different behaviors in a consistent way.
Then, the momentum dependence of the inter-layer hopping matrix plays
an essential role \cite{rf:yanaseTR,rf:ioffe}.
The result of the band calculation has shown the following behavior
\cite{rf:okanderson}, 
\begin{eqnarray}
  \label{eq:okanderson}
  t_{\perp}(\k) = t_{\perp} ({\rm cos} k_{x} - {\rm cos} k_{y})^{2}.
\end{eqnarray}
In short, the inter-layer hopping matrix vanishes at the cold spot and 
the {\it c}-axis transport is mainly determined by the hot spot. 
Therefore, the {\it c}-axis conductivity is significantly reduced
by the effect (1).
Actually, the incoherent nature of the {\it c}-axis optical conductivity
has been obtained in the pseudogap state within the FLEX+T-matrix
approximation, together with the coherent nature of the
in-plane optical conductivity \cite{rf:yanaseISSP}.
This is an experimentally observed behavior
\cite{rf:homes,rf:basov,rf:tajima}.

Naively, it may be expected that the AL-term is important in the c-axis
transport instead of the quasi-particle transport.
However, the AL-term is higher order with respect to the inter-layer hopping,
{\it i.e.}, $\sigma_{{\rm s}} \propto t_{\perp}^{4}$ while
$\sigma_{{\rm n}} \propto t_{\perp}^{2}$ \cite{rf:dorin,rf:IAV}.
Therefore, the AL-term is negligible in the quasi-2D system.
We stress that the qualitative difference between the {\it c}-axis and
in-plane transport is not attributed to the difference of the AL-term
\cite{rf:dorin,rf:IAV}, but to the difference of the normal part
$\sigma_{{\rm n}}$. 
In order to arrive at this conclusion, we have to consider
the spin and SC fluctuations simultaneously and to take account of the
momentum dependent inter-layer hopping matrix.

Finally let us discuss the Hall coefficient.
The Hall coefficient increases owing to the effect (1),
because the momentum dependence of the lifetime is enhanced by (1). 
Actually, the SC fluctuation was proposed to be the origin of the 
enhancement of the Hall coefficient \cite{rf:ioffe}. 
However, the situation is quite altered by taking account of the SPMT-term.
The kernel of the Bethe-Salpeter equation (Eq.~(\ref{eq:currentvertex}))
includes the four-point vertex arising from the spin fluctuation and
the absolute value of the Green function.
Therefore, the vertex correction is reduced by the feedback effect (2)
and furthermore by the effect (1). 
As a result, the Hall coefficient is reduced by the SC fluctuation 
through the SPMT-term. 
We have confirmed that the feedback effect (2) is dominant also for
the Hall coefficient. 
The calculated results in Fig.~\ref{fig:SC-HallU=2.0} 
clearly show the pseudogap behavior of the Hall coefficient. 
The Hall coefficient shows the peak around $T=0.006$ and decreases
with decreasing the temperature.
This phenomenon becomes moderate with increasing the hole-doping.
These results qualitatively explain the experimental results
in the pseudogap state including the doping dependence
\cite{rf:ito,rf:mizuhashi,rf:satoM,rf:xiong,rf:ong}. 

\begin{figure}[t]
\begin{center}
\vspace{2.5cm}
\caption{(a) The temperature dependence of the Hall coefficient
$R_{{\rm H}}$ at $U/t=4.0$ without the effect (3).
The circles and squares correspond to $\delta=0.09$ and $\delta=0.15$,
respectively. (b) The effects of the vertex correction. The meanings of
the lines are the same as in Fig.~\ref{fig:SC-ResistivityU=2.0}(b).}
\label{fig:SC-HallU=2.0}
\end{center}
\end{figure}

The vertex correction arising from the SCMT-term enhances
the momentum dependence of the angle $\varphi(\k)$ \cite{rf:yanaseTRPG}.
Therefore, the Hall coefficient is enhanced by the vertex correction
arising from the SC fluctuation.
It is notable that this enhancement does not occur without the SPMT-term.
In other wards, the SCMT-term indirectly enhances the Hall coefficient
through the combination with the SPMT-term.
Figure \ref{fig:SC-HallU=2.0}(b) shows that this enhancement is not
so significant to alter the qualitative behaviors. 
This is also because of the wave function of the $d$-wave superconductivity.
Including the AL-term in the longitudinal conductivity, the suppression of
the Hall coefficient becomes clearer.

Thus, the transport coefficients in the pseudogap state are explained by
simultaneously taking into account the spin fluctuation and SC fluctuation.
It is confirmed that characteristic behaviors of the electric transport
are mainly caused by the feedback effect through the pseudogap in
the spin fluctuation. 
This effect is not outstanding in the resistivity, but rather apparent
in the Hall coefficient.
This difference reflects the importance of the SPMT-term for these
quantities. We stress that the conventional theory on the fluctuation
conductivity is not satisfactory, but detailed knowledge on the
electronic structure are required for the understanding of
high-$T_{\rm c}$ cuprates.

Recently, the theory including the spin and SC fluctuations has been
extended to the thermal transport \cite{rf:kontaninernst}.
It is shown that the vertex correction from the SCMT-term combined
with the SPMT-term significantly enhances the Nernst coefficient.
This feature is also consistent with the experiments \cite{rf:ongnernst}.
Although the vortex excitation has been considered as an origin of
this behavior \cite{rf:ongnernst}, we consider that the contribution from
quasi-particles is dominant in the wide temperature region like
for the electric conductivity.
That is, the understanding within the \'Eliashberg theory will be relevant also
for the thermal transport.
The importance of the superconducting part in the vicinity
of the critical point is common. The analysis on the superconducting
part under the magnetic field has been recently given
\cite{rf:ikedanernst}.

\begin{figure}[t]
\begin{center}
\vspace{2.5cm}
\caption{The phase diagram obtained by the FLEX and SCFT approximations.}
\label{fig:phasediagram}
\end{center}
\end{figure}

\subsubsection{Phase diagram}

At the last of this section, we show the phase diagram in
Fig.~\ref{fig:phasediagram}.
The SCFT approximation is used to estimate $T_{\rm c}$ suppressed
by the fluctuation. The effects of the SC fluctuation are similarly
classified into the following two parts:
(1) The pseudogap in the single-particle properties.
(2) The feedback effect through the spin fluctuation.
We have concluded that the feedback effect enhances the transition temperature
\cite{rf:yanaseFLEXPG}.
The spin fluctuation has both the pairing and de-pairing effects.
Since the low-frequency component mainly affects as a de-pairing source,
the pseudogap in the spin fluctuation rather enhances the superconductivity
(see also Sec.~4.3.3).
However, the effect (1) is dominant in the present case and significantly suppresses
$T_{\rm c}$.
Note here that strictly speaking, the transition temperature should be
always zero ($T_{\rm c}=0$) in the two-dimension,
which is known as the Mermin-Wagner theorem.
However, the singularity of the two-dimensional system is always removed
in the layered systems. Taking account of the weak inter-layer coupling,
we determine the critical point as $\lambda_{{\rm e}}=0.98$ instead of
$\lambda_{{\rm e}}=1$.
This criterion corresponds to the 2D-3D crossover of the SC fluctuation
with the anisotropy being $\xi_{{\rm ab}}/\xi_{{\rm c}} = 10$ and
$\xi_{{\rm ab}} = 2 \sim 3$.

It is shown that the suppression of $T_{{\rm c}}$ from the mean-field value
becomes remarkable with under-doping.
This is a natural consequence because the SC fluctuation becomes strong
with under-doping (see Sec.~4.3.2). 
In other words, the pseudogap develops with under-doping and
the reduced DOS gives the suppressed $T_{\rm c}$.
It is an interesting result that the transition temperature takes
the maximum value at $\delta \sim 0.11$ and decreases with under-doping
in the SCFT approximation for $U/t=4.4$.
This feature is in sharp contrast to the FLEX approximation where
the $T_{{\rm c}}$ goes on increasing with decreasing $\delta$
(see also Sec.~3.2.4).
Thus, the mean-field transition temperature develops with 
under-doping, however, the strong SC fluctuation decreases the 
$T_{\rm c}$ in the under-doped region. 
This picture is consistent with the experimental results from the 
tunnelling measurements \cite{rf:harris,rf:odatransport}, 
where the SC gap develops with decreasing $\delta$ in the under-doped region. 
We can see from Fig.~\ref{fig:phasediagram} that the decrease of
$T_{\rm c}$ in the under-doped region is only a slight one
in case of $U/t=3.2$. 
Thus, the strong electron correlation plays an essential role
for describing the under-doped region.
It is also important that we can treat the strong correlation
by considering the SC fluctuation which suppresses the AF order.
The stabilization of the metallic state by the SC fluctuation is 
actually expected in the hole-doped cuprates. 

It is worth while to write again that the strength of the SC coupling
is indicated by the ratio $T_{{\rm c}}^{{\rm MF}}/E_{{\rm F}}$,
and not by $T_{{\rm c}}$.
Since the mean-field transition temperature $T_{{\rm c}}^{{\rm MF}}$
increases and the effective Fermi energy $E_{{\rm F}}$ decreases with
under-doping, the superconductivity becomes strong coupling
in spite of the decreasing $T_{{\rm c}}$. 
Thus, the decreasing $T_{{\rm c}}$ in the under-doped region is obtained
by starting from the microscopic Hamiltonian and taking account of the
SC fluctuation. This is an important development for the theory starting
from the Fermi-liquid state.
As has been discussed in Sec.~3.2.5, the SCFT approximation underestimates
the effects of the SC fluctuation. Therefore, the more significant
doping dependence is expected in the higher-order theory,
which is beyond the scope of this review. 

It is quite clear that the SCFT approximation is not sufficient to show the
disappearance of the superconductivity.
If we describe the phase boundary between the SC phase and
the spin glass phase, the loss of the metallic behavior will be an essential
aspect. Then, the validity of the FLEX approximation will be a subject.
The systematic treatment for the electron correlation and the disorder will
be necessary for this issue. This is an interesting and open problem.

\par\vfill
\eject

%
%

\def\Sig{\mathit{\Sigma}}
\def\Gam{\mathit{\Gamma}}
\def\Del{\mathit{\Delta}}
\def\Tr{\mathrm{Tr}}
\def\Vec#1{{\bm #1}}

\section{Heavy-Fermion Systems}

In previous sections, we have discussed the superconductivity in
high-$T_{\rm c}$ cuprates, organic superconductors, and Sr$_2$RuO$_4$.
The pairing mechanism and several physical properties of these
materials have been explained.
In the calculation of the transition temperature within the mean-field
theory, we have adopted the third-order perturbation (TOP) theory with
respect to the on-site Coulomb repulsion $U$ for the weak-coupling case.
When the anti-ferromagnetic (AF) spin fluctuation is strongly enhanced,
the fluctuation-exchange (FLEX) approximation has been applied.
In general, these two methods are useful for the weak and intermediate
coupling regimes in $U$.

Here note that the above approaches stand on a common basis of
the Fermi-liquid state.
As far as the system is in a metallic state, or more strictly,
the existence of Fermi-liquid quasi-particles is insured,
such methods have an established basis.
This is related to the analytic property in $U$ and
the continuity principle stressed by P. W. Anderson \cite{rf:Anderson}.
In this case, the long-lived Fermi-liquid quasi-particles form
the Fermi surface at low temperatures.
The superconductivity is induced by the effective interaction
among quasi-particles and the paring symmetry is determined
by the momentum dependence of the four-point vertex function,
which is naturally caused by the short-range repulsion
through the many-body effect.
Thus, the superconductivity in strongly correlated materials discussed
above can be understood from such a unified view.

On the other hand, the superconductivity in heavy-fermion compounds,
which are the typical SCES materials, has not been explained
from the microscopic point of view, mainly due to the complicated
band  structures and the strong correlation effect.
However, we believe that the key mechanism of the superconductivity
in these compounds should be also explained on the same footing
as described in previous sections.
It is expected that such efforts will extend the universality of
the understanding on the superconductivity.
Then, the identification of the residual interaction between the
quasi-particles is an essential task for this purpose.

In this section, we introduce a strategy to treat heavy-fermion
superconductors and discuss the possible applications to some
Ce-based and U-based heavy-fermion superconductors.
Then, we analyze the effective single $f$-band model by choosing
one dominant band.
Further analysis based on more realistic models
is a desirable future issue, but we believe that
the essential physics will be unchanged and the discussion
in this section will be a basis in the future progress.

\subsection{Experimental view}
\label{sec:introhv}

Most of intermetallic compounds with several localized $f$-electrons
at each atomic site exhibit some kinds of magnetic transition
by the Ruderman-Kittel-Kasuya-Yosida (RKKY) interaction.
Some of Ce-based and U-based compounds, however, form coherent itinerant
electron bands at low temperatures due to the mixing effect
with conduction electrons, while they possess magnetic characters
at high temperatures.
Such electron systems are frequently called ``heavy-fermion systems''
\cite{rf:Grewe}, since the effective mass of the itinerant electron
becomes several hundred times larger than free-electron mass $m_0$
due to the strong electron correlation.
In 1979, the superconductivity in CeCu$_2$Si$_2$, which is one of 
typical heavy-fermion systems, has been reported by Steglich et al.
\cite{rf:Steglich1}.
The discovery of the superconductivity in the material with remarkable
magnetic characters has implied a scenario clearly different from
the electron-phonon mechanism in the conventional BCS theory.
In fact, physical properties in the superconducting phase
at low temperature have shown power-law temperature dependence,
different from exponential decay observed
in the conventional $s$-wave superconductor.
Thus, CeCu$_2$Si$_2$ has become the pioneering discovery of
unconventional superconductivity in SCES.
It is now considered as an even-parity superconductor with line-nodes,
probably $d$-wave pairing symmetry.

Since then, many unconventional superconductors have been discovered
in heavy-fermion systems.
Multi-phase diagrams in UPt$_3$ \cite{rf:stewart,rf:Stewart2,rf:Hasselbach}
and U(Be$_{1-x}$Th$_x$)$_{13}$ \cite{rf:sureview,rf:Ott1,rf:Ott2,rf:Kuramoto}
indicates curious superconductivity with multi-components.
In particular, UPt$_3$ is the odd-parity superconductor
first discovered in electronic systems \cite{rf:Tou1}.
UPd$_2$Al$_3$ and UNi$_2$Al$_3$ \cite{rf:Geibel1,rf:Geibel2} are
unconventional superconductors coexisting with the AF phase,
and are considered to have even- and odd-parity pairing states,
respectively \cite{rf:Kyogaku,rf:Ishida1}.
URu$_2$Si$_2$ indicates coexistence of unconventional superconductivity
and a hidden order \cite{rf:Palstra}.
Furthermore, recent progress in experiments under pressures have
promoted discoveries of new superconductors:
CeCu$_2$Ge$_2$ \cite{rf:Jaccard1,rf:Jaccard2},
CePd$_2$Si$_2$ \cite{rf:Grosche,rf:Mathur},
CeRh$_2$Si$_2$ \cite{rf:Movshovich1},
CeNi$_2$Ge$_2$ \cite{rf:Lister},
and CeIn$_3$ \cite{rf:Mathur,rf:Walker}.
These compounds are AF metals at ambient pressure,
while under high pressures, the AF phases abruptly disappear
accompanied by SC transitions.
Except for cubic CeIn$_3$, all other materials have the same
ThCr$_2$Si$_2$-type crystal structure as in CeCu$_2$Si$_2$.

In addition, quite recently, several kinds of new heavy-fermion
superconductors have been discovered.
One is a family of CeTIn$_5$ (T=Co, Rh, and Ir) with HoCoGa$_5$-type
crystal structure \cite{rf:Hegger,rf:Petrovic1,rf:Petrovic2}.
Since the discovery, a variety of experimental investigations have
rapidly increased.
These compounds possess relatively high transition temperature
such as $T_{\rm c}=2.3$K for CeCoIn$_5$, which is the highest
among heavy fermion superconductors observed yet.
The dominant AF spin fluctuations have been suggested by
the existence of the neighboring AF phase in the pressure-temperature
($P-T$) phase diagram, the power-law ($T^{1\sim 1.3}$) behavior in the
resistivity, and the magnetic behavior observed in the NQR/NMR $1/T_1$.
The situation is quite similar to other Ce-based superconductors
mentioned above.
Another is the coexistence of superconductivity and ferromagnetism in UGe$_2$
at high pressure and URhGe at ambient pressure \cite{rf:saxena,rf:Aoki}.
Much attention has been attracted to features of the superconductivity,
since it is naively believed that a triplet pairing state coexists with
the ferromagnetic phase.
In fact, the unconventional $T^3$-behavior has been observed
in the NQR/NMR-$1/T_1$ \cite{rf:Kotegawa}.
In addition, other new superconductors, including neither Ce nor U atom,
have been also discovered and attracted attentions.
In the filled skutterudite compound PrOs$_4$Sb$_{12}$ ($T_{\rm c}=1.85$K),
the possibility of the double transition has been indicated \cite{PrOs4Sb12}.
In the Pu-based compound PuCoGa$_5$ with the same crystal structure as
a family of CeTIn$_5$, a very high transition temperature $T_{\rm c}=18.5$K
has been reported \cite{PuCoGa5}.

As introduced above, heavy-fermion superconductors show a great variety
of ground states and offer rich examples to investigate unconventional
superconductivity in SCES. We cannot review here in detail the
superconductivity in each heavy-fermion compound.
Alternatively, in the following subsections, we overview characteristic
features in Ce-based and U-based heavy-fermion superconductors,
relevant to the key mechanism of the superconductivity.
For more details of each material, readers can consult the review
articles by Stewart \cite{rf:Stewart2}, Grewe and Steglich \cite{rf:Grewe},
and Sigrist and Ueda \cite{rf:sureview}.

\subsubsection{Ce-based compounds}

Ce-based heavy-fermion compounds possess the typical nature of
what is known as ``Kondo effect'' in the impurity case \cite{rf:Hewson1}.
The impurity Kondo problem is a typical example of the Fermi-liquid
formation in many-body systems.
The local spin fluctuation at impurity (Ce) site, observed
at high temperatures, is quenched at low temperature by the mixing effect
with the conduction electrons.
The system forms the local Fermi-liquid state and has the singlet
ground state as a whole.
In the process, the resistivity shows the $\log T$-behavior with
decreasing $T$ and eventually arrives at the value of the unitarity limit.
The change of the behavior is marked by a characteristic temperature
$T_{\rm K}$, which is called the Kondo temperature.
In the actual case, there remains degeneracy in $f$ orbitals.
For instance, assuming that the crystal-field (CF) ground state is
$\Gamma_7$ doublet and the excited state is $\Gamma_8$ quartet,
we obtain the characteristic Kondo temperature as \cite{rf:Hanzawa}
\begin{equation}
  T_{\rm K}=\left( \frac{D}{T_{\rm K}+\Delta} \right)^2
  D\exp(-N/\rho |J|),
\end{equation}
where $D$, $\Delta$, and $\rho |J|/N$ represent the conduction band-width,
the CF splitting, and the Kondo coupling constant, respectively.
For $T>\Delta$, the higher Kondo temperature
$T_{\rm K}^H \simeq D \exp(-N/3\rho |J|)$ is effective,
while for $T<\Delta$, the lower Kondo temperature
$T_{\rm K}^L \simeq (D/\Delta)^2 D \exp(-N/\rho|J|)$
becomes effective.
Then, the resistivity shows the broad peak around $T_\mathrm{K}^H$
and the typical $\log T$ dependence around $T_\mathrm{K}^L$.
This is the case also in the periodic lattice system.
The different point from the impurity case is that
in the periodic lattice systems, the resistivity smoothly changes into
a power-law decay (typically $T^2$) with a large coefficient $A$
at lower temperature.
This corresponds to the formation of coherent quasi-particle states
with a large effective mass $m^*$, typically $m^* \simeq 100 m_0$.
Here note that there exists a heuristic relation between
the effective mass and the coefficient of the $T^2$-resistivity.
Actually, the Kadowaki-Woods' relation, $A \propto \gamma^2$, has been
found in many heavy-fermion compounds \cite{rf:Kadowaki},
where $\gamma$ denotes the Sommerfeld coefficient
in the electronic specific heat.
For instance, the resistivity in CeCu$_2$Si$_2$ exhibits two hump structures
at $T \sim 100$K and $\sim 20$K. Then, it smoothly changes into $T^2$ 
dependence at low temperatures \cite{rf:Steglich1,rf:Grewe,rf:Kuramoto}.
These two characteristic temperatures ($T\sim 100$K and $\sim 20$K)
correspond to $T_\mathrm{K}^H$ and $T_\mathrm{K}^L$, respectively.
The metallic behavior with the $T^2$-resistivity at low temperatures indicates 
the formation of the coherent itinerant band of heavy quasi-particles.
The coefficient $A$ is also consistent with the large enhanced value of
the electronic specific-heat coefficient, $\gamma=0.7-1.1$J/K$^2$mol.

The compound CeCu$_2$Si$_2$ undergoes the transition into
the SC phase at $T_{\rm c}=0.7$K \cite{rf:Steglich1},
marked by the discontinuity in the specific heat.
The ratio of the jump to the normal-state specific heat
$\Delta C/\gamma T_{\rm c}$ is of the order of unity ($\sim 1.4$),
which could be explained in terms of the conventional BCS theory.
This fact is a strong evidence for the fact that the superconductivity
in this compound is a cooperative phenomenon in the heavy quasi-particles.
It is natural to consider that the heavy quasi-particles on the Fermi
surface form the Cooper-pairs with use of the residual interaction.
In this case, a couple of quasi-particles prefer to form some anisotropic
pairing state, not an $s$-wave one, in order to avoid
the strong on-site repulsion.
Actually, CeCu$_2$Si$_2$ in the SC phase exhibits the power-law dependence
in physical quantities at low temperatures, such as $T^3$-dependence
in the NMR/NQR $1/T_1$ \cite{rf:Kitaoka1,rf:Ishida2} and
$T^2$-dependence in the thermal conductivity $\kappa$ \cite{rf:Steglich2}.
All these facts imply zeros of the gap along lines on the Fermi surface.
In addition, the Knight shift decreases irrespective of directions
of the crystallographic axis \cite{rf:Ueda}.
Thus, the pairing symmetry is considered as an even-parity state with
line nodes, and presumably $d$-wave symmetry from many similarities
with the cuprates and the organic superconductors.

One of the remarkable similarities is the existence of the neighboring
AF phase in the $P-T$ phase diagram and the phase diagram of
chemical substitutions in a series of the isostructural compounds.
The ground state of CeCu$_2$(Si$_{1-x}$Ge$_x$)$_2$, with increasing $x$,
changes from the SC phase to the AF phase through the coexistent phase,
owing to negative chemical pressures by Ge substitution for Si
\cite{rf:YKawasaki,rf:Kitaoka2}.
This property is similar to that of the organic superconductor,
as is explained in Sec.~2.2.
An isostructural compound CeCu$_2$Ge$_2$ is the incommensurate
spin-density-wave (SDW) state with $\Vec{Q}=(0.28,0.28,0.54)$ at ambient
pressure \cite{rf:Knopp}. Under high pressures around 7GPa,
the magnetic phase abruptly disappears, accompanied by the SC transition
with $T_{\rm c}=0.5$K \cite{rf:Jaccard1,rf:Jaccard2}.
CePd$_2$Si$_2$ \cite{rf:Grosche,rf:Mathur},
CeRh$_2$Si$_2$ \cite{rf:Movshovich1},
and CeNi$_2$Ge$_2$ \cite{rf:Lister}
also exhibit the transition from the AF metals to the SC phases 
($T_{\rm c}\simeq 0.5$K) under high pressures.
This is also the case in CeIn$_3$ \cite{rf:Mathur,rf:Walker}
($T_{\rm c}=0.2$K) with the cubic AuCu$_3$-type structure.
The remarkable AF spin fluctuation in these compounds is actually implied
by some measurements.
The power smaller than two of temperature ($T^{1\sim 1.5}$) is observed
in the resistivity \cite{rf:Mathur}, and the NMR/NQR $1/T_1T$ increases
at low temperatures \cite{rf:YKawasaki,rf:SKawasaki1,rf:SKawasaki2}.
These properties are typical ones in the nearly AF Fermi-liquid
in quasi-2D and 3D systems.
The relationship of such AF spin fluctuation with the unconventional
superconductivity has been intensively investigated,
especially in the cuprates,
from the microscopic point of view (see Sec.~3.2).

A series of CeTIn$_5$ (T=Co, Rh, and Ir) discovered recently
\cite{rf:Hegger,rf:Petrovic1,rf:Petrovic2} possess many similarities
with the cuprates and the organic superconductors.
Thus, these compounds offer a great opportunity to bridge our understanding
in Ce-based heavy-fermion superconductors, high-$T_{\rm c}$ cuprates,
and organic superconductors.
These compounds are quasi-2D materials with the layered structure,
which has been confirmed by the de Haas-van Alphen (dHvA) measurements
\cite{rf:Haga,rf:Settai,rf:Hall,rf:Shishido,rf:Onuki}. 
The $P-T$ phase diagram in CeRhIn$_5$ \cite{rf:Kitaoka2} and
the phase diagram in the alloy system CeRh$_{1-x}$Ir$_x$In$_5$
\cite{rf:Pagliuso1} indicate that the superconductivity appears
in the neighborhood of the AF phase.
The power-law behavior of the resistivity in the normal phase with
$T^{1.3}$ in CeIrIn$_5$ ($T_{\rm c}=0.4$K) \cite{rf:Petrovic1} and
$T$ in CeCoIn$_5$ ($T_{\rm c}=2.3$K) \cite{rf:Petrovic2}
implies the strong quasi-2D AF spin fluctuation.
In the SC phase, no coherence peak just below $T_{\rm c}$,
$T^3$-behavior at low temperatures in the NMR/NQR $1/T_1$,
and the decrease in the Knight shift irrespective of directions
indicate the anisotropic even-parity superconductivity,
similar to other Ce-based heavy-fermion superconductors
\cite{rf:Kohori1,rf:Zheng,rf:Kohori2,rf:Curro,rf:Kohori3,rf:Mito}.
In addition, the four-fold symmetry in the thermal conductivity 
in CeCoIn$_5$ \cite{rf:Izawa} strongly suggests
the $d_{x^2-y^2}$-wave singlet pairing.
Furthermore, the pseudogap phenomenon has been reported
in the {$^{115}$In}-NQR measurement in CeRhIn$_5$
in the range of pressures ($P=1.53 \sim 1.75$GPa),
under which the coexistent phase exists \cite{rf:SKawasaki1}.
These similarities, especially the AF spin fluctuation and
$d_{x^2-y^2}$-wave pairing, indicate that the underlying physics
in the superconductivity in these Ce-based heavy-fermion superconductors
is in common with the cuprate superconductors.
This is also inferred from the microscopic electronic state,
where one $f$-electron exists per Ce site and the $f$-electron band itself
is almost half-filled when the hybridization with the conduction electron
band is not taken into account.

In the latter section, we verify the validity of the argument by
estimating $T_{\rm c}$ in the Ce-based heavy-fermion superconductors
on the same framework as in the cuprates.
We introduce neither the coexistent phase in the boundary between
the SC and AF phases nor the $A$ phase with
anomalous magnetic features \cite{rf:Kitaoka3}.
Although these are one of interesting issues of $f$-electron systems,
here we focused our interests on the superconductivity itself.

\subsubsection{U-based compounds}

Next we briefly review U-based heavy-fermion superconductors.
They exhibit various kinds of SC phases probably owing to multi $f$-band
structures with two or three $f$ electrons per U site.
We will show a possibility of our argument even in such complicated systems.

First we note that U-based compounds do not necessarily display
clear $\log T$ dependence in the resistivity.
This is attributed to the multi $f$-orbitals and relatively large
hybridization terms with the conduction electrons.
These properties also provide the complicated multi-band structures and
many Fermi surfaces with dominant $f$ character.
The Fermi surfaces obtained by the band calculations well explain
the dHvA measurements.
One of the characteristics of U-based compounds different from
Ce-based compounds is the clear coexistence between
the SC phase and some kind of magnetic ordered phase.
UPt$_3$ is the odd-parity superconductor ($T_{\rm c}=0.5$K) discovered
for the first time in the electronic systems
\cite{rf:Tou1,rf:stewart,rf:Stewart2,rf:Hasselbach}.
This superconductor coexists with an unusual AF order below
$T_{\rm N}=5$K \cite{rf:Aeppli},
which is observed in the neutron scattering.
Note that it has not been observed by the static and/or dynamically
slow probes. This unusual AF order at $\Vec{Q}=(0.5,0,1)$ becomes
the long-range order below 20mK within the resolution
of the neutron scattering \cite{rf:Koike}.
In URu$_2$Si$_2$, the unconventional superconductivity below 1.5K
also coexists with a hidden order with a clear jump at 17.5K
in the specific heat \cite{rf:Maple,rf:Matsuda}.
UPd$_2$Al$_3$ is an AF metal with $\Vec{Q}=(0,0,0.5)$
below $T_{\rm N}=14.5$K \cite{rf:Krimmel}
and coexists with the anisotropic even-parity superconductivity
below $T_{\rm c}=2$K \cite{rf:Geibel1,rf:Kyogaku}.
The isostructural compound UNi$_2$Al$_3$ is in the SDW state
with $\Vec{Q}=(0.5 \pm \tau,0,0.5)$ and $\tau=0.11 \pm 0.003$
below $T_{\rm N}=4.6$K \cite{rf:Schroder,rf:Lussier}.
This SDW state coexists with the unconventional superconductivity
below $T_{\rm c}=1.2$K \cite{rf:Geibel2},
which may be the odd-parity state as indicated
by the NMR/NQR measurement \cite{rf:Ishida1}.
Recently, in UGe$_2$ and URhGe, the superconductivity
in the ferromagnetic phase has been reported at $T_{\rm c}=0.75$K
(for $P$=11.4kbar) and 0.25K (ambient pressure),
respectively \cite{rf:saxena,rf:Aoki}. 

Although such a variety of interesting phenomena have been observed
experimentally, there are little theoretical progress
from the microscopic point of view.
One of the causes is the complicated band structure, since
U-based compounds possess several $f$ electrons per U site
and many Fermi-surface sheets originating from multi $f$-orbitals.
On the one hand, the complicated nature is unfavorable for
the theoretical effort, but on the other hand,
it is related to a great variety of the ground states.
It seems to be natural to consider that part of the Fermi surfaces
stabilizing the magnetic phase is different from that
leading to the SC state,
even though we cannot completely decouple them.
Then, we can discuss possibility of the superconductivity by
investigating the remaining Fermi surfaces in the magnetic phase.
We will introduce the simple application on the unconventional
superconductivity in the AF phase
in UPd$_2$Al$_3$ and UNi$_2$Al$_3$ later.
The former is considered as an even-parity superconductor,
while the latter is an odd-parity one.
We will discuss how this difference occurs. 

\subsection{Microscopic theory}
\label{sec:microhv}

First we introduce a theoretical treatment of the heavy-fermion
superconductors from the microscopic point of view.
As indicated in a variety of experiments for heavy-fermion compounds,
the superconducting transition occurs after the formation of the coherent
quasi-particle state with heavy effective mass.
Here we formulate a way to calculate $T_{\rm c}$ on the quasi-particle
description based on the Fermi-liquid theory.

Let us start our discussion on the periodic Anderson model (PAM)
as a typical model for heavy-fermion systems.
This model well describes the dual nature of $f$-electron systems.
The perturbation expansion in terms of the hybridization matrix element
leads to the RKKY interaction between the localized $f$-electron spins.
This is the origin of the magnetic transition in $f$-electron systems.
On the other hand, if no magnetic transition occurs, the system goes to
a singlet ground state as the whole.
This is the Fermi-liquid state of quasi-particles with heavy effective mass.
In this case, we should treat the PAM by the perturbation theory
with respect to the on-site Coulomb repulsion $U$ between $f$ electrons,
expecting the analyticity about $U$ as long as no phase transition occurs.
The analytic property has been exactly proved in the impurity Anderson
model \cite{rf:Tsvelisk,rf:Kawakami,rf:Okiji,rf:Zlatic}.
Although such an exact proof does not exist for the PAM,
the continuity principle is believed to justify
the applicability of the perturbation theory with respect to $U$.

From such a point of view, Yamada and Yosida have developed
the Fermi-liquid theory for the heavy-fermion state
based on the PAM \cite{rf:Yamada}.
The momentum independent part of the four-point vertex functions
$\Gamma_{\sigma\sigma'}^{\rm loc}$ between $f$ electrons
with spin $\sigma$ and $\sigma'$ plays an important role.
This dominant $s$-wave scattering part leads to the nearly momentum
independent large mass enhancement factor $\tilde\gamma=a^{-1}$.
In fact, when the momentum dependence of the mass enhancement factor
can be ignored, the $T$-linear coefficient of the specific heat
in the PAM, is given by
\begin{eqnarray}
  \gamma \simeq (2/3)\pi^2k_{\rm B}^2{\rho^f(0)}^2
  \Gam_{\uparrow\downarrow}^{\rm loc}
  =(2/3)\pi^2k_{\rm B}^2\rho^f(0)\tilde\gamma,
\end{eqnarray}
where $k_{\rm B}$ is the Boltzmann's constant and $\rho^f(0)$ is
the $f$-electron density of states at the Fermi level.
This indicates that the large mass enhancement factor
$\tilde\gamma(=a^{-1})$ is represented as
$\rho^f(0)\Gam_{\uparrow\downarrow}^{\rm loc}$
with use of the $s$-wave scattering part of the four-point vertex.
This result implies that the interaction between the quasi-particles
$a^2\Gam_{\uparrow\downarrow}^{\rm loc}$ holds the order of magnitude
of the effective band-width of quasi-particles $T_0 \simeq a/\rho^f(0)$,
where $T_0$ is the characteristic energy scale of the heavy-fermion state.
For $T<T_0$, the low-energy excitation can be described by
the Fermi-liquid theory.

On the other hand, the imaginary part of the $f$-electron self-energy,
which is proportional to the electric resistivity, is given by
\begin{eqnarray}
  \Delta_\Vec{k}=(4/3)(\pi\rho^f(0))^3
  {\Gam_{\uparrow\downarrow}^{\rm loc}}^2 T^2,
\end{eqnarray}
when the momentum dependence of the vertices can be ignored.
Thus, the coefficient $A$ of the $T^2$-term of the electrical
resistivity is proportional to $\gamma^2$ through the vertex
$\Gam_{\uparrow\downarrow}^{\rm loc}$.
This relation $A \propto \gamma^2$, which is well known
as the Kadowaki-Woods' relation \cite{rf:Kadowaki},
holds when the momentum dependence of the vertices is sufficiently
weak to be ignored.
Note that the relation does not hold when the momentum dependence of
the quasi-particle interaction is remarkable.
The high-$T_{\rm c}$ cuprate is the typical case,
as is explained in Sec.~4.3.
In most of heavy-fermion systems, the Kadowaki-Woods' relation
has been confirmed, indicating that the four-point vertex function,
i.e., the interaction among quasi-particles, possesses
rather weak momentum dependence.

The heavy-fermion superconductivity appears under this situation
for $T<T_0$.
The large $s$-wave repulsive part $\Gam_{\uparrow\downarrow}^{\rm loc}$
prevents an appearance of the $s$-wave singlet superconductivity.
In this case, anisotropic pairing states such as $p$- or $d$-wave
will be formed due to the remaining momentum dependence of
the four-point vertices in the particle-particle channel.
Here we discuss such anisotropic paring on the quasi-particles.

First we divide the four-point vertex function
$\Gam_{\sigma\sigma'}(p_1,p_2;p_3,p_4)$ into
the large $s$-wave scattering part $\Gam_{\sigma\sigma'}^{\rm loc}$
and the non-$s$-wave part $\Del\Gam_{\sigma\sigma'}(p_1,p_2;p_3,p_4)$;
\begin{equation}
  \Gam_{\sigma\sigma'}(p_1,p_2;p_3,p_4)=\Gam_{\sigma\sigma'}^{\rm loc}
  +\Del\Gam_{\sigma\sigma'}(p_1,p_2;p_3,p_4),
\end{equation}
where $p_1$ and $p_2$ are the incident momenta,
while $p_3$, $p_4$ the outgoing.
The second term $\Del\Gam_{\sigma\sigma'}(p_1,p_2;p_3,p_4)$ has
remarkable momentum dependence leading to
the the anisotropic Cooper-pairing.
Our purpose is to formulate how to calculate
$\Del\Gam_{\sigma\sigma'}(p_1,p_2;p_3,p_4)$
for the heavy-fermion quasi-particles, which are renormalized
by the large local part $\Gam_{\uparrow\downarrow}^{\rm loc}$.
Corresponding to these two terms of the vertex function, we can also
separate the self-energy into the local and non-local parts as
\begin{equation}
   \Sig(\Vec{k},\omega)=\Sig_{\rm loc}(\omega)+\Del\Sig(\Vec{k},\omega).
\end{equation}
In this case, the $f$-electron Green's function below $T_0$ is given by
\begin{eqnarray}
   \label{eq:G}
   G(\Vec{k},\omega)&=&
   \frac{1}{\omega-\xi_\Vec{k}-\Sig^{\rm loc}(\omega)
   -\Del\Sig(\Vec{k},\omega)-\frac{V_\Vec{k}^2}{\omega-\epsilon_\Vec{k}}},
   \nonumber \\
   &=&\frac{a}{\omega-\tilde{E}_\Vec{k}-\tilde{\Sig}(\Vec{k},\omega)
   -\frac{\tilde{V}_\Vec{k}^2}{\omega-\epsilon_\Vec{k}}}+G_{\rm inc}(\omega),
\end{eqnarray}
where $a$ is the mass renormalization factor,
$\tilde{E}_\Vec{k}=a(\xi_\Vec{k}+{\rm Re}\Sig^{\rm loc}(0))$
is the renormalized $f$-electron dispersion,
$\tilde{V}_\Vec{k}=\sqrt{a}V_\Vec{k}$
is the renormalized hybridization term,
$\tilde{\Sig}(\Vec{k},\omega)=a\Del\Sig(\Vec{k},\omega)$
is the renormalized $f$-electron self-energy,
and $G_{\rm inc}(\omega)$ denotes the momentum independent
incoherent part of the Green's function.
The imaginary part of $\Sig^{\rm loc}$ in proportion to $\omega^2+(\pi T)^2$
can be included in the remaining renormalized self-energy
$\tilde{\Sig}(\Vec{k},\omega)$, as indicated by Hewson for
the impurity Anderson model \cite{rf:Hewson2}.
From the Ward-Takahashi identity,
we can obtain the large mass enhancement factor as
\begin{equation}
  \tilde{\gamma}=a^{-1}=
  1-\left.\frac{d\Sig_{\rm loc}(\omega)}{d\omega}\right|_{\omega=0}
  =1-\frac{i}{2}\int\Gam_{\uparrow\downarrow}^{\rm loc}
  G_{\rm inc}^2(\omega)d\omega.
\end{equation}
In addition, we can set the large local vertex
$\Gam_{\uparrow\downarrow}^{\rm loc}$ as a constant value
at zero frequencies,
as far as the frequency of external lines is less than $T_0$.
In this case, the effective on-site interaction
$a^2\Gam_{\uparrow\downarrow}^{\rm loc}$ works on the quasi-particles.
In the impurity Anderson model,
$a^2\Gam_{\uparrow\downarrow}^{\rm loc}=a\pi\Delta=4T_{\rm K}$,
where $\Delta$ and $T_{\rm K}$ are the width of the virtual
bound state and the Kondo temperature, respectively.
Thus, we can assume that $a^2\Gam_{\uparrow\downarrow}^{\rm loc}$
is of the order of $T_0$ in the periodic system.
This is also consistent with the above-mentioned discussion
of the specific-heat coefficient.
It should be noted again that the renormalized $s$-wave interaction
among quasi-particles becomes comparable with the quasi-particle
band-width, while the on-site repulsion among the bare $f$-electrons 
is sufficiently larger than the bare $f$-electron band-width. 

\begin{figure}[t]
\begin{center}
\vspace{2.5cm}
\end{center}
\caption{(a) The third-order self-energy $\Sigma^{(3)}(\Vec{k},\omega)$
can be divided into three parts;
$\Sig^{(3)}_0=U^3\sum \tilde{\chi}^2 a\tilde{G}$, $\Sig^{(2)}_{\rm v}=
2U^2\sum_\Vec{q}\sum_\nu(U\chi_{\rm inc})\tilde{\chi} a\tilde{G}$
and the remaining terms with no remarkable momentum dependence.
The second term $\Sig^{(2)}_{\rm v}$ has the same momentum
dependence as that in the second-order diagram in $U$.
(b) The renormalized self-energy can be rewritten
as the expansion with respect to the quasi-particle Green's function
$\tilde{G}(p)$ and the effective s-wave interaction
$\tilde{\Gamma}_{\rm loc}=\tilde{\Gamma}_{\uparrow\downarrow}^{\rm loc}=
a^2\Gamma_{\uparrow\downarrow}^{\rm loc}$}
\label{fighv:sig}
\end{figure}

Now we show that in the region $\omega \le T_0$, the renormalized
self-energy $\tilde{\Sig}(\Vec{k},\omega)$ and vertex function
$\tilde{\Gam}_{\sigma\sigma'}(p_1,p_2;p_3,p_4)=
a^2\Del\Gam_{\sigma\sigma'}(p_1,p_2;p_3,p_4)$
can be discussed with the perturbation scheme with respect to the
effective $s$-wave interaction, which is expected to be in the order of
$\tilde{\Gam}_{\uparrow\downarrow}^{\rm loc}=
a^2\Gam_{\uparrow\downarrow}^{\rm loc}$.
First, let us consider the renormalized self-energy
$\tilde{\Sig}(\Vec{k},\omega)=a\Del\Sig(\Vec{k},\omega)$.
Because we have divided the Green's function into the two parts as
$a\tilde{G}(\Vec{k},\omega)+G_{\rm inc}(\omega)$ in Eq.~(\ref{eq:G}),
we can correspondingly divide all diagrams in the perturbative series
into diagrams with and without the remarkable momentum dependence.
The latter is contained in the local self-energy
$\Sig_{\rm loc}(\omega)$.
On the other hand, the former can be rewritten as the expansion
with respect to the effective $s$-wave interaction
$\tilde{\Gam}_{\uparrow\downarrow}$
by concentrating on diagrams with the same momentum dependence.
For instance, let us consider the third-order self-energy
$\Sig^{(3)}(\Vec{k},\omega)=U^3\sum_q\chi(q)^2 G(k+q)$, as shown
in Fig.~\ref{fighv:sig}.
The particle-hole line $\chi(\Vec{q},\nu)=-\sum_p G(p+q)G(p)$
is divided into $\tilde{\chi}(\Vec{q},\nu)=\sum a^2\tilde{G}\tilde{G}$
and $\chi_{\rm inc}(\nu)$ including $G_{\rm inc}(\omega')$.
The former part contributes mainly for $\nu \le T_0$, while
the latter does not possess remarkable momentum dependence
after the $\Vec{p}$ summation.
Then, the third-order diagram is also divided into three parts:
$\Sig^{(3)}_0=U^3\sum \tilde{\chi}^2 a\tilde{G}$
with only the coherent part,
$\Sig^{(2)}_{\rm v}=
2U^2\sum_\Vec{q}\sum_\nu(U\chi_{\rm inc})\tilde{\chi} a\tilde{G}$,
and the remaining terms with no remarkable momentum dependence.
The last terms only contributes to the local self-energy.
By ignoring the frequency dependence of $\chi_{\rm inc}(\nu)$,
the second term $\Sig^{(2)}_{\rm v}$ has the same momentum
dependence as that in the second-order diagram in $U$.
Then, one of the bare vertex $U$ is replaced by $U^2\chi_{\rm inc}$.
This is one of vertex corrections for the second-order self-energy.
Also in the higher-order diagrams, we can pick out diagrams
with three lines of $a\tilde{G}$ producing the same momentum dependence.
We get together such terms and reduce them to a second order diagram
with respect to the effective coupling constant $\Gam_{\uparrow\downarrow}$.
By counting $a^3$ in three lines of $a\tilde{G}(\Vec{k},\omega)$
and $a$ in the definition of the renormalized self-energy,
the renormalized second-order diagram $\tilde{\Sig}^{(2)}=
a^4{\Gam_{\uparrow\downarrow}}^2\tilde{G}\tilde{G}\tilde{G}$
can be regarded as the second-order diagram with respect to
the effective $s$-wave interaction
$\tilde{\Gam}_{\uparrow\downarrow}=a^2\Gam_{\uparrow\downarrow}$.
Here we replace the effective coupling constant
for each reduced diagram to be $\tilde{\Gam}_{\uparrow\downarrow}$
and approximate it as $\tilde{\Gam}_{\uparrow\downarrow}^{\rm loc}$.
Then, we can estimate the renormalized self-energy in the same manner
as explained in Sec.~3.1.
Likewise, within the low-order diagrams, the vertex function
$\tilde{\Gamma}_{\sigma\sigma'}(p_1,p_2;p_3,p_4)$ 
can be reconstructed as the expansion with respect to the quasi-particle
Green's function $\tilde{G}(\Vec{k},\omega)$ and the effective $s$-wave
interaction $\tilde{\Gam}_{\uparrow\downarrow}^{\rm loc}$.
Strictly speaking, in the higher order, there appear diagrams
which cannot be reduced to the expansion in this manner.
For instance, higher-order vertex generally remains.
Here we simply ignore such diagrams and discuss the renormalized effective
PAM. Then, we take a change of energy scale in the band-width and
the interaction between quasi-particles.
A better convergence is expected for this renormalized scheme rather
than the original perturbation scheme.
This procedure just corresponds to an approximation similar to
the pseudo-potential method, such as the ladder summation of the divergent
hard-core potential in {$^3$He} shown by Galitskii \cite{rf:Galitskii}.
The concept of our scheme corresponds to that
the effective Hamiltonian obtained by the renormalization group method
can be written as the renormalized PAM.
Then, the essential assumption is the locality of the interaction and
the unimportance of the higher order vertex,
when the renormalization is performed to the energy scale $T_{0}$.
Of course, we must directly treat the expansion in the bare interaction $U$
to describe the crossover from the localized feature at high temperatures
to the itinerant feature at low temperatures, which is characteristic
in heavy-fermion systems. This is one of the future problems.

Next we formulate the Gor'kov equation for the SC transition
on the quasi-particle description discussed above.
As usual, the SC transition is marked by divergence of the full vertex
in the Cooper channel.
This is determined by evaluating the linearized Dyson-Gor'kov equation
(see Sec.~3.1) as
\begin{equation} 
  \sum_{p'}\Gam^{(2)}(p,p')|G(\Vec{p'},i\omega'_n)|^2
  \Delta(\Vec{p'},i\omega'_n)=\Delta(\Vec{p},i\omega_n),
\end{equation}
where $\Delta(\Vec{p},i\omega_n)$ is an anomalous self-energy
and the Cooper-pairing effective interaction $\Gam^{(2)}(p,p')$
is the particle-particle irreducible vertex.
This equation includes the integral of $|G(\Vec{p'},i\omega'_n)|^2$.
The most important part of this integral comes from the part mediated
by quasi-particles $a^2|\tilde{G}(\Vec{p'},i\omega'_n)|$.
Thus, the Gor'kov equation can be rewritten as
\begin{equation}
  \sum_{p'}a^2\Gam^{(2)}(p,p')|\tilde{G}(\Vec{p'},i\omega'_n)|^2
  \Delta(\Vec{p'},i\omega'_n)=\Delta(\Vec{p},i\omega_n).
\end{equation}
This $a^2\Gam^{(2)}(p,p')$ is the particle-particle irreducible
vertex between quasi-particles with the external frequencies smaller
than $T_0$.
As discussed above, for such a vertex, we can also apply the perturbation
expansion with respect to the renormalized $s$-wave interaction
$a^2\Gam_{\uparrow\downarrow}^{\rm loc}$ between quasi-particles.

Thus, in order to treat the heavy-fermion superconductivity,
we start by introducing the quasi-particle state renormalized
by the dominant $s$-wave scattering part,
which itself does not yield the stable pairing interaction.
Then, we calculate the momentum dependent interaction between
the quasi-particles, using the perturbation expansion in terms of
the renormalized on-site repulsion.
This leads to the transition into unconventional SC phases
as a cooperative phenomenon of the heavy-fermion quasi-particles.
Note that the extension of the above procedure is
straightforward, even if the AF spin fluctuation is relatively
strong and the effective interaction has rather strong momentum
dependence as seen in a kind of Ce-based heavy-fermion
superconductors and the cuprate superconductors.
Now we can begin by the PAM and the Hubbard model describing
the quasi-particle Fermi surface. 
Note also that such quasi-particle description implicitly includes
the following two important arguments.
The renormalized $s$-wave interaction among quasi-particles is expected
to be moderate as compared with the quasi-particle band-width,
even if the on-site repulsion among the bare $f$ electrons is sufficiently
larger than the bare $f$-electron band-width.
The electron-phonon interaction between quasi-particles is renormalized
more severely, since it does not possess enhancement due to
the vertex corrections as
$\Gam_{\uparrow\downarrow}^{\rm loc}\propto a^{-1}$.
Thus, the conventional superconductivity mediated by the electron-phonon
interaction is excluded at this stage.

\subsection{Application to materials}

We proceed to the results of the perturbation theory based on
the renormalized formula in the typical heavy-fermion superconductors;
CeCu$_2$X$_2$ (X=Si and Ge), CeTIn$_5$ (T=Co, Rh, and Ir), CeIn$_3$,
and UM$_2$Al$_3$ (M=Pd and Ni).
It should be noted that the following calculations do not strictly obey
the above renormalization scheme.
For instance, we do not include explicitly the contribution from the
imaginary part of the local self-energy and the cut-off of the energy.
Rather the perturbative scheme explained in Sec.~3.1 will be
directly applied.
However, these differences do not alter the qualitative results
such as the pairing symmetry of the stable state.
Note also that the multi-band system is frequently reduced to
the effective single-band model by taking into account the most important
band triggering the superconductivity.

\subsubsection{CeCu$_2$X$_2$ (X=Si and Ge)}

As introduced in Sec.~\ref{sec:introhv}, the superconductivity in
CeCu$_2$Si$_2$ with $T_{\rm c}=0.7$K locates around the border
with the AF phase at ambient pressure \cite{rf:Steglich1,rf:Ishida2}.
This is also confirmed in terms of similarities with the $P-T$ phase
diagram in the isostructural compound CeCu$_2$Ge$_2$ \cite{rf:Kitaoka4}
(Sec.~5.1.1).
The effect of the dominant AF spin fluctuation has been observed
as deviation from the conventional Fermi-liquid theory in the physical
quantities, such as the resistivity, the specific heat \cite{rf:Gegenwart},
and the NMR/NQR $1/T_1$ \cite{rf:Kitaoka1,rf:YKawasaki}.
Then, the AF spin fluctuation may be responsible for the anisotropic
even-parity superconductivity, which is observed in these compounds.
In the spin fluctuation theory, the SC transition temperature
usually takes maximum value near the AF phase
(see Figs.~\ref{fig:electron-dope}, \ref{fig:takimoto} and
\ref{fig:kondoFLEX}).
CePd$_2$Si$_2$ and CeIn$_3$ \cite{rf:Mathur} are often cited as
the typical examples.
In CeCu$_2$X$_2$, however, the phase diagram is not so simple.
With increasing pressures, $T_{\rm c}$ displays the curious enhancement
in the range of $P=2 \sim 3$GPa, not the monotonic decrease.
The maximum value of $T_{\rm c}$ is about 2K.
Thus, there exists an optimum condition, such as the band structure,
for the superconductivity.
It indicates that the low-lying part of the quantum-critical AF
fluctuation is not responsible for the formation of the superconductivity,
as indicated in CePd$_2$Si$_2$ and CeIn$_3$ \cite{rf:Mathur}.
On the other hand, recent systematic experiments on CeCu$_2$Ge$_2$ have
shown that the curious behavior seems to correspond to the peak structure
of the residual resistivity and the drastic decrease of
the $T^2$-coefficient $A$ of resistivity \cite{rf:Jaccard2}.
As a model involving these properties, a new pairing mechanism mediating
the critical $f$-valence fluctuations has been proposed \cite{rf:Onishi}.
However, here we investigate which kind of anisotropic superconductivity
can be stabilized in the complicated band structure of CeCu$_2$X$_2$
on the basis of TOP based on the quasi-particle description \cite{rf:Ikeda1}.
In addition, let us verify the validity of the simple unified view.

\begin{figure}[t]
\begin{center}
\vspace{2.5cm}
\end{center}
\caption{The renormalized Fermi surfaces at $U/t=8$, $V/t=2.0$, $t_b/t=0$,
$E^f_0/t=-6$ and $T/t=0.05$, which is calculated by
$(\mu-\xi_\Vec{k}-\Sigma^N(\Vec{k},0))(\mu-\epsilon_\Vec{k})-V^2=0$.
(a) The electron Fermi surface around $\Gamma$-point. 
(b) The hole Fermi surface around $Z$-point. 
(c) The periodic zone at $k_z=0$. 
(a) and (b) correspond to light and dark hatches in (c), respectively.}
\label{fighv:FS}
\end{figure}

The band calculation with the linearized augmented-plane-wave (LAPW)
method indicates that
CeCu$_2$X$_2$ are compensated metals with large and small electron Fermi
surfaces around the $\Gamma$-point and complex hole Fermi surfaces
mainly around the $Z$-point \cite{rf:Harima}.
The band structure near these Fermi surfaces are well represented by
two bands, which are formed by the mixing between a quasi-2D $f$-band
\begin{equation}
  \xi_\Vec{k}=-2t[\cos(k_x)+\cos(k_y)]
  -8t_b\cos(k_x/2)\cos(k_y/2)\cos(k_z/2)+E^f_0,
\end{equation}
and a 3D conduction-band
\begin{equation}
  \epsilon_\Vec{k}=2t_c[\cos(k_x)+\cos(k_y)
  -0.8\cos(k_x)\cos(k_y)-2\cos(k_x/2)\cos(k_y/2)\cos(k_z/2)-0.5].
\end{equation}
Here $t$ and $t_b$ are transfer integrals of $f$ electrons,
which are smaller than that of conduction electrons $t_c=10t$,
and $E^f_0=-6t$ is an $f$-electron site energy.
When we choose $t_c=10t=0.025{\rm Ryd.}=0.34$eV and
the hybridization term without momentum dependence $V=0.068$eV,
the band structure near the Fermi level can be almost explained
except for some small Fermi surfaces as shown in Fig.~\ref{fighv:FS}.
We consider the PAM with this band structure as described by
\begin{equation}
  H=\sum_{\Vec{k}\sigma}
  \epsilon_\Vec{k}c_{\Vec{k}\sigma}^\dagger c_{\Vec{k}\sigma}
  +\sum_{\Vec{k}\sigma}
  V\left( f_{\Vec{k}\sigma}^\dagger c_{\Vec{k}\sigma}+{\rm h.c.}\right)
  +\sum_{\Vec{k}\sigma}
  \xi_\Vec{k}f_{\Vec{k}\sigma}^\dagger f_{\Vec{k}\sigma}
  +U \sum_{i\sigma} n_{i\uparrow}^f n_{i\downarrow}^f,
\end{equation}
where $f_{\Vec{k}\sigma}$, $c_{\Vec{k}\sigma}$
($f_{\Vec{k}\sigma}^\dagger$, $c_{\Vec{k}\sigma}^\dagger$) are
the annihilation (creation) operators for $f$ and conduction electrons
with the wave-vector $\Vec{k}$ and a pseudo-spin index $\sigma$.
The Green's functions in the PAM are described
in a $2 \times 2$ matrix form as
\begin{equation}
  \Vec{G}(k)=
  \left(
  \begin{array}{cc}
    G^f(k) & G^{fc}(k) \\
    G^{cf}(k) & G^c(k)
  \end{array}
  \right)=
  \left(
    \begin{array}{cc}
    i\omega_n-\xi_\Vec{k}+\mu-\Sig^N(k) & \hspace{-20pt} V \\
    V & \hspace{-20pt} i\omega_n-\epsilon_\Vec{k}+\mu
    \end{array}
  \right)^{-1}.
\end{equation}
Since the bare interaction works only between $f$ electrons,
the Dyson-Gor'kov equation for $G^f(k)$ is the same one as
in single band case introduced in the previous chapters.
$G^c(k)$ does not directly contribute to the equation like the
$p$-electron Green function in the $d$-$p$ model.
Conduction electrons are incorporated into the SC state through
the hybridization with $f$ electrons.

\begin{figure}
\begin{center}
\vspace{2.5cm}
\end{center}
\caption{The zeroth-order spin susceptibilities $\chi_0(\Vec{q},0)$
at $t_b/t=0.0$, 0.3 and 0.5. The other parameters are the same as in
Fig.~\ref{fighv:FS}.
The peak structure around $X$-point is due to the nesting properties
of the Fermi surfaces.
This peak structure is suppressed with increasing $t_b$.}
\label{fighv:chi}
\end{figure}

\begin{figure}
\begin{center}
\vspace{2.5cm}
\end{center}
\caption{$T_{\rm c}$ as a function of the $f$-level $E^f_0$ for several values
of $V$ at $U=8t$ and $t_b=0.0$. $T_{\rm c}$ decreases with increasing $V$,
although it is almost unchanged for $E^f_0 \lesssim -6t$. This is because
$f$-band character becomes stronger with decreasing $E^f_0$.}
\label{fighv:CeCuX}
\end{figure}

Within the simple TOP, we cannot obtain the large mass enhancement factor
$\bar{\gamma}$, since $U$ cannot be larger than the $f$ band-width.
In CeCu$_2$Si$_2$, most parts of the large mass enhancement factor
comes from almost momentum independent self-energy,
as indicated in the Kadowaki-Woods' relation.
Thus, the renormalization treatment introduced in
Sec.~\ref{sec:microhv} will be valid.
In this case, the large mass enhancement is phenomenologically
taken into account.
Hereafter we consider the on-site Coulomb repulsion $U$
as the renormalized quasi-particle interaction
$a^2\Gamma_{\uparrow\downarrow}^{\rm loc}$ as is discussed in Sec.~5.2.

The most favorable pairing state within the TOP is $d_{x^2-y^2}$-wave
symmetry.
The dominant attractive part originates from the RPA-type diagrams
including the particle-hole bubble-type diagrams.
As shown in Fig.~\ref{fighv:chi}, the bare $f$-electron susceptibility
possesses the peak structure around $(\pi,\pi,q_z)$ for any $q_z$.
Thus, the AF spin fluctuation basically induces
the unconventional superconductivity.
Figure \ref{fighv:CeCuX} illustrates $E^f_0$ dependence of $T_{\rm c}$.
If assuming $\bar{\gamma}\sim 10$, we obtain the renormalized hopping
integral $t \sim 39$K, and the maximum value of
$T_{\rm c} \sim 0.06\times 39 \sim 2.3$K.
This is a reasonable value.
Furthermore, since pressures relatively raise up the $f$-level $E^f_0$
and reduce the $f$-electron number, we can consider that
Fig.~\ref{fighv:CeCuX} shows pressure dependence of $T_{\rm c}$.
The region of $E^f_0\lesssim -6t$, where the hybridization dependence
is weak, corresponds to the AF phase in CeCu$_2$X$_2$.
In the region of $E^f_0 \gtrsim -6t$, on the other hand, $T_{\rm c}$
shows a hump structure around $E^f_0 \sim 2t$.
Suppression of $T_{\rm c}$ for $E^f_0\lesssim 2t$ originates from
the mass renormalization of the normal self-energy due to
the relatively strong correlation, while for $E^f_0\gtrsim 2t$,
from the weak pairing interaction due to
the relatively weak correlation.
This hump structure corresponds to the curious enhancement
of $T_{\rm c}$ under high pressures in CeCu$_2$X$_2$.
If we strictly follow the renormalization procedure in Sec.~5.2,
the most of the mass enhancement is not included in the calculated
normal self-energy.
In fact, the renormalized quasi-particle band itself, which is
phenomenologically introduced here, is sensitive to pressures.
Pressures make the mass enhancement $\tilde{\gamma}$ smaller
and the effective band-width larger.
This will enhance the hump structure in $T-E^f_0$ phase diagram
illustrated in Fig.~\ref{fighv:CeCuX}.
Thus, pressure dependence of $T_{\rm c}$ in CeCu$_2$X$_2$ can be
naturally explained by taking the mass renormalization into account.

\subsubsection{CeTIn$_5$ (T=Co, Rh, and Ir)}

Recently, superconductivity in a series of CeTIn$_5$ has been
discovered \cite{rf:Hegger,rf:Petrovic1,rf:Petrovic2}.
CeRhIn$_5$ is an AF state with $T_{\rm N}=3.8$K at ambient pressure,
and coexists with a SC state under pressures $P>1.5$GPa.
For $P>1.8$GPa, the AF phase vanishes and only the SC state
with $T_{\rm c}=2.1$K appears.
CeIrIn$_5$ and CeCoIn$_5$ are superconducting at ambient pressure with
$T_{\rm c}$=0.4K and 2.3K, respectively.
CeCoIn$_5$ has the highest $T_{\rm c}$ among heavy-fermion superconductors
discovered up to now.

These compounds are worthy of note, since the nature is reminiscent
of the cuprates and the organic superconductors, and therefore,
these compounds will bridge between our understanding of
Ce-based heavy-fermion superconductors and that of the cuprates
and organic superconductors.
Such a viewpoint is suitable for the purpose of this review.
The crystal structure of these compounds is HoCoGa$_5$-type tetragonal
one with a layered structure, in which alternating layers of CeIn$_3$
and TIn$_2$ stack sequentially along the $c$-axis.
In the dHvA measurements, the quasi-2D Fermi surfaces consistent with
the band calculations have been observed
\cite{rf:Haga,rf:Settai,rf:Hall,rf:Shishido,rf:Onuki}.
In the SC phase, the results of the NMR/NQR measurements
indicate the anisotropic even-parity pairing
\cite{rf:Kohori1,rf:Zheng,rf:Kohori2,rf:Curro,rf:Kohori3,rf:Mito}.
In addition, the $d_{x^2-y^2}$-wave pairing
was suggested from the thermal conductivity in CeCoIn$_5$
under the parallel magnetic field  \cite{rf:Izawa}.
In the normal phase, both the transport \cite{rf:Petrovic1,rf:Petrovic2}
and magnetic properties imply the strong quasi-2D AF spin fluctuation.
From the detail analysis of NMR/NQR $T_1$ with the application of
SCR theory \cite{rf:moriyatext}, it has been shown that
these compounds locate close to the quantum critical point (QCP).
In particular, CeCoIn$_5$ is considered to be just above the QCP.
This is also inferred from the behavior $C/T \propto -\ln T$
in the normal-state specific heat under the magnetic field
above $H_{\rm c2}$ \cite{rf:Petrovic2,rf:Sparn1,rf:Sparn2}.
The associated low-temperature entropy is consistent with
the huge zero-field specific-heat jump at $T_{\rm c}$,
$\Del C/\gamma T_{\rm c}=4.5$.
Although this remarkable behavior in the specific heat has not been observed
in the cuprates, the nature can be well understood as the Fermi-liquid state
with the quasi-2D AF spin fluctuation.
Thus, the superconductivity in these compounds should be explained
on the same footing as that in the cuprates.
We show below that $x$-dependence of $T_{\rm c}$ in
CeIr$_x$Co$_{1-x}$In$_5$ \cite{rf:Pagliuso2} is consistent with change of
the carrier number in the main Fermi surface triggering the superconductivity,
and that the large enhanced value of the specific-heat jump can be explained
by the strong-coupling theory of superconductivity including
the quasi-2D AF spin fluctuation.

\begin{figure}
\begin{center}
\vspace{2.5cm}
\end{center}
\caption{$T_{\rm c}$ as a function of $U$ for several $n$. 
For comparison, we illustrate $T_{\rm c}$ corresponding to the transition
temperatures of CeCoIn$_5$ and CeIrIn$_5$.}
\label{fighv:CeTIn}
\end{figure}

Let us discuss the $x$-dependence of $T_{\rm c}$ in
CeIr$_x$Co$_{1-x}$In$_5$ within TOP on the basis of the quasi-particle
description \cite{rf:Nisikawa1}.
The main part of the band structure can be approximated by the
tight binding fitting in a square lattice,
$\xi_\Vec{k}=2t(\cos k_x + \cos k_y)$,
which reproduces the large Fermi surface with the heaviest cyclotron mass
\cite{rf:Haga,rf:Settai,rf:Hall,rf:Shishido,rf:Onuki}.
The main Fermi surfaces in CeIrIn$_5$ and CeCoIn$_5$ are well reproduced
for the electron number $n$=0.69 and 0.77, respectively.
Thus, we consider the Hubbard model with such quasi-particle band
and the renormalized on-site repulsion.
The resulting model is the same one as that was discussed
in Sec.~3.2.3 with $t'/t=0$.
We have already shown that TOP, in this case, gives the qualitatively
same results with the spin fluctuation theory (Sec.~3.2).
If using the PAM as discussed above, we may reproduce more parts of
the Fermi surfaces, but here we simply consider the single-band model
based upon a belief that quasi-particles on the main Fermi surface
play an essential role for the pair formation.
Multi-band Hubbard model for $f$-electron systems has been studied
\cite{rf:Takimoto1}, but the multi-band effect is not discussed here.

The result of TOP is illustrated in Fig.~\ref{fighv:CeTIn}.
The third-order perturbation expansion with respect to $U$ leads
to the effective pairing interaction for the $d_{x^2-y^2}$-wave
superconductivity as expected from the results in Sec.~3.2.
$T_{\rm c}$ provides proper values for the moderate $U$.
We see that $T_{\rm c}$ increases with approaching to the half-filling.
This tendency is also consistent with experimental results
in CeIr$_x$Co$_{1-x}$In$_5$ \cite{rf:Pagliuso2}.

\begin{figure}[t]
\begin{center}
\vspace{2.5cm}
\end{center}
\caption{(a)$1/\chi(Q)$ as a function of $T/T_\mathrm{c}$ and
$\Delta/T_{\rm c}=\Sig_a^\mathrm{max}(\Vec{k},\pi T)/T_\mathrm{c}$
as a function of $T/T_\mathrm{c}$.
(b) The entropy $S$ and (c) the specific heat $C$ in the normal and
SC states. As $U/t$ is larger, the jump of $C/T$ at $T_\mathrm{c}$
is enhanced.}
\label{fighv:C}
\end{figure}

Next we discuss the huge enhanced value of the specific-heat jump
at $T_{\rm c}$ in CeCoIn$_5$, by analyzing the same model Hamiltonian
within FLEX \cite{rf:Ikeda2}.
The thermodynamic potential $\Omega(T,\mu)$ in FLEX is given by
the scheme of Baym and Kadanoff, since FLEX is a kind of the conserving
approximations.
Then, the entropy $S=-\left(\partial \Omega/\partial T \right)_\mu$
is given by explicit derivative of $\Omega(T,\mu)$ with respect to $T$,
since its implicit derivative through the self-energy vanishes
owing to the stationary conditions (Eq.~\ref{eq:variational-condition})
\cite{rf:luttinger}. As a result, we obtain the entropy
\begin{equation}
  S=-\frac{\partial}{\partial T}\left[
  \frac{T}{N}\sum_\Vec{k}\sum_{\omega_n}\Tr\ln\hat{G}(k) \right].
\end{equation}
After the analytic continuation on the real axis,
we obtain in the normal state
\begin{equation}
  S=2\frac{1}{2\pi{\rm i}T}\frac{1}{N}\sum_\Vec{k}\int d\epsilon
  \epsilon\left(-\frac{\partial f}{\partial \epsilon} \right)
  \left[\ln\mathcal{G}_{\rm n}^{\rm R}-\ln\mathcal{G}_{\rm n}^{\rm A} \right],
\end{equation}
and in the superconducting state
\begin{equation}
  S=\frac{1}{2\pi{\rm i}T}\frac{1}{N}\sum_\Vec{k}\int d\epsilon
  \epsilon\left(-\frac{\partial f}{\partial \epsilon} \right)
  \left[ \ln G_{\rm sc}^{\rm R}-\ln G_{\rm sc}^{\rm A} \right],
\end{equation}
where $G_{\rm sc}^{\rm R,A}=%
[\omega_\pm^2 Z_{k}^2-\bar{\xi}_\Vec{k}^2-\Sig_a^2]^{-1}$ with
$\bar{\xi}_\Vec{k}=\xi_\Vec{k}+\chi$ and
$\Sig_n(\Vec{k},\pm\omega)=\pm\omega(1-Z_{k})+\chi$.
The results are illustrated in Fig.~\ref{fighv:C}.
Figure \ref{fighv:C}(a) illustrates the maximum of the anomalous
self-energy $\Sig_a(\Vec{k},\pi T)$ as a function of $T/T_{\rm c}$
for $U/t$=4.0, 4.5, and 5.0.
The rapid increase of the anomalous self-energy below $T_{\rm c}$
is the characteristic behavior in the strong coupling theory,
which has been already obtained in the FLEX approximation
\cite{rf:monthouxFLEX,rf:paoFLEX,rf:takimotoFLEX2}.
This is because the de-pairing effect arising from the normal self-energy
is suppressed below $T_{\rm c}$.
We can see that superconductivity becomes strong coupling,
as $U/t$ is larger and equivalently the largest $\chi_s(\bm{Q})$
at $\bm{Q} \sim (\pi, \pi)$ is larger.
In this case, the entropy $S$ is evaluated as in Fig.~\ref{fighv:C}(b).
It shows the convex behavior from high temperatures,
which originates from the AF spin fluctuation.
Too small entropy below $T/t=0.005$ may be due to the numerical errors.
With use of the fitting by the polynomial function of the entropy
in the vicinity of $T_{\rm c}$, we obtain the specific heat $C$ and
the enhanced jump of the specific heat $\Del C/\gamma T_{\rm c}$
in Fig.~\ref{fighv:C}(c).
As $U/t$ is larger, $T_{\rm c}$ and $\Del C/\gamma T_{\rm c}$
are also larger.
At $U/t=5.0$, $T_{\rm c}/t=0.017$ and $\Del C/\gamma T_{\rm c}=4.6$.
It is not easy to obtain larger value than this,
since the system becomes the AF phase.
Increase of $C/T$ in the normal state at lower temperatures can be
considered as the precursor of $-\ln T$ dependence,
which is predicted by the SCR theory for 2D AF spin fluctuations.
Furthermore, if we set $T_{\rm c}=0.017t$ as 2.3K, then $t \simeq 135$K
and $C/T$ just above $T_{\rm c}$ corresponds to $\sim$ 200 mJ/mol.K$^2$
(290 mJ/mol.K$^2$ in the experimental data \cite{rf:Petrovic2}).
Since the bare $t$ is expected to be $2 \sim 3$ times larger than
the present value 135K, the renormalization inherent in the quasi-particle
description is not large in this case.
Thus, the anomalous behavior of the specific heat in CeCoIn$_5$
can be almost explained by the quasi-2D AF spin fluctuation.

\subsubsection{CeIn$_3$}

CeIn$_3$ is only one heavy-fermion superconductor with the cubic symmetry.
At ambient pressure, it is the AF state with an ordering vector
$\Vec{Q}=(0.5,0.5,0.5)$ and $T_{\rm N}=10$K, while at the critical pressure
$P_c=2.55$GPa, it becomes superconducting with $T_{\rm c} \simeq 0.2$K
\cite{rf:Mathur,rf:Walker}.
At $P=2.65$GPa, $1/T_1$ displays a significant decrease below $T^*=30$K,
and $T_1T=$const. below $T_{\rm FL}=5$K,
which is the typical Fermi-liquid behavior.
Recently, it has been confirmed that the superconductivity possesses
the unconventional nature from no coherence peak
in the {$^{115}$In}-NQR measurement \cite{rf:SKawasaki1,rf:SKawasaki2}.
We note again that the relative material CeRhIn$_5$,
which is the quasi-2D material, has relatively high $T_{\rm c}=2.1$K
under pressure.
This comparison indicates that the dimensionality is
one of important factors for occurrence
of the unconventional superconductivity.
Effect of dimensionality on $T_{\rm c}$ has been already discussed
for the spin-fluctuation mediated superconductivity.
Both in the phenomenological models \cite{rf:moriyaST1996,rf:Monthoux}
and in the microscopic calculations on the basis of FLEX
\cite{rf:Arita,rf:Takimoto2}, it has been shown that the magnitude
of $T_{\rm c}$ is higher in quasi-2D systems than in 3D systems.
With increasing three-dimensionality on the energy dispersion,
the AF phase is more stabilized owing to the suppression of the
fluctuation, and then the SC phase shrinks.
Moreover, the total weight of the spin fluctuation decreases with
increasing the three-dimensionality.
This is a general feature of the fluctuation theory as explained in
Sec.~4.2.2. Then, $T_{\rm c}$ decreases because the pairing interaction
becomes effectively weak. We consider that CeIn$_3$ is a clear example
for this suppression of $T_{\rm c}$ due to the three-dimensionality.
We here estimate the superconducting $T_{\rm c}$ in cubic CeIn$_3$
and its variation for the dimensionality within TOP \cite{rf:Fukazawa}.
We discuss relation between $T_{\rm c}$ and dimensionality. 

\begin{figure}
\begin{center}
\vspace{2.5cm}
\end{center}
\caption{$T_\mathrm{c}/t_1$ as a function of the anisotropy $t_z$.
In the inset, the ratio $T_\mathrm{c}/T_\mathrm{c}^\mathrm{RPA}$ is
illustrated. $T_{\rm c}$ in 3D systems becomes one order smaller
than that in 2D systems.}
\label{fighv:CeIn}
\end{figure}

For simplicity, we consider the Hubbard Hamiltonian
with the energy dispersion
\begin{eqnarray}
  \epsilon_\Vec{k}&=&-2t_1(\cos k_x+\cos k_y+ t_z\cos k_z) \nonumber \\
  &+&4t_2(\cos k_x\cos k_y+t_z\cos k_y\cos k_z+t_z\cos k_z\cos k_x),
\end{eqnarray}
where $t_1$ and $t_2$ denote the nearest-neighbor and
the next-nearest-neighbor hopping integrals.
Note that dimensionality is controlled by a parameter $t_z$.
The cases of $t_z$=0 and 1 correspond to the 2D square lattice and
the 3D cubic lattice, respectively.
In order to describe the main Fermi surface with a large volume in CeIn$_3$,
we choose $t_2=-0.2t_1$ and the electron density $n=0.9$ for $t_z=1$
\cite{rf:Betsuyaku}.
The Fermi surface possesses the nesting property, and the bare susceptibility
exhibits a peak structure in the vicinity of $\Vec{Q}=(0.5,0.5,0.5)$.
The qualitative nature of the superconductivity is the same as in the
quasi-2D systems.
The main pairing interaction originates from the RPA-type diagrams.
The most favorable pairing state has $d_{x^2-y^2}$-wave symmetry.
In 3D cubic systems, this state degenerates with $d_{3z^2-r^2}$-wave
symmetry.
According to the weak coupling theory, time-reversal-symmetry-breaking
state, which is the linear combination of the two pairing states,
is expected in the SC state.
However, one or another may be preferred because of the feedback effect.
This subject will be clarified in the NQR/NMR $1/T_1$
at very low temperatures.
The obtained $T_\mathrm{c}=0.003t_1$ for the moderate value of
$U/W=3/4$ with the band-width $W^{3D}\simeq 12t_1$ can explain
$T_\mathrm{c} \simeq 0.2$K in CeIn$_3$ with $t_1\simeq 100$K.
We show the role of dimensionality in Fig.~\ref{fighv:CeIn}.
It can be seen that $T_{\rm c}$ in 3D systems becomes one order
smaller than that in 2D systems, although $T_{\rm c}$ in quasi-2D systems
is robust for $t_z\lesssim 0.5$.
These features are consistent with the results in the spin fluctuation
theory and with the experimental fact that $T_{\rm c}=0.2$K
in CeIn$_3$ and $T_{\rm c}=2.1$K in CeRhIn$_5$.
It is generally expected that the superconducting $T_{\rm c}$ is small
in 3D systems, since the momentum dependence of the effective interaction
is relatively weak.
This is the underlying physics in common with the spin fluctuation theory.

Finally, let us comment on the quasi-1D system.
In this case, $T_{\rm c}$ for the $d$-wave superconductivity is generally
low, because the quasi-1D momentum dependence of the effective interaction
is not suitable for the $d$-wave superconductivity and the de-pairing
effect is enhanced by the nested Fermi surface.
Thus, the quasi-2D systems like high-$T_{\rm c}$ cuprates,
$\kappa$-(ET)$_2$X and CeTIn$_5$ are the most favorable for the appearance
of the $d$-wave superconductivity.

\subsubsection{UM$_2$Al$_3$ (M=Pd and Ni)}

\begin{figure}[t]
\begin{center}
\vspace{2.5cm}
\end{center}
\caption{(a) $T_{\rm c}$ as a function of the anisotropy $t_m/t$ at $U=7.5t$.
(b) and (c) illustrate, respectively, the eigen values $\lambda$ of
the \'Eliashberg equation for $n=1.200$ and $n=1.144$ at $T=0.003t$.
$\Delta_{B_{1g}}$ is stable in a wide range of $t_m/t$,
while $\Delta_{B_{2u}}^2$ only in the vicinity of $t_m/t=1$.
Although the region of $\Delta_{B_{2u}}^2$ becomes wider as $n$ decreases,
the transition temperature is suppressed abruptly.}
\label{fighv:UMAl}
\end{figure}

UPd$_2$Al$_3$ and UNi$_2$Al$_3$ \cite{rf:Geibel1,rf:Geibel2} exhibit
clear coexistence between an AF phase and an unconventional superconductor.
UPd$_2$Al$_3$ is an AF metal \cite{rf:Krimmel} and coexists with
the anisotropic even-parity superconductivity below $T_{\rm c}=2$K
\cite{rf:Geibel1,rf:Kyogaku}.
UNi$_2$Al$_3$ is the SDW state with $\Vec{Q}=(0.5 \pm \tau,0,0.5)$
\cite{rf:Schroder,rf:Lussier} and coexists with
the odd-parity superconductivity at $T_{\rm c}=1.2$K \cite{rf:Geibel2}.
The AF state in UPd$_2$Al$_3$ possesses relatively large ordered moments
0.85$\mu_{\rm B}$/U as compared with 0.2$\mu_{\rm B}$/U in UNi$_2$Al$_3$,
and the AF transition seems to be that of the localized $f$-electron system.
The ordered moments are aligned parallel to the [11\=20] direction
on the c-plane, and alternate along the c-axis with $\Vec{Q}=(0,0,0.5)$.
For $T_{\rm c} \lesssim T \ll T_{\rm N}$,
both systems exhibit typical behaviors of heavy-fermion systems;
the large enhanced coefficient of the electronic specific heat
({$\gamma=140$mJ/K$^2$mol} in UPd$_2$Al$_3$ and {$\gamma=120$mJ/K$^2$mol}
in UNi$_2$Al$_3$) and $T^2$ behavior in the resistivity
with the large coefficient \cite{rf:Geibel1,rf:Geibel2}.
Thus, it has been indicated that, especially in UPd$_2$Al$_3$,
two separated subsystems, a localized part leading to the magnetic
long-range order and an itinerant part forming the heavy-fermion state,
seem to coexist in momentum space \cite{rf:Caspary,rf:Feyerherm}.
On the other hand, the inelastic peak observed in the neutron scattering
measurement in UPd$_2$Al$_3$ implies a sizable interaction
between two subsystems \cite{rf:Sato1,rf:Metoki,rf:Bernhoeft}.
At $T$=4.2K, the spectrum at the AF zone center $\Vec{Q}=(0,0,0.5)$ exhibits
a quasi-elastic peak at $\omega=0$ and an inelastic peak (magnetic exciton)
at $\omega=1.5$meV.
After the SC transition, the former peak shifts to the high-energy side,
and develops into the inelastic (`resonance') peak centered
at $\omega=0.4$meV at $T=0.4$K.
This is considered as a circumstantial evidence for the scenario that
the dispersive magnetic exciton by localized $5f$ electrons is responsible
for the unconventional superconductivity \cite{rf:Sato2,rf:Miyake}.
This mechanism indicates the horizontal line nodes on the AF zone boundary.
However, it has been recently shown by Thalmeier that this mechanism rather
favors an odd parity state \cite{rf:Thalmeier}.
In addition, it is suspicious whether such an indirect interaction
as mediated by magnetic-excitons dominates the many-body effect
originating from the Coulomb repulsion between itinerant electrons.
Here we consider that the Coulomb interaction between the itinerant
electrons should be responsible for the pairing mechanism in this compound.
In this case, the band structure of the itinerant electrons is important
for the superconductivity, according to the renormalization procedure
in Sec.~5.2.
In UPd$_2$Al$_3$, the dHvA effect in the AF phase is in good agreement
with the Fermi surfaces calculated by the band calculation \cite{rf:Inada}.
The two dominant Fermi surfaces (`party hat' and `column') with heavy
cyclotron mass have quasi-2D  nature to some extent.
These quasi-2D Fermi surfaces will prefer the vertical line node
rather than the horizontal one in the SC state,
if their own Coulomb repulsion is a dominant interaction.
This is another probable candidate for the unconventional
superconductivity in UPd$_2$Al$_3$.
This mechanism is suitable for unified viewpoint on the unconventional
superconductivity in SCES as frequently stressed in this review.
Here we investigate a possible scenario for the appearance of the
different SC states in UPd$_2$Al$_3$ and UNi$_2$Al$_3$ on the basis
of the above standpoint.

We investigate the 2D Hubbard model on an anisotropic triangular lattice,
which corresponds to the c-plane in UPd$_2$Al$_3$ and UNi$_2$Al$_3$.
Note that it is essentially the same model as investigated for
organic superconductors (Sec.~3.3).
For simplicity, here we introduce only an anisotropy of
the Fermi surfaces as an effect of the AF order.
Then, the dispersion of the quasi-particles is represented by
\begin{equation}
  \epsilon_\Vec{k}=-4t\cos(\frac{\sqrt{3}}{2}k_x)\cos(\frac{1}{2}k_y)
  -2t_m\cos(k_y),
\end{equation}
reflecting that the Brillouin zone is reduced by the AF arrangement
to a c-based-center orthorhombic, although the chemical unit cell
is hexagonal \cite{rf:Nisikawa2,rf:Nisikawa3}.
Here $t$ and $t_m$ are the nearest and next nearest hopping integrals,
respectively.
Note that the dispersion possesses the $D_{2h}$ symmetry for
$t_m \ne t$, while at $t_m=t$, it is reduced to that of the triangular
lattice with the $D_{6h}$ symmetry.
Now we evaluate \'Eliashberg equation within TOP to study stable
superconducting states for the anisotropy $t_m/t=0.75 \sim 1.0$,
the electron density $n=1.0 \sim 1.4$, and
the Coulomb repulsion $U/t=3.5 \sim 7.5$.
Among irreducible representations of $D_{2h}$ symmetry,
we investigate the following probable four kinds of pairing symmetry,
except $A_{1g}$ and ones including $k_z$-dependence:
$\Delta_{B_{1g}}=\sin(\frac{\sqrt{3}}{2}k_x)\sin(\frac{1}{2}k_y)$, 
$\Delta_{B_{2u}}^1=\cos(\frac{\sqrt{3}}{2}k_x)\sin(\frac{1}{2}k_y)$, 
$\Delta_{B_{2u}}^2=\sin(k_y)$, and 
$\Delta_{B_{3u}}=\sin(\frac{\sqrt{3}}{2}k_x)\cos(\frac{1}{2}k_y)$.
We have found two kinds of stable state in a finite range of parameters,
irrespective of $n$ and $U$:
One is the even-parity pairing state $\Delta_{B_{1g}}$ and
another is the odd-parity pairing state $\Delta_{B_{2u}}^2$.
As shown in Fig.~\ref{fighv:UMAl}, the former is stable in a wide range
of $t_m$, while the latter only in the vicinity of the symmetric point
$t_m=t$ ($D_{6h}$ symmetry).
Although the range of $\Delta_{B_{2u}}^2$ becomes wider as $n$ decreases,
the transition temperature is suppressed abruptly.
We propose these two stable states as the even-parity pairing in
UPd$_2$Al$_3$ and the odd-parity pairing in UNi$_2$Al$_3$, respectively.
This proposal is consistent with a feature in each compound,
because we can expect that the distortion from the $D_{6h}$ symmetry
under the magnetic order is larger in UPd$_2$Al$_3$ than in UNi$_2$Al$_3$,
according to the magnitude of the ordered magnetic moment.
Thus, the result obtained by TOP for this simplified model can well
explain the emergence of the different SC states in UPd$_2$Al$_3$ and
UNi$_2$Al$_3$.
Note that the simplification of the model is obviously hypothetical
and some future works will be required.
For instance, the effects of the AF order, three-dimensionality,
multi-band effect, and the relation with the `resonance' peak
at $\Vec{Q}=(0,0,0.5)$ should be investigated in more details.

\par\vfill
\eject

%
%
\section{Concluding Remarks and Discussion}

In this review, we have discussed systems possessing the strong
electron correlation.
They are cuprate superconductors, organic superconductors,
Sr$_2$RuO$_4$, and heavy-fermion superconductors.
We have analyzed the superconductivity realized in these substances
on the basis of the single- and multi-orbital Hubbard Hamiltonian.
In the single-band Hubbard Hamiltonian, the on-site electron correlation
$U$ is the only many-body interaction.
Although the Coulomb repulsion suppresses the $s$-wave superconductivity,
it induces various types of anisotropic superconductivity
through the momentum dependence of quasi-particle interaction,
which originates from the many-body effect.
While some inter-orbital interactions exist in the multi-orbital Hubbard
Hamiltonian, anisotropic superconductivity is induced by essentially
the same mechanism, namely the momentum dependence of
quasi-particle interaction.

This is an important understanding of the mechanism of superconductivity
in strongly correlated electron systems (SCES).
Only the above understanding is believed to be the unified one. 
In other words, some simplified ideas will be insufficient
for the purpose to provide a unified picture.
For example, the pairing mechanism is frequently attributed to
fluctuation of some order parameter.
The spin fluctuation theory is a typical one.
While this theory has obtained great success as is reviewed in Sec.~3.2,
the unified understanding cannot be derived from this theory.
We can easily understand this fact from the result on Sr$_2$RuO$_4$,
where the approach from the fluctuation theory clearly fails.
As is explained in Sec.~3.4, the pairing mechanism of Sr$_2$RuO$_4$ is
appropriately clarified on the basis of the understanding proposed above.

Now we emphasize that the quasi-particle in the Fermi-liquid theory 
is an essentially important concept to describe the SCES.
The basis of this fact is the continuity principle existing in the
Fermi liquid. 
This widely accepted idea not only ensures the applicability of
the perturbative methods but also promises some universal
understandings in the SCES.
While the superconducting materials in SCES frequently show
the non Fermi-liquid behaviors, the usefulness of the Fermi-liquid
theory as a starting point is robust.
In the general statement, the existence of Fermi-liquid quasi-particle
is justified when the damping rate is smaller than its energy.
If we adopt the nearly anti-ferromagnetic (AF) Fermi-liquid theory,
the energy width
of single-particle spectrum sometimes approaches to their energies,
as seen in the vicinity of hot spots
in the under-doped cuprates (Sec.~4.3.1).
However, it is still possible to consider the quasi-particle states
with strong damping effects.
The usefulness of the approach from the Fermi-liquid theory has been
confirmed by the arguments on the pseudogap phenomena (Sec.~4).
In the argument of the pseudogap, we have calculated the normal
self-energy due to superconducting (SC) fluctuations.
The self-energy shows anomalous behaviors:
The real part possesses positive slope around Fermi energy and
the imaginary part shows a large peak around Fermi energy
in its absolute value.
These behaviors are in sharp contrast to the normal Fermi liquid.
We have clarified that these anomalous behaviors of the self-energy
are the origin of the pseudogap.
Because the superconducting long-range order violates
the continuity principle, it is natural for its precursor to destroy
the picture of quasi-particles by degrees.

On the basis of the above standpoint, we have reviewed several topics.
We have clarified the mechanism of superconductivity for each system
with the corresponding pairing symmetry.
They are the $d$-wave pairing for cuprates and organic superconductors,
$p$-wave pairing for Sr$_2$RuO$_4$, and $d$- or $p$-wave pairing
for heavy-fermion superconductors.
These results have been obtained by solving the Dyson-Gor'kov equation
derived by perturbation calculation or FLEX approximation.

In particular, we should stress that the physical properties related to
the cuprate high temperature superconductivity have been explained
appropriately in the normal and superconducting states.
The hole-doped and electron-doped systems have been explained in common
on the basis of the same Hubbard Hamiltonian by only adjusting
the carrier number to that in real systems.
Moreover, pseudogap phenomena have been also explained by taking
the SC fluctuations into account without any other assumptions.

In this review we have shown the following unified view for
strongly correlated electron systems: 

{\bf 1.}
The anisotropic superconductivity in SCES originates from the momentum
dependence of the quasi-particle interaction, stemming from
the on-site Coulomb repulsion.
The triplet $p$-wave pairing in Sr$_{2}$RuO$_4$ also can be explained
on the basis of this unified understanding.
The obtained mechanism is a new one, which is different from
the paramagnon mechanism.
In general, the $d$-wave superconductivity is stabilized
near the half-filling, while the $p$-wave superconductivity
is stabilized apart from the half-filling.

{\bf 2.} 
The pseudogap phenomena arise from the SC fluctuation
in the quasi-two-dimensional system.
In addition to the quasi-two-dimensionality, the strong-coupling
superconductivity possessing short coherence length $\xi$ is
necessary for the appearance of the pseudogap phenomena.
The SC fluctuation itself originates from attractive
interactions induced by AF spin fluctuation.
Starting with the repulsive Hubbard Hamiltonian, we have derived
the pseudogap phenomena and succeeded in explaining not only
the single particle properties but also the magnetic and transport ones
consistently in the pseudogap region.
It is important that the explanation of pseudogap is the natural
extension of our theory applied to the other regions
with different hole- and electron-dopings.
A kind of resonance between quasi-particle states in Fermi-liquid
and Cooper-pairing states give rise to the pseudogap.
This is an essentially new physical phenomenon discovered in
quasi-two-dimensional SCES.
The pseudogap above $T_{\rm c}$ gains the energy without
superconducting long-range order.

{\bf 3.}
Among the anomalous properties in high-$T_{\rm c}$ cuprates,
we have made an important progress in the analysis of
transport phenomena.
It is shown that the vertex correction in the linear response theory
(Kubo formula) sometimes plays an essential role in SCES.
In other words, not only the properties of quasi-particles but also
those of the residual interaction is necessary to understand the
transport phenomena. 
For example, the Hall coefficient, the magnetic penetration depth,
and the Nernst coefficient in high-$T_{\rm c}$ cuprates are the cases. 
This is one of the characteristic features in the SCES with 
significant momentum dependence. 
The seemingly anomalous behaviors in cuprates are 
reasonably explained on the basis of the Fermi-liquid theory. 

{\bf 4.}
To describe the superconductivity in SCES, the following
renormalization procedure offers an important perspective.
First we include the on-site Coulomb interaction and then,
band energy and effective interaction are renormalized.
The momentum dependence of the effective interaction between
quasi-particles gives rise to the anisotropic superconductivity.
This perspective gives the explanation on the enhanced jump
of the specific heat at $T_{\rm c}$ in heavy fermion materials.
Furthermore, important knowledge for the superconductivity
in SCES is obtained.
For instance, by considering the renormalization effect due to electron
correlation, we can show that the electron-phonon interaction is
generally reduced by the renormalization factor $z$ compared with
the electron-electron interaction.
Therefore, the $s$-wave pairing is almost impossible in heavy-fermion
systems. The above argument is mainly related to a strategy
to understand the superconductivity in heavy-fermion systems.
By following this strategy,
we have analyzed the superconductivity in the Ce-based and U-based
heavy fermion superconductors.
Concerning the systems possessing not so heavy effective mass such as
the $d$-electron and organic materials, we can treat simultaneously
both the mass enhancement and the superconductivity, as has been
done in this review.

{\bf 5.}
Recently, the higher-order correction beyond the third-order
perturbation theory, which is mainly used in this review,
has been discussed.
The fourth order perturbation with respect to $U$ has been performed
by Nomura and Yamada (Appendix B).
It is shown that the effective interaction for $d$-wave pairing
converges very smoothly.
The fourth-order terms considerably cancel each other and as a result,
their contribution becomes small in total.
The superconducting critical temperature also shows a good convergence.
These results qualitatively justify the understanding reviewed in Sec.~3.2.
On the other hand, the obtained critical temperatures for the
$p$-wave pairing case show an oscillatory behavior with respect to
the calculated order. This behavior originates from the oscillation
in particle-particle scattering terms.
For the $p$-wave case, the correction from the particle-particle ladder
diagrams induces the important contribution to the quasi-particle
interaction.
As is well known, the particle-particle ladder terms
can be collected up to the infinite order to give a smooth function.
It is shown that rather convergent results are obtained by
applying this procedure.
Thus, it is expected that the higher-order terms can be treated so as
not to change the result obtained within the third-order perturbation
theory.
Of course, the theory to treat the electron correlation in a closed form 
is highly desirable.
The results given in Appendix B imply a possibility to develop
the theory of unconventional superconductivity in a closed form
by starting with an appropriate renormalized form.

New superconducting systems are continuously found in the various
systems such as heavy-fermion systems.
The superconductivity in heavy-fermion systems provides many interesting
subjects which deserve theoretical efforts.
We close this review by noting that the heavy-fermion superconductivity
will be an important and attractive future issue.

%
%
\section*{Acknowledgement}

The authors would like to thank E. Dagotto, K. Deguchi, S. Fujimoto,
H. Fukazawa, R. Ikeda, K. Ishida, K. Kanki, H. Kohno, S. Koikegami,
H. Kondo, H. Kontani, K. Kuroki, Y. Maeno, T. Moriya, Y. Nisikawa,
M. Ogata, T. Ohmi, M. Sigrist, Y. Takada, T. Takimoto, K. Ueda,
H. Yasuoka, and K. Yosida for discussions on the present subject.
Figures \ref{fig:takimoto}, \ref{fig:kurokifig1}, and
\ref{fig:kondoFLEX} were presented
by courtesy of T. Takimoto, K. Kuroki, and H. Kondo, respectively.
Y. Y. is supported by the Grant-in-Aid for Scientific Research from
Japan Society for the Promotion of Science.
T. N. is supported from the Japan Society for the Promotion of Science
for Young Scientists.
T. H. has been supported by the Grant-in-Aid for Scientific Research
Priority Area from the Ministry of Education, Culture, Sports,
Science, and Technology of Japan. He is also supported by the Grant-in-Aid
for Scientific Research from Japan Society for the Promotion of Science.

%
%
\appendix
\section{Fermi-liquid theory on the London constant}

As an interesting topic, in this Appendix,
we discuss the London constant in high-$T_{\rm c}$ cuprates.
The London constant is described by the magnetic penetration depth as
$\Lambda=1/4\pi\lambda_{\rm L}^{2}$ and often described by the 
``superfluid density'' as $n_{\rm s}/m^{*}$. 
The experimental results for high-$T_{\rm c}$ cuprates have been reviewed
in Sec.~2.1.6.
The so-called ``Uemura plot'' \cite{rf:uemura} has particularly stimulated
interests, because the relation $T_{\rm c} \propto \Lambda(0)$ indicates
the Kosterlitz-Thouless (KT) transition which is established
in the exactly two-dimensional system \cite{rf:KT}.
From this insight, the phase disordered state was proposed as
a possible pseudogap state \cite{rf:emery,rf:franz,rf:kwon}.
The situation is different for the quasi-two-dimensional system where
the long-range order occurs at finite temperature.
Actually, the characteristic behavior for the KT transition,
such as the Nelson-Kosterlitz jump, is not observed experimentally.
Here, we do not enter into the validity of the phase-only model,
but focus on the microscopic origin of the doping and
temperature dependence of $\Lambda$.

Not only the Uemura plot but also the doping independence of $a=-d\Lambda/dT$
have attracted theoretical interests.
Lee and Wen have shown that the doping and temperature dependences of
$\Lambda$ are explained by the SU(2) formulation for the $t$-$J$ model
\cite{rf:leewen}.
Another understanding was proposed on the basis of the phenomenological
description for the Fermi-liquid theory
\cite{rf:millismagpen,rf:paramekantiFermi}.
The conventional Fermi-liquid theory for the isotropic system
\cite{rf:leggettmagpen} is not available in this case.
In the following, we derive the microscopic description for the anisotropic
Fermi-liquid and propose an understanding based on the Fermi-liquid theory
\cite{rf:jujomagpen1,rf:jujomagpen2}.
It is shown that the expected behaviors of the Fermi-liquid parameters
are successfully reproduced by the FLEX approximation.


The Fermi-liquid description for the London constant is obtained by the
Kubo formula. 
The superconducting (SC) state is no longer the Fermi-liquid
in a strict sense.
However, the description based on the Bogoliubov quasi-particle
is similarly possible. 
The London constant $\Lambda$ is obtained by subtracting the 
paramagnetic contribution from the diamagnetic contribution.
Then, the London constant is derived from the current-current correlation
function and therefore described by the quasi-particles
as well as the quasi-particle interaction.
The quasi-particle interaction is expressed by the vertex correction
for the correlation function.
At $T=0$, the interaction between the quasi-particles is taken into account
in the same way as in the Drude weight \cite{rf:okabe,rf:maebashi}
and in the cyclotron resonance frequency \cite{rf:kankicyc}.
This is natural because the Drude weight at $T$=0 is equal to the London
constant according to the $f$-sum rule.
A careful treatment for the many-body effect is needed at finite temperature.
A systematic discussion is presented in
Refs.~\cite{rf:jujomagpen1} and \cite{rf:jujomagpen2}.
Hereafter, we use the mean-field theory for the superconductivity.
We ignore the SC fluctuation because it is not important at low temperature.
In particular, it has been concluded that the thermal fluctuation does not
contribute to the $T$-linear coefficient
$a=-\partial \Lambda/\partial T$ \cite{rf:paramekantiFL}.
While the importance of the quantum fluctuation has been pointed out
\cite{rf:paramekantiFL}, we do not take it into account.
The doping dependence of the quantum fluctuation cooperatively affects 
on the London constant together with the effect discussed below. 

From the results in Ref.~\cite{rf:jujomagpen1},
the London coefficient is written at finite temperature as
\begin{equation}
 \label{eq:London-constant}
 \Lambda_{\mu\nu}(T) = e^{2} 
 \int_{\rm FS}\frac{{\rm d}S_k}{4\pi^3|{\vec{v}}^*(\k)|}
 v^*_{\mu}(\k)(1-Y(\k_{\rm F};T))\bar{v}^*_{\nu}(\k;T),
\end{equation}
where $\int_{\rm FS}{\rm d}S_k$ means the integral over the Fermi surface,
$\mu$ and $\nu$ are the spatial indices, and 
$Y(\k_{\rm F};T)$ is the Yosida function, given by
\begin{equation} 
 \label{eq:yoshida-function}
 Y(\k_{\rm F};T)=\int{\rm d}\varepsilon^*(\k)\left(
 -\frac{\partial f(E^*(\k))}{\partial E^*(\k)}\right).
\end{equation}
Here $E^*(\k)=\sqrt{\varepsilon^{*2}(\k)+\Delta(\k)^2}$ and
$\Delta(\k)$ is the momentum dependent excitation gap renormalized
by the many-body effect.
$v^*_{\mu}(\k)$ is the renormalized velocity of quasi-particles
($\vec{v}^*(\k)=\partial \varepsilon^{*}(\k)/\partial \k$) and
$\bar{v}^*_{\nu}(\k;T)$ is determined by the following integral equation
\begin{equation}
  \label{eq:renormalization1}
  \bar{v}^*_{\nu}(\k;T)=j^*_{\nu}(\k)
  -\int_{\rm FS}\frac{{\rm d}S_{\k'}}{4\pi^3|{\vec{v}}^*(\k')|}
  f(\k,\k')Y(\k'_{\rm F};T)\bar{v}^*_{\nu}(\k';T).
\end{equation}
Here $f(\k,\k')=z(\k) \Gamma^{\omega}(\k,\k') z(\k')$ is the interaction
between quasi-particles, where $\Gamma^{\omega}$ is obtained from the
$\omega$-limit of reducible four-point vertex.
The notation about $\omega$-limit and $k$-limit follows Ref.~\cite{rf:AGD}.
The current vertex $j^*_{\mu}(\k)$ in the collision-less region
is usually written as
\begin{equation}
  \label{eq:renormalization2}
  j^*_{\mu}(\k)=v^*_{\mu}(\k)+\sum_{\smallkk}
  f(\k,\k')\delta(\varepsilon^*(\k'))v^*_{\mu}(\k').
\end{equation}
It is easily understood that $\bar{v}^*_{\nu}(\k;0)=j^*_{\nu}(\k)$
because $Y(\k;T=0)=0$.
Note that the current vertex $j^*_{\mu}(\k)$ is different from
$J_{\mu}(\k)$ in Sec.~4.3.4.
The latter is defined in the hydrodynamic region.

The above expression for the current vertex is rewritten
by more convenient form as
\begin{equation}
  \label{eq:renormalization3}
  j^*_{\mu}(\k)=v_{\mu}(\k)+z(\k)w_{\mu}(k)|_{\epsilon=0},
\end{equation}
where $\vec{w}(k)$ is obtained from the following integral equations:
\begin{eqnarray}
  \label{eq:renormalization4}
  \vec{w}(k) &=& \vec{u}(k) + \int_{k'} dk' I(k,k')[G(k')^2]^{\omega} 
  \vec{w}(k'),
  \\
  \label{eq:renormalization5}
  \vec{u}(k) &=& \int_{k'} dk' I(k,k')[G(k')^2]^{\omega}
  \left(1-\frac{\partial \Sigma(k')}{\partial \epsilon'}\right)
  (\vec{v}(\k')-\vec{v}(\k)),
\end{eqnarray}
where $I(k,k')$ is the irreducible four-point vertex which satisfies
the relation $I(k,k'):=\delta\Sigma(k)/\delta G(k')$.
We can show that $u_{\mu}(k)=0$ and therefore $j^*_{\mu}(\k)=v_{\mu}(\k)$
in the Galilei invariant system.
Thus, the renormalization for the current vertex $j^*_{\mu}(\k)$ generally
results from the Umklapp scattering.

\begin{figure}[t]
\begin{center}
\vspace{2.5cm}
\caption{Results of the self-consistent second order perturbation
for $\int_{\rm FS}\frac{{\rm d}S_k}{|\vec{v}^*(\k)|}v_x^*(\k) X$.
Here $X=j_x^*(\k)$ (circles), $z(\k)v_x(\k)$ (triangles),
and $v_x(\k)$ (squares).
This quantity is proportional to $\Lambda_{xx}(0)$ in
the case of $X=j_x^*(\k)$.}
\label{fig:jujofig13}
\end{center}
\end{figure}

\begin{figure}[t]
\begin{center}
\vspace{2.5cm}
\caption{The schematic figure for the current vertex $\vec{j}^*(\k)$
in the under-doped region. The renormalized velocity
$\vec{v}^*(\k)$ is perpendicular to the deformed Fermi surface,
while $\vec{j}^*(\k)$ is not.}
\label{fig:HotColdstatic}
\end{center}
\end{figure}

Finally we derive the coefficient of the $T$-linear term from
Eqs.~(\ref{eq:London-constant}) and (\ref{eq:renormalization1}) as
\begin{equation}
  \label{eq:t-linear-term}
  a=- \frac{{\rm d }\Lambda_{\mu\nu}(T)}{{\rm d} T}|_{T=0}
  =e^{2} \int_{\rm FS}\frac{{\rm d} S_k}{2\pi^2|{\vec{v}}^*(\k)|}
  j^*_{\mu}(\k)\left(\frac{\partial Y(\k_{\rm F};T)}{\partial T}\right)_{T=0}
  j^*_{\nu}(\k).
\end{equation}
This coefficient is determined by the current vertex around the node
as is expected. This is contrasted from the behavior of $\Lambda(0)$
which is determined by the quasi-particles around the whole Fermi surface.
We obtain the expression
\begin{equation}
  \label{eq:London-constant-T=0}
  \Lambda_{\mu\nu}(0) = e^{2} 
  \int_{\rm FS}\frac{{\rm d}S_k}{4\pi^3|{\vec{v}}^*(\k)|}
  v^*_{\mu}(\k) j^*_{\nu}(\k). 
\end{equation}
This is an underlying origin of the qualitatively different doping 
dependence between $\Lambda(0)$ and $a$.


Before discussing the realistic situation,
we comment on the failure of the Fermi-liquid theory for the
isotropic system \cite{rf:leggett,rf:nozieresbook}.
If we apply the above expressions for the two-dimensional isotropic system
where $\varepsilon^{*}(\k)=\k^{2}/2m^{*}$, the London constant is obtained
in the low-temperature region as \cite{rf:leggettmagpen}
\begin{eqnarray}
  \label{eq:isotropic-case}
  \Lambda(T)=(1+F_{1{\rm s}}/2) \frac{n}{m^*}-(1+F_{1{\rm s}}/2)^{2}\alpha T,
\end{eqnarray}
where $\alpha$ is a factor in the order of unity.
The simplified expression $\Lambda(0) = n/m^*$ is obtained
if $F_{1{\rm s}} \ll 1$.
If we apply Eq.~(\ref{eq:isotropic-case}) to the high-$T_{\rm c}$ cuprates,
the Uemura plot requires the scaling behavior $m^* \propto \delta^{-1}$ or
$1+F_{1{\rm s}}/2 \propto \delta$ or $n \propto \delta$.
The first candidate clearly contradicts the ARPES measurement
\cite{rf:marshall}, where the mass-renormalization does not
significantly depend on the doping.
The second candidate leads to the relation $-d\Lambda/dT \propto \delta^{2}$
and clearly contradicts the experimental results (see Sec.~2.1.6).
Thus, we can not resolve the anomalous behaviors by taking account of
the electron correlation on the basis of the isotropic Fermi-liquid theory.
Contrary to that, the last candidate gives an appropriate result.
This observation is sometimes suggested as an evidence for that
the under-doped cuprate is a low-carrier system. 
This is one of the reasons why the NSR theory has been applied to 
the pseudogap phenomena (Sec.~4.1). 
However, this argument is too naive, because the appearance of the
particle number $n$ is an accidental result in the isotropic case.
That is, the relation $\Lambda \propto n$ is derived from the replacement
$n=k_{\rm F}^{2}/2\pi$ which is justified only in the isotropic case.
From the expressions
Eqs.~(\ref{eq:London-constant})-(\ref{eq:London-constant-T=0}),
the London constant is generally determined by the quasi-particles
near the Fermi surface.
The excitation deeply below the Fermi level is not important.
The similar observation has been obtained for the Hall coefficient
(Sec.~4.3.4).

\begin{figure}[t]
\begin{center}
\vspace{2.5cm}
\caption{Results of the FLEX approximation 
for $\int_{\rm FS}\frac{{\rm d}S_k}{|\vec{v}^*(\k)|}v_x^*(\k) X$. 
Here $X=j_x^*(\k)$ (circles) and $z(\k)v_x(\k)$ (triangles).}
\label{fig:jujofig14}
\end{center}
\end{figure}

In the following, we show that the generalized Fermi-liquid theory resolves
the anomalous behaviors.
The Hubbard model is adopted as a microscopic Hamiltonian.
We start from the usual case where the momentum dependence of
the quasi-particle interaction is not essential.
The over-doped cuprate is probably included in this case.
Then, the current vertex is given as
\begin{eqnarray}
  j^*_{\mu}(\k) \simeq v^*_{\mu}(\k) \simeq  z(\k) v_{\mu}(\k).
\end{eqnarray}
Thus, the London constant is reduced by the renormalization factor $z(\k)$.
If the momentum dependence of the velocity $|\vec{v}^*(\k)|$ is not
so significant, the conventional behavior is expected
for the London constant.
In order to study this case, namely, the over-doped region,
we use the perturbation theory.
This choice is in common with the discussion on the pairing
mechanism (Sec.~3.2).
In this case, the momentum dependence of the quasi-particle interaction
exists, but is not strong.
Figure \ref{fig:jujofig13} shows the results of the self-consistent
second-order perturbation.
It is shown that the London constant at $T=0$ decreases with doping.
This is an observed behavior in the over-doped region
\cite{rf:tallonmagpen}.
We see that the vertex correction (quasi-particle interaction)
is not important.

Note that the above situation is irrelevant in the under-doped region,
because the strong momentum dependence is an important property
of the under-doped cuprates (see Sec.~4.3).
Here, the most effective momentum dependence should be derived from
AF spin fluctuations.
If we use the phenomenological description for the quasi-particle 
interaction, which is represented by the Feynmann diagram
in Fig.~\ref{fig:phenomenovertex}, a rough estimation \cite{rf:jujomagpen2}
shows that the current vertex behaves as in the schematic figure
(see Fig.~\ref{fig:HotColdstatic}).
The current vertex is significantly reduced by the quasi-particle
interaction $f(\k,\k')$ which is enhanced around $\k = \k'+\Q$.
The reduction is especially significant around the ``hot spot''.

Figure \ref{fig:HotColdstatic} indicates that the London constant
at $T=0$ decreases with the development of the spin fluctuation.
On the other hand, the coefficient of the $T$-linear term $a$ is not
so doping-dependent because the ``cold spot'' is not directly
affected by the AF spin fluctuation at $\q=\Q$.
These behaviors are expected in the phenomenological proposal
on this subject \cite{rf:millismagpen,rf:paramekantiFermi},
while the origin of the quasi-particle interaction has not been
identified there.
Note here that the Fermi-liquid parameter assumed in
Ref.~\cite{rf:paramekantiFermi} is qualitatively different from that 
in our microscopic treatment.

\begin{figure}[t]
\begin{center}
\vspace{2.5cm}
\caption{The doping dependence of the current vertex $j^{*}_{\rm x}(\k)$
obtained by the FLEX approximation. The horizontal axis represents
the momentum along the Fermi surface (see Fig.~\ref{fig:HotColdstatic}).}
\label{fig:jujofig15}
\end{center}
\end{figure}

The FLEX approximation well reproduces the above estimation for the current
vertex and the London constant \cite{rf:jujomagpen2}.
Figure \ref{fig:jujofig14} shows that the London constant at $T=0$ decreases
in the under-doped region like the Uemura plot.
The development of the AF spin fluctuation is essential for this behavior.
Note that the quasi-particle interaction $f(\k,\k')$ derived from $I(k,k')$
is related with the real part of the spin susceptibility,
while the imaginary part is important in the hydrodynamic response
(Sec.~4.3.4).
Therefore, the vertex correction arising from the AF spin fluctuation is
not so suppressed by the excitation gap.
If we neglect the quasi-particle interaction, qualitatively different
doping dependence is obtained.
We conclude that the Fermi-liquid correction in the nearly AF Fermi-liquid
is essential for the Uemura plot.

Figure \ref{fig:jujofig15} shows that the current vertex around the gap node
is almost independent of the doping concentration.
Thus, the Fermi-liquid correction is not important for the
doping dependence of $a$. 
We stress that this result is in sharp contrast to
Eq.~(\ref{eq:isotropic-case}).
The detailed information of the SC gap is needed for the explicit
estimation of $a$. 
The deformation of the gap function $\Delta(\k)$, which is observed
experimentally \cite{rf:mesot}, may play a role to explain
the doping independence of $a$.
This tendency is consistent with the FLEX approximation
(Fig.~\ref{fig:PhiFS}).
We consider, however, that the essential origin of the doping
independence of $a$ is not this deformation but also the character
of the Fermi-liquid correction.

Thus, the anomalous behaviors of the London constant are explained
on the basis of the general formulation for the Fermi-liquid theory.
The vertex correction arising from the AF spin fluctuation plays
an essential role, as in the hydrodynamic transport
in the normal state (Sec.~4.3.4).
These results excellently contribute to the coherent understanding of
high-$T_{\rm c}$ cuprates.

So far, we have used the two-dimensional model and discussed
the in-plane transport.
We briefly comment on the qualitatively different behavior of the
$c$-axis London constant $\Lambda_{\rm c}$
\cite{rf:bonnmagpen,rf:panagopoulosmagpen}.
Owing to the the momentum dependence of the inter-layer hopping
(see Eq.~(\ref{eq:okanderson})),
the coherent transport gives the $T^{5}$-law of $\Lambda_{\rm c}$
in the clean limit \cite{rf:xiangcaxis}.
However, the power is easily affected by the randomness and then,
the incoherent nature should be taken in the under-doped region.
The lower power is observed in many cases \cite{rf:xiangcaxis}.
The Fermi-liquid correction is probably not important for the $c$-axis
London constant, because the $k_{\rm z}$-dependence of the quasi-particle
interaction is very weak.

Contrary to the above discussion on the low-temperature behaviors,
the temperature dependence around $T_{\rm c}$ is dominated by
the SC fluctuation.
The critical fluctuation generates the rapid growth of the London constant,
which is more remarkable along the $c$-axis \cite{rf:yanaseSC}.
These behaviors are also observed in the experimental result which has
indicated the scaling behavior corresponding to the universality class
of the 3D XY model \cite{rf:kamal}.

%
%
\section{Effects of higher-order perturbation terms}

In Sec.~3.1, we have provided the perturbation expansion for
the effective pairing interaction or the anomalous self-energy
up to third order of the on-site Coulomb repulsion $U$.
However, the value of $U$ leading to the realistic $T_{\rm c}$ is not
always sufficiently small compared with the kinetic energy of the system.
Thus, the convergence of the perturbation expansion should be examined.
Furthermore, it is interesting to investigate the momentum dependence of
the higher-order perturbation terms, since the momentum dependence of 
the effective interaction among quasi-particles is important 
for unconventional superconductivity in SCES.
For these problems, recently two of the present authors, Nomura and Yamada,
have performed the fourth-order expansion for the pairing interaction and
investigated the momentum dependence of fourth-order contributions
\cite{rf:NomuraA}. In this Appendix, following their work,
we briefly discuss the effects of the higher-order terms.

Nomura and Yamada have considered two typical cases in the two-dimensional
Hubbard model on the square lattice, given by Eqs.~(\ref{eq:Hubbard-model})
and (\ref{eq:high-tc-dispersion}) in Sec.~3.2.3:
One is the case (I) similar to the high-$T_{\rm c}$ cuprates,
where the system is near the half-filling and AF spin fluctuations are
strong due to the Fermi surface nesting.
Another is the case (II) similar to the $\gamma$ band in the
spin-triplet superconductor Sr$_2$RuO$_4$,
where the system is away from the half-filling.
In the both cases, the Fermi level is set close
to the van Hove singularity.

\begin{figure}[t]
\begin{center}
\vspace{2.5cm}
\caption{The fourth-order diagrams for the effective pairing interaction.
The vertex points and the solid lines represent the bare interaction $U$
and the bare Green's function $G^{(0)}(k)$, respectively.}
\label{Fig:4thgraphs}
\end{center}
\end{figure}

Note here that the higher-order expansions have been carried out
by Efremov et al. for a fermion gas model with repulsive interaction
in the isotropic three-dimensional space \cite{rf:Efremov}.
They have evaluated the effective interactions perturbatively up to
the fourth order with respect to the $s$-wave scattering component
of the bare interaction.
Their discussions are applicable to the diluted liquid $^3$He.

The irreducible vertex function
$V_{\sigma_1\sigma_2,\sigma_3\sigma_4}(k,k')$ is
expanded in terms of $U$ as 
\begin{eqnarray}
  V_{\sigma_1\sigma_2,\sigma_3\sigma_4}(k,k')
  &=& V^{(1)}_{\sigma_1\sigma_2,\sigma_3\sigma_4}(k,k') U
  + V^{(2)}_{\sigma_1\sigma_2,\sigma_3\sigma_4}(k,k') U^2
  + V^{(3)}_{\sigma_1\sigma_2,\sigma_3\sigma_4}(k,k') U^3 \nonumber \\
  && + V^{(4)}_{\sigma_1\sigma_2,\sigma_3\sigma_4}(k,k') U^4 + \cdots.
\end{eqnarray}
Then, we define the following quantities:
\begin{eqnarray}
  && V_{\sigma_1\sigma_2,\sigma_3\sigma_4}^{(\leq 2)}(k,k')=
  V^{(1)}_{\sigma_1\sigma_2,\sigma_3\sigma_4}(k,k') U
  + V^{(2)}_{\sigma_1\sigma_2,\sigma_3\sigma_4}(k,k') U^2,\\
  && V_{\sigma_1\sigma_2,\sigma_3\sigma_4}^{(\leq 3)}(k,k')=
  V_{\sigma_1\sigma_2,\sigma_3\sigma_4}^{(\leq 2)}(k,k')
  + V^{(3)}_{\sigma_1\sigma_2,\sigma_3\sigma_4}(k,k') U^3,\\
  && V_{\sigma_1\sigma_2,\sigma_3\sigma_4}^{(\leq 4)}(k,k')=
  V_{\sigma_1\sigma_2,\sigma_3\sigma_4}^{(\leq 3)}(k,k')
  + V^{(4)}_{\sigma_1\sigma_2,\sigma_3\sigma_4}(k,k') U^4.
  \label{Eq:Sumto4th}
\end{eqnarray}
These functions are used as perturbative approximate forms
for the effective pairing interaction
$V_{\sigma_1\sigma_2,\sigma_3\sigma_4}(k,k')$. 
Note that in the Sec.~3, we have taken
$V_{\sigma_1\sigma_2,\sigma_3\sigma_4}^{(\leq 3)}(k,k')$ 
as the approximate form.

The diagrammatic expressions for the fourth-order perturbation terms
are shown in Fig.~\ref{Fig:4thgraphs}.
The analytic expressions for the diagrams are found in
Ref.~\cite{rf:NomuraA} (see also Ref.~\cite{rf:Efremov}).
Since the weak coupling region is considered here, 
the normal self-energy is simply ignored and 
the bare Green's function $G^{(0)}(k)$ is assigned to 
the internal lines in Fig.~\ref{Fig:4thgraphs}. 
When we include the normal self-energy correction in the internal lines,
the transition temperature should be decreased due to the increase of
quasi-particle damping.
However, the momentum dependence of the effective pairing interaction 
would not be affected at least qualitatively and thus,
the most probable pairing symmetry would not be changed. 

Now all fourth-order terms are at our hands, but unfortunately,
it is not feasible task to perform faithfully the summation in momentum and
frequency for the fourth-order terms.
Up to the third order, we could carry out the summation numerically both
in momentum and frequency by exploiting the fast Fourier transformation (FFT)
algorithm, but the fourth-order terms include contributions which cannot be
summed up in the same manner.
In order to proceed to further calculations, we consider the momentum
dependence $only$ on the Fermi surface for
$V_{\sigma_1\sigma_2,\sigma_3\sigma_4}(k,k')$ and
the anomalous self-energy $\Delta(k)$.
This simplification is justified as far as we discuss low-energy phenomena
including superconductivity, for which the quasi-particles in the vicinity
of the Fermi level play the most important roles.
We define $N_{\rm F}$ points, $\bm{k}_{{\rm F}i}$ ($1 \leq i \leq N_{\rm F}$),
on the Fermi circle.
Then, the effective interaction $V(\bm{k}_{\rm F}, \bm{k}'_{\rm F})$ and
the gap function $\Delta(\bm{k}_{\rm F})$ are expressed as functions of
the positions $\bm{k}_{\rm F}$ and $\bm{k}'_{\rm F}$ on the Fermi surface.
By fixing momenta $\bm{k}_{\rm F}$ and $\bm{k}'_{\rm F}$ on the Fermi surface,
we can numerically perform the summation with respect to the internal momenta
and frequencies up to the fourth order with the use of FFT algorithm.
The above simplification is summarized as follows:
\begin{eqnarray}
 V_{\sigma_1\sigma_2,\sigma_3\sigma_4}(k,k')
 &\rightarrow&
 V_{\sigma_1\sigma_2,\sigma_3\sigma_4}({\bm k},{\bm k}')
 \equiv \lim_{\omega=\omega' \rightarrow +0}
 V_{\sigma_1\sigma_2,\sigma_3\sigma_4}
 (k=({\bm k},{\rm i}\omega),k'=({\bm k}',{\rm i}\omega'))\\
 V_{\sigma_1\sigma_2,\sigma_3\sigma_4}^{(p)}(k,k') 
 &\rightarrow& 
 V_{\sigma_1\sigma_2,\sigma_3\sigma_4}^{(p)}({\bm k},{\bm k}')
 \equiv \lim_{\omega=\omega' \rightarrow +0}
 V_{\sigma_1\sigma_2,\sigma_3\sigma_4}^{(p)}
 (k=({\bm k},{\rm i}\omega),k'=({\bm k}',{\rm i}\omega'))\\
 \Delta_{\sigma_1\sigma_2}(k)
 &\rightarrow& \Delta_{\sigma_1\sigma_2}({\bm k}), 
\end{eqnarray}
where $p$ symbolically denotes $p$=$\leq 2$, $\leq 3$, $\leq 4$, ... etc.
The quantity $V_{\sigma_1\sigma_2,\sigma_3\sigma_4}(\bm{k},\bm{k}')$
is regarded as the scattering amplitude in the pair scattering process,
where the two electrons (or Fermi-liquid quasi-particles) with
the momenta and spins, $(\bm{k},\sigma_1)$ and $(-\bm{k},\sigma_2)$,
are scattered to the states characterized by the momenta and spins,
$(\bm{k}',\sigma_4)$ and $(-\bm{k}',\sigma_3)$, respectively.
As a result of the simplification for Eq.~(16),
we obtain the BCS-like gap equation as 
\begin{eqnarray}
 \lambda \cdot \Delta_{\sigma_1\sigma_2}({\bm k}_{{\rm F}i})
 &=& -\frac{1}{(2 \pi)^2} \frac{L_{\rm F}}{N_{\rm F}}
 \ln \Bigl( \frac{2e^{\gamma}W}{\pi T} \Bigr)
 \sum_{j=1}^{N_{\rm F}}
 \Bigl| \frac{\partial \varepsilon(\bm{k})}
 {\partial {\bm k}} \Bigr|_{{\bm k}={\bm k}_{{\rm F}j}}^{-1}
 \nonumber \\
 && \times \sum_{\sigma_3\sigma_4} V_{\sigma_1\sigma_2,\sigma_3\sigma_4}
 ({\bm k}_{{\rm F}i},{\bm k}_{{\rm F}j}) 
 \Delta_{\sigma_4\sigma_3}({\bm k}_{{\rm F}j}), 
 \label{Eq:Gapeq2}
\end{eqnarray}
where the summation $\sum_{j=1}^{N_{\rm F}}$ is performed only
with respect to the Fermi surface points ${\bm k}_{{\rm F}j}$,
and $N_{\rm F}$ is the total number of the Fermi surface points
at which the vertex function
$V_{\sigma_1\sigma_2,\sigma_3\sigma_4}({\bm k}_{{\rm F}i},{\bm k}_{{\rm F}j})$
and the gap function $\Delta({\bm k}_{{\rm F}i})$ are calculated.
$L_{\rm F}$ is the circumference of the Fermi circle,
$W$ is the cut-off energy, and $\gamma(=0.5772 \cdots$)
is the Euler's constant.
The cut-off energy is taken as $W=0.700$ throughout the present calculation.
We can determine the transition temperature
at which the maximum eigenvalue $\lambda_{\rm max}$ is equal to unity
and the eigenfunction giving $\lambda_{\rm max}$ is related to
the most probable pairing symmetry.
Details of the formulation are given in Ref.~\cite{rf:NomuraA}.

\begin{figure}[t]
\begin{center}
\vspace{2.5cm}
\caption{(a) Fermi surface for the nearly half-filled case (I).
(b) The gap function $\Delta_{\rm singlet}({\bm k}_{\rm F})$
as a function of the position ${\bm k}_{\rm F}$ on the Fermi circle.
Note that ${\bm k}_{\rm F}$ traces the Fermi circle following the arrows
in (a).
The symmetry is $d_{x^2-y^2}$-wave state for all cases.
The parameters are $T=0.0100$ and $U=2.066$.}
\label{Fig:afhf}
\end{center}
\end{figure}

Now we show the numerical results of the gap function
$\Delta_{\sigma_1\sigma_2}({\bm k})$ and the transition temperatures
as functions of the repulsion $U$.
For the case (I), we have taken the hopping parameters
$t=1.00$ and $t'=-0.100$ and set the electron number as $n=0.98$.
The Fermi surface is depicted in Fig.~\ref{Fig:afhf}(a).
The obtained gap function $\Delta_{\sigma_1\sigma_2}({\bm k})$
exhibits the $d_{x^2-y^2}$-wave state in all the cases
of the second-, third-, and fourth-order perturbation theories.
This may be natural if we recall that the momentum dependence of
the pairing interaction is well approximated by using the susceptibility
as $g^2\chi(k-k')$ in Sec.~3.2.2 for the superconductivity induced by
the strong anti-ferromagnetic spin fluctuations.

\begin{figure}[t]
\begin{center}
\vspace{2.5cm}
\caption{Transition temperature as a function of the repulsion $U$ for
the nearly half-filled case (I).
The most probable pairing state is spin-singlet $d_{x^2-y^2}$-wave.}
\label{Fig:4thdTc}
\end{center}
\end{figure}

We show the transition temperatures as functions of the repulsion $U$
for the case (I) within the second-, third-, and fourth-order
perturbation theories in Fig.~\ref{Fig:4thdTc}.
The values of $T_{\rm c}$ for the third- and fourth-order theories are
quantitatively close to each other, although $T_{\rm c}$
for the second-order theory is very low.
If we further take account of the perturbation terms beyond fourth order,
the curve of $T_{\rm c}$ as a function of $U$ is considered to converge
to a single curve, leading to a better estimation for $T_{\rm c}$ than
that of the present fourth-order calculation.

Let us discuss the validity of the perturbation expansion by comparing
the magnitude of the third- and the fourth-order vertex functions,
$V^{(3)}(\bm{k},\bm{k}')$ and $V^{(4)}(\bm{k},\bm{k}')$.
This naive analysis justifies the expansion in $U$ up to
$U \sim |V^{(3)}/V^{(4)}| \sim 3.5$ \cite{rf:NomuraA}.
Although there are much more perturbation terms in the fourth order 
than in the second or third order, most of the fourth-order contributions
cancel each other, and therefore the total magnitude of 
$V^{(4)}(\bm{k},\bm{k}')$ is smaller than that of $V^{(3)}(\bm{k},\bm{k}')$.
Thus, the convergence of the perturbation expansion in $U$ becomes good
up to moderately strong $U$ within the fourth-order perturbation theory.
In addition, the momentum dependences of the functions,
$V^{(2)}(\bm{k},\bm{k}')$, $V^{(3)}(\bm{k},\bm{k}')$,
and $V^{(4)}(\bm{k},\bm{k}')$ are all similar to each other.
Therefore the perturbation expansion would give a good estimation of
the effective pairing interaction and a reasonable analysis of the pairing
symmetry up to moderately strong $U$ in this case.

\begin{figure}[t]
\begin{center}
\vspace{2.5cm}
\caption{The strong contributions with particle-particle ladder in the vertex.
The contributions enclosed by the dotted lines are added to
$V^{(\leq 4)}(k,k')$ in order to suppress ill-convergence.}
\label{Fig:vertcor}
\end{center}
\end{figure}

Next we consider the other case (II).
According to the recent work \cite{rf:NomuraA}, the convergence of
the perturbation expansion is not good for the spin-triplet channel,
in contrast to the spin-singlet channel near the half-filled case (I).
In one word, this is due to an oscillatory behavior in terms of $U$.
As shown in Sec.~3.4.2, the third-order terms give the momentum dependence
much favorable for the $p$-wave pairing.
After the momentum dependences of all fourth-order terms are examined,
it is found that dominant contributions have the momentum dependence
similar to that of the third-order term, but their signs are opposite.
Further analysis on higher-order terms has revealed that
the contributions shown in Fig.~\ref{Fig:vertcor} provide 
the dominant momentum dependence in each order. 
These contributions in each higher-order term have
positive (negative) sign for odd (even) order of $U$ and thus,
these contributions can be summed up to the infinite order.
For the pairing interaction, then we use $V^{(\leq 4 \,{\rm corr.})}(k,k')$,
which is obtained by adding the higher-order corrections
shown in Fig.~\ref{Fig:vertcor} to $V^{(\leq 4)}(k,k')$.
Due to this improved procedure, the most probable pairing symmetry is
triplet $p$-wave. 

\begin{figure}[t]
\begin{center}
\vspace{2.5cm}
\caption{(a) Fermi surface for the case (II) away from the half-filled.
(b) The gap function $\Delta_{\rm triplet}({\bm k}_{\rm F})$ as a function
of the position ${\bm k}_{\rm F}$ on the Fermi circle.
Note that ${\bm k}_{\rm F}$ traces the Fermi circle following the arrows
in (a). The parameters are $T=0.0100$, $U=3.428$.}
\label{Fig:afahf}
\end{center}
\end{figure}

\begin{figure}
\begin{center}
\vspace{2.5cm}
\caption{Transition temperature as a function of the repulsion $U$ for the case (II).
The most probable pairing state is spin-triplet $p$-wave pairing state.}
\label{Fig:pTc}
\end{center}
\end{figure}

In Fig.~\ref{Fig:afahf}, we show the Fermi surface and the gap function
for the case (II), where we have taken the hopping parameters
$t=1.00$ and $t'=-0.375$ and set the electron number as $n=1.334$.
The gap functions obtained for the third-order and the above mentioned
improved fourth-order perturbation theories show the highly anisotropic
$p$-wave symmetry with the nodes on the line $k_y$=0, 
while the gap function obtained in the second-order theory
has eight additional nodes on the Fermi surface.
We show the transition temperature as functions of the repulsion $U$
in Fig.~\ref{Fig:pTc}. Note that the second-order perturbation theory 
gives only low $T_{\rm c}$, because the momentum dependence of
the effective interaction $V^{(\leq 2)}(\bm{k}, \bm{k}')$ is less
favorable for the triplet $p$-wave pairing than
$V^{(\leq 3)}(\bm{k}, \bm{k}')$ and
$V^{(\leq 4 \, {\rm corr.})}(\bm{k},\bm{k}')$.

Finally we again emphasize that the momentum dependence induced by the
electron correlations is important for the anisotropic pairing in SCES. 
In some cases, the picture that some magnetic fluctuations induce
the anisotropic superconductivity would be valid.
However, we could $not$ always consider that the essential momentum
dependence of the pairing interaction originates from some strong
fluctuations.
The most essential point for anisotropic pairings is what momentum
dependence the effective pairing interaction could acquire
as a result of many-body effects or correlation effects,
as we have stressed throughout this review article.
In general, we could expect that the perturbation theories developed
in this review capture qualitatively well the essential momentum dependences
of the pairing interaction.
It is highly believed that the anisotropic pairing deduced within
the weak-coupling perturbation theory evolves smoothly up to
the realistically strong $U$. 

\par\vfill
\eject

%
%

\end{document}